\documentclass[a4paper,11pt]{article}
\pdfoutput=1 

\usepackage{jinstpub} 
\usepackage{multicol}
\usepackage{multirow}
\usepackage{gensymb}
\usepackage{array}
\newcolumntype{x}[1]{>{\centering\arraybackslash\hspace{0pt}}p{#1}}
\usepackage[utf8]{inputenc}
\usepackage{graphicx}
\usepackage[labelformat=simple]{subcaption}

\usepackage{xcolor} 
\usepackage{soul}
\usepackage{lineno}
\usepackage{float}
\usepackage{mathrsfs}
\usepackage{amsmath}
\usepackage{array,tabularx}   
\usepackage{siunitx}
\usepackage{comment}
\def\Put(#1,#2)#3{\leavevmode\makebox(0,0){\put(#1,#2){#3}}}
\usepackage{sectsty}
\allsectionsfont{\boldmath}

\newenvironment{conditions*} 
  {\par\vspace{\abovedisplayskip}\noindent
   \tabularx{\columnwidth}{>{$}l<{$} @{${}={}$} >{\raggedright\arraybackslash}X}}
  {\endtabularx\par\vspace{\belowdisplayskip}}

\title{First results on ProtoDUNE-SP liquid argon time projection chamber performance from a beam test at the CERN Neutrino Platform}

\author[142]{B.~Abi,}
\author[21,118]{A.~Abed Abud,}
\author[61]{R.~Acciarri,}
\author[8]{M.~A.~Acero,}
\author[65]{G.~Adamov,}
\author[61]{M.~Adamowski,}
\author[17]{D.~Adams,}
\author[21]{P.~Adrien,}
\author[16]{M.~Adinolfi,}
\author[182]{Z.~Ahmad,}
% ...
\affiliation[8]{Universidad del Atl{\'a}ntico, Atl{\'a}ntico, Colombia}
% ...
\affiliation[16]{University of Bristol, Bristol BS8 1TL, United Kingdom}
\affiliation[17]{Brookhaven National Laboratory, Upton, NY 11973, USA}
% ...
\affiliation[21]{CERN, The European Organization for Nuclear Research, 1211 Meyrin, Switzerland}
% ...
\affiliation[61]{Fermi National Accelerator Laboratory, Batavia, IL 60510, USA}
% ...
\affiliation[65]{Georgian Technical University, Tbilisi, Georgia}
% ...
\affiliation[118]{University of Liverpool, L69 7ZE, Liverpool, United Kingdom}
% ...
\affiliation[142]{University of Oxford, Oxford, OX1 3RH, United Kingdom}
% ...
\affiliation[182]{Variable Energy Cyclotron Centre, 700 064 West Bengal, India}
% ...

\emailAdd{F. Cavanna (cavanna@fnal.gov)}
\emailAdd{T. Junk (trj@fnal.gov)}
\emailAdd{T. Yang (tjyang@fnal.gov)}

\abstract{The ProtoDUNE-SP detector is a single-phase liquid argon time projection chamber with an active volume of $7.2\times 6.1\times 7.0$~m$^3$.  It is installed at the CERN Neutrino Platform in a specially-constructed beam that delivers charged pions, kaons, protons, muons and electrons with momenta in the range 0.3~GeV$/c$ to 7~GeV/$c$.  Beam line instrumentation provides accurate momentum measurements and particle identification.  The ProtoDUNE-SP detector is a prototype for the first far detector module of the Deep Underground Neutrino Experiment, and it incorporates full-size components as designed for that module.  This paper describes the beam line, the time projection chamber, the photon detectors, the cosmic-ray tagger, the signal processing and particle reconstruction. It presents the first results on ProtoDUNE-SP's performance, including noise and gain measurements, $dE/dx$ calibration for muons, protons, pions and electrons, drift electron lifetime measurements, and photon detector noise, signal sensitivity and time resolution measurements.  The measured values meet or exceed the specifications for the DUNE far detector, in several cases by large margins.
ProtoDUNE-SP's successful operation starting in 2018 and its production of large samples of high-quality data demonstrate the effectiveness of the single-phase far detector design.  
}


\arxivnumber{2007.06722 } 

\begin{document}
\hfill FERMILAB-PUB-20-059-AD-ESH-LBNF-ND-SCD

\hfill CERN-EP-2020-125

\maketitle
\flushbottom

\section{Introduction}
\label{sec:intro}

The Deep Underground Neutrino Experiment (DUNE)~\cite{Abi:2020wmh} is a next-generation, long-baseline neutrino oscillation experiment, with a near detector at Fermilab and a far detector located at the 4850 ft level of Sanford Underground Research Facility (SURF), in Lead, South Dakota, USA, 1285 km from the neutrino production target. The neutrino detectors at the DUNE far site will be housed inside four cryostats, each of which will contain 17.5 kt of liquid argon (LAr). The first detector to be constructed will be a single-phase time projection chamber (TPC), similar to, but a factor of 25 more massive than the pioneering T600 detector built by the ICARUS Collaboration~\cite{Amerio:2004ze}. ProtoDUNE-SP, the ProtoDUNE single-phase apparatus (the NP04 experiment at CERN)~\cite{Abi:2017aow}, assembled and tested at the CERN Neutrino Platform~\cite{Pietropaolo:2017jlh}, is designed as a test bed and full-scale prototype for the elements of the first far detector module of DUNE~\cite{Abi:2020loh}. NP04 is a CERN-approved experiment to explore large volume LArTPCs and forms an integral part of the DUNE Collaboration.


In addition to its role as a demonstration prototype and engineering test bed, the ProtoDUNE-SP TPC was exposed to a tagged and momentum-analyzed particle beam with momentum settings ranging from 0.3~GeV/$c$ to 7~GeV/$c$. This beam enabled the acquisition of large samples of data on the behavior of charged pions, kaons, protons, muons and positive electrons (positrons) in LAr. The beam was set to deliver only positively-charged particles for the data samples used in this paper, although future runs will also include negatively-charged particle beams.  These data serve as templates for understanding how these particles will appear when produced in neutrino interactions in DUNE, and they will be an important reference in the analysis of interactions in DUNE. These data also provide a real-world test bed for the development of algorithms for pattern recognition, event reconstruction and analysis, and they will be used  to measure the cross sections of interactions of charged particles in LAr. 

The ProtoDUNE-SP apparatus is designed to satisfy the stringent new requirements and achieve the improved levels of performance required by DUNE~\cite{Abi:2020evt}. The membrane cryostat and its associated cryogenic system are the largest LAr systems ever constructed. The argon purification system is the largest constructed to date. As compared to previous devices, such as ICARUS~\cite{Amerio:2004ze}, ArgoNeuT~\cite{Anderson:2012vc}, LongBo~\cite{Bromberg:2015bqs}, MicroBooNE~\cite{Acciarri:2016smi}, and the 35-ton prototype~\cite{Adams:2019wxc} which had shorter maximum drift distances, the 3.6~m drift distance in ProtoDUNE-SP makes higher demands on argon purity. Under the nominal electric field of 500~V/cm, the maximum electron drift time is 2.25~ms. The long drift distance also requires higher voltages in the HV system used to provide the drift field, and the stored energy which may be released in a discharge is also higher than in previous devices. To allow for higher voltages and to reduce the chance of discharges, ProtoDUNE-SP incorporates specially-chosen materials for the cathode and the field cage structure, and new shapes for the field rings. The sense-wire assemblies, known as Anode Plane Assemblies (APAs), contain three planes of readout wires on both faces and are of a novel design and construction. To improve the signal-to-noise ratio, the sense wire readout amplifiers and analog-to-digital converters (ADCs) are placed inside the LAr close to the wires. Furthermore, the data acquisition system accommodates a higher data rate and larger event sizes than previous LArTPC systems. 

ProtoDUNE-SP includes a novel photon-detector design which embeds the photon detectors within the APAs in order to collect scintillation light from ionized LAr.  Due to the small available area, the photon detectors are required to be highly efficient for detecting single photons.  The performance of the photon detectors in ProtoDUNE-SP is a primary topic of this paper.

Cosmic-ray interactions with the detector cause a buildup of positive ions that drift very slowly towards the cathode.  The accumulated space charge is proportional to the rate of incident cosmic rays and it depends strongly on the drift distance. The space charge alters the electric field in the detector, changing both its strength and its direction, causing distortions in both the measured positions of particles traversing the detector and their apparent ionization densities. However, the effects of space charge buildup are expected to be largely absent in the DUNE Far Detector due to its low cosmic-ray rate, a consequence of its deep underground location.  In the analyses presented in this paper, corrections for the effects of space charge are applied where appropriate in order for the results of these studies to be generally applicable.

The ProtoDUNE-SP Technical Design Report~\cite{Abi:2017aow} contains a detailed description of the design. A description of the apparatus as built, plus a description of the installation, testing and commissioning is given in~\cite{protodunetechnicalpaper}. 

This paper is organized as follows. Section~\ref{sec:detector} describes different components of the ProtoDUNE-SP detector. Section~\ref{sec:beamline} describes details of the CERN beam line instrumentation. Sections~\ref{sec:tpc}-\ref{sec:pdresponse} summarize results on TPC characterization, photon detector characterization, TPC response, and photon detector response. Section~\ref{conclusions} concludes the paper. This paper summarizes the initial results from analyzing the ProtoDUNE-SP data. More in-depth studies will be reported in future publications.

\section{The ProtoDUNE-SP detector}
\label{sec:detector}
The ProtoDUNE-SP apparatus, shown in figure~\ref{fig:tpc}, is described by a right-handed coordinate system in which the $y$ axis is vertical (positive pointing up) and the $z$ axis is horizontal and points approximately along the beam direction.  The $x$ axis is also horizontal and points along the nominal electric field direction and is perpendicular to the wire planes\footnote{Throughout this paper, the charge and energy deposited per unit track length are conventionally referred to as $dQ/dx$ and $dE/dx$ respectively.  The $dx$ in these expressions is not oriented along the detector coordinate $x$ but rather it is a differential step along the track path.}.

\subsection{Cryostat}
\label{sec:cryostat}

The TPC is installed in a membrane cryostat~\cite{Montanari:2015pxz} with internal dimensions of 8.5~m in both the $x$ and $z$ directions, and 7.9~m in $y$. The cryostat is filled to a height of about 7.3~m and its pressure is maintained to 1050~mbar (absolute).  The TPC is suspended by the detector support system, which is a network of steel beams held in place by nine penetrations in the roof of the cryostat. 

A detailed description of the design, construction, leak-checking, testing and validation of the cryostat is given in~\cite{protodunetechnicalpaper}. The cryogenics control system is also described there. We give a summary here of the argon purification system system that has played a crucial role in the detector performance achieved.

\begin{figure}[!ht]
\begin{center}
\includegraphics[width=0.8\textwidth]{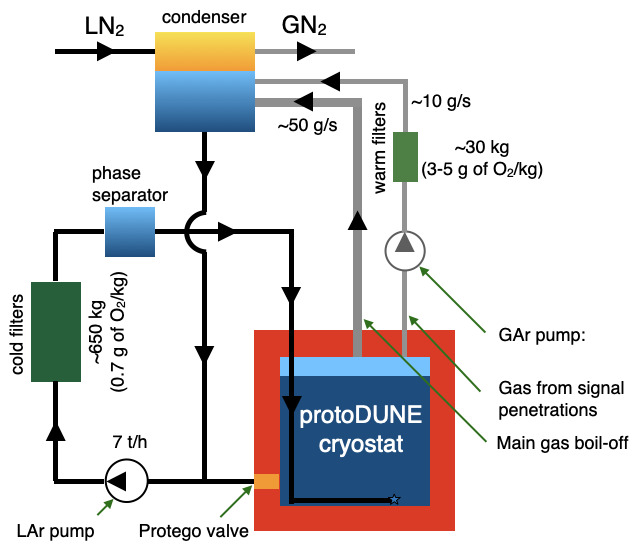}
\end{center}
\caption{A schematic of the argon purification system at NP04.}
\label{fig:cryoschematic}
\end{figure}

The argon received from the supplier has contaminants of water, oxygen and nitrogen at the parts per million level each. Water and oxygen will capture drifting electrons and the concentration of these contaminants needs to be reduced by a factor of at least $10^{4}$ and maintained at this level to allow operation of the TPC. Purification of argon in the liquid phase, as required for the mass of argon involved here, is reported in~\cite{Benetti:1993mf}. The present system builds on  purification systems developed for ICARUS and most recently at Fermilab~\cite{Adamowski:2014daa} and~\cite{Acciarri:2016smi}, including the use of the same filter materials. The main features of the system are indicated in figure~\ref{fig:cryoschematic}. There are three circulation loops. In one, liquid leaves the cryostat via a penetration in the side.  It is pumped as liquid through a set of filters, and it is reintroduced to the cryostat at the bottom. The pump can drive about 7~t/hr giving a volume turnover time of about 4.5 days. In the second loop, argon gas from the purge pipes with which each signal penetration is equipped is purified directly while warm and it is recondensed to join the liquid flow out of the cryostat. In the third loop, the main boil-off from the argon is recondensed directly and it then joins the liquid flow out of the cryostat. When the argon is first circulated,  the contamination level falls following a perfect mixing model with a time constant of the turnover time until a steady state is reached in which the rate of contamination from leaks and outgassing from impurities balances the clean-up rate. In the NP04 cryostat, thanks to the rate of recirculation and the avoidance of leaks, this state is equivalent to an oxygen contamination \cite{bettini1991study} of a few parts per trillion resulting in essentially full-strength signals from the furthest parts of the TPC. 

Instrumentation for monitoring the state of the argon is distributed outside the TPC near the inner walls of the cryostat. Three purity monitors, formerly used with the ICARUS T600 detector~\cite{Amerio:2004ze}, were refurbished with new gold photocathodes and quartz fibers.  They are deployed in ProtoDUNE-SP, each at a different height.  They monitor and give fast feedback on the drift electron lifetime in the liquid argon. Two vertical columns of resistance temperature detectors (RTDs) measure the temperature gradient of the liquid argon.  Computational fluid dynamics (CFD) calculations have been performed that predict the temperature distribution and the internal flow pattern of the argon~\cite{sdsu_protodune_cfd}. The temperature is predicted to vary by 15~mK total over the height of the liquid.  The RTDs have been cross calibrated in situ to better than 2~mK and their measurements agree with the predictions within $\pm 3.7$~mK.  A set of cameras and LED lights in the liquid and in the ullage provides monitoring of the mechanical state of the apparatus during filling and operation.

\subsection{Time projection chamber}
\label{sec:det_tpc}

\begin{figure}[!ht]
     \centering
         \includegraphics[width=0.7\textwidth]{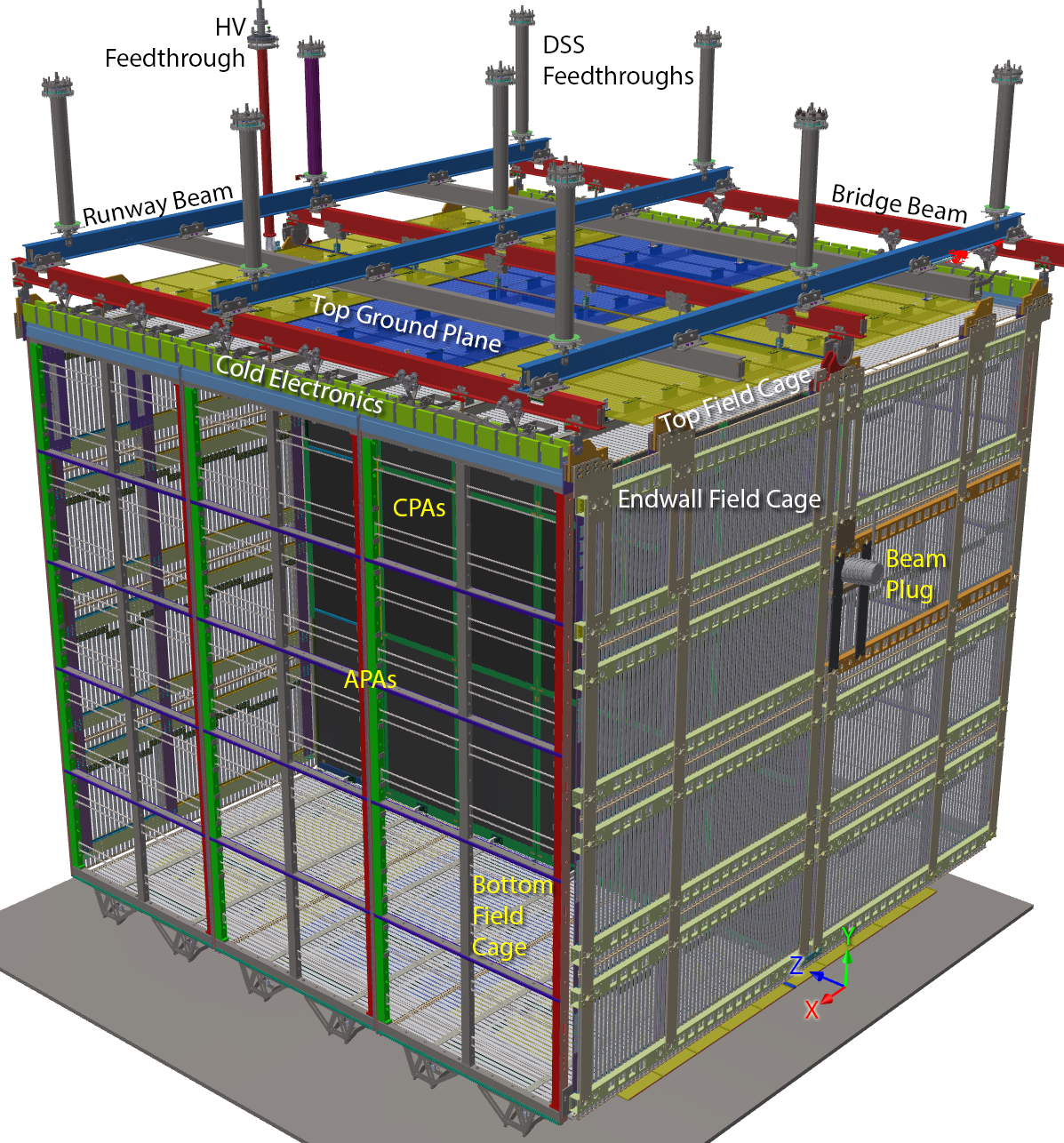}
         \includegraphics[width=0.7\textwidth]{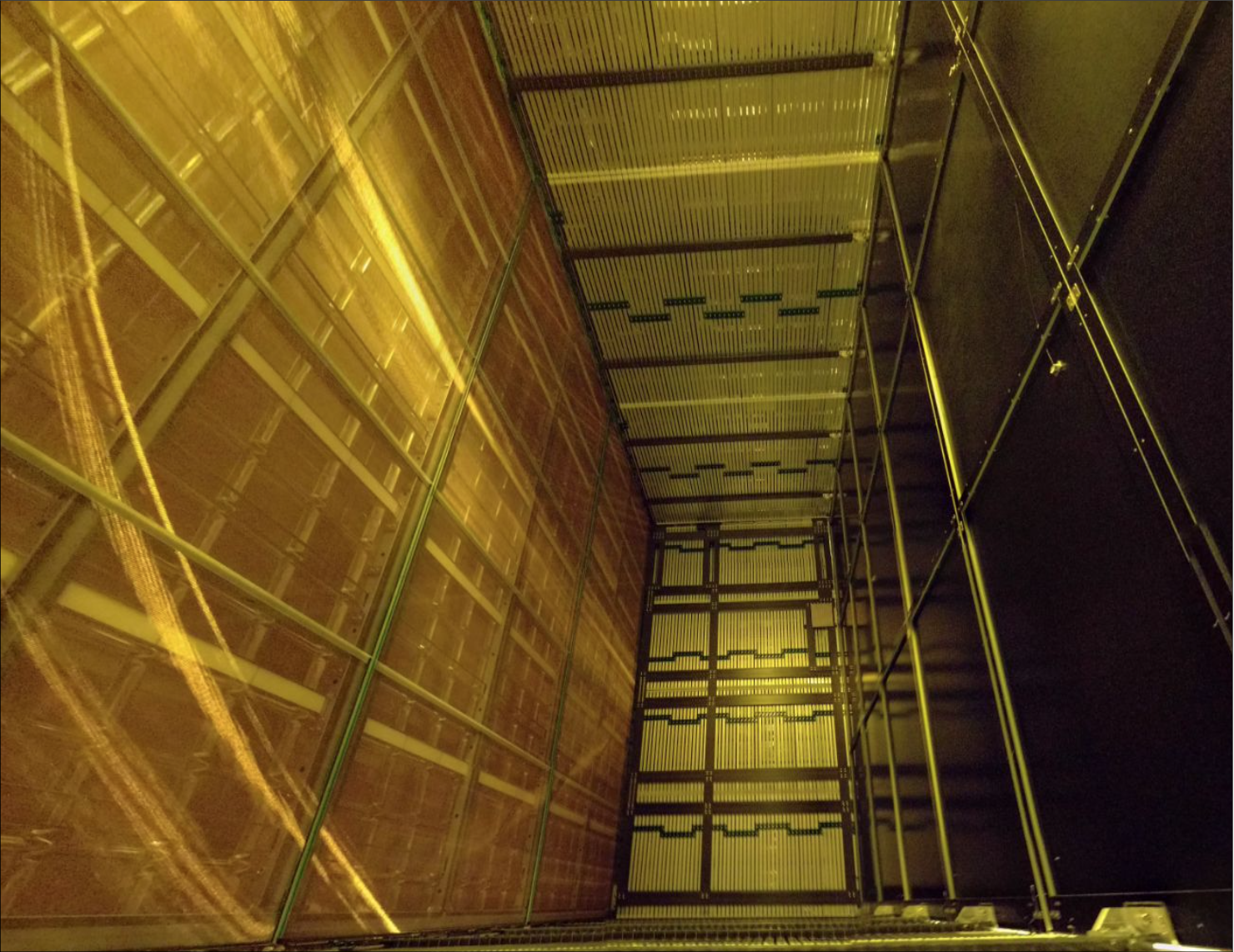}
     \caption{Top: a view of the TPC with its major components labeled; bottom: a photo of one of the two drift volumes, where three APAs are on the left side and the cathode is on the right side.}
     \label{fig:tpc}
\end{figure}

The time projection chamber is divided into two separate half-volumes with a solid, planar cathode in the center, at $x=0$ in the $yz$ plane, with three APAs forming an anode plane opposite to the cathode on either side. The two active regions of the TPC, permeated by an electric field, enclose a volume 6.1~m high along the $y$ direction, 7.0~m along $z$, and 3.6~m in the positive and negative $x$ direction. The entire active volume, except very thin regions at the boundaries, is instrumented for ionization charge readout at the APA end of the drift.   
Table~\ref{TPC} summarizes the nominal TPC parameters and features. Figure~\ref{fig:tpc} shows a view of the TPC with its major components labeled and a photo of one of the two drift volumes. 

\begin{table}[htp!]
\centering
\caption{Nominal LArTPC parameters and features. Here, X refers to all collection planes, and
  Z and C refer to the collection planes on the sides of the TPC and cryostat, respectively.}
\vspace{0.2cm}
\begin{tabular}{l|l}
	\hline\hline
TPC configuration  &  Anode-Cathode-Anode (2~active~volumes)  \\ 
TPC dimensions (active volumes) &  $6.086~{\rm{(h)}}\times 3.597~{\rm{(w)}}\times 7.045~{\rm{(l)}}~$m$^{3}$  \\ 
~~~~~~~~~~~~~~~~~(instrumented volumes) &  $5.984~{\rm{(h)}}\times 3.597~{\rm{(w)}}\times 6.944~{\rm{(l)}}~$m$^{3}$  \\ 
Total active volume (nominal, at room T) & 2 $\times 154 {\rm~m}^3$ \\
Total instrumented LAr mass (87.65 K) & 419 t \\ \hline
Number of TPC wire planes & 4 (G, U, V, X)  \\ 
Number of wires (total) & 15360 (instrumented) \\ 
~~~~~~G: Grid plane      &  2 $\times$ 2880  (non-instrumented)\\ 
~~~~~~U: 1$^{\rm{st}}$ induction plane &  2 $\times$ 2400 (instrumented, wrapped) \\ 
~~~~~~V: 2$^{\rm{nd}}$ induction plane &  2 $\times$ 2400 (instrumented, wrapped) \\ 
~~~~~~Z: TPC-side collection plane & 2 $\times$ 1440 (instrumented)\\ 
~~~~~~C: Cryostat-side collection plane & 2 $\times$ 1440 (instrumented)\\ 
Wire orientation (w.r.t. vertical) & G: $0^{\circ}$, U: $+35.7^{\circ}$, V: $-35.7^{\circ}$, X: $0^{\circ}$ \\ 
Wire pitch (normal to wire direction) & 4.79~mm (G, X); 4.67~mm (U, V)  \\ 
Wire type & Cu-Be Alloy \#25, diam. 150 $\mu$m  \\ 
Gap width between planes & 4.75~mm \\ \hline
 E-Field (nominal) in drift volume &  500~V/cm  \\ 
~~~~~~ Cathode plane voltage  & $-180$ kV  \\
~~~~~~ Anode plane bias voltages  & G: -665~V, U: -370~V, V:~0~V, X: +820~V    \\
~~~~~~ Ground mesh  & 0~V  \\
Max. drift length &  \multirow{2}{*}{3572~mm} \\
 (Cathode-to-G-plane distance at 87.65 K) &   \\ 
Drift velocity (nominal field, 87.65 K) & 1.59 mm/$\mu$s   \\ 
Max. drift time (nominal field, 87.65 K)  & 2.25~ms \\ \hline
\hline
\end{tabular}
\label{TPC}
\end{table}

The cathode plane of the TPC is formed from six cathode plane assemblies (CPAs). Each CPA is 1.15~m wide and 6.1~m high, and consists of three vertically stacked cathode panels.  The stored electrical energy in the TPC when fully charged presents a challenge.  If the cathode were electrically conducting, an electrical breakdown can discharge it rapidly, endangering the front-end electronics.  Instead, the cathode is constructed out of resistive materials which give it a very long discharge time constant, reducing the risk.  The CPA panels are constructed from FR4, a fire-retardant fiberglass-epoxy composite material.  These panels are laminated on both sides with a commercial Kapton film with a resistivity of $\sim$3.5~M$\Omega$/sq.  The cathode plane is biased at -180~kV to provide a 500~V/cm drift field. 
A field cage with 60~voltage steps on each side of the cathode ensures the uniformity of the nominal drift field between the cathode plane and the sense planes.  The electric field differs from the nominal prediction due to space-charge effects, which are described in section~\ref{sec:SCE}.

\begin{figure}[!ht]
\begin{center}
\includegraphics[width=0.9\textwidth,trim = 0mm 10mm 0mm 0mm,clip]{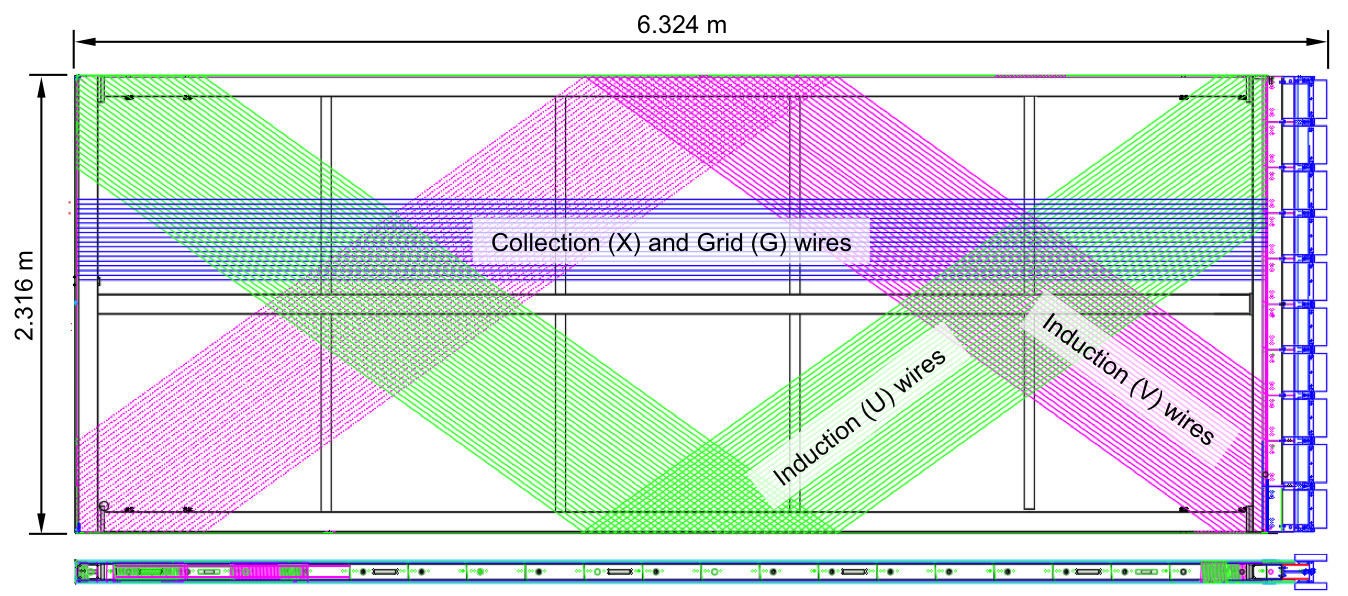}
\end{center}
\caption{Sketch of a ProtoDUNE-SP APA, showing some of the  U, V and X wires, to accentuate their angular relationships to the frame and to each other.  The induction layers are connected electrically across both sides of the APA.  The grid layer (G) wires run parallel to the X wires.  Separate sets of G and X wires are strung on the two sides of the APA.  The blue boxes on the end house the cold electronics.  The APA is shown turned on its side.  Each APA is installed vertically in the detector.  From ref.~\cite{Abi:2020loh}. }
\label{fig:apasketch}
\end{figure}

\begin{figure}[!ht]
\begin{center}
\includegraphics[width=0.9\textwidth]{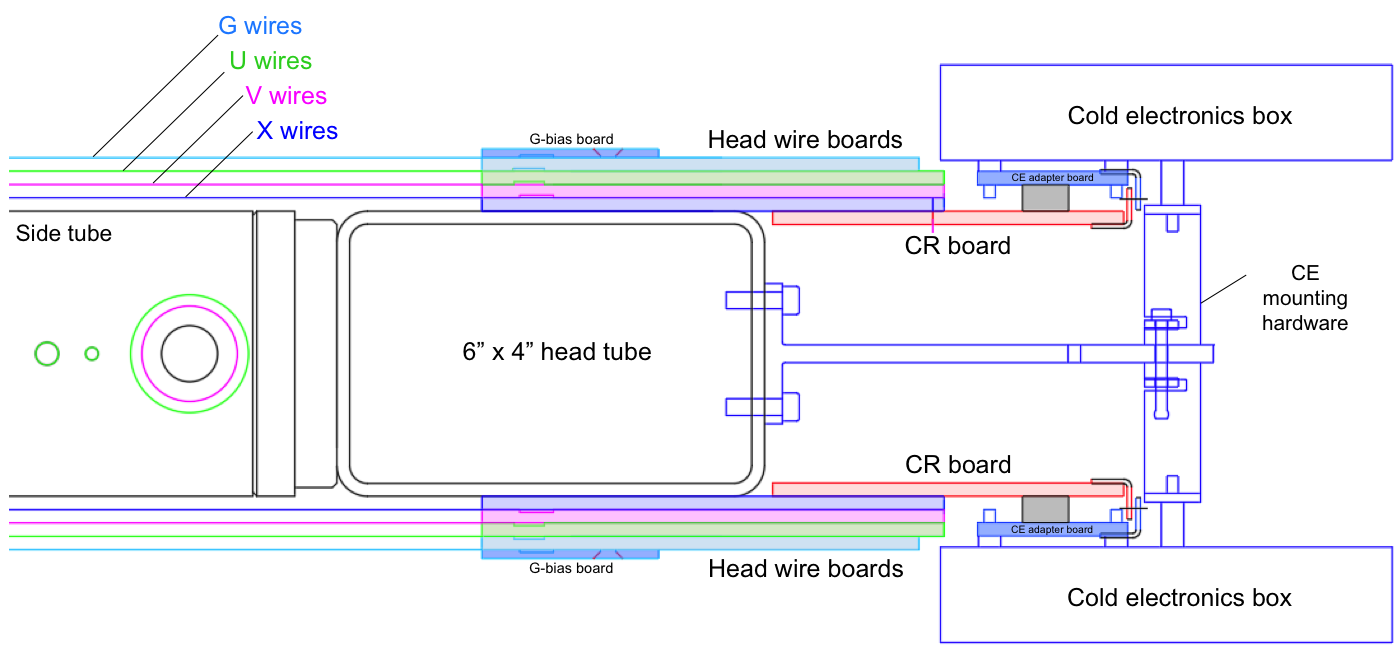}
\end{center}
\caption{Side view of the end of a ProtoDUNE-SP APA showing the mounting of the front-end electronics and the arrangement of the G (light blue), U (green), V (magenta), the induction layers; and X (blue) wire planes.  Wires are mounted on the head wire boards and bias voltages and signals are brought through the capacitance-resistance (CR) boards.  From ref.~\cite{Abi:2020loh}. }
\label{fig:apasketchside}
\end{figure}

Fig.~\ref{fig:apasketch} is a diagram of an APA viewed from the front, and fig.~\ref{fig:apasketchside} shows one end of an APA viewed from the side. Each APA has a rectangular stainless steel frame 6.1~m high, 2.3~m wide, and 76~mm thick. There are four layers (planes) of wires bonded on each side of the frame.  Following the notation of ref.~\cite{Abi:2017aow}, the wire planes and their wire orientations are (from outside in) the Grid (G) layer (vertical), the U layer (+35.7$^\circ$ from vertical), the V layer (-35.7$^\circ$ from vertical), and the X layer (vertical). A bronze wire mesh with 85\% optical transparency is bonded directly over each side of the APA frame to provide a grounded shield plane for the four wire planes mentioned above. Each successive wire plane is built 4.75~mm above the previous layer, including the wire mesh.  The wires are terminated on wire boards which are stacked on the short ends the APA. The G and X layers have the same wire pitch of 4.79~mm, but are staggered by half a wire pitch in relative position. The U and V wires have a pitch of 4.67~mm. Wires on the two induction planes are helically wrapped around the frame from the head, to the sides, and then to the foot. Wires are held in place with FR-4 boards with teeth cut in them as they wrap around the sides.  The wire angle is chosen such that the wires do not wrap more than one revolution to avoid creating ambiguities in track reconstruction.  Four wire support combs made out of 0.5~mm-thick G10 (a fiberglass-epoxy composite material) are installed on each side of each APA uniformly spaced along the $y$ direction, in order to hold the long wires in place, helping to counteract gravitational, electrostatic, and fluid-flow forces that would otherwise cause portions of the wires to be displaced from their nominal positions.
Each wire plane is electrically biased at a different potential such that the primary ionization electrons created in the drift volume pass through the G, U and V planes without being captured, and finally are collected on the X wires. Therefore, the X plane wires are also referred to as the collection wires, and the U, V plane wires as the first and second induction plane wires.  The grid plane wires serve as an electrostatic discharge (ESD) protective shield and are not read out. The nominal wire-plane bias voltages, to ensure electron collection only on collection-plane wires, are $V_{\rm G}= -665$~V, $V_{\rm U} = -370$~V, $V_{\rm V}= 0$~V, $V_{\rm X}= +820$~V.  

The G~plane on the lowest-$z$ APA on the $x<0$ side of the detector was unintentionally not connected to its voltage supply. The break in connectivity was determined to be inside the cryostat and it could not be repaired for the duration of the run.  Groups of four G~plane wires are connected to a 3.9~nF capacitor with the other terminal grounded.  Without a voltage supply, drifting ionization electrons will charge up the G-plane from its initial state to a potential that repels electrons and prevents further charge collection.  This charging process takes approximately 100 hours~\cite{Abi:2020loh}, and the average charge measured by this APA is reduced during the charge-up time. While we observed the transparency of the G plane on this APA to be as low as 10\% immediately after the drift field had been switched on, the total loss in transparency after 100 hours became negligible.

Electron diverters are installed in the two vertical gaps between the APAs on the negative-$x$ side of the cathode, but not between the APAs on the positive-$x$ side.  These diverters consist of two vertical electrode strips, an inner electrode and an outer electrode, mounted on an insulating board that protrudes approximately 25~mm into the drift volume beyond the G plane wires.  Voltages applied to the diverter electrodes modify the local drift field so that electrons drift away from the gaps between the APAs and into the active area.  A diagram showing the field lines and equipotentials in the vicinity of the electron diverters when they are working as designed is given in ref.~\cite{Abi:2017aow}.  High currents were drawn from the electron diverters' power supplies when they were energized, due to one or more electrical shorts in the cold volume. The electron diverters were therefore left unpowered.  A resistive path to ground on each one ensured that the actual voltage on the outer electrode was close to zero, which was not the intended voltage.  The grounded diverter electrodes collected charge near the gaps, and also distorted nearby drift paths.  The impacts on the measured charge values and spatial distortions are discussed in section~\ref{sec:muonchargecalibration} and also in ref.~\cite{Abi:2020loh}.

\subsection{Beam plug}
\label{sec:det_beamplug}

\begin{figure}[!htp]
    \centering
    \includegraphics[width=.48\textwidth]{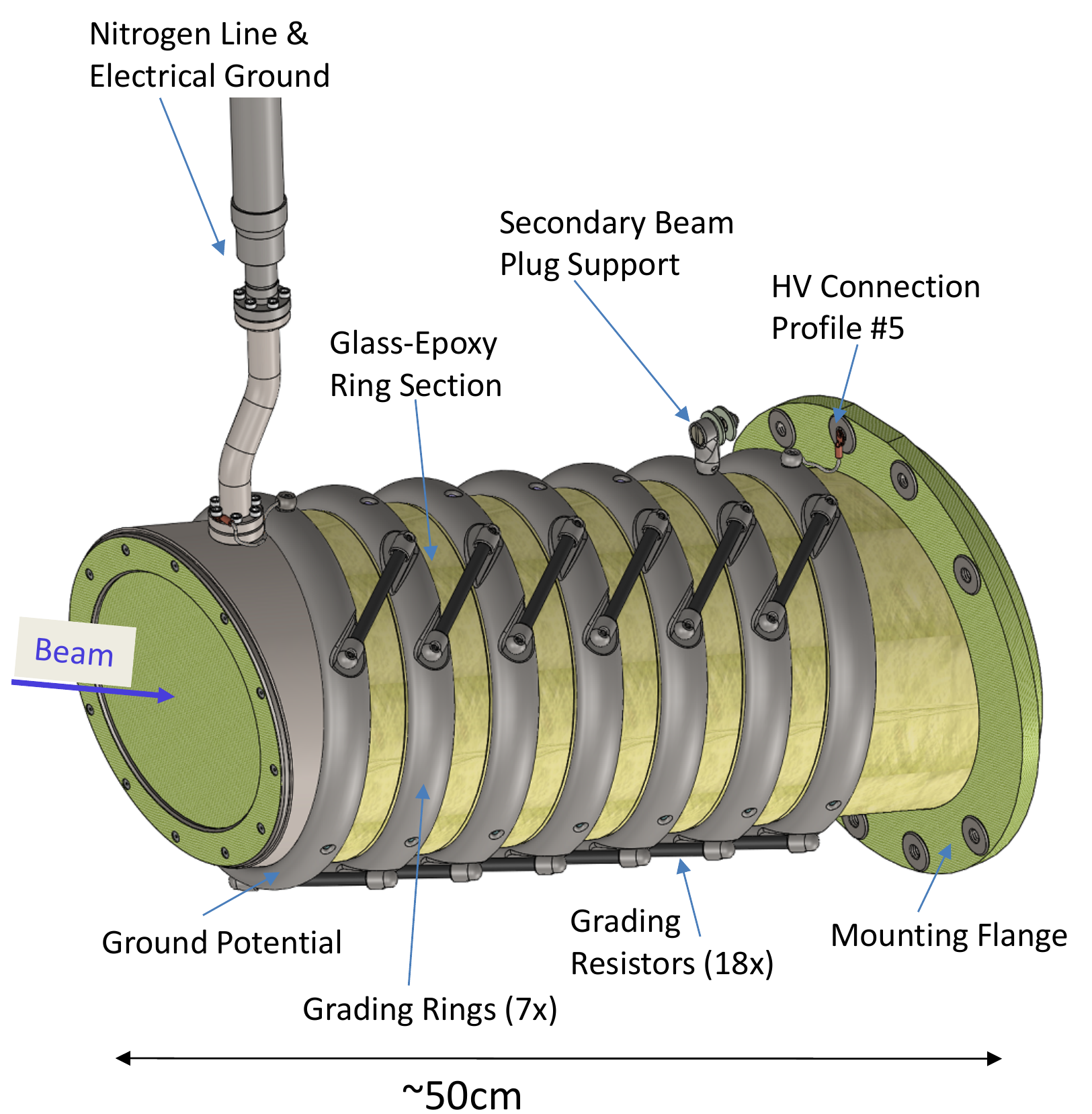}
    \includegraphics[width=.48\textwidth]{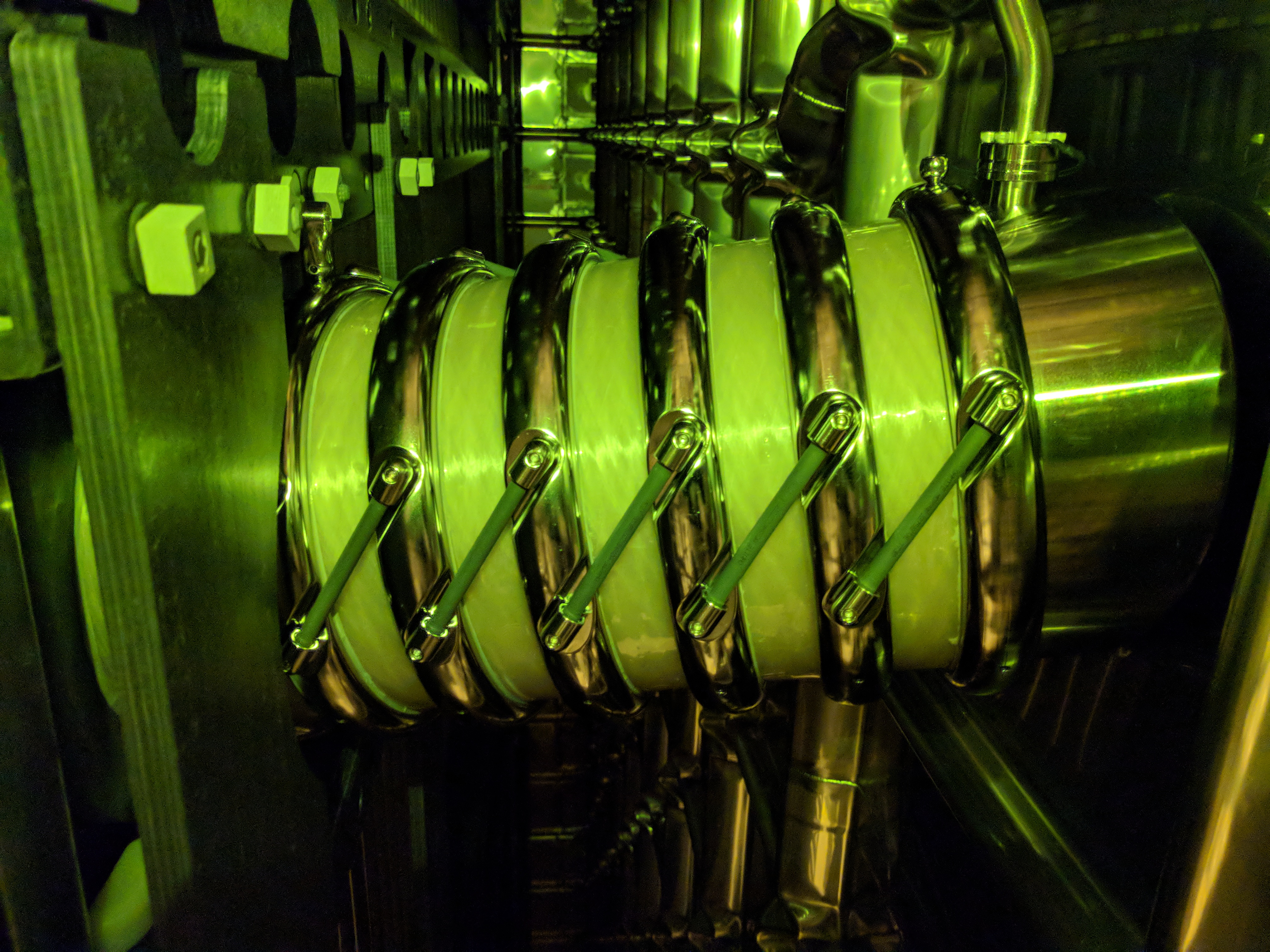}
    \caption{Drawing of the beam plug  (left) and an image of the beam plug installed inside the cryostat (right).}
    \label{fig:beamplug}
\end{figure}

The test beam enters the detector at mid-height and about 30~cm away from the cathode, on the negative $x$ side.  It points down $11^\circ$ from the horizontal, and towards the APA on the negative  $x$ side, $10^\circ$ to the right of the $z$ direction. In order to minimize the energy loss of beam particles prior to their entry in the TPC due to the materials in the cryostat, the 40\,cm of inactive liquid argon in front of the TPC, and the field cage, a ``beam plug''~\cite{Abi:2017aow} is installed on the low-$z$, negative-$x$ side of the end-wall field cage, as shown in figure~\ref{fig:beamplug}. This beam plug is constructed from a series of alternating fiberglass and stainless steel rings to form a cylinder, and capped at entrance and exit ends with low mass fiberglass plates. The stainless steel rings are connected to three sets of resistors to regulate the voltage from the field cage to the grounded cryostat membrane. The beam plug extends through an opening in the field cage about 5~cm inside the field cage boundary.  The inside face of the beam plug is covered with a mini field cage made from 0.8~mm thick printed circuit board to reduce the drift field distortion introduced by this opening. The beam plug is filled with nitrogen at a nominal pressure of 1.3~bar (absolute pressure) to balance the hydrostatic pressure of the liquid argon at this height and also to maintain high dielectric strength to avoid HV breakdown. Besides, the cryostat warm structure and the insulation are also modified to reduce the beam interaction with passive materials.

\subsection{Cold electronics}
\label{det-coldelectronics}

The U, V and X wire planes on both sides of an APA are read out by 20 front-end motherboards (FEMBs) installed close to the wire boards on top of each APA.  The FEMBs amplify, shape, digitize, and transmit all 15,360 TPC channels' signals to the warm interface electronics through cold data cables, which are up to 7~m in length. Each FEMB contains one analog motherboard, which is assembled with eight 16-channel analog front-end (FE) ASICs~\cite{fe-asic}, to provide amplification and pulse shaping, and eight 16-channel Analog to Digital Converter (ADC) ASICs for a total of 128 channels readout per FEMB. 

Each FE ASIC channel has a dual-stage charge amplifier circuit with a programmable gain selectable from 4.7, 7.8, 14 and 25~mV/fC, and a 5th-order anti-aliasing shaper with a programmable time constant with peaking times of 0.5, 1, 2, and 3\,$\mathrm{\mu}$s.  The FE ASIC also has an option to enable AC coupling and a baseline adjustment for operation at either 200~mV for the unipolar pulses on the collection wires or 900~mV for the bipolar pulses on the induction wires. Under normal running conditions the ASIC gain is set at 14~mV/fC and the peaking time is set at 2~$\mathrm{\mu}$s for all channels.  On October 11, 2018, the internal ASIC baseline was changed to 900~mV for both induction and collection channels in order to mitigate ASIC saturation with large input charge. Each FE ASIC also has an adjustable pre-amplifier leakage current selectable from 100, 500, 1000, and 5000~pA. The default leakage current is 500~pA. The estimated power dissipation of a FE ASIC is about 5.5~mW per channel at 1.8\,V. Each FE ASIC contains a programmable pulse generator with a 6-bit DAC for electronics calibration, which is connected to each channel individually via an injection capacitor.   The ADC ASIC has 16 independent 12-bit digitizers performing at speeds up to 2 megasamples per second (MS/s). 

A commercial Altera Cyclone IV FPGA, assembled on a mezzanine card that is attached to the analog motherboard, provides clock and control signals to the FE and ADC ASICs. The FPGA also serializes the 16 data streams from the ADCs into four 1.25~Gbps links for transmission to the warm interface electronics over the cold data cables. The FPGA can also provide a calibration pulse to each FE ASIC channel via the same injection capacitor used for the internal FE ASIC DAC, as a cross-check for the electronics calibration.  The production, commissioning and performance of the cold electronics components are described in~\cite{Adams:2020tqx}.

The number of TPC channels that do not respond to charge signals from cosmic-ray muons evolved over the course of the data-taking period.  Twenty-nine channels never showed any sensitivity to signals, from September 2018 to January 2020. An additional seven became solidly unresponsive during the run, making the total unresponsive channel count 36 in January 2020. Approximately 30 additional channels were found to be intermittently unresponsive during the run.  During initial cold-box testing before installation, 34 channels were identified as non-responsive; this includes the 29 initially dead channels and five intermittent ones that happened to be non-responsive during the test.


\subsection{Photon detectors}

\label{sec:det_photon}

Liquid argon is a prolific emitter of scintillation light.  Approximately 2.4$\times 10^4$ vacuum ultraviolet (VUV) photons are created per MeV of energy deposited by ionization in LAr at the nominal electric field of 500~V/cm.  Photon detectors are installed in ProtoDUNE-SP in order to detect a fraction of these photons to measure interaction times and to get an independent measurement of deposited energy.  These photon detectors, however, cannot be placed outside of the field cage because it blocks the scintillation light, and so the photon detectors are integrated in the APAs, occupying the space between the two mesh planes.  Ten bar-shaped photon detectors with dimensions of 8.6~cm (height) 2.2~m (length)  and 0.6~cm (thickness)  are embedded at equally spaced heights within each APA.  A number of different designs of photon-detector technologies are implemented within this size constraint. In each design, silicon photomultipliers~\cite{Acerbi:2019qgp} are used to convert the light to electrical signals, which are brought out of the cryostat on copper cables. Most of the photon detectors sense the light that reaches the ends of the bars -- the exception is the so-called ARAPUCA design, which collects light at several positions along the bar. More details on the photon detector system are provided in section~\ref{sec:pd}.

\subsection{Cosmic-ray tagger}
\label{sec:det_crt}

The CRT is a system of scintillation counters that covers almost the entire upstream and downstream faces of the TPC.  It was installed in order to provide triggers to read out the detector for a set of cosmic-ray muons that pass through with known timing and direction, parallel to the TPC readout planes. Since the ProtoDUNE-SP detector is on the surface, it is exposed to 20 kHz of cosmic-ray muons.  Most of these muons are not tagged before entry into the TPC. Both untagged muons and muons tagged by the CRT are exploited to provide important calibration data and performance indicators.

\begin{figure}
    \centering
    \includegraphics[width=.58\textwidth]{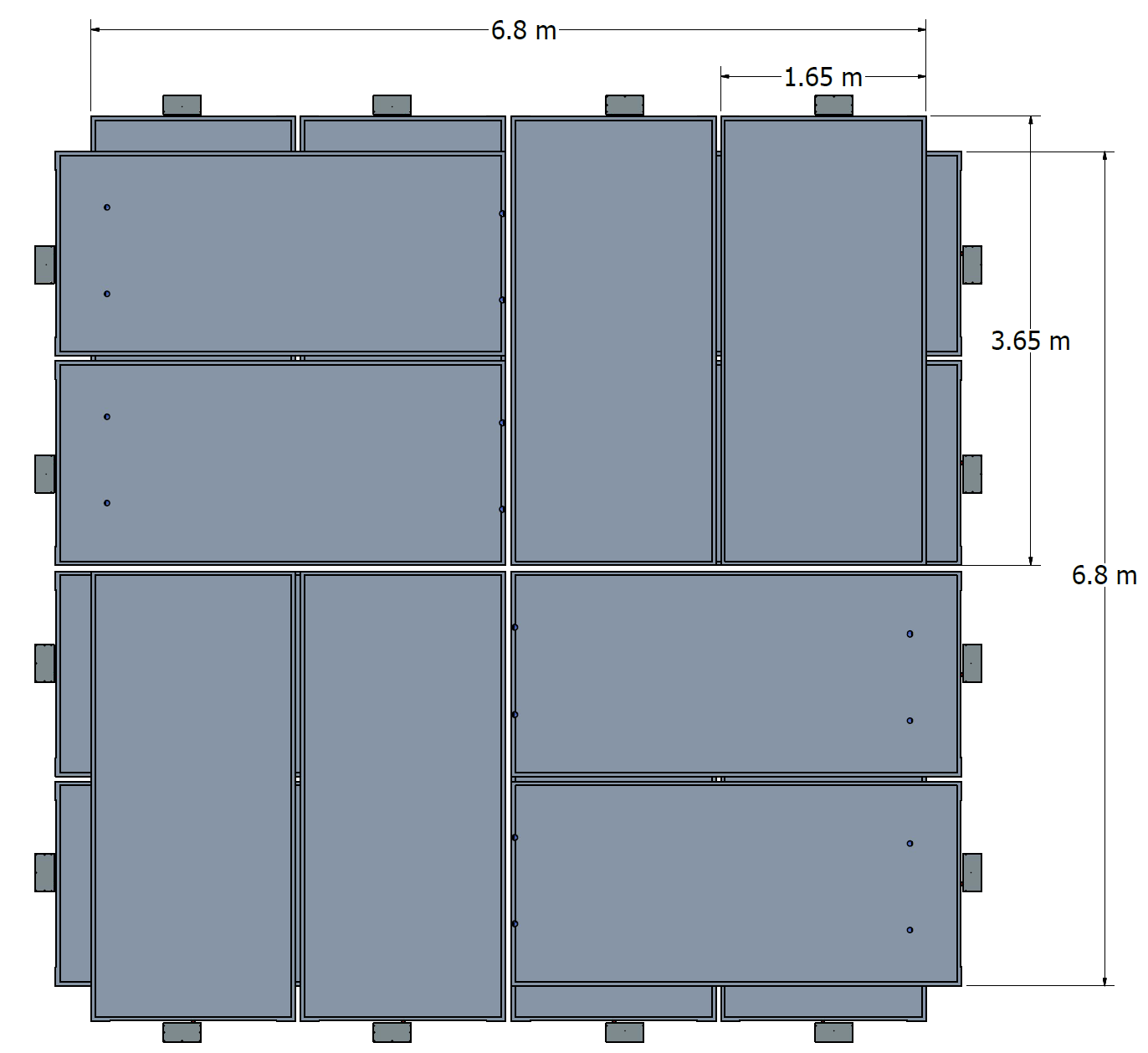}
    \includegraphics[width=.4\textwidth]{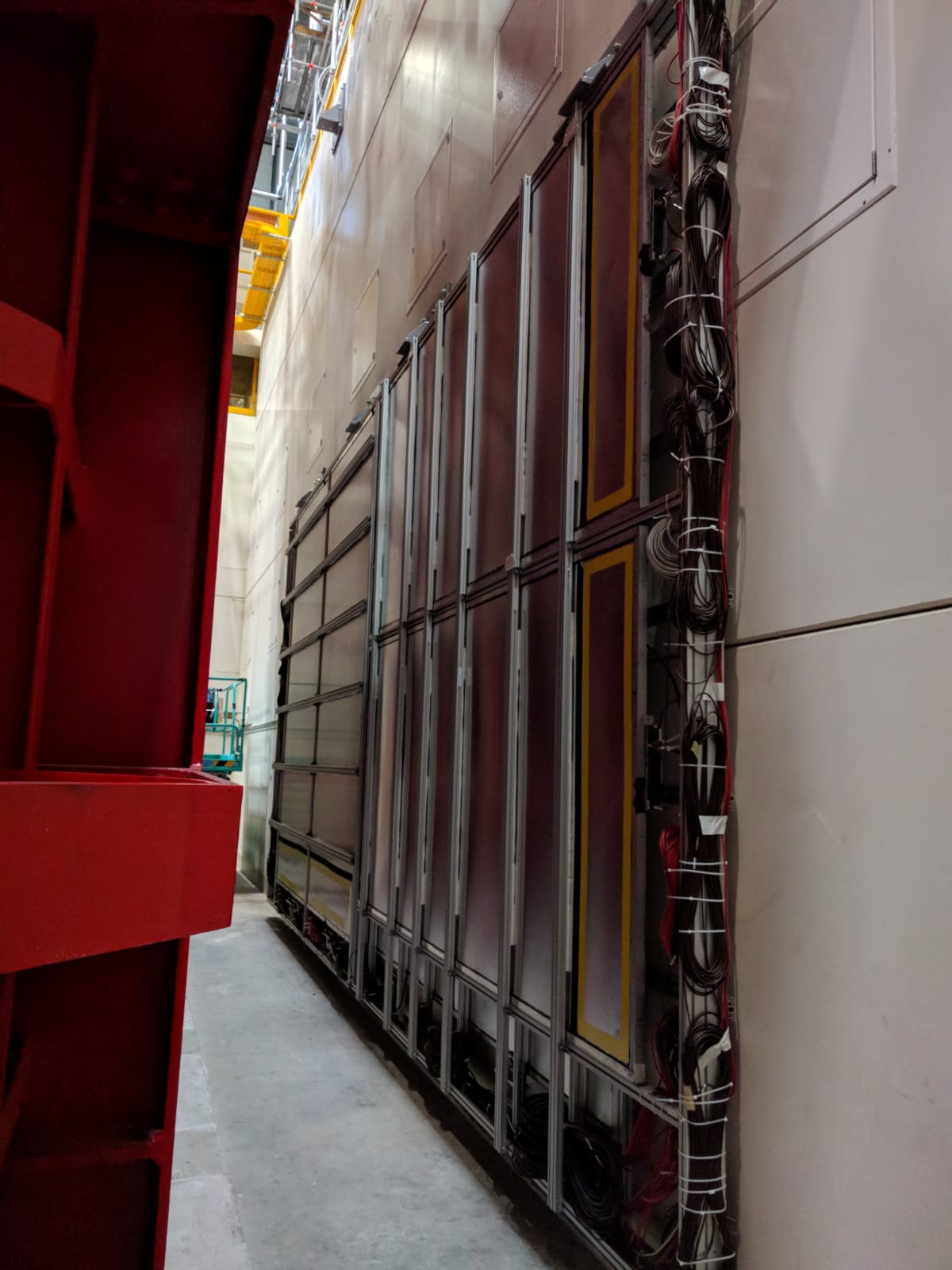}
    \caption{Drawing of CRT modules overlaid (left) and an image of CRT modules installed downstream (right). Two CRT modules measure the $x$~coordinate and two CRT modules are rotated to measure the $y$~coordinate.}
    \label{fig:crtModule}
\end{figure}

The CRT uses scintillation counters recycled from the outer veto of the Double Chooz experiment~\cite{2014PhDT.......100C}. It is constructed in four large assemblies, two mounted upstream and two mounted downstream of the cryostat.  Each assembly covers an area approximately 6.8~m high and 3.65~m wide. 
The CRT uses 32 modules containing 64 scintillating strips each. The strips are 5~cm wide and 365~cm long. The strips in each module are parallel to each other, and thus a module provides a one-dimensional spatial measurement for each track at a given position along $z$. In order to enable two-dimensional sensitivity in $x$ and $y$, four modules are placed together into eight assemblies with two modules being rotated by 90 degrees to create an assembly of 3.65~m by 3.65~m in size, as shown in figure~\ref{fig:crtModule}. Four of these units are placed to cover the upstream (front) face of the detector and the other four placed against the downstream (back) face. Hamamatsu~M64 multi-anode photomultiplier tubes detect the scintillation light and the resulting electrical pulses are digitized by ADCs and recorded by the data acquisition system along with timestamps with 20~ns resolution. A digitized pulse and its timestamp are called a ``one-dimensional hit.''  Two-dimensional hits are reconstructed when two one-dimensional hits are recorded in overlapping CRT modules within a coincidence window of 80~ns.  A cosmic-ray muon track is reconstructed in the CRT by drawing a line from hits in the front modules to hits in the back modules within a coincidence window dictated by the estimated time of flight to travel from the front modules to the back modules.

\begin{figure}[!ht]
\centering
\includegraphics[width=0.8\textwidth]{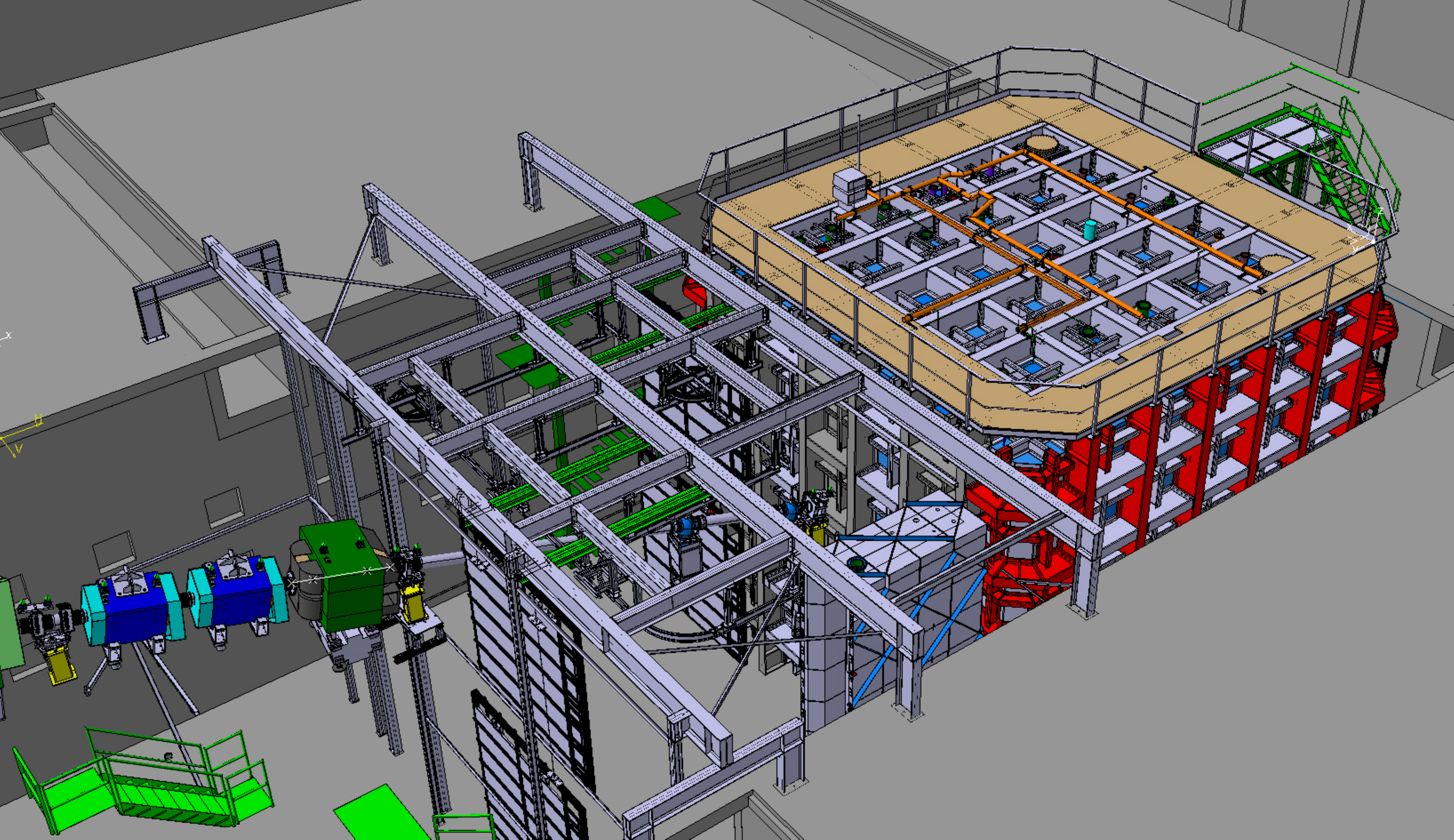}
\includegraphics[width=0.49\textwidth]{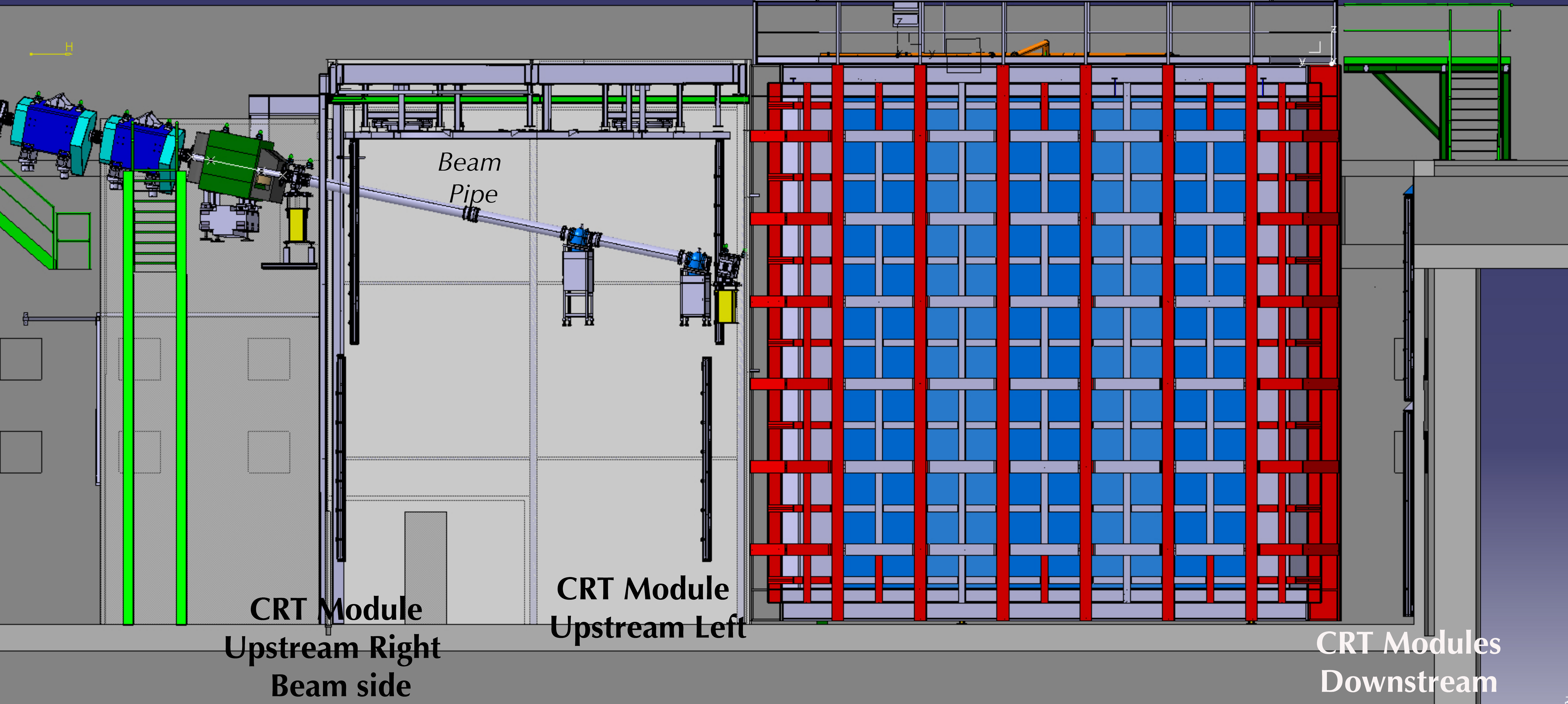}
\includegraphics[width=0.49\textwidth]{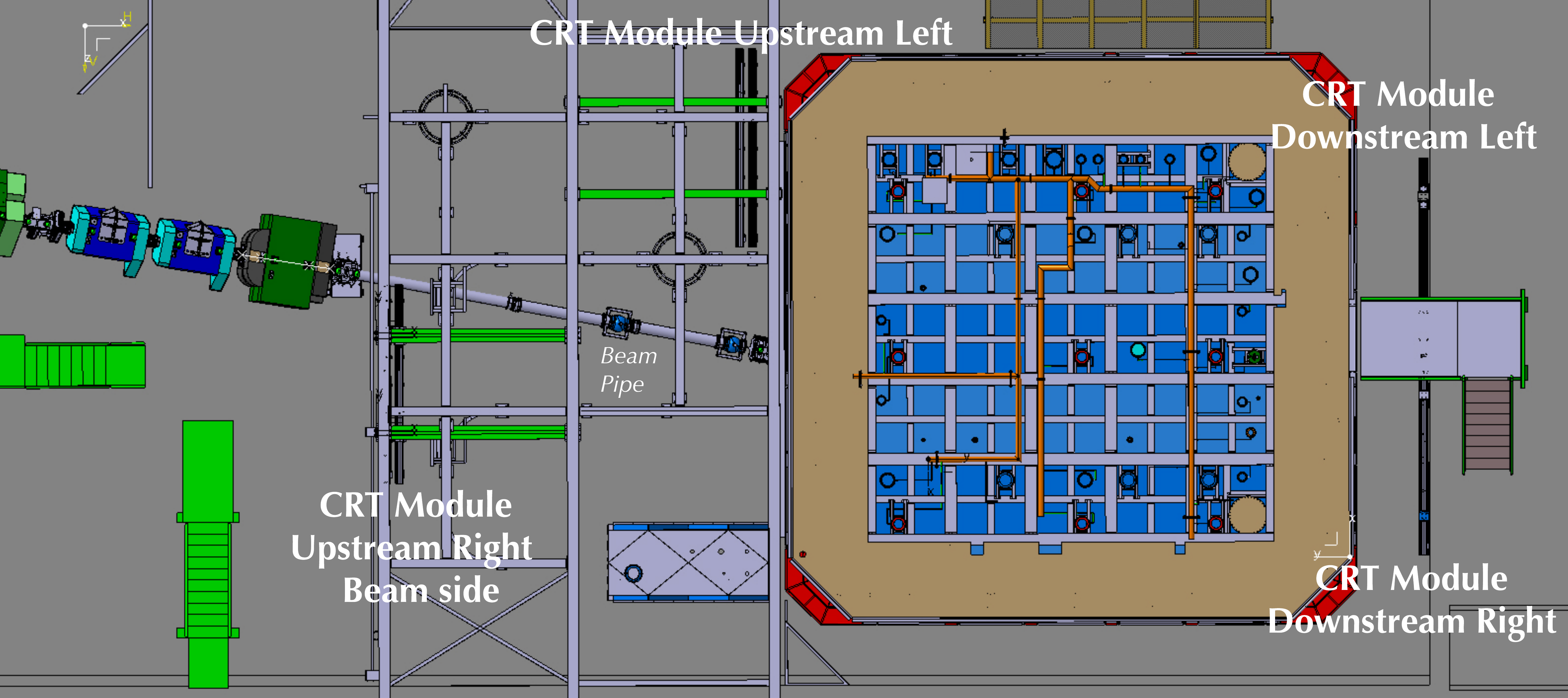}
\caption{Installation of the Cosmic Ray Tagger (CRT): (top) 3D view with staggered upstream modules visible in front of the cryostat, (bottom-left) side view, (bottom-right) top view, with positions of the staggered upstream modules and downstream parallel modules indicated by labels.
}
\label{fig:CRT-CAD-views}
\end{figure}

Half of the 32 upstream CRT modules cover the upstream face of the detector and the other half of the CRT modules cover the downstream face of the detector as seen in figure~\ref{fig:CRT-CAD-views}. The upstream CRT modules are offset due to the beam pipe, which enters the cryostat at an angle. Because of this, eight CRT modules cover the area near the cathode along the $x$ direction, but 9.5~m upstream from the front face of the TPC. The other eight upstream CRT modules sit to the left of the beam pipe with an offset of 2.5~m upstream from the from face of the TPC. The downstream CRT modules are centered with respect to the center of the TPC in $x$ and sit 10~m downstream from the front face of the TPC. 

\subsection{Data acquisition, timing and trigger system}
\label{subsec:DAQTimingTrigger}

The ProtoDUNE-SP data acquisition system (DAQ) is responsible for reading the data from the TPC, the photon detector and the CRT.  It also reduces the data volume using online triggering and compression techniques and formats the data into trigger records\footnote{The word ``event'' is customarily used for a triggered detector readout in many high-energy physics experiments.  Due to the need to refer to interactions as events and the presence of multiple interactions per detector readout, we standardize on ``trigger record'' as the name of a unit of data produced by the DAQ.} for storage and offline processing. The TPC has two candidate readout solutions under test in ProtoDUNE-SP: RCE (ATCA-based)~\cite{Herbst:2016prn,Tsang:2019okh} and FELIX (PCIe-based)~\cite{Anderson:2016lfn,Borga:2018uqw}.  Both of these systems ran simultaneously.  For the beam runs, five out of the six APAs were read out using RCEs and one APA was read out using FELIX. After the beam runs, four APAs were converted from RCE readout to FELIX readout. Fermilab's artDAQ~\cite{Biery:2013cda} is used as the data-flow software.

The ProtoDUNE-SP timing system provides a 50~MHz clock multiplexed on an 8b10b encoded data stream that is broadcast to all endpoints. The timing system interfaces to the CERN SPS beam presence signals and can be used to switch modes for data taking with and without beam. The timing system data stream also provides the trigger distribution. The timing system is partitionable, a feature that allows parts of the experiment to run independently.  A clock synchronized to the Global Positioning System  provides 64-bit timestamps that are used to mark the trigger and data times irrespective of file name, run, or trigger record numbers.



A hardware triggering system was designed in order to perform event selection in ProtoDUNE-SP.
The core element of this system is the Central Trigger Board (CTB) which is a custom printed circuit board (PCB) in charge of processing the status of the auxiliary detectors to aid in making prompt readout decisions.
The readout decisions are ultimately made by the timing system which communicates with the CTB through various commands.
The CTB hosts a MicroZed, which is a commercial PCB with an onboard System-On-a-Chip (SoC).
The SoC contains both Programmable Logic (PL) and a Processing System (PS) and serves as an interface between the auxiliary detectors (photon detectors, beam instrumentation, and CRT) and the DAQ through the timing system.
The CTB has 32 individual CRT pixel inputs (a pixel being a unit of two overlapping panels), 24 optical inputs for the photon detection system, and seven inputs for beam instrumentation signals, all of which are translated into digital pulses and forwarded to the PL for further processing.

\begin{figure}[!htp]
    \centering
    \includegraphics[width=\textwidth]{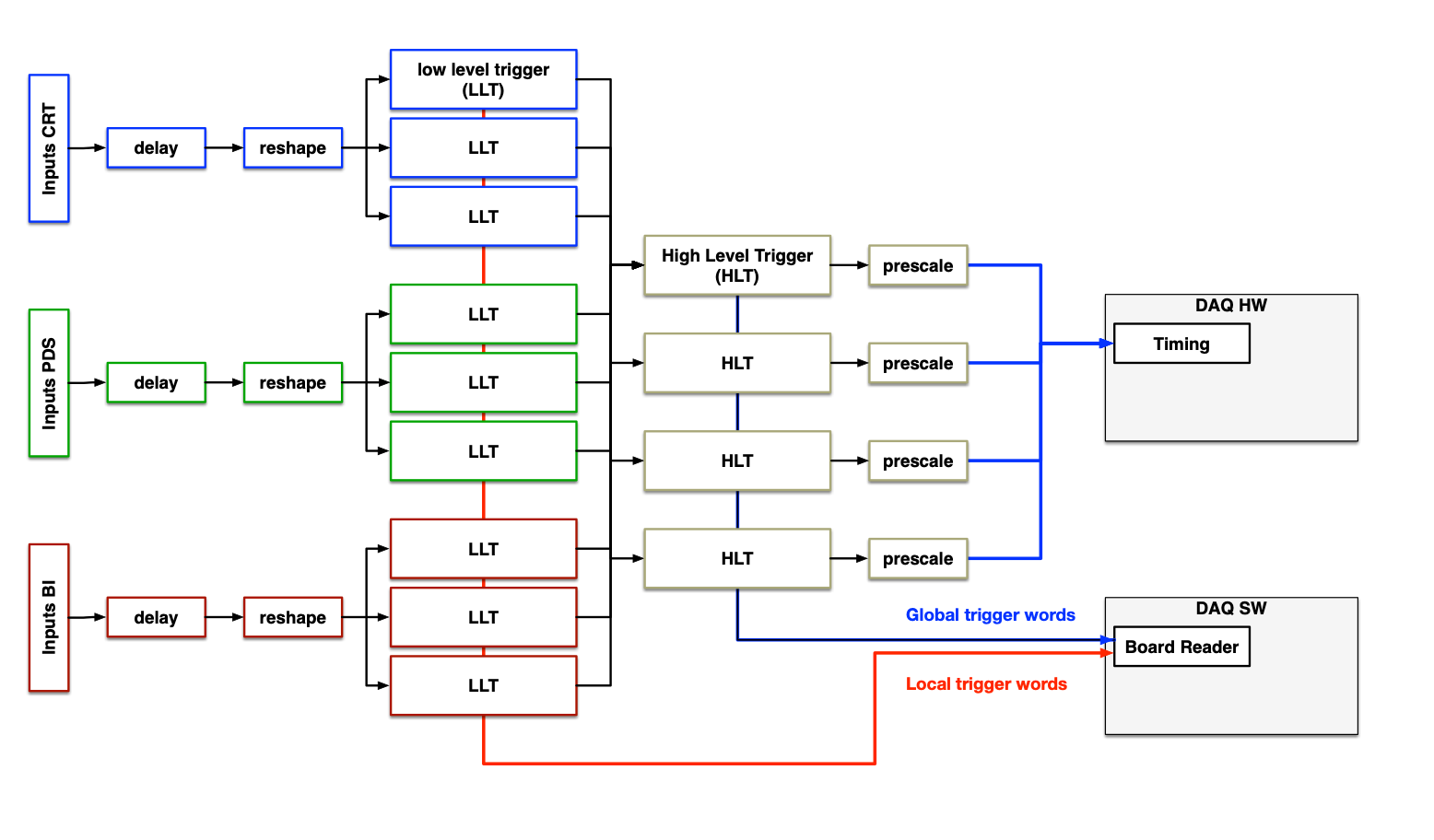}
    \caption{CTB trigger hierarchy.}
    \label{fig:ctbfw}
\end{figure}

The CTB triggering firmware operating in the PL is organized into a two-level hierarchy of low-level and high-level triggers (LLTs and HLTs) which are configurable at run-time by the DAQ system. 
LLTs are defined for inputs from a single subsystem while HLTs can be defined using the various LLTs and can therefore span any or a combination of the subsystems.
Several trigger conditions can be set up; each one is uniquely identified by a bitmask and is embedded into a trigger word issued to the DAQ. An overview of the CTB trigger scheme and its interface with the DAQ is depicted in figure~\ref{fig:ctbfw}.

Additionally, multiple trigger conditions can be satisfied during a single triggered detector readout.  To distinguish between these, the CTB timestamps all LLTs and HLTs generated with the 50 MHz system clock. This (64-bit) timestamp is also included in the trigger word along with the bitmask.



In the HLT, trigger conditions can be configured to require coincidences or anti-coincidences between the various LLTs.
Only when all the required conditions for an HLT are satisfied is a trigger command passed to the timing system. 
The timing system is then responsible for validating or vetoing\footnote{In case a trigger has already been issued by the timing system.} the issued trigger. 
If accepted, the timing system forwards the readout decision to the DAQ software and to the individual readout systems (i.e. TPC, photon detectors, and CRT).
However, for accountability\footnote{If beam pile-up occurs, the timing system vetoes any additional beam triggers if it has issued one in the last 10~ms. However, the CTB still reports multiple beam triggers in this case.}, the CTB sends a data word directly to the DAQ to be stored regardless of whether or not the trigger is validated by the timing system.


All HLTs can be classified as beam-on or beam-off triggers.
The former relies mostly on the beam instrumentation inputs and requires the conditions to be satisfied during a beam spill while the latter requires that the conditions are satisfied outside the beam spill.
The most common examples of beam-on triggers include those aimed at tagging electron, proton and kaon events.
By requiring different signal combinations from the beam instrumentation inputs, one can identify specific particles for a relevant energy range, which will be discussed in section~\ref{sec:beamline}.



The most common examples of beam-off triggers are those arising from cosmic-ray activity. Several of these triggers are in place to select events with specific topologies, requiring CRT pixels from specific regions to register hits in coincidence with pixels from another region. 
For example, by requiring that at least one upstream CRT pixel is hit in coincidence with a downstream CRT pixel, one can select throughgoing muon candidates. 
Another trigger is set up for cathode-crossing muon candidates, which is achieved by requiring coincidence hits on CRT pixels on opposing drift volumes and sides of the cryostat.
In addition to the logic-specific triggers, an aperiodic random trigger is provided to read out the detector  without regard to trigger conditions.


For each triggered readout of the detector, the TPC data consists of 6000 consecutive samples of each ADC, which are digitized at a rate of 2~MHz, for a total of 3~ms of time.  Each time period of 500~ns between ADC samples is called a ``tick."  The data readout starts 250~$\mu$s (500 ticks) before the trigger time in order to collect charge deposited by particles that arrive earlier than the trigger but cause charge to arrive at the anodes during time periods that overlap those of triggered events. Corresponding data from the photon detectors and the CRT are saved in the output data stream for analysis.  Compressed raw data trigger records have a typical size of 60~MB, and trigger rates of 40~Hz were reliably sustained by the data acquisition system. A typical physics run lasts several hours.


\section{CERN beam line instrumentation}
\label{sec:beamline}
The ProtoDUNE-SP TPC is located in the CERN North Area in a tertiary extension branch of the H4 beam line. The 400 GeV/$c$ primary proton beam is extracted from the CERN Super Proton Synchrotron (SPS) and is directed towards a beryllium target, producing a mixed hadron beam with a momentum of 80~GeV/$c$. This secondary beam is then transported to impinge on a secondary target, producing a tertiary, very low energy (VLE) beam in the 0.3~-~7~GeV/$c$ momentum range. The H4-VLE beam line then accepts, momentum-selects and transports these particles to the ProtoDUNE-SP detector. The secondary target material can be changed between copper and tungsten.  The latter is chosen for momenta below 4~GeV/$c$ in order to increase the hadron content of the beam. However, the copper target was unintentionally used for the 2~GeV/$c$ run instead of the tungsten target.

\subsection{Beam line instrumentation components}\label{sec:hardware}
\begin{figure}[!ht]
\begin{center}
\includegraphics[width=\textwidth]{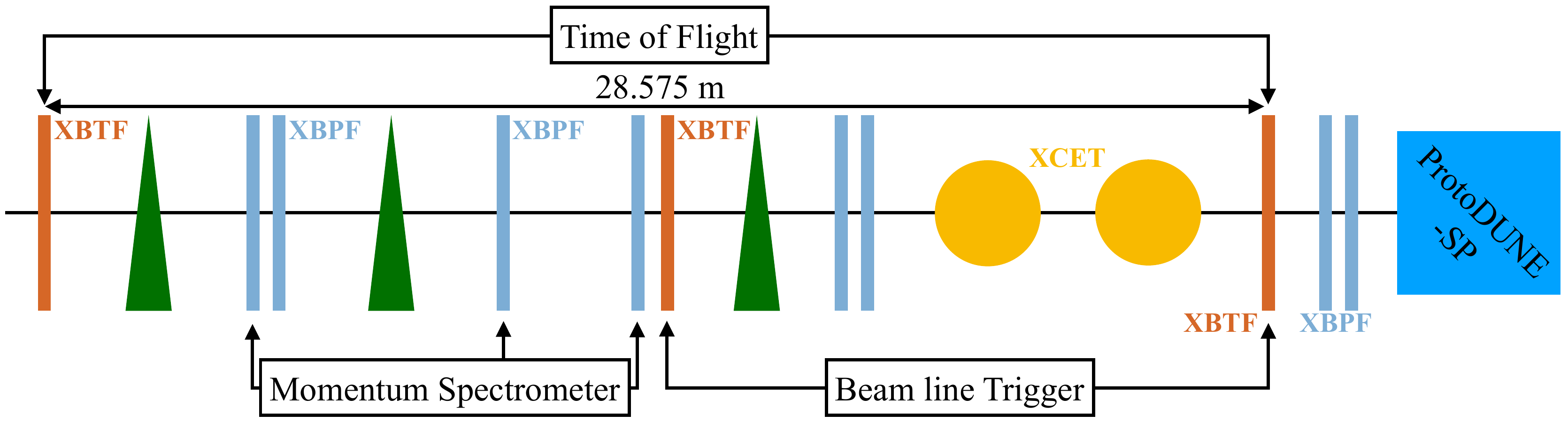}
\end{center}
\caption{A schematic diagram showing the relative positions of the trigger counters (XBTFs), bending magnets (triangles), profile montiors (XBPFs) and Cherenkov detectors (XCETs) in the H4-VLE beam line. Combining data from different pieces of instrumentation can be used for triggering, reconstructing momentum and measuring time of flight.}\label{fig:bischem}
\end{figure}

The H4-VLE beam line is instrumented with three types of detectors that provide particle identification and a trigger for the TPC. There are eight profile monitors (``XBPF''), three trigger counters (``XBTF'') and two threshold Cherenkov counters (``XCET''). There are also three bending magnets that direct the beam toward the ProtoDUNE-SP detector. The second of these magnets is also used as part of a momentum spectrometer. The relative positions of each of these features can be seen in figure~\ref{fig:bischem}. A description of the beam line design has been reported elsewhere~\cite{PhysRevAccelBeams.20.111001}, while an in-depth discussion of the instrumentation can be found in~\cite{PhysRevAccelBeams.22.061003}.

The XBPFs, described in detail in \cite{OrtegaRuiz:253109}, are scintillating fiber detectors, each containing 192~square fibers, approximately 1~mm thick. The fibers are arranged in a planar configuration and cover an area of approximately $20 \times 20\ \textrm{cm}^2$. Each device contains a single plane of fibers and therefore measures one spatial coordinate. Pairs of these detectors, rotated by 90\degree ~with respect to each other, are placed at several points along the beam line. This arrangement allows the beam position to be tracked on a particle-by-particle basis. The XBPF data is also used in the reconstruction of a particle's momentum, discussed in section \ref{subsec:momSpec}. 
Hits in the last two sets of XBPF devices are used to measure the trajectories of the beam particles that are then extrapolated to the face of the ProtoDUNE-SP TPC. The XBTFs are designed in a similar way. However, instead of each fiber being read out separately, they are gathered into two bundles and therefore offer no position resolution. Instead, the signals from upstream and downstream planes, which are separated by 28.575~m, are connected to a time-to-digital converter (FMC-TDC \cite{ANSI}). The TDC signals from these two planes provide a particle's time of flight (TOF). The resolution of this measurement has been measured to be approximately 900~ps \cite{PhysRevAccelBeams.22.061003}.  

Coincident signals from the middle and downstream XBTFs act as a ``general trigger.'' These general triggers are sent to the CTB serving as conditions for HLTs as described in section \ref{subsec:DAQTimingTrigger}. During data taking across the momentum regime of interest, the measured efficiencies of the XBPFs with respect to these triggers are greater than 95\% for all chambers \cite{PhysRevAccelBeams.22.061003}.

The two Cherenkov counters used in the H4-VLE are of similar design \cite{ Charitonidis:2202366,CHARITONIDIS2017134}, although one is able to sustain a higher radiator gas pressure. The internal pressures of the two devices were tuned to tag different particle species at various momenta. A combination of the TOF and the two Cherenkov signals (high and low pressure), offers particle identification for analysis across the whole momentum spectrum of interest. During the beam run, signals from these devices were sent to the CTB to form HLTs tagged as various beam particle species. 




\subsection{Beam line simulation and optimization}\label{sec:beamline_simulations}
The beam transported in the H4-VLE beam line is produced by the collision of the secondary mixed hadron beam of 80~GeV/$c$ with the secondary fixed target. To limit the contributions from the decays of unstable low-energy hadrons such as pions and kaons, a beam line length of less than 50~m is required. Low-energy beam particles need to be sufficiently separated from the high-energy background in this distance, and enough space for the beam line  instrumentation is required.  Detailed simulation studies were carried out in order to meet these specifications.

The performance of the initial layout was calculated with the beam optics code {\tt Transport}~\cite{osti_97260} and refined by a comprehensive {\tt MAD-X}~\cite{MADX} (and {\tt MAD-X-PTC}~\cite{Schmidt:2007zzd}) simulation~\cite{Chatzidaki:2666202}.
The Monte Carlo simulations use two frameworks, {\tt G4beamline}~\cite{Roberts:2008zzc} and {\tt FLUKA}~\cite{BOHLEN2014211,Ferrari:898301}.
Different target lengths and materials were investigated to satisfy the experimental needs of rate and beam composition.
The target choice (either copper or tungsten) and the different field strengths of the beam line's dipoles and quadrupoles are incorporated into the {\tt G4beamline}  and {\tt FLUKA} models. Based on these studies, estimates of the beam rates, compositions and background rates at the experiment location are obtained.
The background suppression was improved by optimizing the shielding using the {\tt FLUKA} simulation~\cite{PhysRevAccelBeams.22.061003}.


\subsection{Beam line event reconstruction and particle identification}\label{sec:pid}
    
    Information from the three types of beam line instruments (discussed in section~\ref{sec:hardware}) is combined in order to perform particle identification on an event-by-event basis. A search window in time of 500~ns is defined around each general trigger; timestamps associated with data packets from each device are then matched within this interval.  
    \subsubsection{Momentum spectrometer technique/calculation}\label{subsec:momSpec}

\begin{figure}[!ht]
\begin{center}
\includegraphics[width=\textwidth]{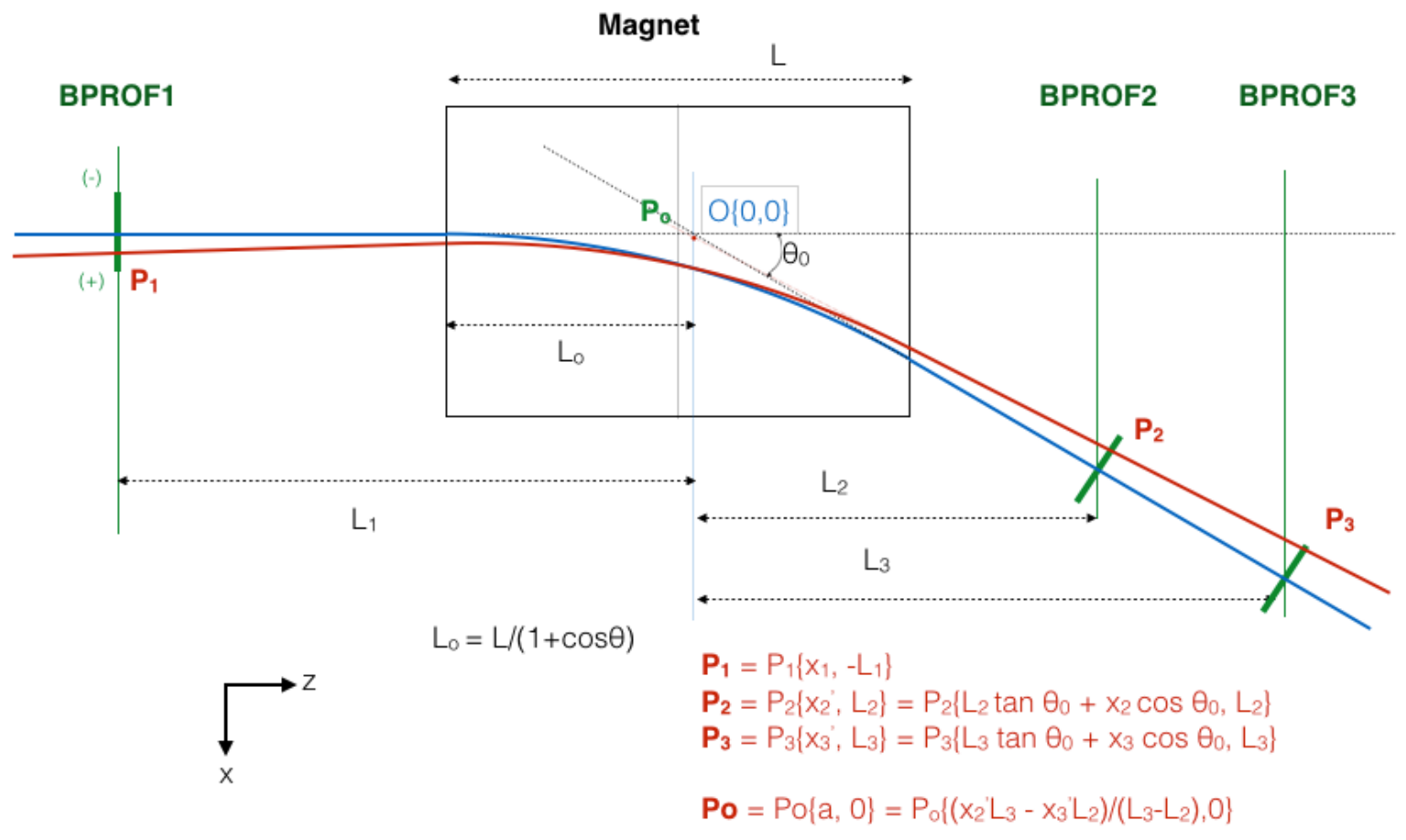}
\end{center}
\caption{A schematic diagram showing the method by which momentum is reconstructed for a given beam particle (red), as discussed in the text. Taken from ~\cite{Charitonidis:2202366}. The direction of the $x$ axis is opposite to the convention used in this paper.}\label{fig:momspecschem}
\end{figure}

The three XBPF detectors surrounding the middle bending magnet provide a measurement of each particle's momentum. This is illustrated in figure \ref{fig:momspecschem} \cite{Charitonidis:2202366}. 
The lateral position of the particle at each XBPF detector ($\chi_1$, $\chi_2$, $\chi_3$) is provided by the index of the activated fibers in the profile monitors.  These measurements, along with the known distances between the monitors ($L_1$, $L_2$, $L_3$) and the measured magnetic field are used with equations~\ref{eq:bending_angle} and~\ref{eq:beam_momentum} to determine a particle's bending angle $\theta$ and momentum $p$.

n\begin{equation}\label{eq:bending_angle}
\cos\theta = \frac{ M [\Delta L \tan \theta_0 + \Delta\chi\cos\theta_0] + L_1\Delta L}{\sqrt{[M^2 + L_1^2][(\Delta L\tan\theta_0 + \Delta\chi\cos\theta_0)^2 + \Delta L^2]}}
\end{equation}

\begin{equation}\label{eq:beam_momentum}
p = \frac{299.7924}{\theta} \times \int_0^{L_{\tiny{\textrm{mag}}}} (Bdl)
\end{equation}

Here, $M \equiv \alpha + \chi_1$, $\alpha = \frac{\chi_3 L_2 - \chi_2 L_3}{L_3 - L_2}\cos\theta_0$, $\Delta L \equiv L_3 - L_2$, and $\Delta \chi \equiv \chi_2 - \chi_3$. $\theta_0$ is the nominal bending angle of the beam and is equal to 120.003~mrad~\cite{PhysRevAccelBeams.22.061003}. 

\subsubsection{Particle identification logic}\label{subsec:pid}
\begin{figure}[!htp] 
  \begin{subfigure}[t]{0.45\linewidth}
    \centering
    \includegraphics[width=0.99\linewidth]{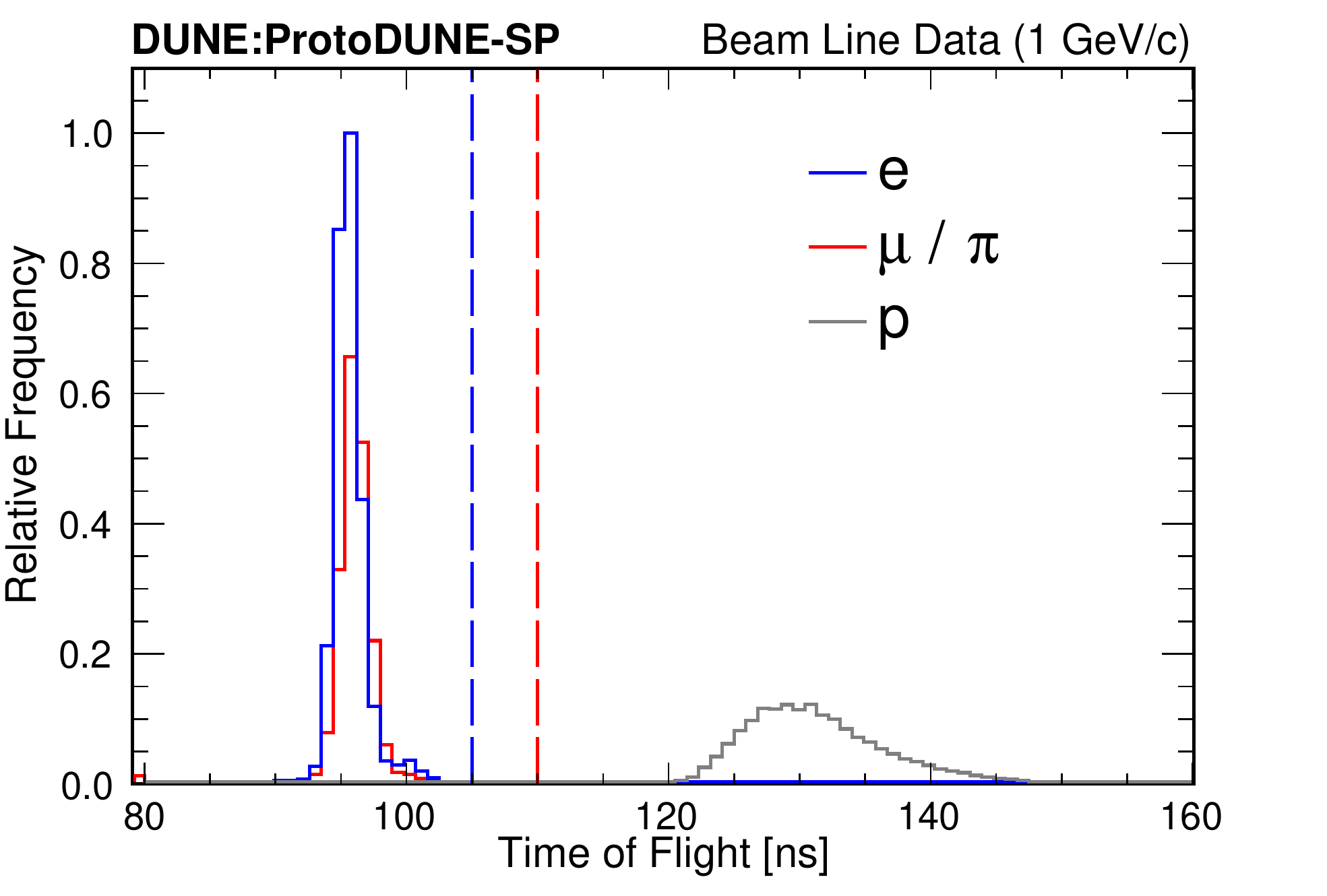} 
    \caption{Nominal beam momentum = 1~GeV/$c$. Vertical lines represent the time of flight cuts used for electrons (blue), and muons/pions (red).} 
    \label{fig:TOF_1} 
    \vspace{1ex}
  \end{subfigure}\hspace{0.1\linewidth}
  \begin{subfigure}[t]{0.45\linewidth}
     \centering
     \includegraphics[width=0.99\linewidth]{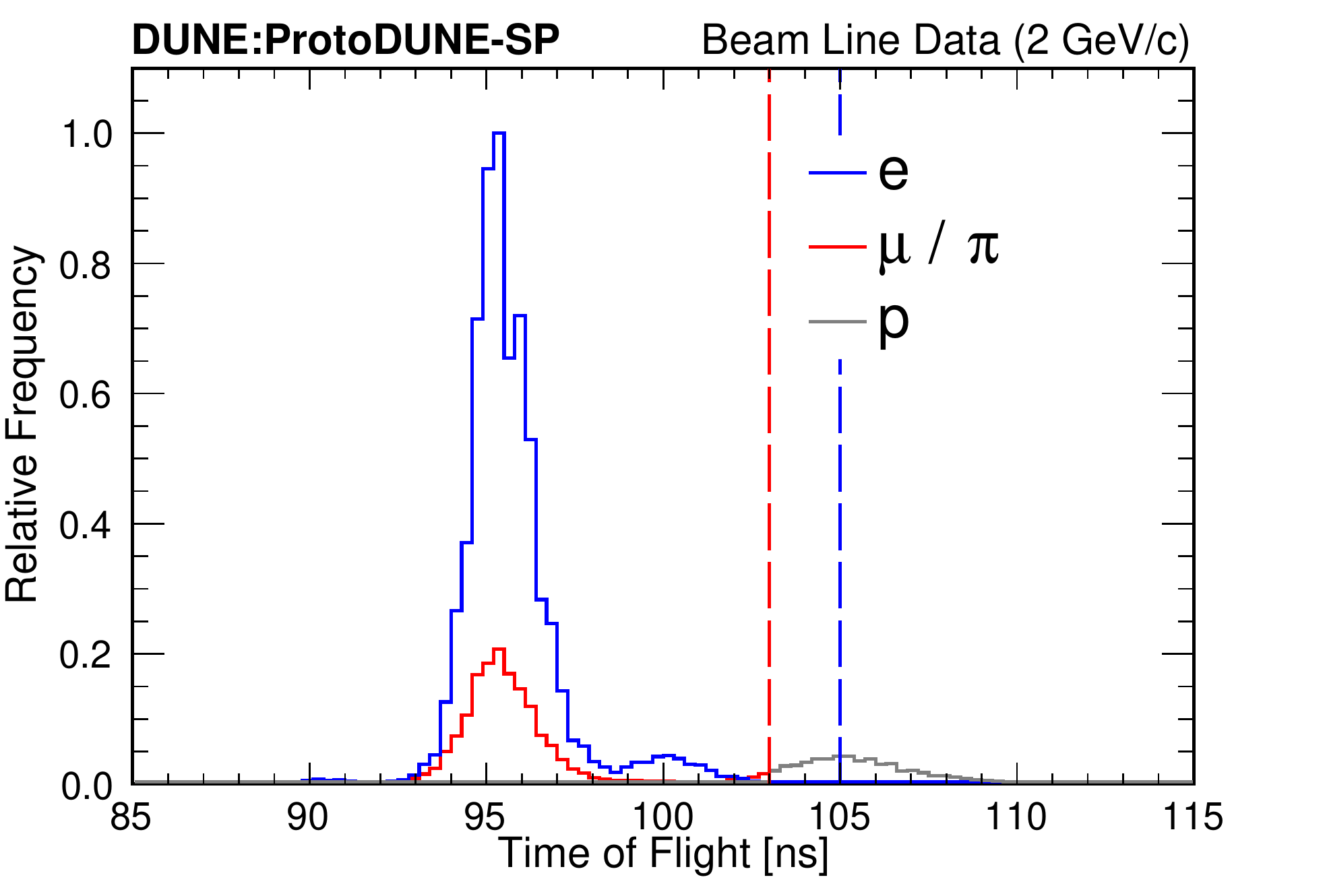}
     \caption{Nominal beam momentum = 2~GeV/$c$. Vertical lines represent the time of flight cuts used for electrons (blue), and muons/pions (red).} 
     \label{fig:TOF_2} 
     \vspace{1ex}
   \end{subfigure} 
   \begin{subfigure}[t]{0.45\linewidth}
     \centering
     \includegraphics[width=0.99\linewidth]{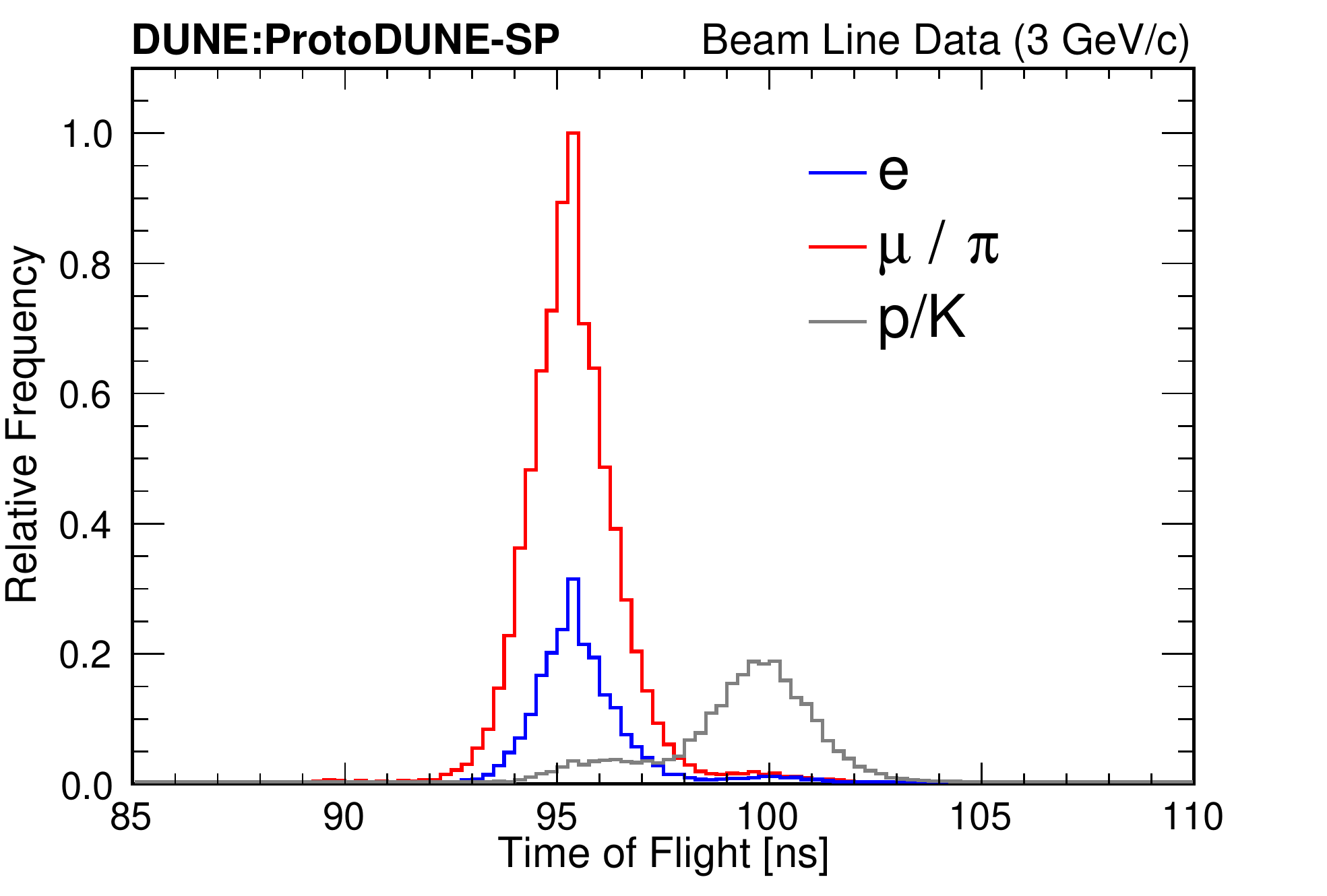} 
     \caption{Nominal beam momentum = 3~GeV/$c$.} 
     \label{fig:TOF_3} 
   \end{subfigure}\hspace{0.1\linewidth}
     \begin{subfigure}[t]{0.45\linewidth}
    \centering
    \includegraphics[width=0.99\linewidth]{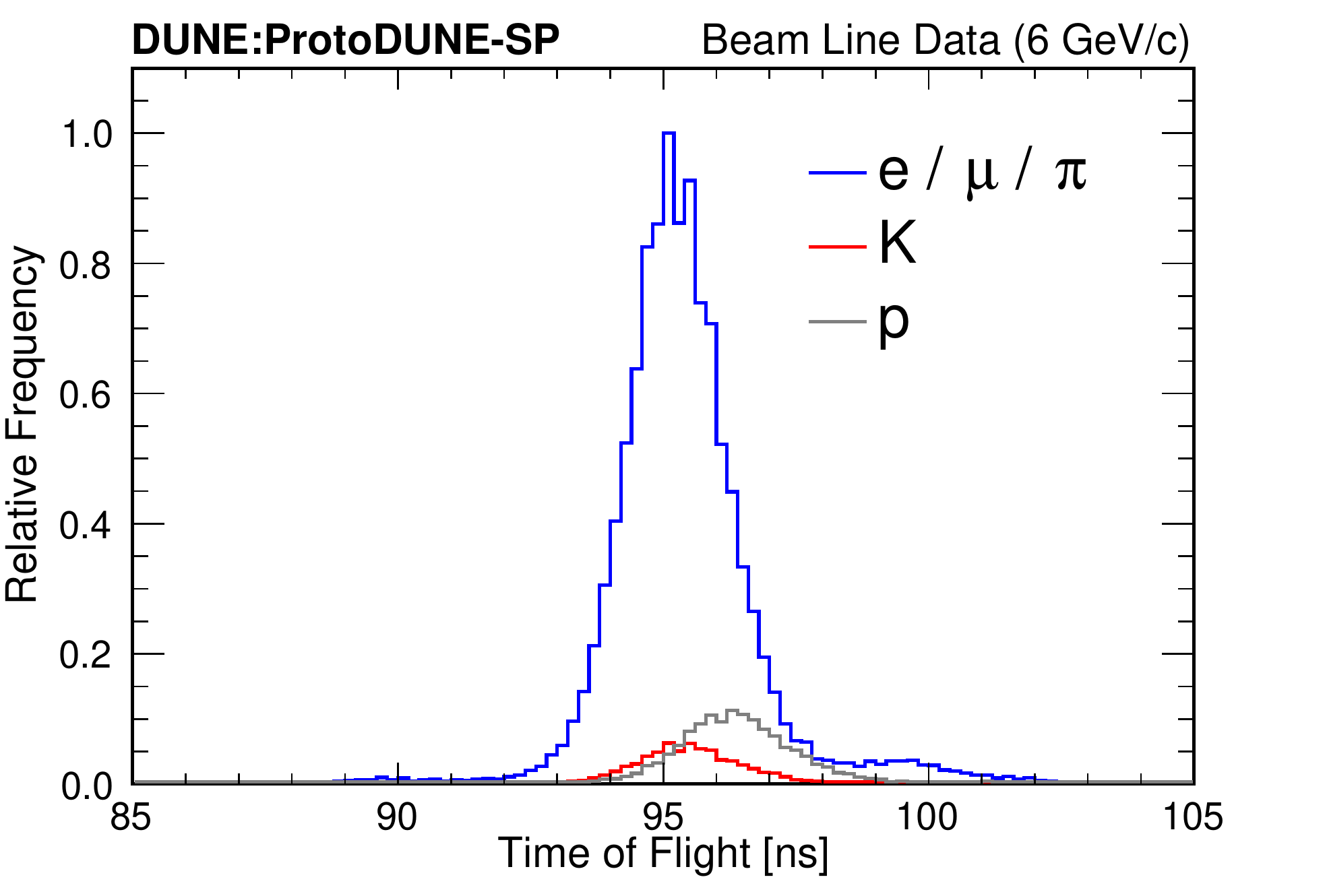} 
    \caption{Nominal beam momentum = 6~GeV/$c$.} 
    \label{fig:TOF_6} 
  \end{subfigure} 
  \caption{Time of flight distributions for different reference momenta, separated by particle using the PID techniques listed in table~\ref{tab:PID}. The distributions are normalized such that the maximum height is equal to 1.}
  \label{fig:TOF} 
\end{figure}

\begin{figure}[ht]
    \centering
    \includegraphics[width=\textwidth]{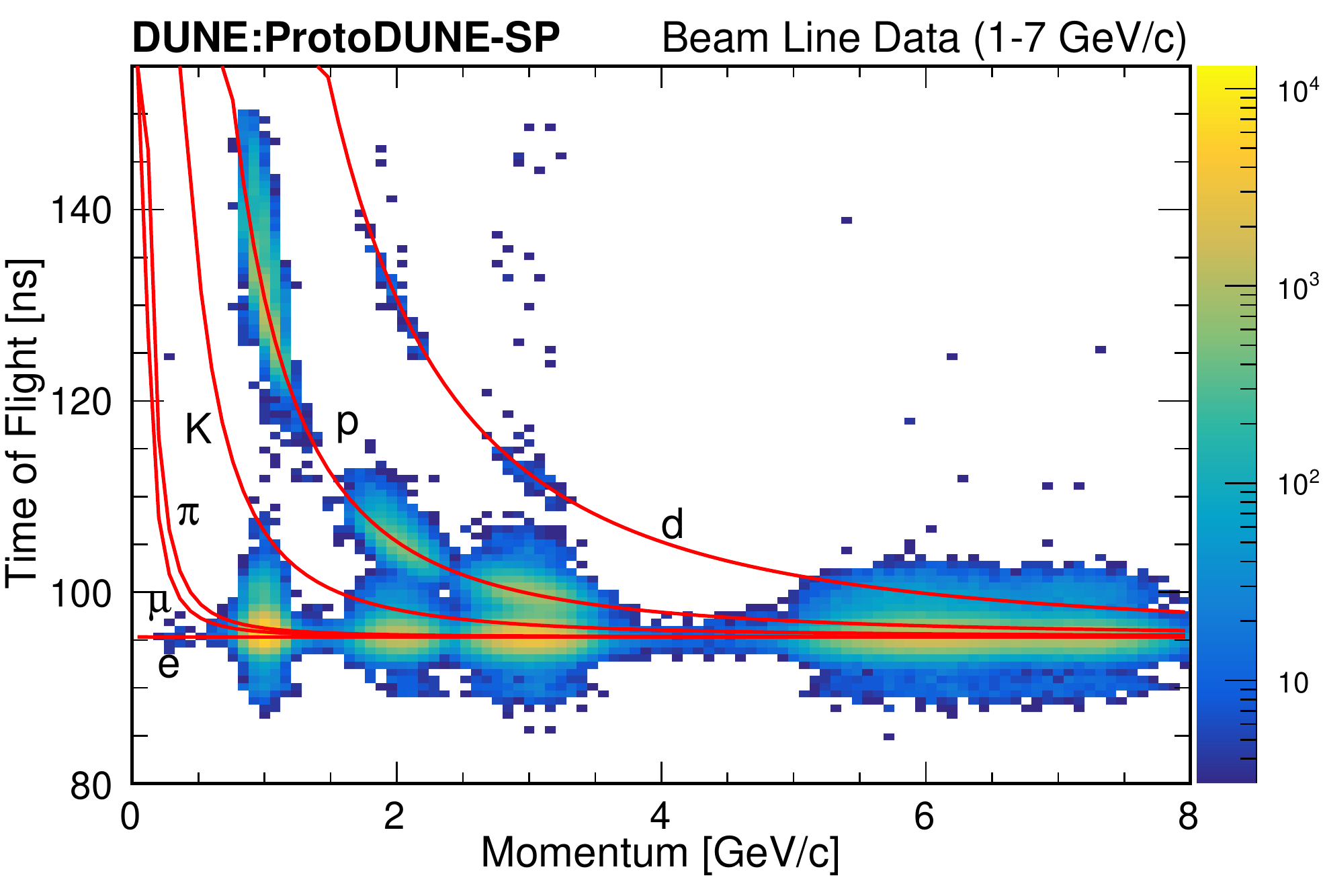}
    \caption{The distribution of particles' time of flight against reconstructed momentum from several runs at various beam reference momenta. The red curves are predictions for $e$, $\mu$, $\pi$, $K$, $p$ and deuterons ($d$) in order of increasing time of flight. }
    \label{fig:tof_mom}
\end{figure}

The beam line is designed to provide particle identification (PID) for the various particle types ($p$, $\mu$, $\pi$, $e$, $K$) comprising the beam. Depending on the beam momentum settings, different conditions are applied to the data from the beam line instrumentation to extract the particle types. These conditions are listed in table~\ref{tab:PID}. This technique is demonstrated for selected runs at various beam momenta in figures~\ref{fig:TOF_1} -- \ref{fig:TOF_6}. Figure \ref{fig:tof_mom} shows the measured momentum and TOF distribution throughout the selected runs. The red curves show expected TOF for several particle types ($e$, $\mu$, $\pi$, $K$, $p$ and $d$) given the particle's momentum, its mass, and assuming a distance of 28.575~m between the TOF monitors. 


\begin{table}[!htp]
\center
\caption{A summary of beam line instrumentation logic used in the identification of particle types. 
Each cell reflects how a particular type of instrumentation is used at a given reference momentum. When time of flight is used, the values of the lower and upper cuts are given in nanoseconds. In the case of the high-pressure Cherenkov (XCET-H) and the low-pressure Cherenkov (XCET-L), zero and one represent the absence and presence of a signal respectively. When a given piece of instrumentation is not involved in a logic decision at a given momenta, a dash is used.
\label{tab:PID}}
\begin{tabular}{|c|c|x{15mm}|x{15mm}|x{15mm}|x{15mm}|}
	\cline{3-6} 
    \multicolumn{2}{c|}{}  & \multicolumn{4}{c|}{\textbf{Momentum (GeV/{\boldmath $c$})}}  \\ 
    \cline{3-6} 
	\multicolumn{2}{c|}{}  & \textbf{1} & \textbf{2} & \textbf{3} & \textbf{6 - 7}  \\
	\hline
	\multirow{3}{*}{$e$}  & TOF (ns) & 0, 105 & 0, 105 & -- & --  \\ 
	\cline{2-6} 
	 & XCET-L  & 1 & 1 & 1 & 1 \\
	\cline{2-6}   
	 & XCET-H & -- & -- & 1 & 1  \\ 
		\hline 
	\multirow{3}{*}{$\mu \ /\ \pi$}& TOF (ns) & 0, 110 & 0, 103 & -- & -- \\
	\cline{2-6} 
	& XCET-L & 0 & 0 & 0 & 1 \\
	\cline{2-6}   
	& XCET-H & -- & -- & 1 & 1 \\ 
		\hline 
	\multirow{3}{*}{$K$}& TOF (ns) & -- & -- & -- & -- \\ 
	\cline{2-6} 
	& XCET-L & -- & -- & 0 & 0 \\
	\cline{2-6}   
	& XCET-H & -- & -- & 0 & 1 \\ 
		\hline 
	\multirow{3}{*}{$p$}& TOF (ns) & 110, 160 & 103, 160 & -- & -- \\ 
	\cline{2-6} 
	& XCET-L & 0 & 0 & 0 & 0 \\
	\cline{2-6}   
	& XCET-H & -- & -- & 0 & 0 \\ 
	\hline 
\end{tabular}
\end{table}


\section{TPC characterization}
\label{sec:tpc}

The large quantities of high-quality data collected by ProtoDUNE-SP enable many studies of the performance  of the TPC.  This section describes the offline data preparation and noise suppression, charge calibration, noise measurement, signal processing, event reconstruction, signal-to-noise performance, and a measurement of the electron lifetime.

\subsection{TPC data preparation and noise suppression}
\label{tpcnoisesupp}

The ProtoDUNE-SP detector is typically triggered at a rate of 1-40~Hz where each trigger record includes
synchronized contiguous samples from all TPC channels, typically with a length of 3~ms
corresponding to 6000 ticks (ADC samples).
Trigger records are processed independently of one another, beginning with data preparation which 
converts the ADC waveform (ADC count for each tick) for each channel to a charge waveform.
The data preparation comprises evaluation of pedestals, charge calibration, mitigation of
readout issues, tail removal and noise suppression.  These operations are necessary in order to optimize the performance of subsequent stages of event reconstruction.
The data preparation steps are described in detail in the following subsections.

\subsubsection{Pedestal evaluation}

Voltage offsets are introduced at the inputs to the amplifier and ADC for each channel
to keep the signals in the appropriate range for each of these devices.
These offsets and the gains of both devices vary from channel to channel and so there
are channel-to-channel variations in the ADC pedestal, i.e.~the mean ADC count that would
be observed in the absence of signal.
In addition, the pedestal is observed to have significant variation from one trigger record to another, presumably due to low-frequency (compared to the 3~ms readout window) noise pickup before the amplifier.
To cope with this, the pedestal is evaluated independently for each channel and each trigger record.

The pedestal is evaluated by histogramming the ADC count for all (typically 6000) ticks and
fitting the observed peak with a Gaussian whose mean is used as the pedestal.
The RMS of the fit Gaussian provides an initial estimate of the noise in the channel and
is typically around four to six ADC counts.
Due to the sticky-code issues described in section~\ref{sec:stickycodeidentification}, these ADC count
distributions are sometimes observed to have spikes at the offending sticky codes which can bias the pedestal estimate.
To reduce this bias, the peak bin is excluded from the fit if it holds more than 20\% of the samples.

\subsubsection{Initial charge waveforms}

Initial charge waveforms are obtained for each channel by subtracting the pedestal from
each of the ADC counts and multiplying this difference by the gain assigned to the channel.
These gains may be set to 1.0 to obtain a charge waveform in units
of ADC counts or they may be taken from a charge calibration.
The standard ProtoDUNE-SP reconstruction makes use of the calibration discussed in section~\ref{sec:gain_calib}.

\begin{figure}[!ht]
  \centering
  \begin{subfigure}[t]{0.49\textwidth}
    \centering
    \includegraphics[width=\textwidth]{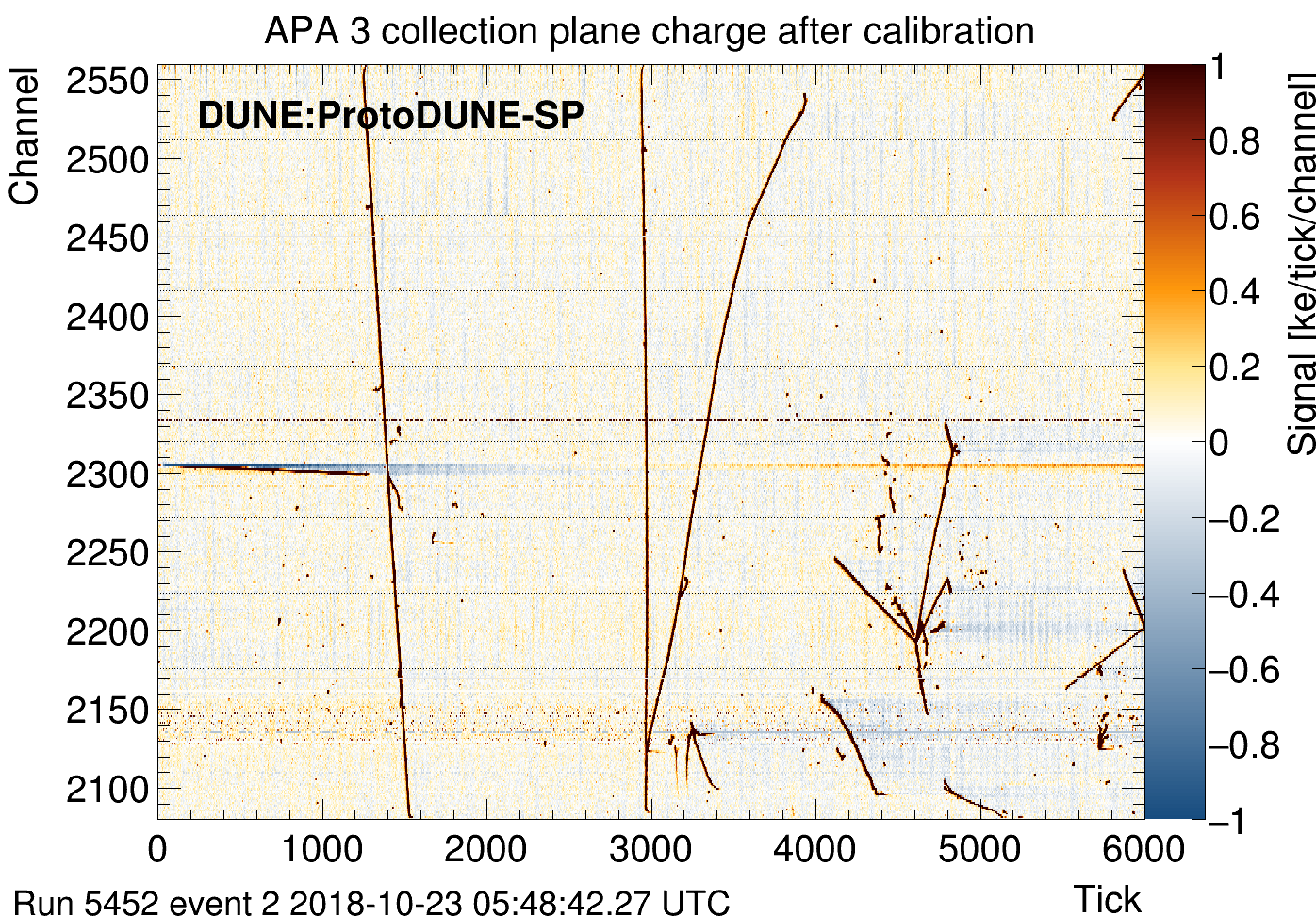}
    \caption{After pedestal subtraction and calibration.}
    \label{fig:evdcal}
  \end{subfigure}
  \hfill
  \begin{subfigure}[t]{0.49\textwidth}
    \centering
    \includegraphics[width=\textwidth]{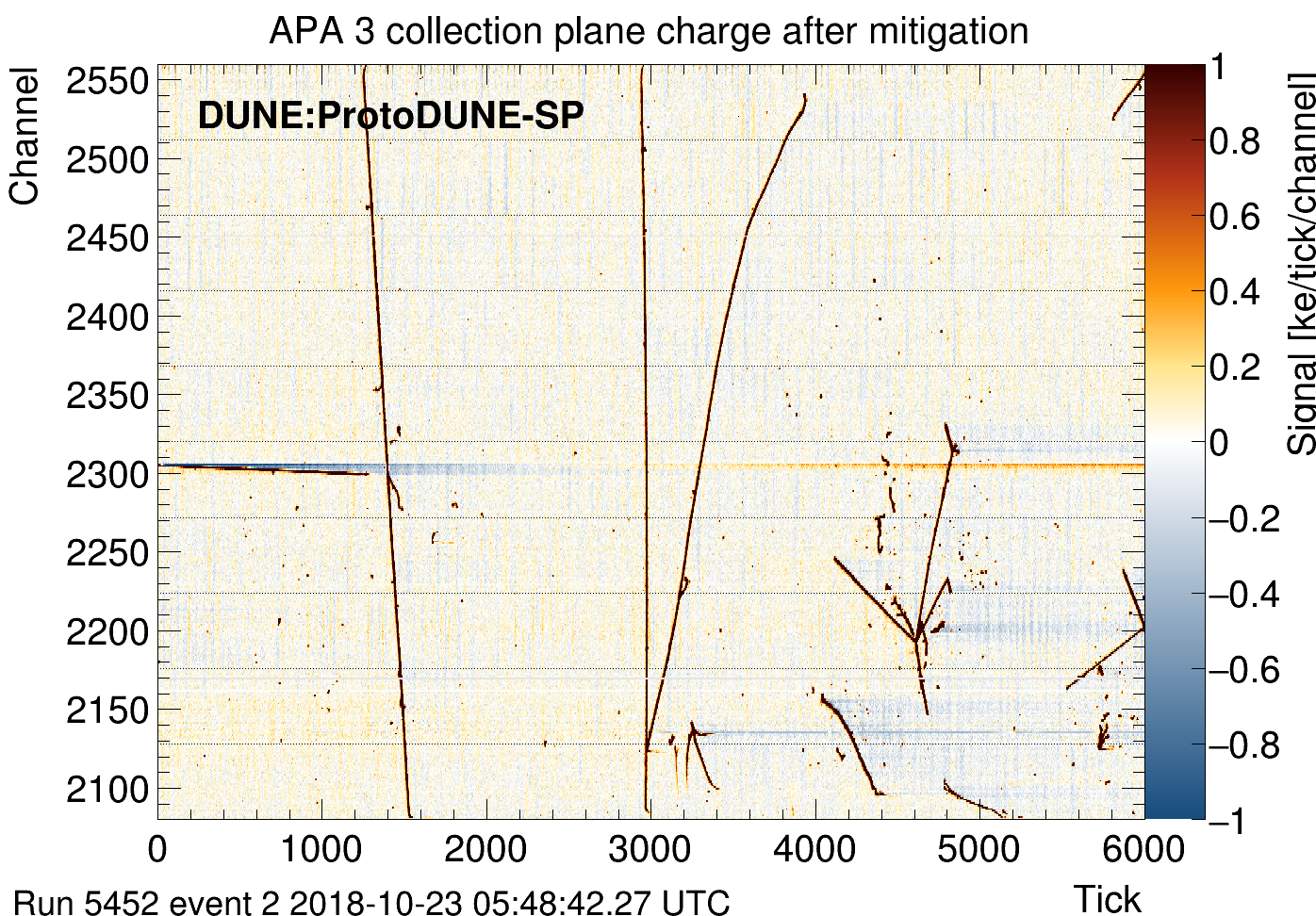}
    \caption{After ADC sticky code and timing mitigation.}
    \label{fig:evdmit}
  \end{subfigure}
  \hfill
  \vspace{3mm}
  \begin{subfigure}[t]{0.49\textwidth}
    \centering
    \includegraphics[width=\textwidth]{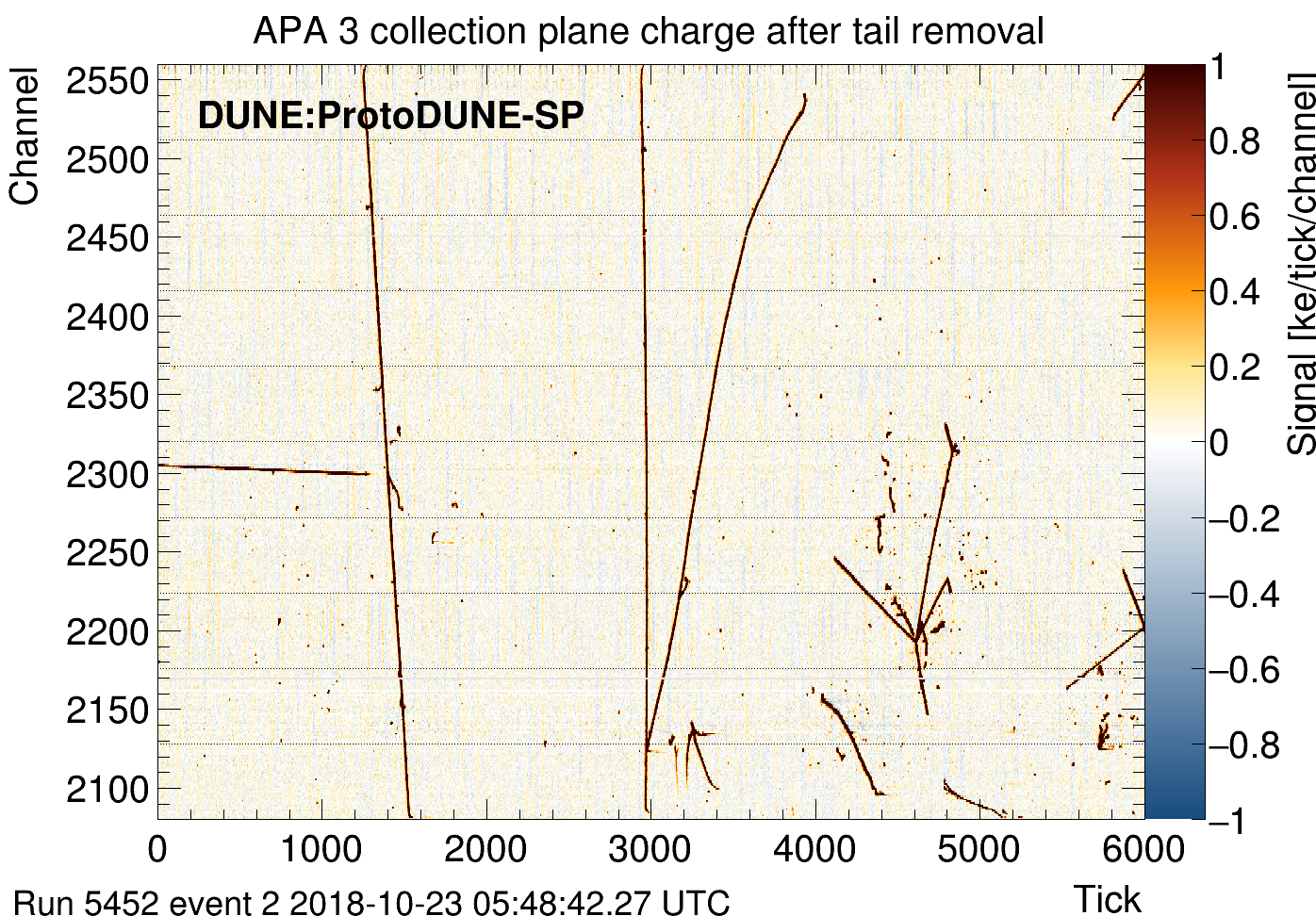}
    \caption{After tail removal.}
    \label{fig:evdtai}
  \end{subfigure}
  \hfill
  \begin{subfigure}[t]{0.49\textwidth}
    \centering
    \includegraphics[width=\textwidth]{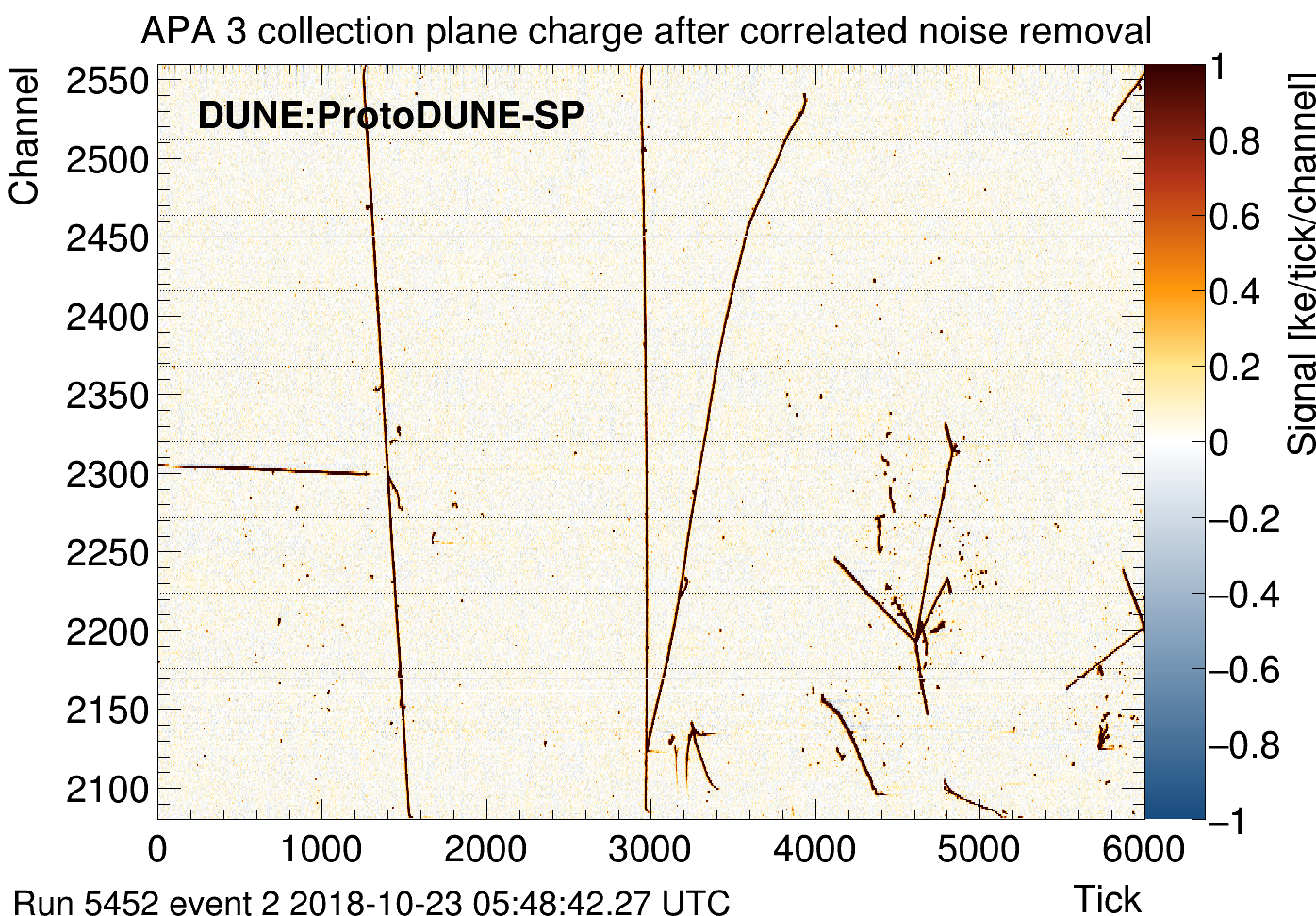}
    \caption{After correlated noise removal.}
    \label{fig:evdcnr}
  \end{subfigure}
  \hfill
  \caption{
  Example event displays for a collection plane showing background reduction in successive stages of data processing.
  The horizontal axis is the tick and vertical axis is the channel number.
  The color scale represents the charge for each channel averaged over five ticks
  with the range chosen to make the noise visible.
  Signals from charged tracks appear mostly in black and are off scale, well
  above the noise level.
  Horizontal dashed lines indicate the boundaries between the ten FEMBs used
  to read out the channels for this plane.
  The second from the bottom is FEMB~302 referenced in the text.
  }
  \label{fig:evdall}
\end{figure}

Figure~\ref{fig:evdcal} shows an example event display consisting of waveforms on wires in the collection plane of APA~3 shown side by side in a two-dimension color plot.  Pedestal subtraction and charge calibration have been applied.  APA~3 instruments the upstream drift volume on the side of the cathode on which the beam enters.

\subsubsection{Sticky code identification}
\label{sec:stickycodeidentification}

A few percent of ADC ASIC channels suffer from an issue known as ``sticky code,'' in which certain ADC values would be preferentially produced by the ADC independent of the input voltage causing the readout channel to appear to ``stick'' at a particular value. The flaw in this ADC design is a failure of transistor matching at the transition from digitizing the six most significant bits to the six least significant bits. The sticky codes therefore tend to prefer values of zero or 63 plus a multiple of 64, though other sticky codes have been observed in the data as well.  Sticky codes were observed in test-bench measurements in advance of installation, where the dynamic range of the ADC was tested with a calibrated source of charge. 

The pedestal histograms and a few waveforms for all channels were scanned by eye to obtain
an initial list of sticky codes
and this list is extended when other problematic channels are uncovered.
A total of 498 codes in 312 channels (of 15360) have been identified as sticky and are mitigated
as described in the following section.

Approximately 70 channels are flagged as bad due to very high fractions of sticky codes
or population of multiple widely-separated sticky values.
Another 35 are flagged as noisy due to serious but less-severe sticky-code issues.  The solidly unresponsive channels mentioned in section~\ref{det-coldelectronics} are also flagged as bad.  A total of 133 channels are flagged as bad or noisy.
The data for these channels are prepared like any others, but downstream
processing such as deconvolution
or track finding may choose to ignore these channels or treat them in a special manner.

\subsubsection{ADC code mitigation}

For channels that have known sticky codes, if the ADC value on a particular sample is at one of the sticky values, it is replaced with a value interpolated from the nearest-neighboring non-sticky codes.
If the two neighbors on either side exhibit a significant jump (20 ADC counts), the interpolation
uses a quadratic fit.
Otherwise, linear interpolation is used.

\begin{figure}[!ht]
\centering
\includegraphics[width=\textwidth]{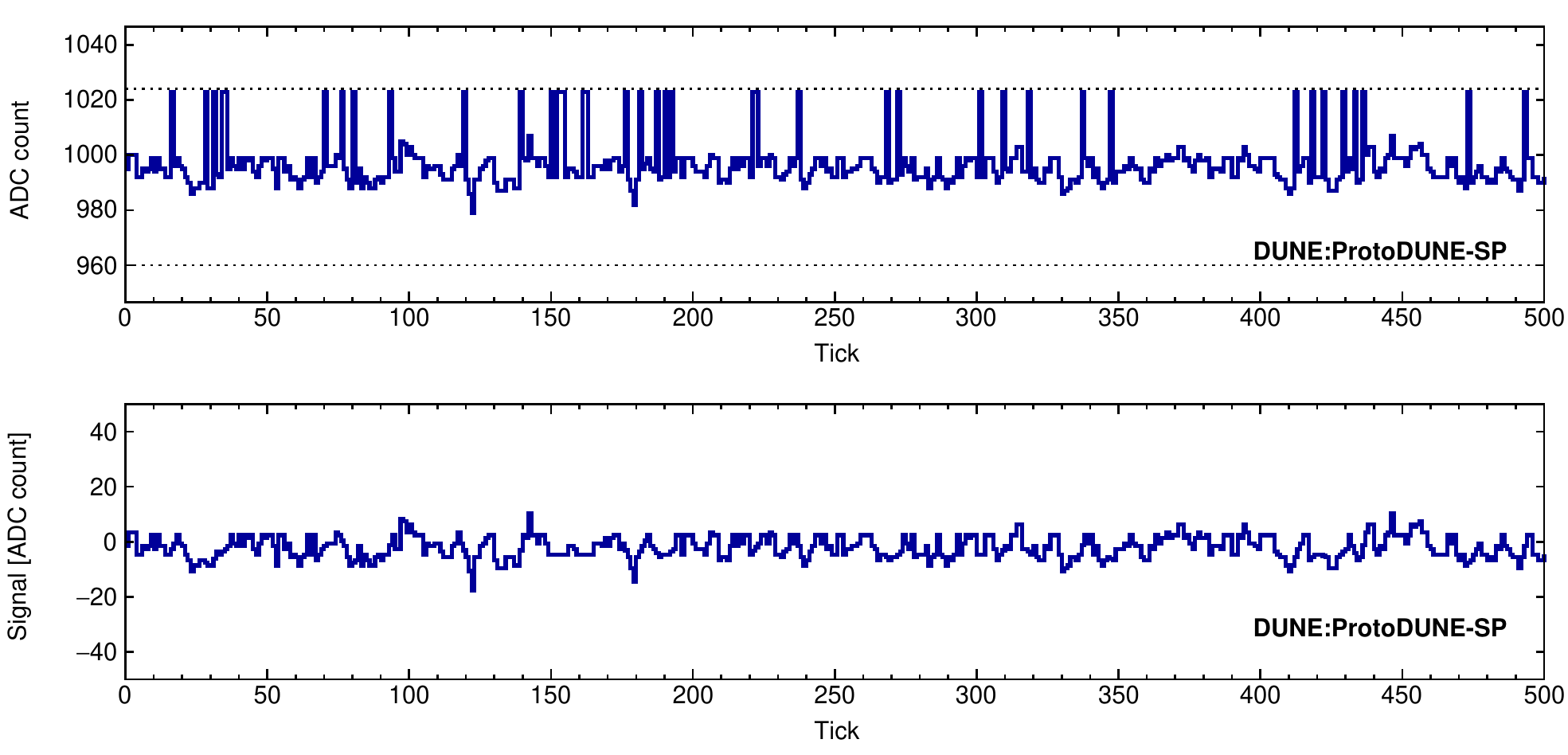}
\caption{Example of a raw ADC waveform with sticky codes (top) and the corresponding
         waveform after pedestal subtraction and ADC sticky code mitigation (bottom).
         The dashed lines in the top plot show the six-bit boundaries.}
\label{fig:sticky_bit}
\end{figure}

Figure~\ref{fig:sticky_bit} shows an example charge waveform before and after mitigation.
This is from FEMB~302 where sticky codes are particularly prevalent.
A sticky code has been identified and the waveform is significantly
improved after mitigation. Sticky codes very close to the pedestal are not flagged.

\subsubsection{Timing mitigation}
One of the 120 FEMBs (FEMB 302) does not receive the master timing
signal used to clock the ADCs.
The ADCs on that FEMB make use of a backup clock that resides on the FEMB.
Although the master and FEMB clocks both nominally run at 2~MHz, reconstructed signals show that the FEMB clock runs 0.07\% slower than the master clock.
The charge waveforms for FEMB 302 are corrected to match the sampling rate and offset
for the other channels.
The charge for each sample is replaced with a linear interpolation of the original charges
of two samples nearest in time.

Figure~\ref{fig:evdmit} shows an event display made from 
mitigated waveforms for the same data with the same
scale and binning as figure~\ref{fig:evdcal}.
Both sticky-code and timing mitigations are added.
The shift in the FEMB~302 timing and reduction in noise are discernible.
Channels flagged as bad or noisy are zeroed and appear white in the display.

\subsubsection{Tail removal}
In each TPC channel, the amplifier and ADC are AC-coupled using a high-pass
RC filter with a time constant of approximately $\tau_{\rm{RC}} = $1.1~ms (2200 ticks) for collection-plane channels
and 3.3~ms (6600 ticks) for induction-plane channels.
The typical signal from a charged track will be much faster, 10-20~$\mu$s (20-40 ticks)
and this AC coupling implies the observed signal will be followed by
a long tail of opposite sign whose area cancels that of the initial signal.
The tails are much smaller and are neglected in induction-plane
channels, where the signals are bipolar and thus integrate to zero.

The decay time is comparable to the mean time between cosmic-ray signals, about 1500 ticks,
and it is a significant fraction of the data readout time used to evaluate the pedestal, typically 6000 ticks.
Variations in cosmic arrival time and charge deposit per channel (in particular due to varying
angle of incidence) imply that all signals are superimposed on a fluctuating background of accumulated 
tails from preceding signals, many of which arrive before the readout window starts.
Tails from these fluctuations are clearly visible in collection-plane waveforms and event displays.
Large charge deposits in figure~\ref{fig:evdmit} are followed by blue regions indicating negative tails.
Some other regions are positive (orange) because the initial pedestal estimate is biased by the
negative tails.
The tails are removed in collection-plane channels using a time-domain correction, chosen because of the large fraction of channels that start each trigger record with a significant tail from charge that has arrived before the readout window starts.

Before tail removal, the (nominal) pedestal-subtracted value (ADC count or calibrated charge)
in sample $i$, $d_i$, is the sum of signal, $s_i$, and tail, $\xi_i$, contributions
and a pedestal offset, $p_{o}$:
\begin{equation}
\label{eqn:tailsum}
d_i = s_i + \xi_i + p_o
\end{equation}
The pedestal offset is not zero because the method used to evaluate the pedestal includes
contributions from tails.
Because the tails are exponential, the tail in any sample may be expressed in terms of the
signal and tail in the preceding sample:
\begin{equation}
\label{eqn:taili}
\xi_i = \beta \xi_{i-1} + \alpha s_{i-1}
\end{equation}
where $\beta = e^{-1/\tau_{\rm{RC}}}$ ($\tau_{\rm{RC}}$ is the time constant in ticks) and
$\alpha = 1 - 1/\beta$ obtained by requiring the integral of the tail cancel
that of the signal.
Equations~\ref{eqn:tailsum} and~\ref{eqn:taili} may be solved to obtain the
signal and tail from the input data:
eq.~\ref{eqn:tailsum} is used to obtain $s_i$ from $d_i$ and $\xi_i$ and
eq.~\ref{eqn:taili} to obtain $\xi_{i+1}$ from $s_i$ and $\xi_i$ for $i = 0, 1, 2, ...$.
The tail correction replaces the input data with the evaluated signal: $d_i \rightarrow s_i$.

In addition to $\alpha$, $\beta$, this solution depends on two parameters: the pedestal
offset and the tail in the first sample, $\xi_0$.
The latter is unknown because the data preceding the start of the triggered readout have not been recorded.
These parameters are obtained by identifying signal-free regions and choosing the values
that minimize the sum of $s_i^2$ over those regions.
This is done by iteratively applying the signal finder first to the input data and then to
each signal estimate.

\begin{figure}[!ht]
\centering
\includegraphics[width=0.49\textwidth]{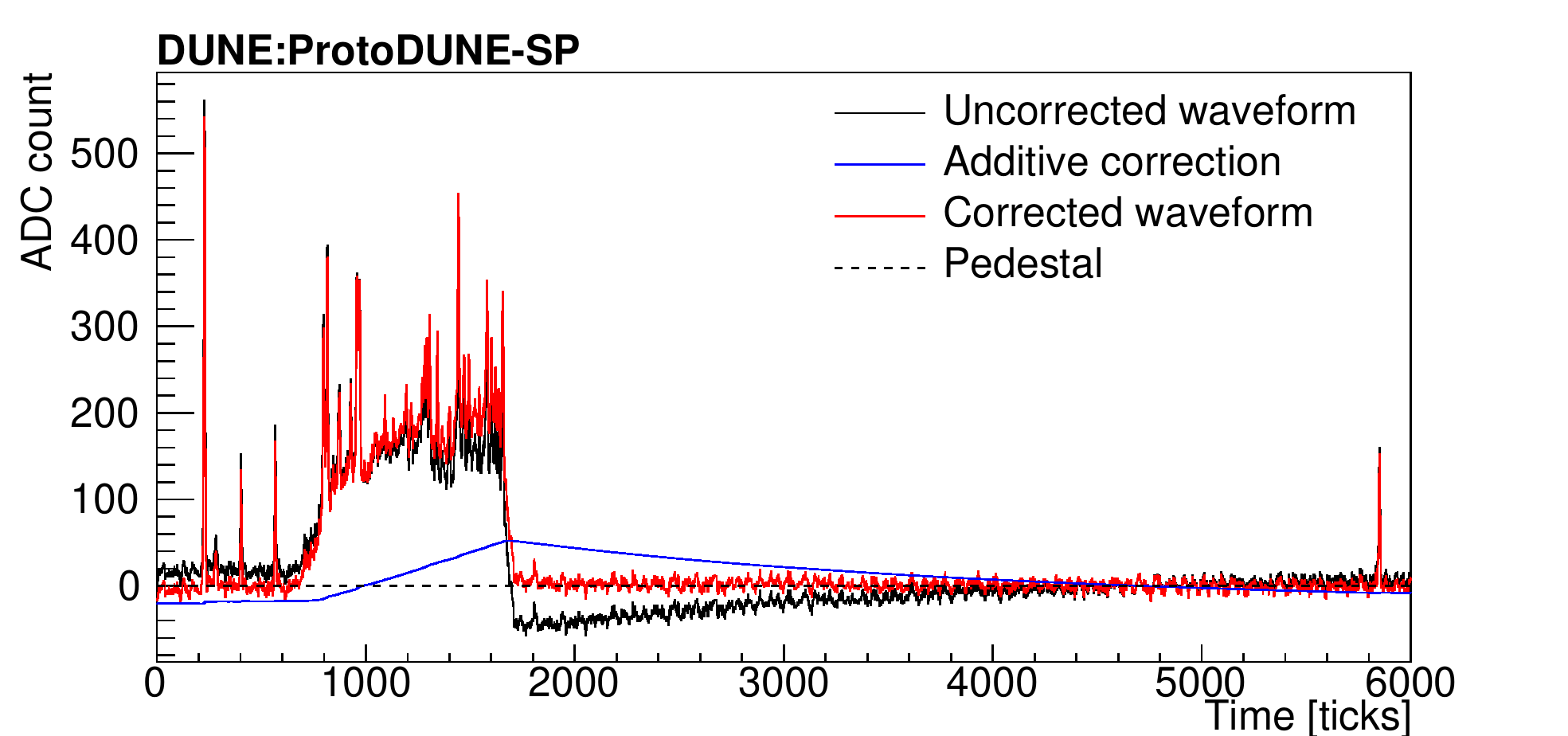}
\includegraphics[width=0.49\textwidth]{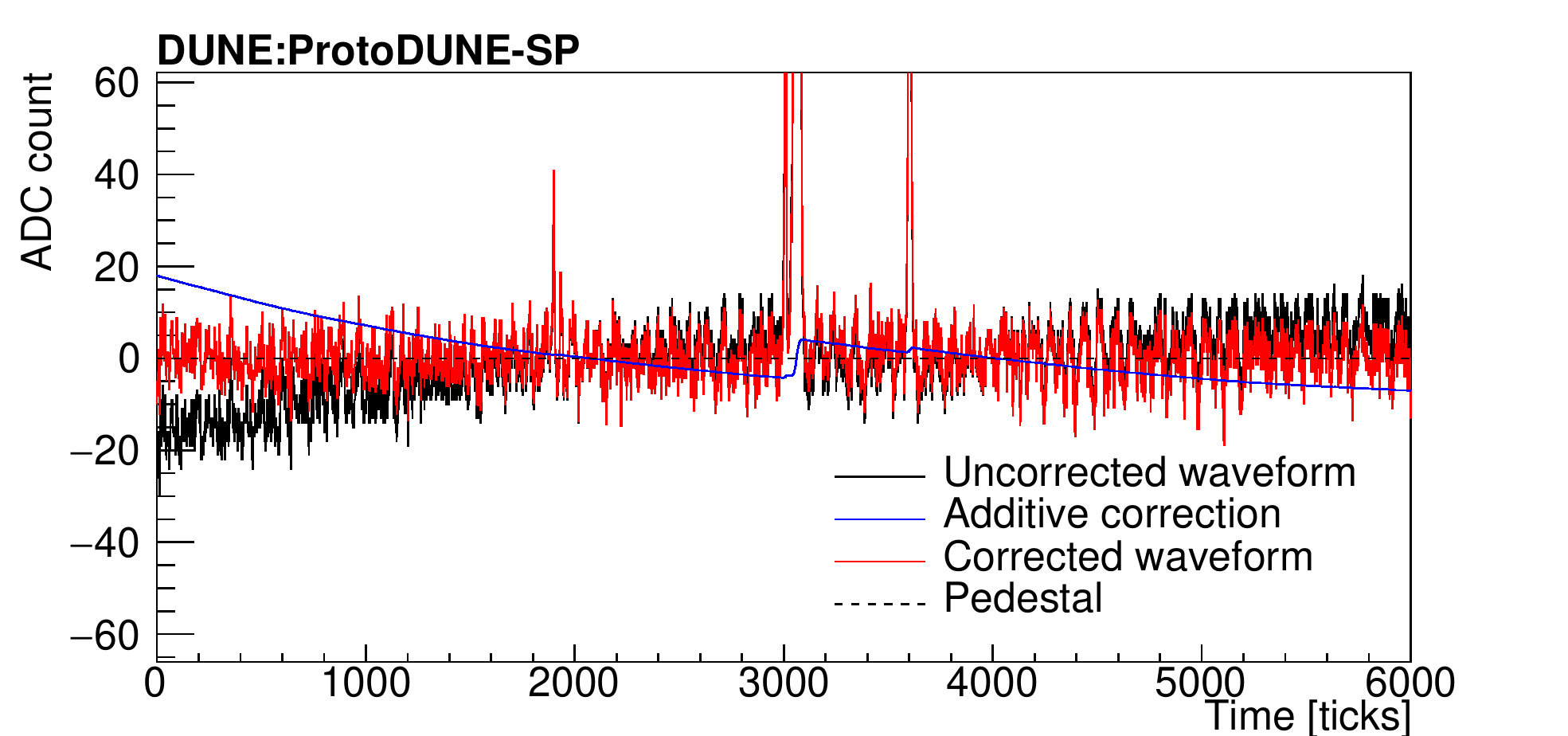}
\caption{
Examples of waveforms with tail removal.
The black curve shows the original data $\{d_i\}$ and the red curve shows the corrected data,
i.e.~the signal $\{s_i\}$.
The blue curve is the correction added to the data, i.e.~pedestal offset $p_o$ minus
estimated tail $\{\xi_i\}$.
On the left, a very long pulse produces a clearly visible tail.
On the right, there is clear evidence of a tail from charge that preceded the readout window
as well as tail from signal within the window.
}
\label{fig:undershoot1}
\end{figure}

Figure~\ref{fig:evdtai} shows an event display composed of waveforms after tail removal for the same data with the same
scale and binning as figure~\ref{fig:evdmit}.
Examples of two corrected waveforms from ProtoDUNE-SP data are shown in figure~\ref{fig:undershoot1}.


\subsubsection{Correlated noise removal}
One of the most significant sources of excess noise observed in the ProtoDUNE-SP detector
has a frequency distribution with a peak around 45~kHz. This noise
source is found to be highly correlated among a group of channels that share the
same low-voltage regulator in the same FEMB.  
Following Ref.~\cite{Acciarri:2017sde}, a mitigation method is developed by dividing
channels into groups. 
Each FEMB amplifies and digitizes 128 channels: 40 adjacent U-plane channels, 40 adjacent V-plane channels,
and 48 adjacent collection-plane channels.   The U-plane channels form a group, as do the V-plane channels and the collection-plane channels.  For each of the three groups, a correction waveform is constructed based on the median value of samples from the group
at every time tick and it is subtracted from each channel's waveform in that group.
However, if the majority of waveforms contain signals of ionization electrons, it is necessary
to protect this time region to avoid signal suppression. 
A region of interest (ROI) is defined as the ADC counts above an expected threshold as well
as 8 (20) ticks before (after).  
As an example, event displays consisting of waveforms before and after the correlated noise removal (CNR) are
shown in figure~\ref{fig:evdtai} and~\ref{fig:evdcnr}, respectively.  The correlated noise is visible as vertical bands in figure~\ref{fig:evdtai}, and it is suppressed in figure~\ref{fig:evdcnr}.  Some sources of correlated noise remain in some portions of the detector, specifically those for which the spatial correlation does not coincide with FEMB boundaries.




\subsection{Charge calibration}
\label{sec:gain_calib}

The ProtoDUNE-SP electronics provide the capability to inject a known
charge in short-duration ($<1\mu$s) pulses into each of the amplifiers connected to the TPC wires.
The level of that charge is controlled by a six-bit voltage digital-to-analog converter
(DAC) and is nearly linear with $Q = S Q_s$ where $S$ is the DAC setting (0, 1, ..., 63)
and the step charge $Q_s =$~3.43~fC~=~21.4~ke, which is comparable to the charge deposition of a minimum ionizing particle traveling parallel to the wire plane and perpendicular to that plane's wire direction.

A charge calibration is carried out so that ADC counts read out for
each channel may be converted to collected charge.
The calibration is expressed as a gain for each channel normalized such
that the product of the gain and the integral of the ADC signal over the
pulse in a collection channel gives the charge in the pulse,
i.e.~$Q = g A$.
The evaluation of charge for the bipolar TPC signals in induction channels
is more complicated but also proportional to the gain derived here.

Special runs were taken with injected voltage regularly alternating
between ground and the DAC level (one setting for each run)
producing charge pulses of alternating sign.
Fifty trigger records with typically 12 pulses of each sign are processed for
each channel at each DAC setting.
For each channel, the pedestal is evaluated for each event and a distribution
of approximately 600 pedestal-subtracted areas in units of (ADC count)-ticks is obtained for each charge sign.
The mean of these signal area measurements are plotted as a function of
DAC setting using data from many runs, and a line constrained to pass through
the origin is fit to DAC settings 1-7.
The step charge divided by the slope of this line provides the calibrated gain for each channel.

\begin{figure}
  \includegraphics[width=0.90\textwidth,angle=0,trim=0 0 0 0, clip]{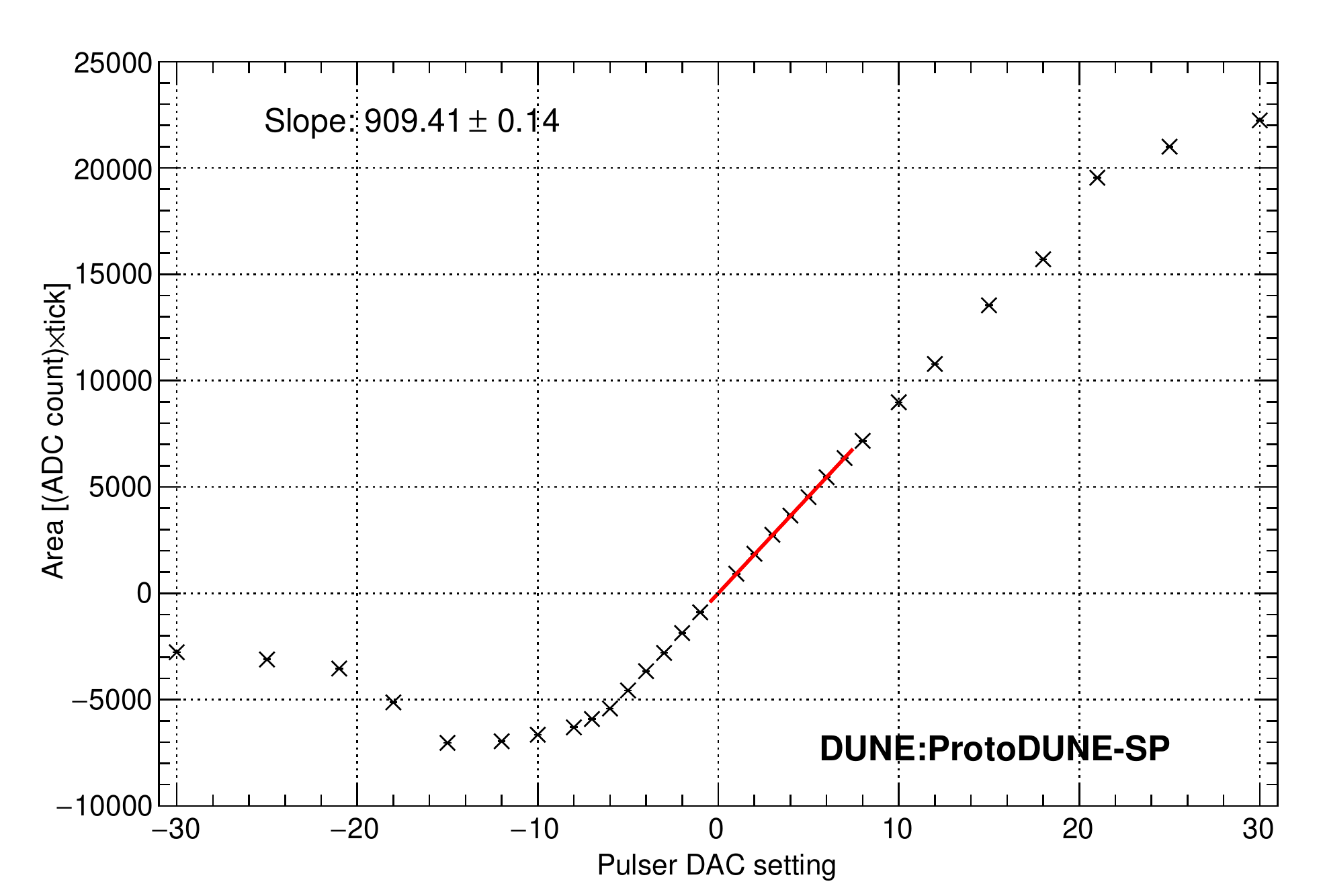}
\caption{
Measured pulse area vs.\ DAC setting for a typical collection channel.
The red line shows the fit used to extract the gain.
}
\label{fig:tpcarea}
\end{figure}

Figure~\ref{fig:tpcarea} shows the uncalibrated area vs.\ DAC setting and
the fit for a typical collection channel.
The response is fairly linear over the DAC setting range (-5,~20) with saturation setting in
outside this range.
Typical track charge deposits are one to four times the step charge
and this saturation is only an issue for very heavily ionizing tracks.
The gain for this channel is $g = (21.4$~ke)/(909.4~(ADC~count)-tick) =~23.5~e/((ADC~count)-tick).

\begin{figure}
  \includegraphics[width=0.90\textwidth,angle=0,trim=0 0 0 0, clip]{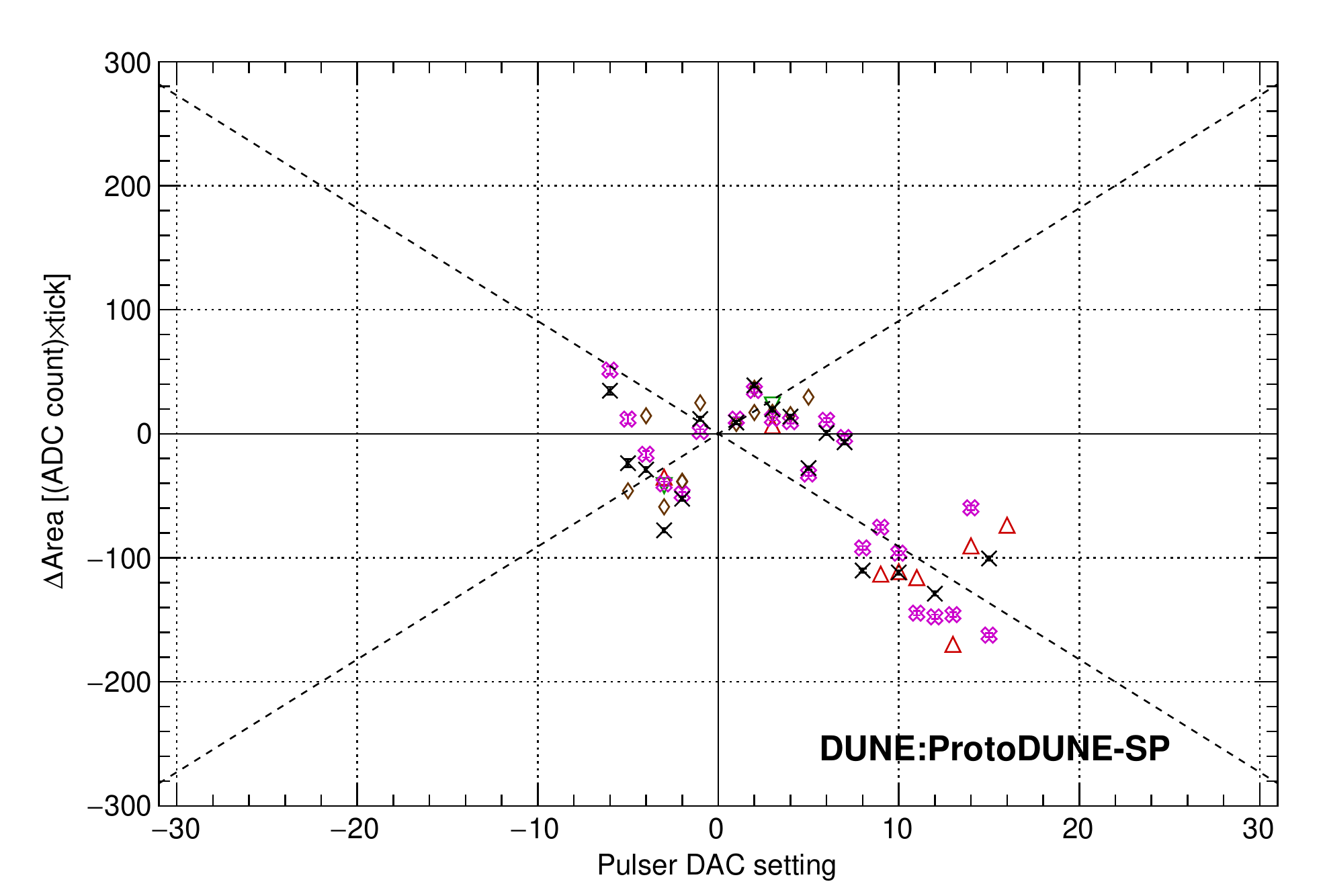}
\caption{
Measurement residuals (data - fit) for the same data as the preceding figure (black)
plus the same for data taken in the following months (colors).
The dashed lines indicate deviations of $\pm 1\%$.
}
\label{fig:tpcares}
\end{figure}

Figure~\ref{fig:tpcares} shows the residuals for the same data, i.e.~the measured area
minus that expected for the fitted gain $\Delta A = A - S Q_{s}/g$.
These results are typical---most of the measured areas for positive smaller
($S \le 7, Q \lesssim 160 $~ke) pulses are within 1\% of their fitted values.
The systematic shift for higher values is also typical and presumably reflects
non-linearity in the DAC.

\begin{figure}
\centering
  \includegraphics[width=0.90\textwidth,angle=0,trim=0 0 0 0, clip]{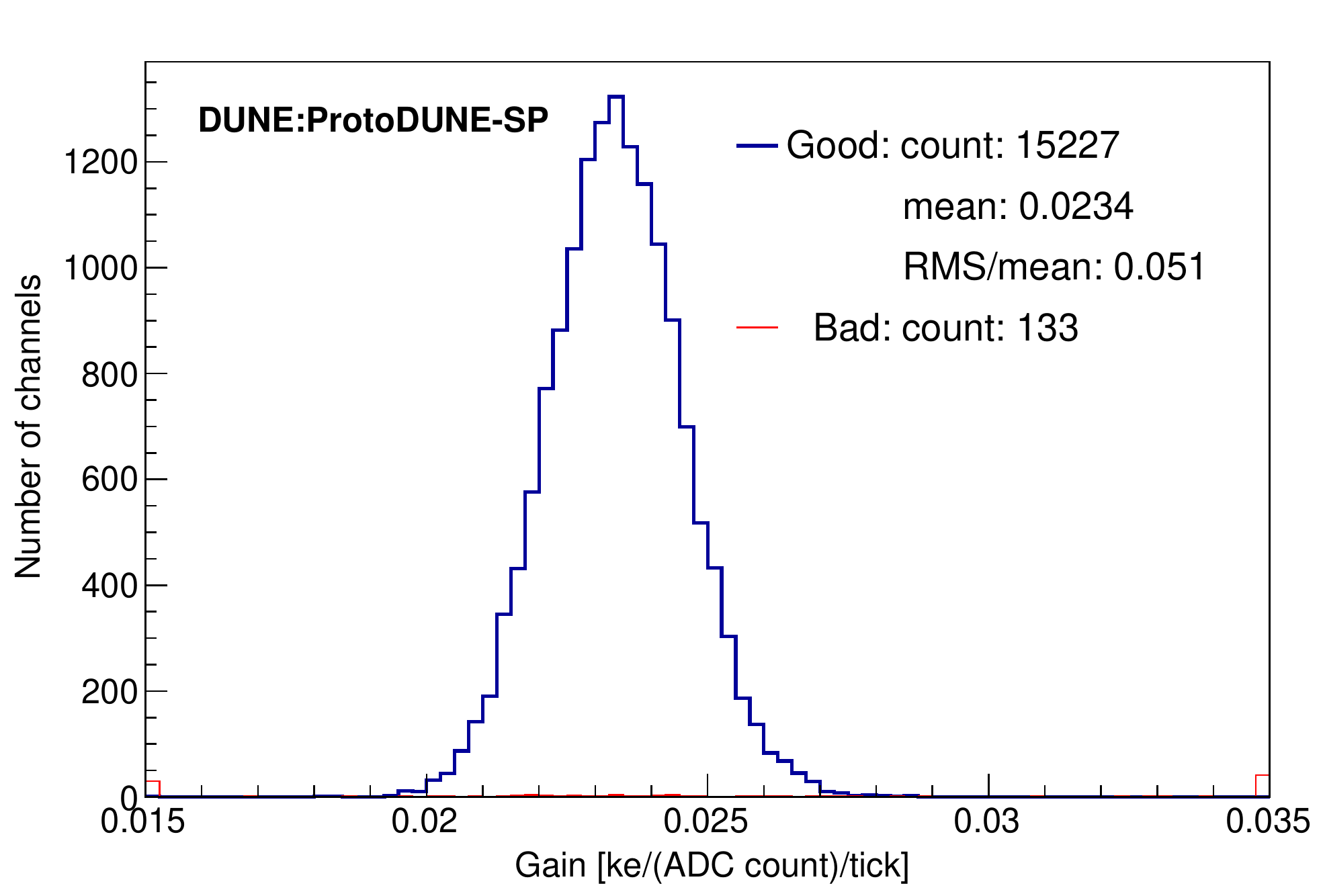}
\caption{
Distribution of fitted gains for good (blue) and bad/noisy (red) channels.
The legend indicates the number of channels in each category and gives the mean (23.4~e/(ADC~count)/tick)) and RMS/mean (5.1\%) for the good channels.
}
\label{fig:tpcgains}
\end{figure}

All 15,360 ProtoDUNE channels were calibrated in this manner, and those gains are applied
early (before the mitigation and noise removal) in the typical processing of data from the detector.
Figure~\ref{fig:tpcgains} shows the distribution of these gains for all channels.
Channels flagged as bad or especially noisy in an independent hand scan are shown
separately.
The gains for the remaining good channels are contained in a narrow peak with an
RMS of 5.1\% reflecting channel-to-channel response variation in the ADCs
and gain and shaping time variations in the amplifiers.

\subsection{TPC noise level}
\label{noiselevel}

One very important goal for ProtoDUNE-SP is to demonstrate that noise levels
are well below signals from charged tracks; this is found to be the case
for nearly all of the channels in the detector.
The noise is evaluated both for single ADC samples (\emph{sample noise})
and for a contiguous range of 50 samples (\emph{integrated noise}).
The latter range is chosen to be sufficient to obtain the area of the
signal from a charged track in the detector traveling in the $yz$ plane.  Tracks with other angles with respect to the electric field will leave longer pulses on the sense wires.

The noise is measured after initial data preparation.
As discussed in section~\ref{tpcnoisesupp}, the pedestal is evaluated for each trigger record,
and the charge calibration is applied to the ADC count minus pedestal
to obtain the initial charge measurement for each channel.
Sticky codes are mitigated and the AC-coupling tails are removed
in the collection channels.
The noise is evaluated both at this point and after applying the CNR.  The noise is expressed in units of collected electrons. 
To set the scale, the bulk of the observed distribution of signal
areas in each of the channels starts (as expected) at about 30~ke.

The high rate of cosmic-ray signals---an average of one every 1500 ticks (0.75~ms) for the TPC-side collection-plane channels---complicates the measurement of the noise.
To avoid contamination from these and radioactive (e.g.\ $^{39}$Ar) signals,
a signal finder is applied and the noise is defined to be the RMS ADC value
outside the signal regions.
For the integrated noise measurement, integration regions start every 50 ticks
(i.e. at ticks 0, 50, 100, ...) and regions are discarded if they have any overlap
with signal regions.

The signal finder used for this study makes use of a variable sample threshold and retains a
region of (-30, +50) ticks around any tick with signal magnitude above
that threshold.
The threshold is evaluated independently for each channel in every trigger record.
The threshold starts at 300~e and, if it is below five times the sample noise,
is increased until it reaches that level.
This allows efficient removal of signals in quiet channels while retaining the noise
in those that are noisier.

Visual inspection of raw waveforms were performed to identify bad channels in the detector, mostly those with
no signal or exceptionally high noise typically from sticky ADC codes.
The number of such channels is 90, i.e.\ 0.6\% of the channels in the detector.
These are excluded from the noise summary plots below.

\begin{figure}
  \includegraphics[width=0.99\textwidth]{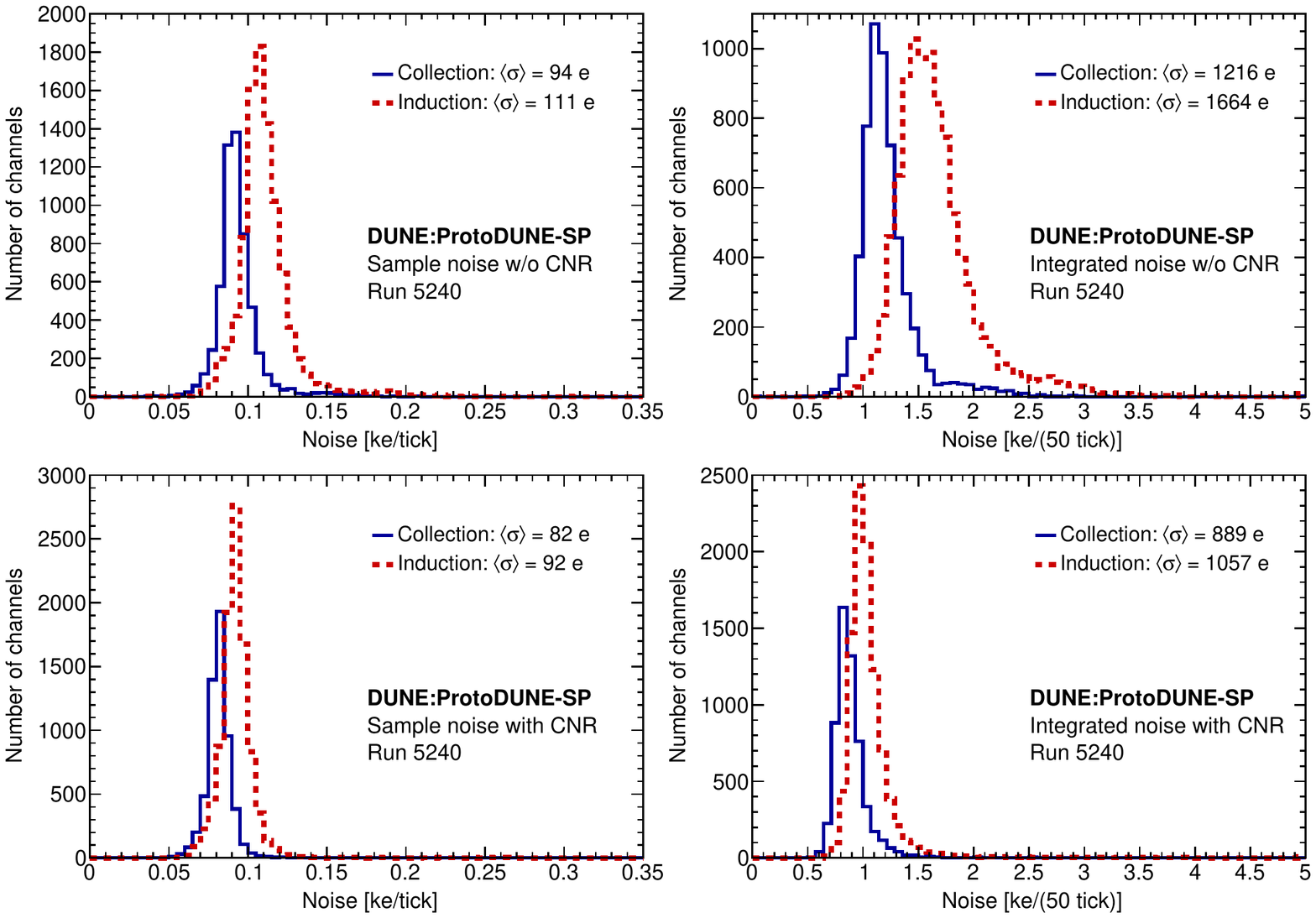}
\caption{
TPC sample (left) and 50-sample integrated (right) noise distributions before (top) and after (bottom)
correlated noise removal.
Each plot has one entry for each channel excluding bad channels with overflows shown
in the last bin.
Collection and induction channels are shown separately.
}
\label{fig:tpcnoise}
\end{figure}

Figure~\ref{fig:tpcnoise} shows the distributions of sample and integrated noise levels
before and after correlated noise removal
for trigger records 1-1000 of run 5240 taken October~12, 2018.
The collection-plane and induction-plane channels are shown separately and, as expected, the
noise levels are higher for the induction-plane channels as the wires are longer.
For the collection channels, the sample noise is around 100\,e before correlated noise removal
falling to 80\,e after the channel correlations are removed.
The corresponding values for the integrated noise are 1200\,e and 900\,e.

For charge deposits much faster than the nominal 2~$\mu$s shaping time of the amplifier, the area $A$ of the resulting signal pulse is proportional to the height $h$ and shaping time $\tau$: $A = K h \tau$.
The shape is well understood~\cite{Acciarri:2017sde} and has been verified with fits of the ProtoDUNE-SP pulser signals.
Numerical integration gives $K = 1.269 /$tick~$= 2.538 /\mu$s.
For such fast signals, the charge may be deduced directly from the pulse height and its standard deviation is called
the ENC (equivalent noise charge)~\cite{Acciarri:2017sde}.
The ProtoDUNE-SP signals are slower than this but the ENC is a standard metric and is presented here to allow comparison with results from other detectors.

The ratio of ENC to sample noise defined here is $A/h = K\tau$.
The actual shaping time varies from channel to channel but has central value around 2.2~$\mu$s
which gives a ratio of ENC to sample noise of 5.58.
With this factor and the above values for the sampling noise, the mean ENC for the
collection channels is 530\,e before correlated noise removal and 430\,e after.
The corresponding numbers for the induction channels are 620\,e and 500\,e.  These noise values are similar to the 500-600\,e values obtained from bench measurements with a prototype FEMB at LN2 temperature~\cite{Abi:2017aow}.


\begin{figure}
  \includegraphics[width=0.99\textwidth,angle=0,trim=0 0 0 0, clip]{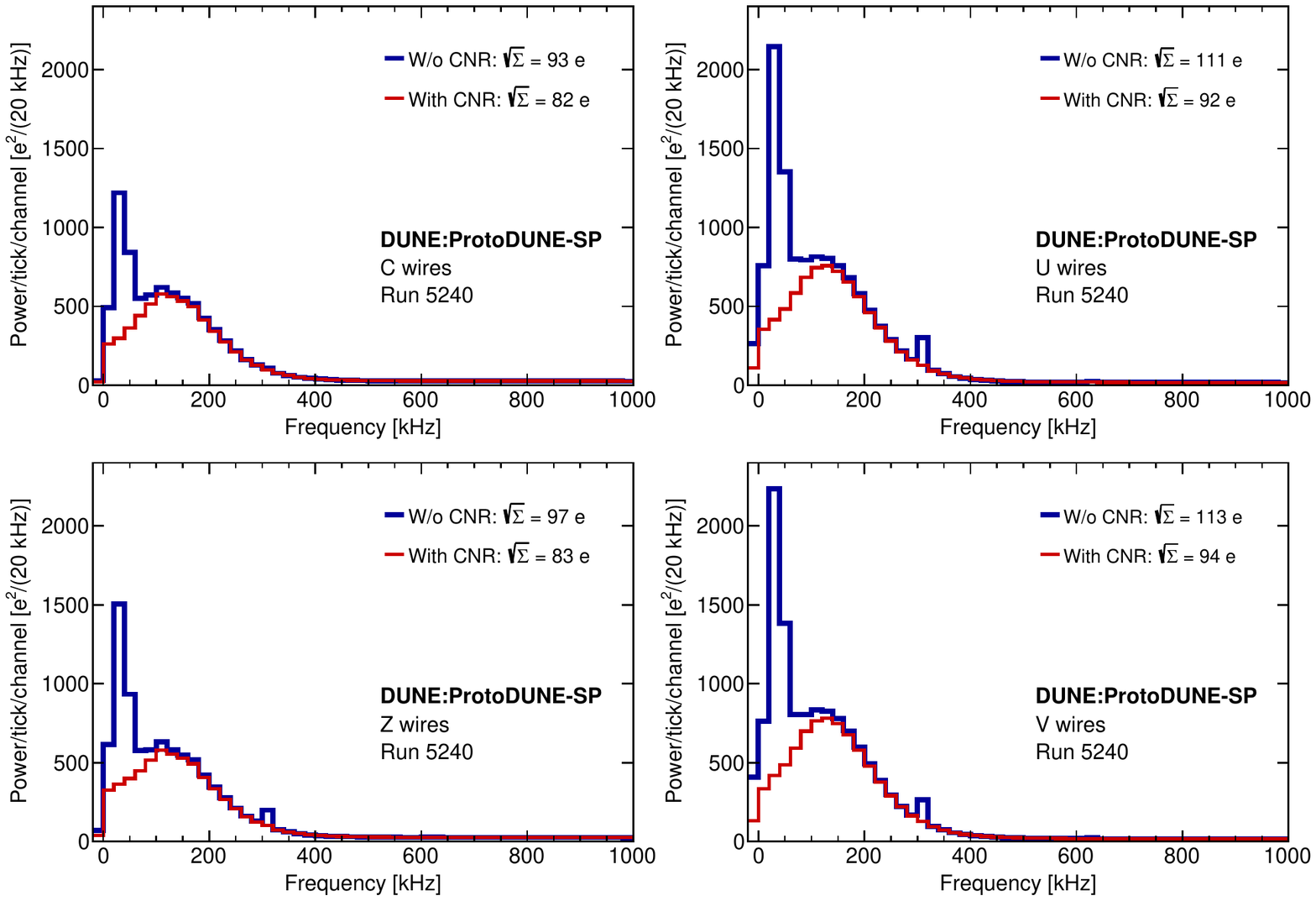}
\caption{
Noise frequency power spectra before and after CNR.
Each plot is evaluated from 1000-sample blocks in non-signal regions.
Upper and lower left respectively show the cryostat- and TPC-side collection planes.
The induction planes are on the right: top is U and and bottom is V.
The first bin in each plot includes only the zero frequency component.
The others are sums over 20~kHz bins.
}
\label{fig:dftall}
\end{figure}

Figure~\ref{fig:dftall} shows the noise frequency power spectra for data collected at the same time as those  used for figure~\ref{fig:tpcnoise}.
For each channel, a signal finder with a dynamic threshold of five times the sample RMS in the non-signal region is used to identify signal samples.
A discrete Fourier transform is performed on non-overlapping blocks of 1000 contiguous samples selected from the non-signal regions.
The power spectra for good channels are averaged separately for each of the four wire types: TPC-side collection,
cryostat-sided collection and the two induction orientations, U and V.
These are normalized so that the sum over power terms or histogram entries is equal to the RMS charge per sample, i.e.\ the sample noise shown on the left side of figure~\ref{fig:tpcnoise}.

As expected for effective removal of signals from the TPC, the power distributions are very similar for the TPC-side and cryostat-side collection channels.
The induction wire distributions have similar shapes.
The small spike around 300~kHz is due to pickup noise in one of the APAs.
The CNR effectively suppresses both that and the excess noise below 100 kHz.

\subsection{Signal processing} 
\label{signalproc}

The recorded waveform on each TPC readout channel is a linear transformation of the current on the connected wire as a function of time.  This transformation includes the effect of induced currents due to drifting and collecting charge, as well as the response of the front-end electronics.  The goal of the signal-processing stage of the offline data processing chain is to produce distributions of charge arrival times and positions given the input waveforms.  These charge arrival distributions are used in subsequent reconstruction steps, such as hit finding.  Because the response is linear in the arriving charge distribution, a deconvolution technique forms the core of the signal processing.

A charge moving in the vicinity of an
electrode can induce electric current. The Shockley-Ramo theorem
~\cite{Ramo} states that the instantaneous electric
current $i$ on a particular electrode (wire) which is held at constant voltage, is given by
\begin{equation}\label{eq:ramo_theorem}
    i = e\nabla\phi \cdot \vec{v}_e,
\end{equation}
where $e$ is the charge in motion, and $\vec{v}_e$ is the
charge velocity at a given location. The so-called weighting potential $\phi$
of a selected electrode at a given location is determined by virtually
removing the charge and setting the potential of the selected
electrode to unity while grounding all other conductors.

\begin{figure}[!ht]
     \centering
     \begin{subfigure}[b]{0.49\textwidth}
         \centering
         \includegraphics[width=\textwidth]{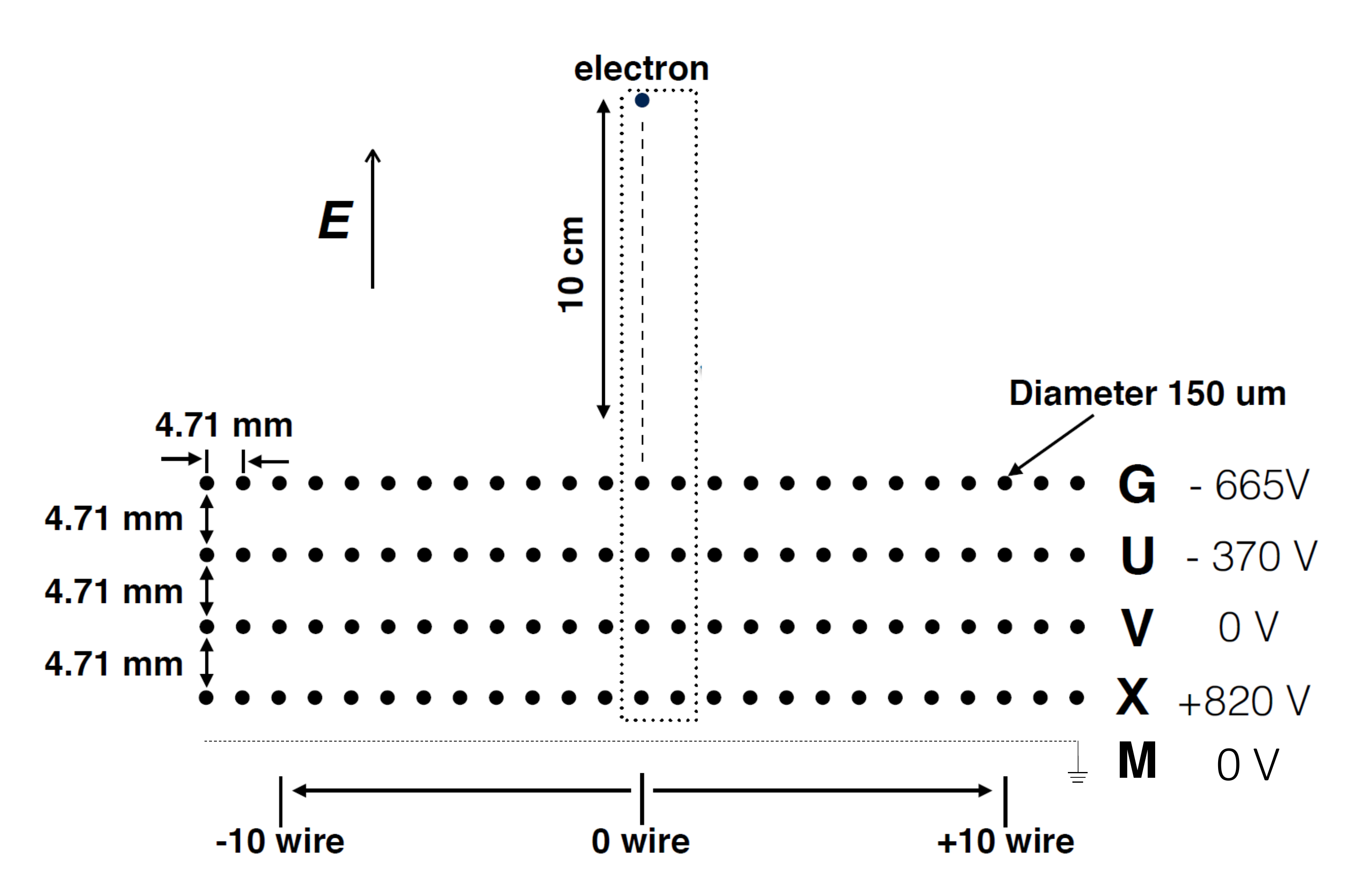}
         \caption{}
         \label{fig:dune_response_scheme}
     \end{subfigure}
     \hfill
     \begin{subfigure}[b]{0.49\textwidth}
         \centering
         \includegraphics[width=\textwidth,  trim={0cm 8cm 3cm 4.5cm},clip]{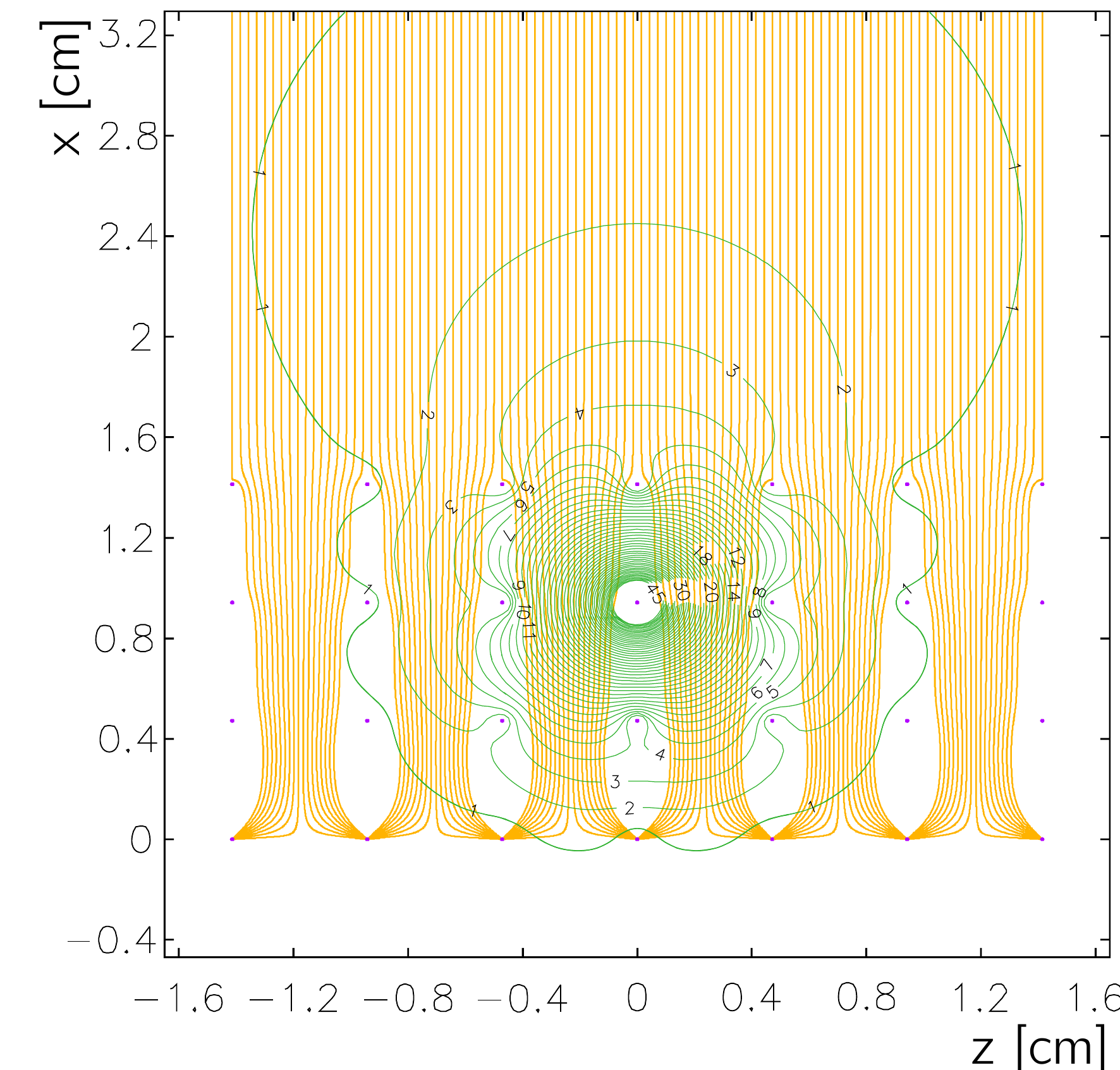}
         \caption{}
         \label{fig:garfield_path}
     \end{subfigure}
        \caption{(a) Illustration of the 2D ProtoDUNE-SP TPC scheme for the Garfield simulation, where $\pm$10 wires (large black dots) are considered for each wire plane and the electrons drift simulation starts 10~cm away from the grid plane. The inset shows the spacing of the starting points of the simulated electrons;
        (b) Garfield simulation of electron drift paths (yellow lines) in a 2D ProtoDUNE-SP TPC scheme and the equal weighting potential lines (green) for a given wire in the first induction plane, where the latter is shown in percentage from 1\% to 45\%. A long-range induction effect is noticed as the weighting potential has significant strength several wire spacings away from any particular wire.}
        \label{fig:garfield_summary}
\end{figure}

\begin{figure}[!ht]
     \centering
     \begin{subfigure}[b]{0.65\textwidth}
         \centering
         \includegraphics[width=\textwidth]{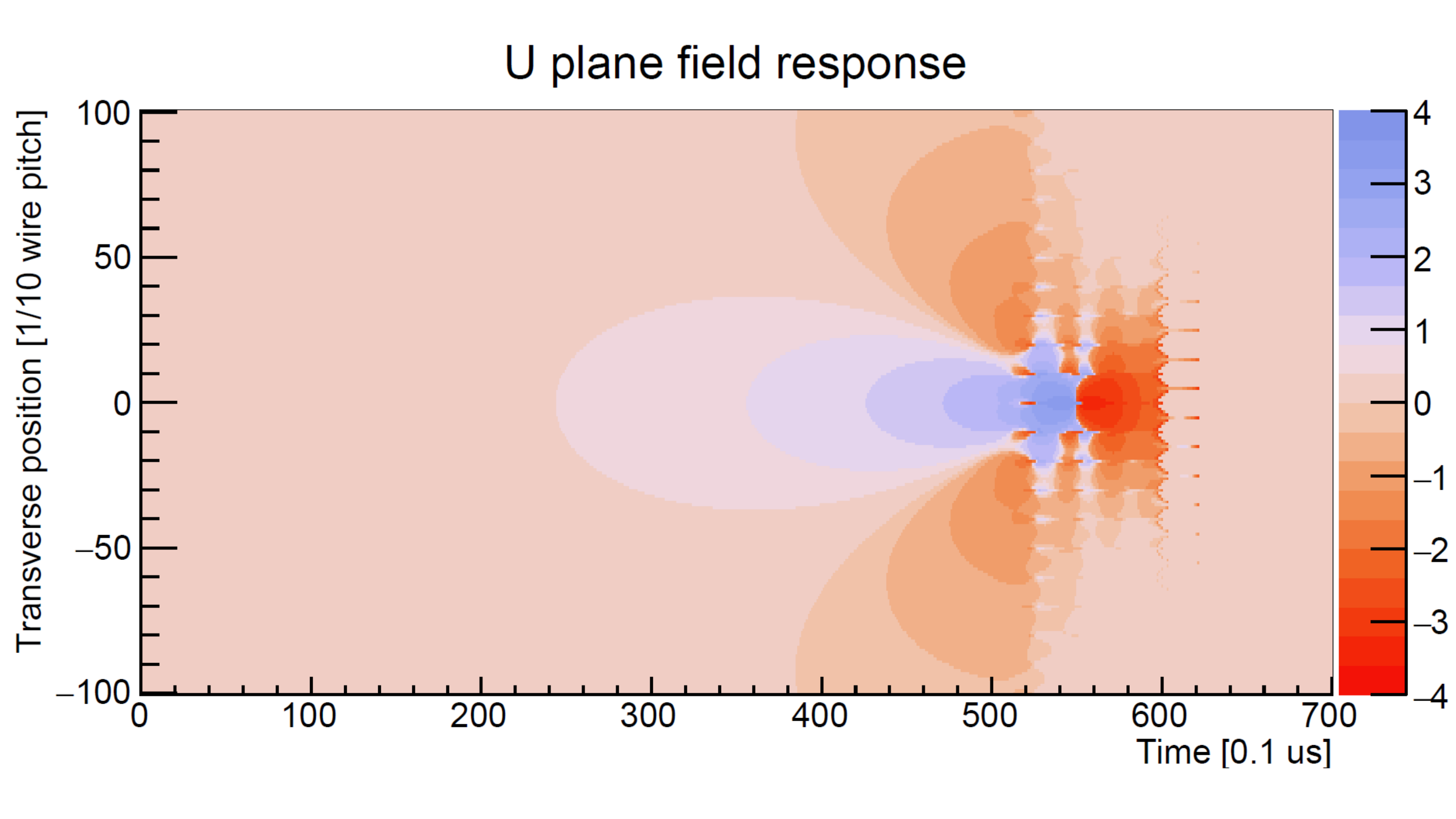}
     \end{subfigure}
     \hfill
     \begin{subfigure}[b]{0.65\textwidth}
         \centering
         \includegraphics[width=\textwidth]{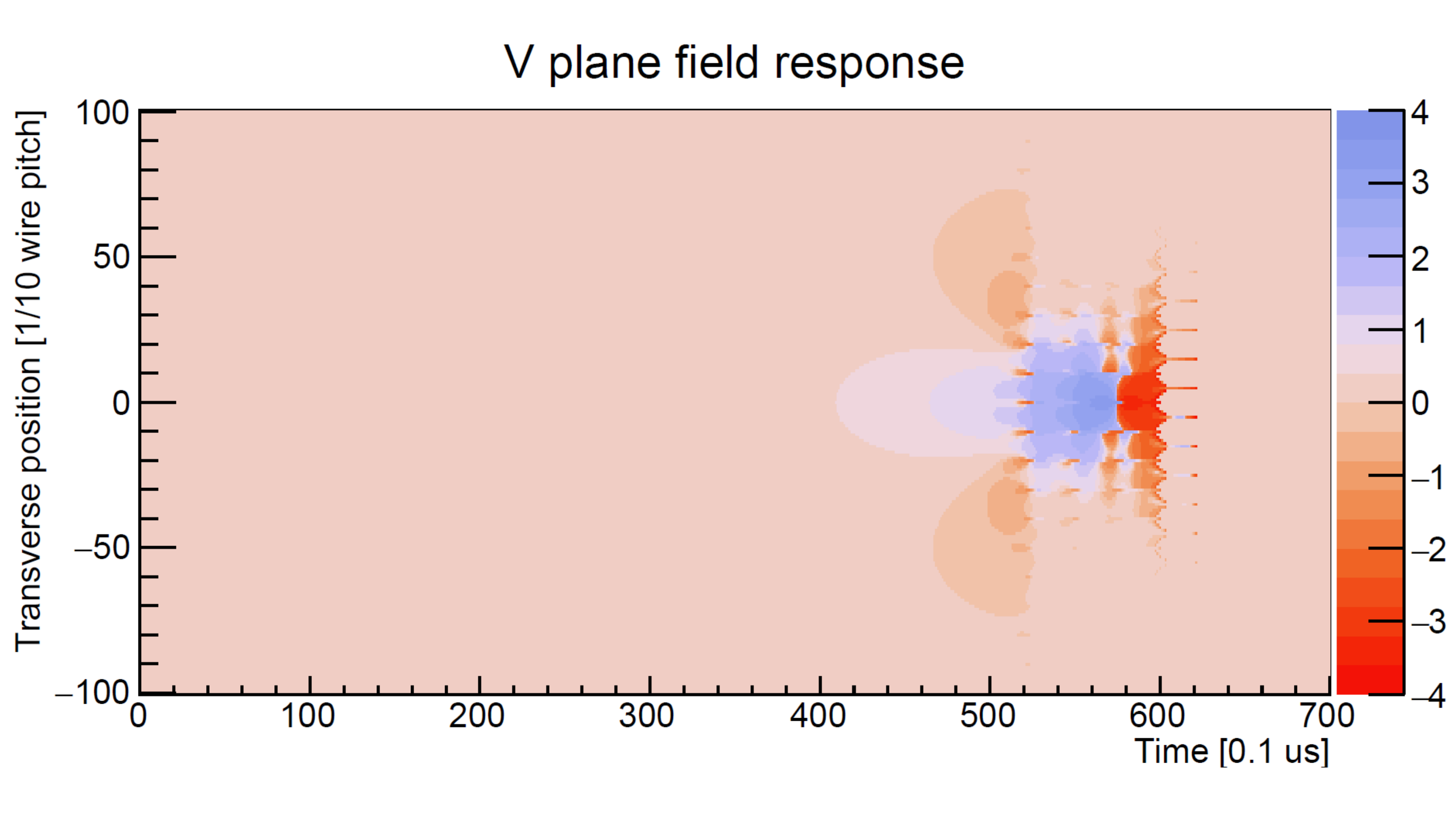}
     \end{subfigure}
     \begin{subfigure}[b]{0.65\textwidth}
         \centering
         \includegraphics[width=\textwidth]{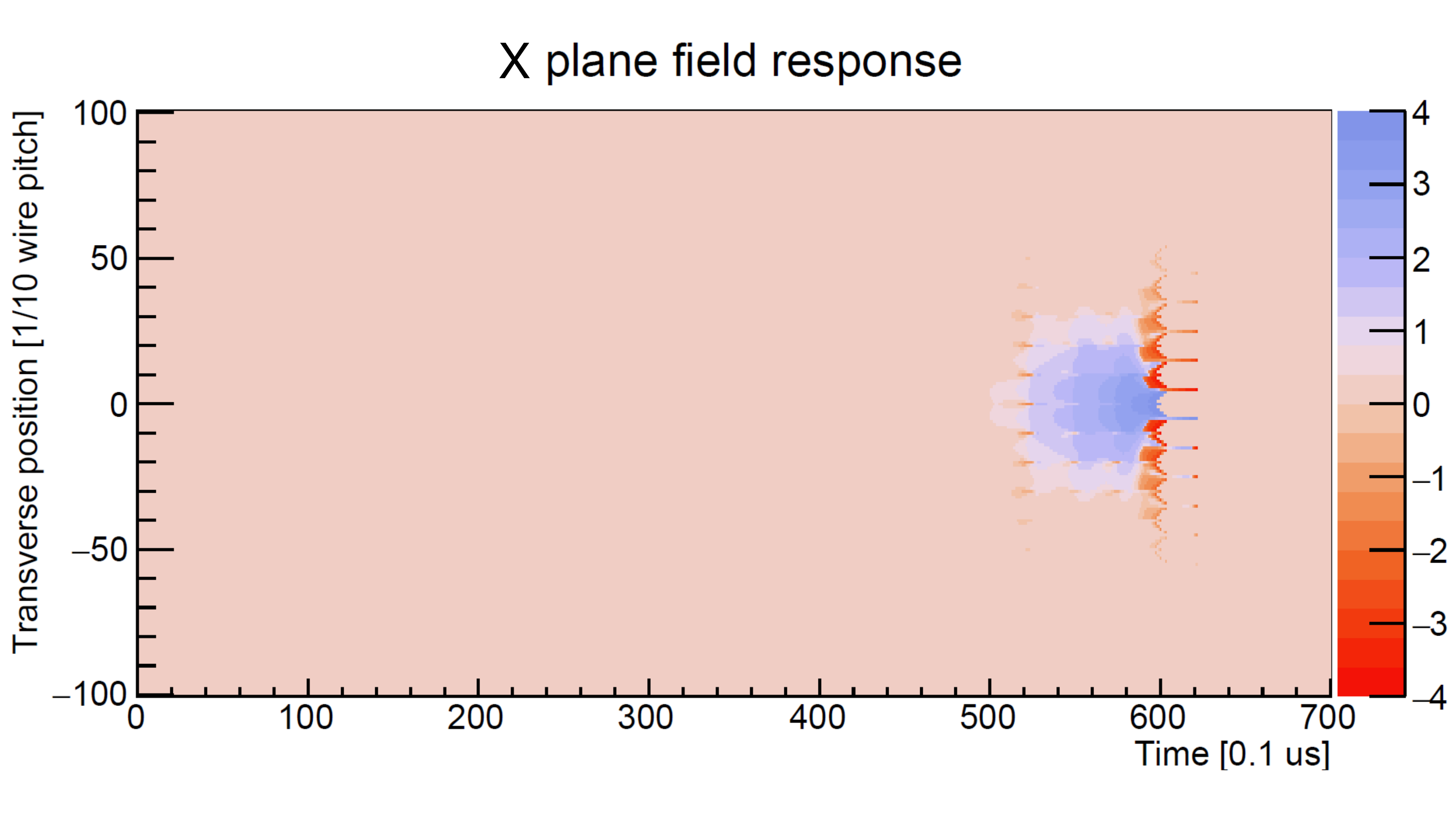}
     \end{subfigure}
        \caption{The overall response function convolved with electronics response.
        The color density is shown in a modified, sign-dependent base 10 log scale, as described in the text.}
        \label{fig:dune_response}
\end{figure}

The field response is defined to be the induced current on different wires due to a moving point charge. The field response is an essential input to the signal processing
procedure as will be discussed below. For ProtoDUNE-SP, the field response is calculated with Garfield~\cite{garfield},
a TPC drift simulation code, in a 2D scheme as illustrated in
figure~\ref{fig:dune_response_scheme}.
During the field response simulation, a point charge is positioned at different positions in a horizontal plane 10 cm away 
from the grid plane and the drift path is recorded from the simulation as shown
in figure~\ref{fig:garfield_path}. The electron drift velocity can be determined from
the electric field~\cite{Li:2015rqa, ref:lar_property}, while
the precomputed weighting potential for a U-plane wire is also shown in
figure~\ref{fig:garfield_path}. With the drift path and the weighting potential,
the field response of the point charge on the sense wire can be calculated according to
eq.~(\ref{eq:ramo_theorem}). This procedure is repeated for a series of point charges
that spans a region of 21 wires with the wire of interest at the center.
In order to sample the rapidly-changing field response functions adequately, point charges
are simulated drifting in from positions on a grid with a spacing of one tenth of the wire pitch.
After convolving the electronics response, the total response as
a function of time and wire pitch is presented in figure~\ref{fig:dune_response}, where
a ``Log10'' color scale is defined for the sake of visibility:
\begin{equation}
    i~\text{in ``Log10''} = 
    \begin{cases}
      \mathrm{log_{10}}(i\cdot10^5), & \mathrm{if}~i > 1\times10^{-5},\\
      0, & \mathrm{if}~-1\times10^{-5} \leq i \leq 1\times10^{-5},\\
      \mathrm{-log_{10}} (-1\cdot i\cdot10^5), & \mathrm{if}~ i<-1\times10^{-5}.\\
    \end{cases}
\end{equation}

As shown in figure~\ref{fig:garfield_path},  the weighting potential of the
first induction (U) plane is significantly different from zero over a region of a few wires even with the presence of the
grid plane. As a result, the current induced on a sense wire contains contributions not only
from charges passing between the wire and its immediate neighbors, but also from moving charges that are farther away. A region that is 10~cm in front of the grid plane and
$\pm$10 wires around the wire of interest in the simulation is sufficient to envelope
the field response. 

In order to deconvolve the ionization electron distribution from the measured signal,
it is natural to mathematically describe this long-range effect as follows:
\begin{equation}\label{eq:decon_2d_1}
  \begin{aligned}
    M(w_{i1}, t_{j1}) & = \int_{w_{i2}}\int_{t_{j2}}\int_{t_{j3}} \
    Q(w_{i2}, t_{j2})\cdot F(w_{i2} - w_{i1}, t_{j3} - t_{j2})\
    \cdot E(t_{j1} - t_{j3}) \cdot dt_{j3} \cdot dt_{j2} \cdot dt_{w_{i2}}\\
    & = \int_{w_{i2}}\int_{t_{j2}} \
    Q(w_{i2}, t_{j2})\cdot R(w_{i2} - w_{i1}, t_{j1} - t_{j2})\cdot dt_{j2} \cdot dt_{w_{i2}}\\
    \end{aligned}
\end{equation}
where measured signal $M(w_{i1}, t_{j1})$ on the $w_{i1}$th wire and time $t_{j1}$
is a convolution of i) the ionization charge distribution as a function
of the position in wire number and the drift time: $Q(w_{i2}, t_{j2})$, ii) the field response that
describes the induced current on the wires when the ionization charge moves:
$F(w_{i2} - w_{i1}, t_{j3} - t_{j2})$, and iii) the electronics response that amplifies
and shapes the induced current on the wire: $E(t_{j1} - t_{j3})$.
For simplicity, one can firstly convolve the field response and the electronics response
into the overall response function: $R(w_{i2} - w_{i1}, t_{j1} - t_{j2})$, in which the
fine-grained position-dependent field response function is averaged within
one wire pitch.

\begin{figure}[!ht]
     \centering
     \begin{subfigure}[b]{0.49\textwidth}
         \centering
         \includegraphics[width=\textwidth, trim={10cm 1cm 3cm 1.5cm},clip]{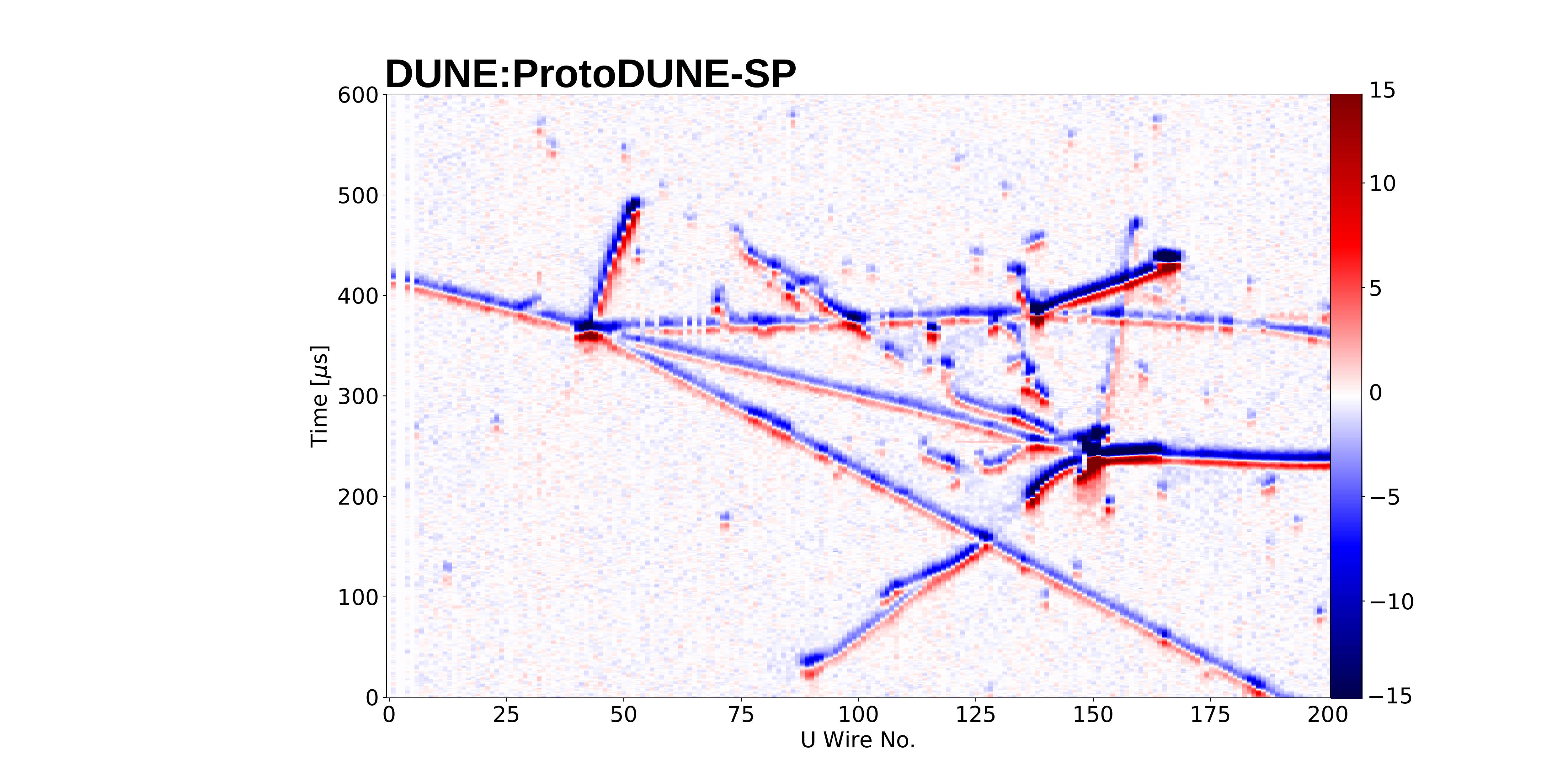}
         \caption{After Noise Filtering}
         \label{fig:example_nf}
     \end{subfigure}
     \hfill
     \begin{subfigure}[b]{0.49\textwidth}
         \centering
         \includegraphics[width=\textwidth, trim={10cm 1cm 3cm 1.5cm},clip]{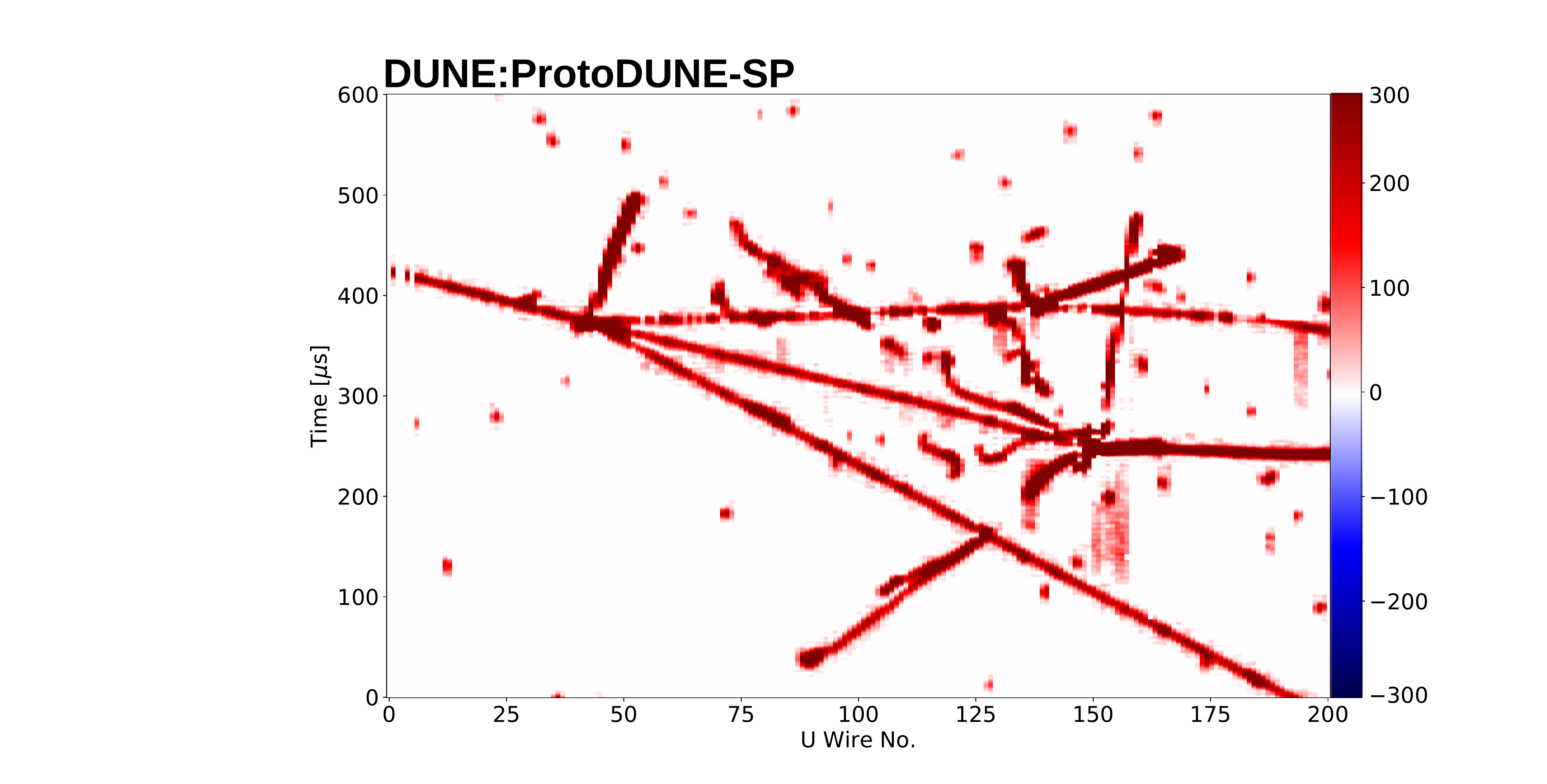}
         \caption{After Deconvolution}
         \label{fig:example_decon}
     \end{subfigure}
     \caption{An interaction vertex from the 7-GeV charged particle beam data
     (run 5152, event 89) measured on the induction U plane:
     \subref{fig:example_nf} Raw waveform in ADC counts after noise filtering;
     \subref{fig:example_decon} Ionization charge in number of electrons, scaled by 200, extracted with the 2D deconvolution technique.}
     \label{fig:example_nfsp}
\end{figure}

Because of the long-range induction effect, instead of a 1D deconvolution involving
only the time dimension, a two-dimensional (2D) deconvolution involving both the time
and wire dimensions is performed to extract the ionization electron distribution.
In practice, the FFT algorithm is used to convert the data from the discrete 2D time and wire domain
to a discrete 2D frequency domain~\cite{Adams:2018dra, Adams:2018gbi}.
To avoid amplifying high-frequency noise in the deconvolution, two Wiener-inspired filters
are applied separately in both dimensions. In addition, to further reduce the
noise contamination and improve the charge resolution, a technique for identifying
the signal regions of interest (SROI) is adopted and adjusted accordingly. 
The application of SROIs are particularly important to process the induction plane signals.
As an example, a raw waveform that has been de-noised as described in section~\ref{tpcnoisesupp} and 
the corresponding extracted charge distribution after the
deconvolution are shown in figure~\ref{fig:example_nf} and figure~\ref{fig:example_decon},
respectively.

\subsection{Event Reconstruction}\label{sec:reco}
There are two distinct steps in the ProtoDUNE-SP event reconstruction chain to go from the deconvolved waveforms to fully reconstructed interactions: hit finding and pattern recognition.  These steps are described in sections~\ref{sec:reco:hitfinder} and~\ref{sec:reco:pandora}, respectively. 

\subsubsection{Hit finding}\label{sec:reco:hitfinder}
\begin{figure}[!ht]
    \centering
    \includegraphics[width=\textwidth]{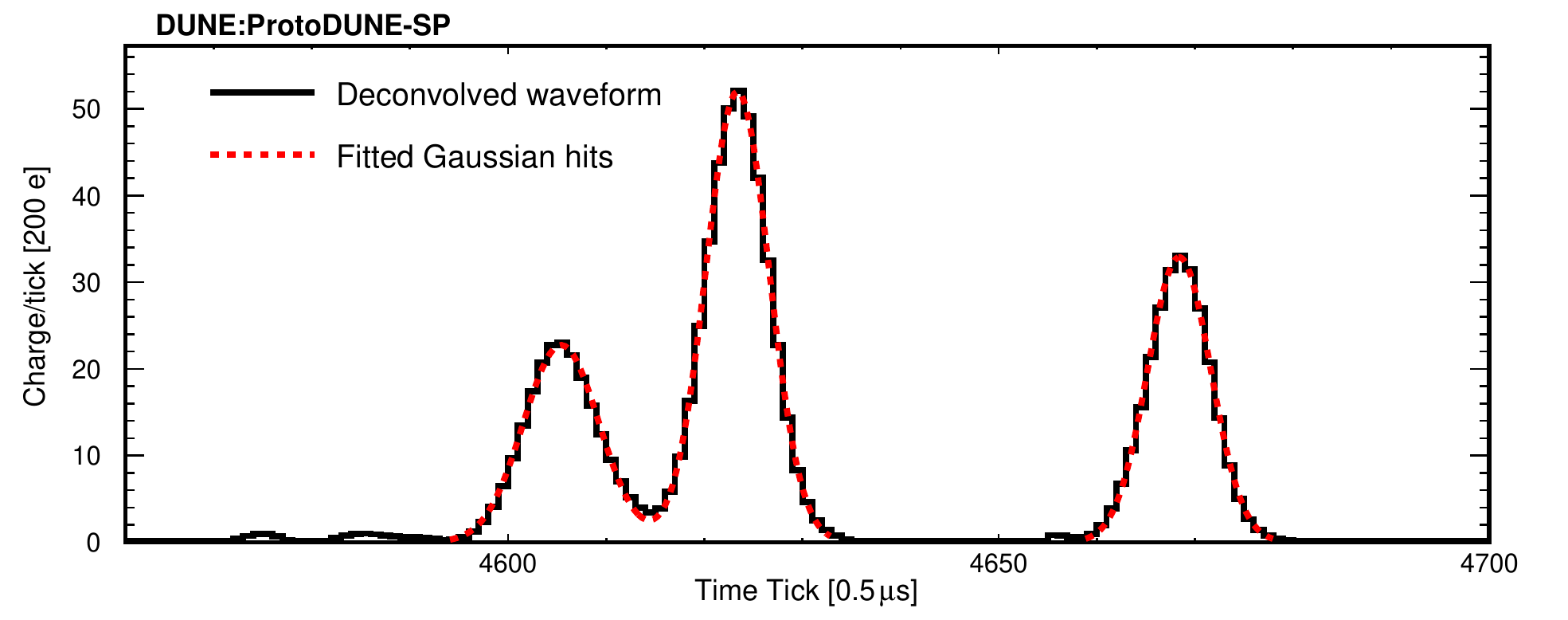}
    \caption{An example of a reconstructed waveform on a single wire from ProtoDUNE-SP data.}
    \label{fig:reco:hits}
\end{figure}

The hit finding algorithm fits peaks in the wire waveforms, where a hit represents a charge deposition on a single wire at a given time.  Each hit corresponds to a fitted peak.  Ideally, after the deconvolution process described in section~\ref{signalproc}, the signals on all wires, regardless of whether they are induction-plane wires or collection-plane wires, will be waveforms containing possibly overlapping Gaussian-shaped peaks. The algorithm searches for candidate hits in the waveform and fits them to a Gaussian shape to produce the hits.  Situations can occur in which charge deposits do not form a simple Gaussian shape, for example when a particle trajectory is close to being in the plane containing the wire under study and the electric field. If, after the candidate peak-finding, a very large number of candidate peaks are found in a given SROI then the algorithm bypasses the hit-fitting step and the pulse is instead divided into a number of evenly-spaced hits. An example of a fitted waveform is shown in figure~\ref{fig:reco:hits}, where three hits have been reconstructed.

The two induction planes consist of wires that are wrapped around the APA. As a result, it must be determined on which segment of the wrapped wire that a given energy deposit was actually measured. Firstly, triplets of wires (one on each plane) are formed using signals within a narrow time window. Often, for a given collection wire, only a single pair of induction wires are matched, thus the hits are disambiguated at this stage. Otherwise, there can be multiple induction wires consistent in time with the collection wire. In this case, the algorithm aims to minimize the difference in charge between the collection wire and the candidate induction wires in a deterministic manner. A full description of the method is given in Ref.~\cite{Abi:2020evt}. Simulation studies show that this technique assigns more than 99\% of hits to their correct wire segments.

\subsubsection{Pattern recognition with Pandora}\label{sec:reco:pandora}

Pattern recognition in ProtoDUNE-SP is performed using the Pandora software package~\cite{Marshall:2015rfa}, which executes multiple algorithms to build up the overall picture of  interactions in the detector. Pandora has been used successfully in other liquid argon time projection chambers (LArTPCs) such as MicroBooNE~\cite{Acciarri:2017hat}. New features have been developed for ProtoDUNE-SP since it differs from MicroBooNE with its multiple TPCs and drift volumes in addition to the need for a testbeam particle specific reconstruction chain. 

Pandora contains chains of reconstruction algorithms that focus on specific topologies, but they all follow a common pattern. The first step involves two-dimensional clustering of the reconstructed hits in each of the three detector readout planes separately.  Dedicated algorithms then match sets of 2D clusters between the three views. If matching ambiguities are discovered, information from all three views is used in order to motivate changes to the original 2D clustering. Once consistent matches between 2D clusters have been made, three-dimensional hits are constructed and particle interaction hierarchies are created.

\begin{figure}[!ht]
    \centering
    \includegraphics[scale=0.6]{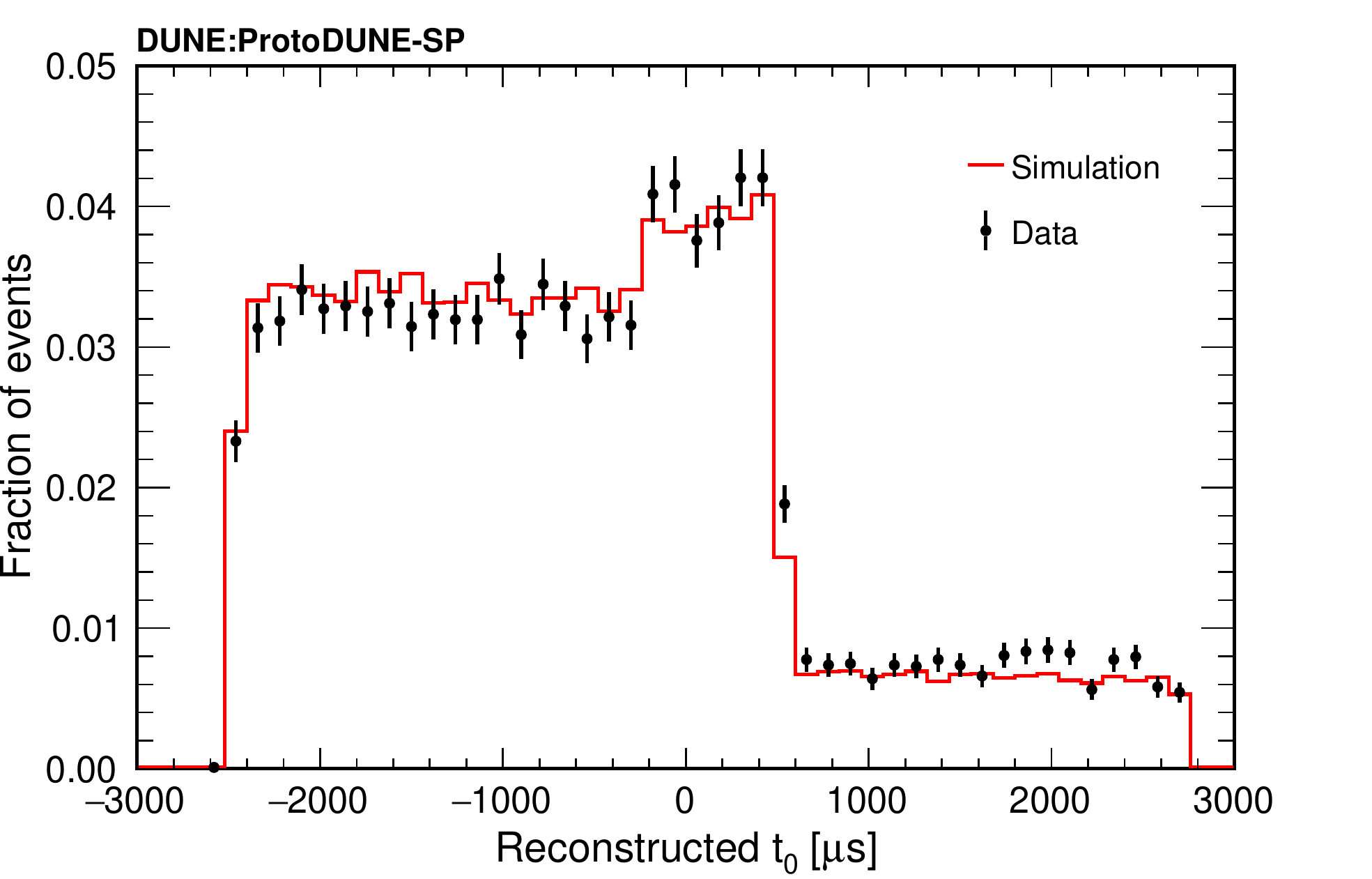}
    \caption{The Pandora reconstructed $t_0$ distribution for cosmic-ray muons that cross either the cathode or anode in data (black points) and simulation (red).}
    \label{fig:reco:t0}
\end{figure}

In the Pandora ProtoDUNE-SP reconstruction, all of the clusters are reconstructed first under the cosmic-ray hypothesis using a set of algorithms designed to reconstruct track-like particles. Cosmic-ray candidates are subsequently identified and removed so that beam-particle analysis can proceed. One important feature of the cosmic-ray reconstruction step is the ``stitching'' of tracks across the boundaries between neighboring drift volumes bounded by a CPA or an APA. The stitching procedure is applied when two 3D clusters have been reconstructed in neighboring drift volumes that have consistent direction vectors and an equal but opposite shift in the drift direction from the CPA or APA. When the clusters are shifted by this amount,  a single collinear cluster with a known absolute position along the drift direction and time $t_0$ relative to the trigger time is produced. Figure~\ref{fig:reco:t0} shows the reconstructed $t_0$ distribution for data and simulation for those cosmic-ray muon tracks that have been stitched at the cathode or anode. Cosmic-ray muons that cross the cathode have $t_0$ values between -2500$\,\mu$s and 500$\,\mu$s, and those that cross the anode have $-250\,\mu$s $< t_0 < 2750\,\mu$s. Tracks satisfying one or more of the following criteria are identified as ``clear'' cosmic-ray candidates:
\begin{itemize}
    \item The particle enters through the top of the detector and exits through the bottom.
    \item The measured $t_0$ for stitched tracks is inconsistent with a particle coming from the beam.
    \item Any of the reconstructed hits appear to be located outside of the detector when assuming $t_0 = 0$, which indicates that the object is inconsistent with the timing of the beam.

\end{itemize}
The hits from these clear cosmic-ray candidates are removed from the trigger record before the Pandora reconstruction chain continues to further process the data. These tracks, and in particular those with a measured $t_0$, form a critical component of the various detector calibrations detailed in section~\ref{sec:tpcresponse}. 

Once the energy deposits from the clear cosmic rays have been removed, the reconstruction continues with a 3D slicing algorithm that divides the detector into spatial regions containing all of the hits from a single parent particle interaction. These 3D slices could contain beam particles or cosmic rays that were not clear enough to be removed in the first-pass cosmic-ray removal process.  Two parallel reconstruction chains are applied to these slices - one is the aforementioned cosmic-ray reconstruction, and the other is a test-beam specific reconstruction.

The test-beam specific reconstruction consists of a more complex chain of algorithms capable of reconstructing the intricate hierarchies of particles seen in hadronic interactions that can produce numerous track-like and shower-like topologies. Included in this reconstruction chain is a dedicated search for the primary interaction vertex of the test-beam particle. As well as being used to inform the clustering, the vertex is essential for constructing the correct particle hierarchy. 

\begin{figure}
    \centering
    \includegraphics[scale=0.3]{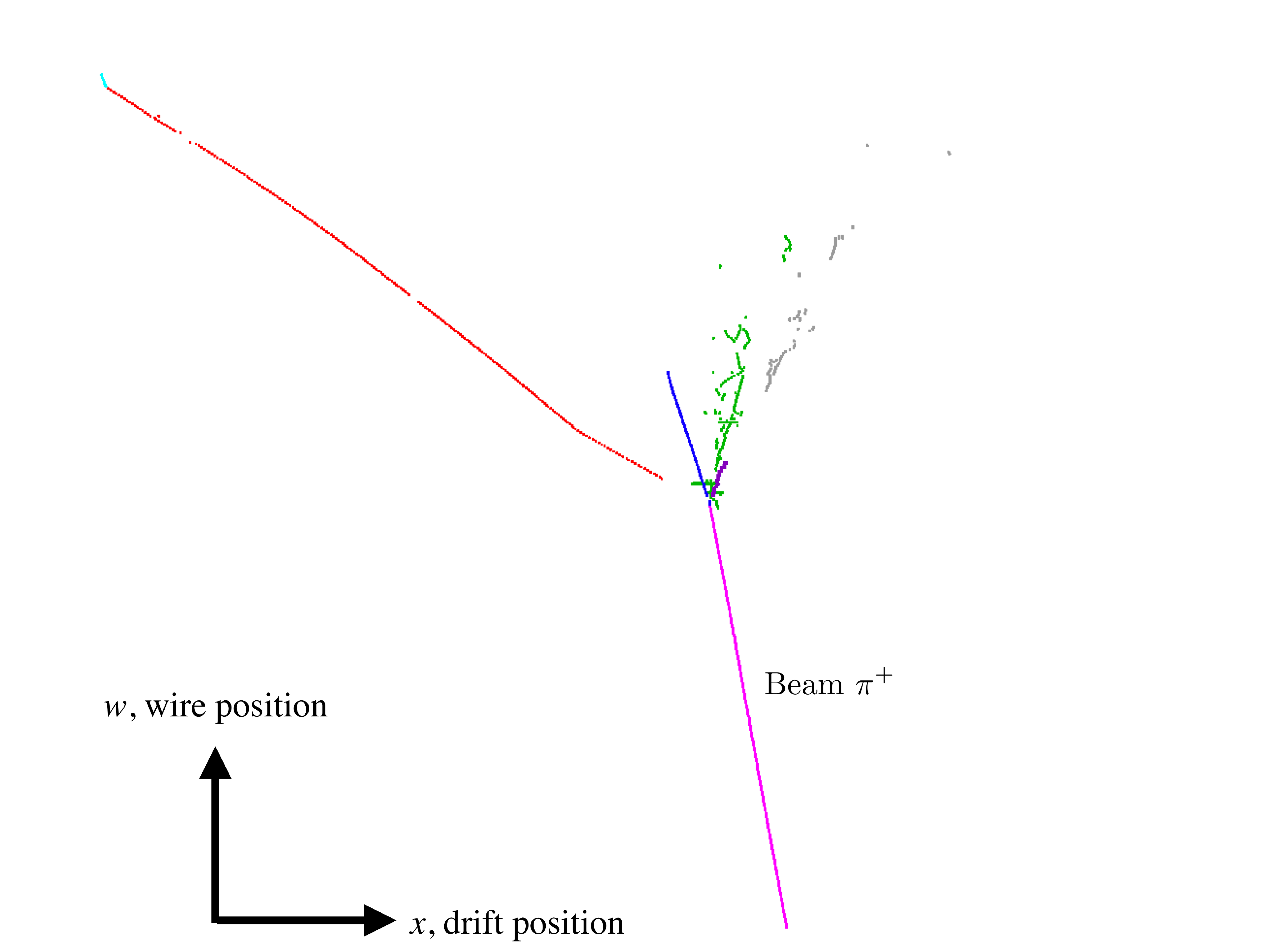}
    \caption{A reconstructed, simulated test-beam interaction showing the incoming beam $\pi^+$ track in magenta and a number of secondary particles created at the interaction vertex. The different colors represent different reconstructed particles.}
    \label{fig:reco:hierarchy}
\end{figure}

Once the slices have been reconstructed under both hypotheses, cosmic-ray and test-beam, a boosted decision tree (BDT) algorithm is used to determine which, if any, of the slices are consistent with being of test-beam origin. The input variables to the BDT primarily use topological information to measure the consistency of the interaction with the test-beam particle hypothesis. The output from the reconstruction is in the form of a particle hierarchy, where links are made between parent and child particles to give the flow of an interaction from the initial beam particle. Figure~\ref{fig:reco:hierarchy} shows an example of a fully reconstructed particle hierarchy for a simulated beam interaction, where the incoming beam $\pi^+$ is shown as the magenta track.
A suite of tools has been produced for ProtoDUNE-SP analysers to easily access this hierarchical information in order to perform the analyses.

The charge deposition per unit length, $dQ/dx$ is reconstructed for track-like objects such as muons, charged pions, kaons, protons and the beginnings of electromagnetic showers. The charge $Q$ is taken as the area of the Gaussian fit to the individual hit. The segment length $dx$ is calculated as the  wire spacing divided by the cosine of the angle between the track direction and the direction normal to the wire direction in the wire plane. The raw $dQ/dx$ is further calibrated to remove nonuniform detector effects and converted to energy loss $dE/dx$ for energy measurement and particle identification. This procedure is described in section~\ref{sec:muonCal}.

\subsection{Signal to noise performance}
The measurement of the signal-to-noise ratio (S/N) for the ProtoDUNE-SP detector is carried out using a selected cosmic-ray muon sample in Run 5432 taken on Oct. 20, 2018.  Muon tracks crossing the LArTPC at shallow angles with respect to the anode plane and large angles with respect to the direction of the wires in the planes are considered for the S/N characterization. To ensure good track quality, track length is required to be at least 1 m. The electron drift lifetime of the sample was approximately 24\,ms as independently measured by the purity monitor. 

\begin{figure}[!htp]
      \centering
      \includegraphics[width=0.6\textwidth]{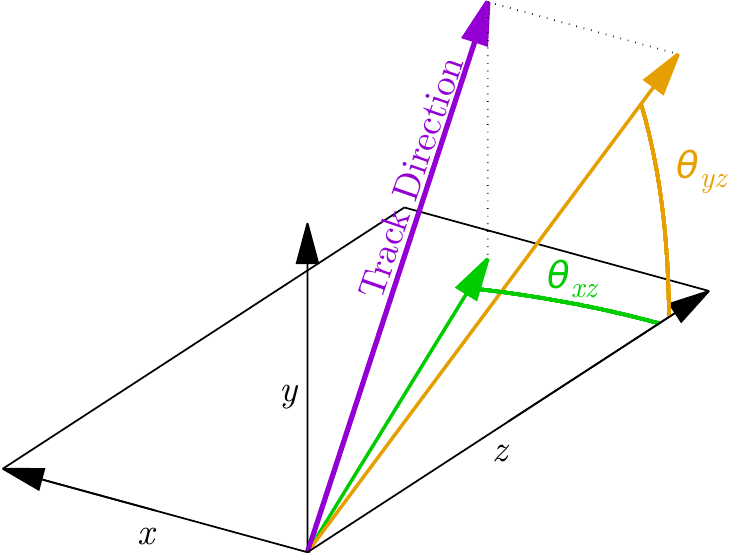}
      \caption{The definitions of the track direction angles $\theta_{xz}$ and  $\theta_{yz}$}
      \label{fig:thetaxzyz_definition}
  \end{figure}

The signal on each wire in a plane is defined to be the maximum pulse height of the raw waveform after subtracting the pedestal. The noise value is defined to be the standard deviation of a Gaussian function fit to the distribution of ADC values in signal-free regions of a channel's waveform.  The signal size depends on the angle of the track with respect to the wire and also with respect to the electric field.  We standardize the signal on a wire to be that from tracks that are perpendicular to the wire and also perpendicular to the electric field.  We define two angles $\theta_{xz}$ (the angle made by the projection of a track on the $xz$ plane with the $z$ direction) and $\theta_{yz}$ (the angle made by the projection of a track on the $yz$ plane with the $z$ direction). The two angles are illustrated in figure~\ref{fig:thetaxzyz_definition}. To minimize the influence of angular dependence of S/N, we select muon tracks that have minimum component in both drift and wire direction. Angle cuts of  $\theta_{xz}$ and $\theta_{yz}$ within 20$^{\circ}$ are applied. A correction is applied to the raw waveform signal to standardize the strength, adjusting for angular effects.

\begin{figure}[!htp]
\centering
\includegraphics[width=\textwidth]{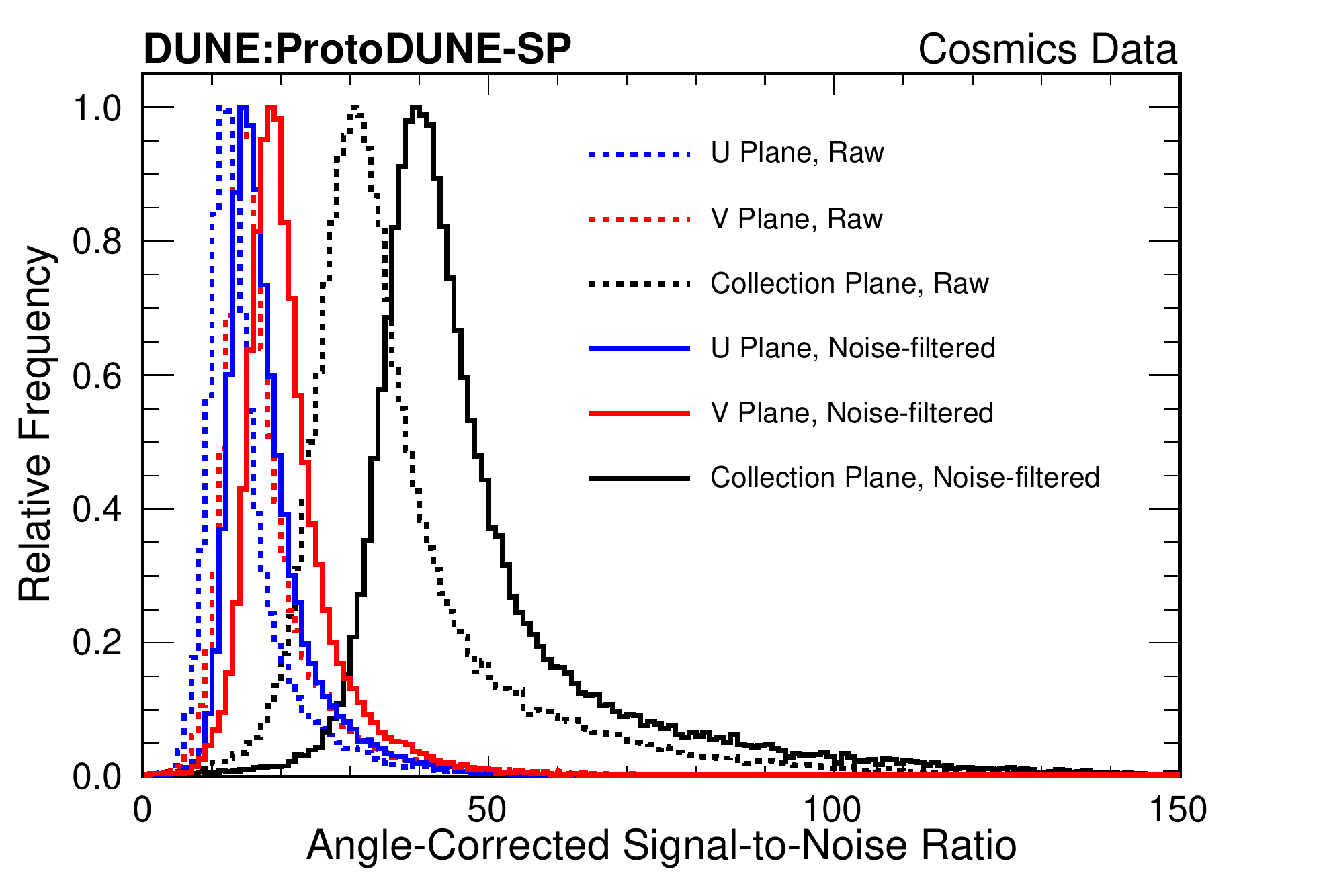}
\caption{Angle-corrected S/N distributions before and after the noise filtering of the three planes
using the cosmic-ray muons for the characterization. The histograms are normalized such that the maximum frequency is one.
}
\label{fig:signal_to_noise_ratio}
\end{figure}

 The angle-corrected S/N distributions are shown in  figure~\ref{fig:signal_to_noise_ratio}. No electron drift lifetime corrections are applied to the angle-corrected S/N calculations. The most probable values (MPVs) of the S/N distributions after the noise mitigation are 40.3, 15.1, and 18.6 for the collection plane, the U plane, and the V plane, respectively. The actual performance of the S/N for the three planes is much better than the expectation in the ProtoDUNE-SP technical design report~\cite{Abi:2017aow} - 9.0 for the three planes. The angle-corrected S/N results with and without the noise filters, together with the estimation using the averaged values of the S/N distributions, are summarized in table~\ref{tab:snr}. 
 
 The differences in the average S/N values for the three planes is explained using the Shockley-Ramo theorem, discussed in section~\ref{signalproc}. The three planes have similar weighting fields but different local drift velocities. Among the three planes, the collection plane has the largest local drift velocity and hence the best S/N performance. The S/N performance is slightly better for the V plane with respect to the U plane. This is because the local drift velocity at the V plane is higher than that of the U plane due to larger bias voltage, while the weighting fields are the same for both.
 


\begin{table}[!htp]
\centering
\caption{Summary table of the angle-corrected S/N performance before and after the noise mitigation for the ProtoDUNE-SP detector. No electron drift lifetime corrections are applied. The most-probable value (MPV) and the average value for each plane are listed. 
}
\begin{tabular}{ccccc}
\hline
\multirow{3}{*}{Plane} & \multicolumn{4}{c}{Peak signal-to-noise ratio} \\
 \cline{2-5}
 & \multicolumn{2}{c}{Raw Data} & \multicolumn{2}{c}{After noise filtering} \\
 \cline{2-5}
 & MPV & Average & MPV & Average\\
 \hline
 Collection & 30.9 & 38.3 & 40.3 & 48.7 \\
 U & 12.1 & 15.6 & 15.1 & 18.2\\
 V & 14.9 & 18.7 & 18.6 & 21.2\\
 \hline
\end{tabular}
\label{tab:snr}
\end{table}


\section{Photon detector characterization}
\label{sec:pd}
\subsection{The photon detector system}
\label{sec:PDSDescription}
The ProtoDUNE-SP photon detector system (PDS) comprises 60 optical modules embedded within the six APA frames of the TPC.  These modules view the LAr volume from each anode side opposite the central cathode.
Three different photon collection technologies proposed for DUNE's far detector modules~\cite{Abi:2017aow} are implemented in ProtoDUNE-SP's PD system.  In each technology, incident LAr scintillation photons, which have wavelengths around 128~nm, are converted into longer-wavelength photons using photofluorescent compounds as wavelength shifters (WLS). Visible light is trapped within the modules, a portion of which is eventually incident on an array of silicon photomultiplier photosensors (SiPMs)~\cite{Acerbi:2019qgp}.

\subsubsection{Light collectors}

\begin{figure}[!htbp]
    \centering
      \includegraphics[height=9.5cm,width=1.05\textwidth]{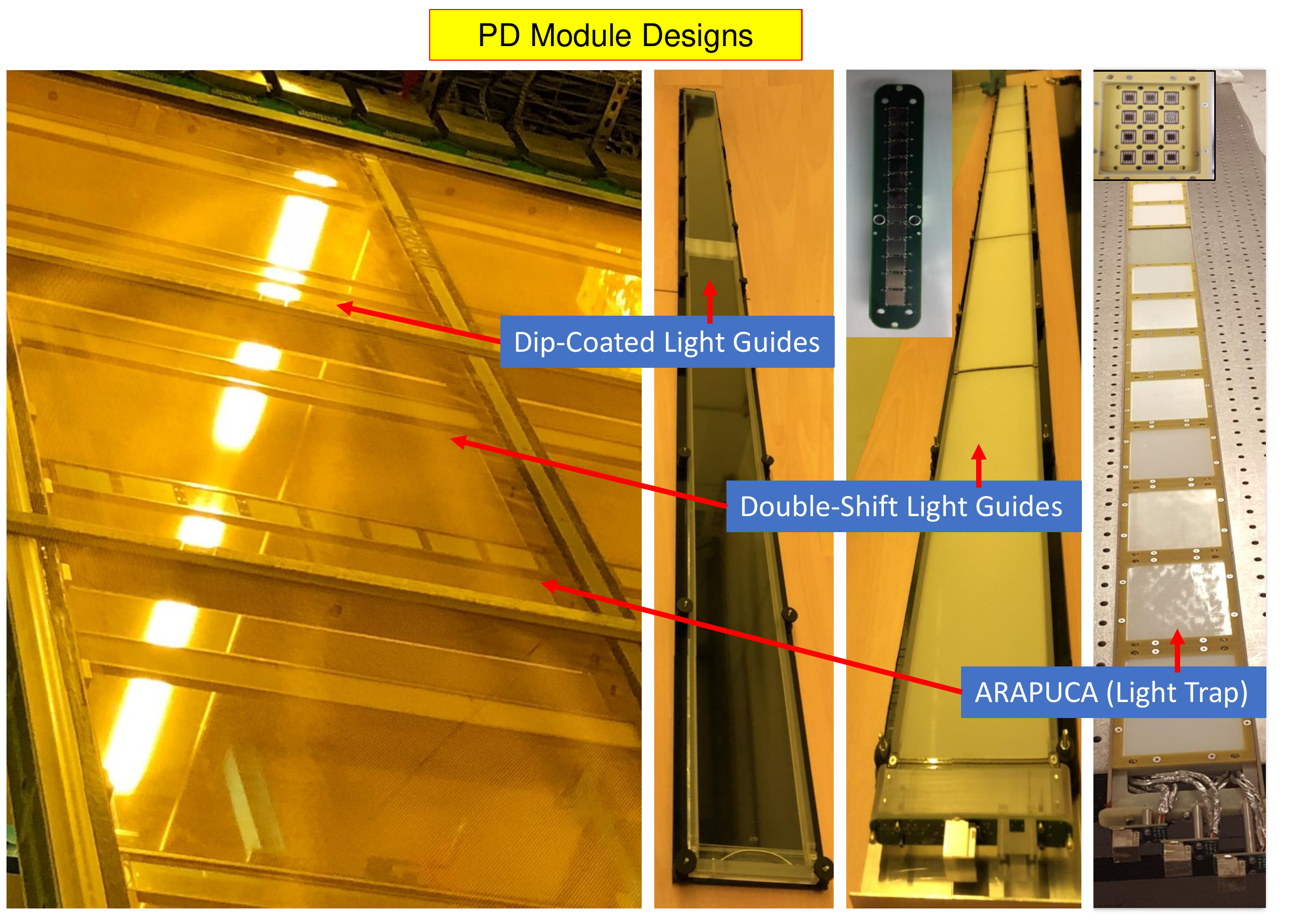}
    \caption{Picture of a ProtoDUNE-SP APA during assembly. Labels indicate the three types of PDS modules  inserted into the APA frame (left). Pictures of the three technologies (right) with details of the photo-sensor arrays equipping the modules (insets).  From left to right, they are a dip-coated light-guide module, a double-shift light-guide module, and an ARAPUCA module.}
    \label{fig:protodune-Three_technologies}
\end{figure}

Each APA contains ten support structures located behind the wire planes for the PDS modules.  Each module is a long, thin bar oriented along the $z$ axis.  The spacing between modules in $y$ is approximately 60~cm, as illustrated in figure \ref{fig:protodune-Three_technologies}. Of the 60 modules, two are based on the ARAPUCA photon detector technology~\cite{Machado:2016jqe}, 29 are dip-coated light guides~\cite{Moss:2016yhb, Bugel:2011xg}, and 29 are double-shift light guides~\cite{Howard:2017dqb}. 
All light-guide modules have the same dimensions.  The optical area of a module is $207.4 \times 8.2$~cm$^2$ in size, and the light is read out on one end of the bar. The ARAPUCA modules are segmented longitudinally (along the $z$ direction) into 12 cells, each with its own readout.  The first eight cells each have an optical area of $9.8\times 7.9$~cm$^2$ and the remaining four cells are double-area cells, each with an optical area of $19.6\times 7.9$~cm$^2$. One ARAPUCA module is located in the top half of the upstream APA in the beam-side drift volume. A second is located in the middle of the APA in the center of the opposite drift volume. The two light-guide designs fill the remaining modules in alternating positions in the APAs. One example of each module type is highlighted in figure~\ref{fig:protodune-Three_technologies}, together with the photo-sensor arrays equipping the module. The installation of the modules within an APA behind the wire planes and a  grounding mesh is also visible in figure~\ref{fig:protodune-Three_technologies}.  

The two light-guide designs convert incident VUV photons into the visible range using tetra-phenyl butadiene (TPB) (emission peak $\sim$430~nm), while the ARAPUCA design uses p-terphenyl (PTP) (emission peak $\sim$340~nm). 
In the dip-coated acrylic light guide, wavelength-shifted photons are confined inside by total internal reflection and they are guided toward the end of the bar that is in optical contact with a photosensor array.
The double-shift light guide contains an internal light guide doped with the second WLS (490~nm emission) to facilitate trapping of double-converted photons within the module and guide them toward the photosensors at the end of the bar.
The ARAPUCA uses a dichroic filter window (400~nm cutoff) to reflect photons from a second WLS (TPB) inside the cell underneath the window and prevent them from exiting before they are absorbed or detected by the photosensors distributed inside the cell. 

\subsubsection{Photosensors}

Three silicon photosensor models, each with an active area of $6\times 6$~mm$^2$, are employed throughout the photon detection system.  They were selected among those that were available commercially or that were newly developed during the PDS material procurement phase: the SensL SiPM MicroFC-60035-SMT (35~$\mu$m pixel size) and two types of Hamamatsu MPPC S13360-6050 (50~$\mu$m pixel size) - the CQ-type (Quartz windowed for Cryogenic application) and the VE-type (VErtical through-silicon via). Arrays of photosensors of the same model are passively ganged together in parallel forming large-area single channels for voltage supply and signal readout. The arrays formed by three SensL SiPMs are indicated in the following as 3-S-SiPM, those formed by 3 Hamamatsu MPPCs (VE-type) indicated as 3-H-MPPC and those with 12 Hamamatsu MPPC (CQ-type) as 12-H-MPPC.
Each of the light-guide modules is read out by four channels with three photosensors each, either 3-S-SiPM or 3-H-MPPC, arranged in a strip of 12 photosensors in total at one end of the module. 
Each cell of each ARAPUCA module is read out by one 12-H-MPPC channel made of 12 Hamamatsu MPPCs distributed in the plane opposite to the cell optical window, for a total of 144 photosensors in the ARAPUCA module (12 cells). The numbers of channels of the different types and their distribution in the PDS modules are summarized in table~\ref{tab:pdmoduletypes}. A 12-H-MPPC array and a strip made by four 3-S-SiPM are shown in figure~\ref{fig:protodune-Three_technologies}. 
\begin{table}[!htp]
\centering
\caption{Numbers of each type of PDS module installed in ProtoDUNE-SP, and the numbers of sensors per channel and channels per module}
\label{tab:pdmoduletypes}
\begin{tabular}{ c|c|c|c|c|c|c}
{\footnotesize Ph. sensors}        &   {\footnotesize Type}   & {\footnotesize  Channels}  & {\footnotesize Dip-coated} &  {\footnotesize Double-shift}  & {\footnotesize ARAPUCA} &  {\footnotesize Total channels}  \\
{\footnotesize  per channel}        &    {\footnotesize of channel}   & {\footnotesize per module}  & {\footnotesize  modules}     &     {\footnotesize modules}     &  {\footnotesize modules}  &  {\footnotesize  in PDS}         \\
\hline
 3  &  3-S-SiPM   &   4  & 21  & 22 & - & 172 \\
\hline
3   & 3-H-MPPC    &    4  & 8  & 7  & - &  60  \\
\hline
12  & 12-H-MPPC   &   12  & -  & -  & 2 &  24  \\
\end{tabular} 
\end{table}

The ratio of the photosensor area to the light collector surface area for the light-guide modules (0.26\%) is lower than that of the ARAPUCA cells (4.2\% and 2.1\% for the cells with double area). 

\subsubsection{Readout DAQ and triggering}

\begin{figure}[!htbp]
\centering
\includegraphics[height=7.5cm,width=1.0\textwidth]{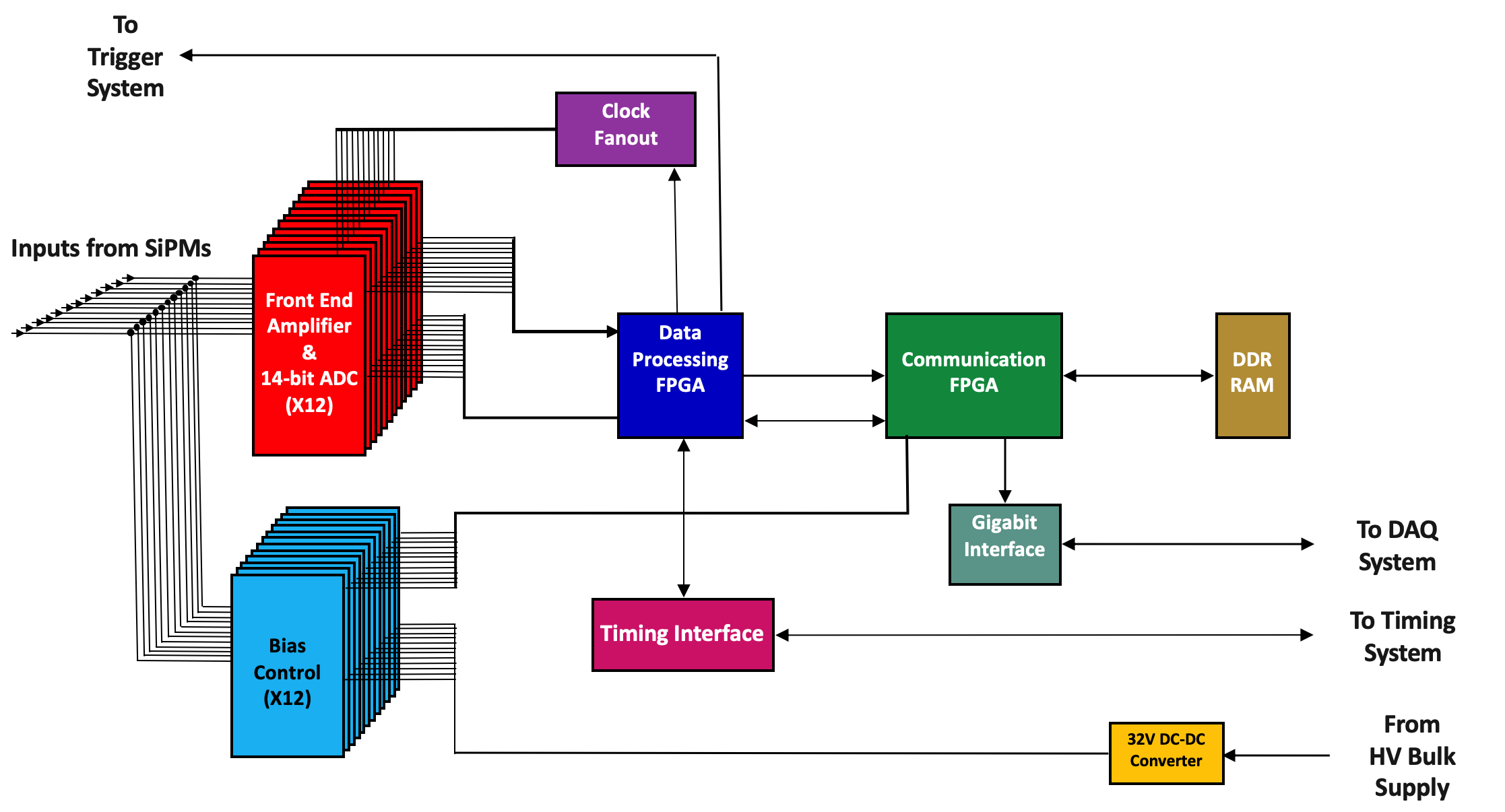}    
\caption{Block diagram of the ProtoDUNE-SP photon-detector readout module (SSP) with interfaces to Trigger and DAQ systems.}
\label{fig:SSP-diagram}
\end{figure}

Unamplified signals from the photosensors in the LAr are transmitted outside the cryostat on copper cables.  A dedicated, custom module was built for receiving and processing silicon photosensor signals for the trigger and DAQ.  The module is called the SiPM Signal Processor (SSP).  A self-contained, 1U SSP module reads out 12 independent PDS channels. Each channel has a voltage amplifier and a 14-bit, 150 megasamples per second ADC that digitizes the (current) output signal from photosensors into analog-to-digital units.
The front-end amplifiers are configured to be fully-differential with high common-mode noise rejection.  Each amplifier has on its input a termination resistor that matches the 100~$\Omega$ characteristic impedance of the signal cable. The least-significant ADC bit corresponds to 66~nA in this configuration.  The digitized data are stored in pipelines in the SSP, corresponding to as much as 13.3~$\mu$s per trigger. The processing is performed by FPGA
(Field-Programmable Gate Array). The FPGA implements an independent data processor for each channel. The processing incorporates a leading-edge discriminator and a constant fraction discriminator for sub-clock timing resolution. 
A block diagram of the system is shown in Fig.~\ref{fig:SSP-diagram}. 

Each channel is individually triggerable.  Triggers can come from a periodic trigger, an internal trigger based on the leading-edge discriminator local to the individual channel, or an external global trigger distributed by the timing system as the general triggers issued by the beam line instrumentation (sections~\ref{sec:hardware},\ref{sec:pid}). If a trigger is present, the channel will produce a data packet consisting of a header and a waveform (a sequence of values in ADU) of predefined length. 
An artDAQ process in the PD DAQ system (SSP boardreader) generates a fragment when the timing system produces a trigger.  This fragment contains all packets received from the SSP with timestamps in a window -2.25+2.75~ms from the timestamp of the trigger.
Each SSP fragment contains 12 packets, one for each channel, with identical timestamps corresponding to the trigger time, and an arbitrary number of additional packets generated for channels that have discriminators that fired.

\subsubsection{Photon detector calibration and monitoring system}

The PDS incorporates a pulsed UV-light monitoring and calibration system to determine the photosensors' gains, linearities, and timing resolution, and to monitor the stability of the system response over time.
\begin{figure}[!htbp]
\centering

\includegraphics[height=7.0cm,width=1.03\textwidth]{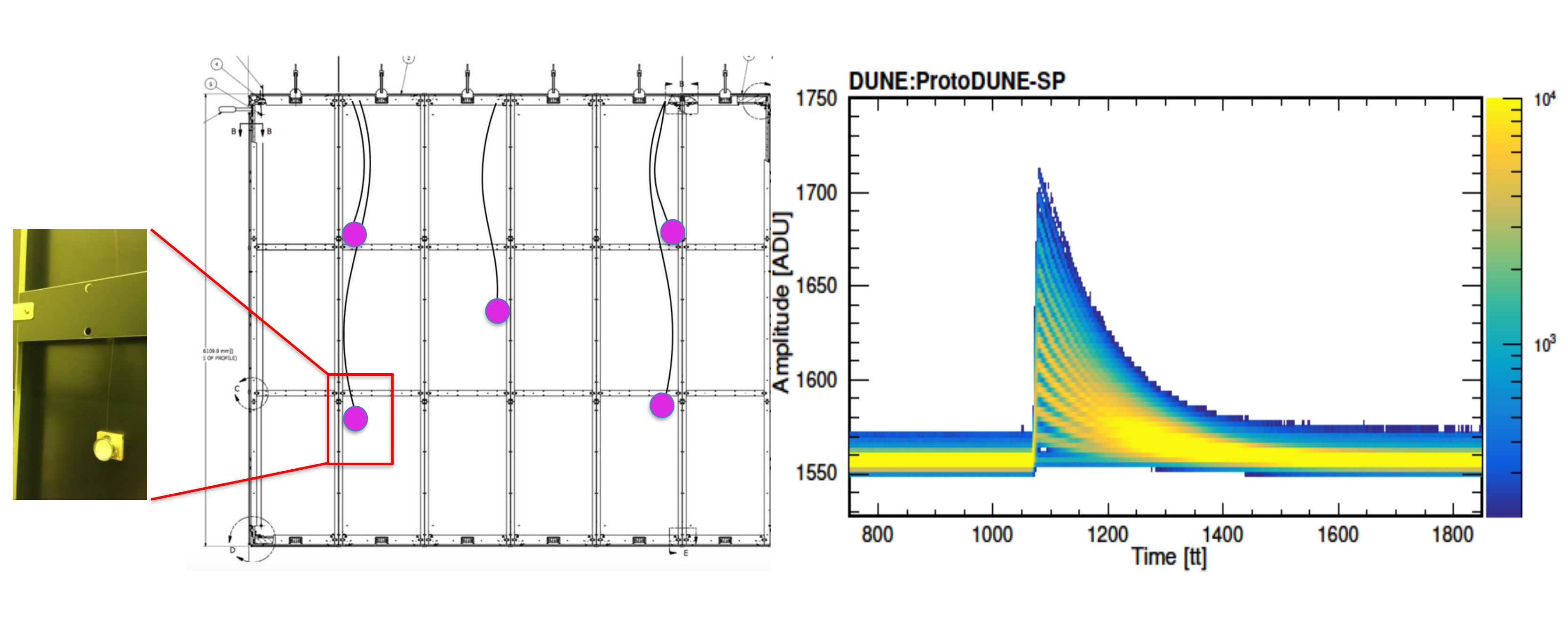}
\caption{The ProtoDUNE-SP photon-detector calibration and monitoring system installation on the CPA panels (left). The inset shows the actual lower left diffuser at time of installation. Waveforms in a LED calibration run displayed in persistence trace mode (right) showing recorded single and multi-photon signals. Each time tick (tt) represents 6.67 ns.}
\label{fig:cal_system_scheme}
\end{figure}

The system hardware consists of both warm and cold components. Figure~\ref{fig:cal_system_scheme} (left) shows a schematic of the ProtoDUNE-SP PD calibration and monitoring system. Diffusers mounted on the cathode-plane assembly panels serve as point light sources that illuminate the APA with the PDS modules on the opposite side of the drift volume.
The location of the diffusers on the CPA panel is indicated by magenta discs.  Also shown are the quartz fibers from the top of the CPA to the diffusers. The other CPA side holds a second set of five diffusers to calibrate the PDS in the opposite APA array.

The active system component is a 1U rack mount Light Calibration Module (LCM) sitting outside the cryostat. The LCM generates light pulses that propagate through the quartz fiber-optic cable to the diffusers at the CPA. The LCM consists of an FPGA-based control logic unit coupled to an internal LED Pulser Module (LPM) and an additional bulk power supply. The LPM utilizes multiple digital outputs from the control board to control the pulse characteristics, and it incorporates DACs to control the LPM pulse amplitude. 

The calibration system produces UV light flashes with predefined pulse amplitude, pulse width, repetition rate, and total number of pulses. 
A typical photosensor response to low amplitude, shortest duration calibration pulses is shown in 
figure~\ref{fig:cal_system_scheme} (right) where many recorded waveforms are overlaid with the color indicating the frequency, as in an oscilloscope persistence trace mode.
The UV light flashes can be produced in pairs with a fixed time difference between the two pulses to study timing properties of the photon system.


\subsection{Photosensor performance}
 
 Silicon photomultipliers convert light flashes into analog electrical pulses.  When a photon hits and is absorbed in a microcell of the avalanche photodiode matrix, an electron is lifted into the conduction band. If the bias voltage exceeds the breakdown voltage, this photoelectron creates an avalanche multiplication, with amplification (gain) typically in the $10^6$ range, which leads to a rising current signal.   Silicon photomultipliers are implemented as photosensors in the ProtoDUNE-SP PDS modules because of their compact design, low operating voltage and sensitivity to single photons. To overcome the limitation due to their small sensitive area (6$~\times~$6 mm$^2$) and to limit the number of channels, arrays of photosensors are passively ganged together in parallel forming a large-area single channel for voltage supply and signal readout. The capacitance of the array however increases with the number of photosensors connected in the array, and correspondingly its recovery time. Therefore, the signal amplitude decreases and the intrinsic noise increases. Operating at cryogenic temperature helps reduce the dark count rate. On the other hand, correlated noise --- afterpulses in the same pixel and optical crosstalk in neighboring pixels, both generated by the primary photoelectron event --- is expected to grow at high signal gain settings. The adopted multiplicity of the arrays in the ProtoDUNE-SP PDS design (three per channel in the 3-S-SiPM and 3-H-MPPC for the electron bars and twelve per channel in the 12-H-MPPC for the ARAPUCA cells) and the working parameters  for operation (bias voltage and gain) were determined with the requirement of acceptable signal-to-noise ratio allowing for sensitivity to single photoelectron discrimination with minimal secondary effects.
 
 Fifteen channels out of a total of $256$ show anomalous readings.  There are two that appear to be disconnected.  They fail their continuity checks and they have been unresponsive since the earliest tests after cabling.  There are two that appear to respond anomalously to light signals.  Their gains are similar to the others, but are somewhat higher, and their peak signal amplitudes are much higher than those of similar channels.  The remaining 11 pass their continuity checks but they do not respond to light signals. These anomalous channels (all of the 3-S-SiPM type) have not been investigated further.  No channels have changed state since detector operation began.
 
 The photosensor response is characterized using dedicated calibration runs that have fast, 
 low-amplitude LED flash triggers.  Data are collected at several different bias voltage settings. 
 These calibration runs are periodically repeated in time. 
 


 
\subsubsection{Single photoelectron sensitivity}
\label{sec:SPESensitivity}
\begin{figure}[tbh]
\begin{minipage}[t]{0.5\textwidth}
\includegraphics[width=1.05\textwidth]{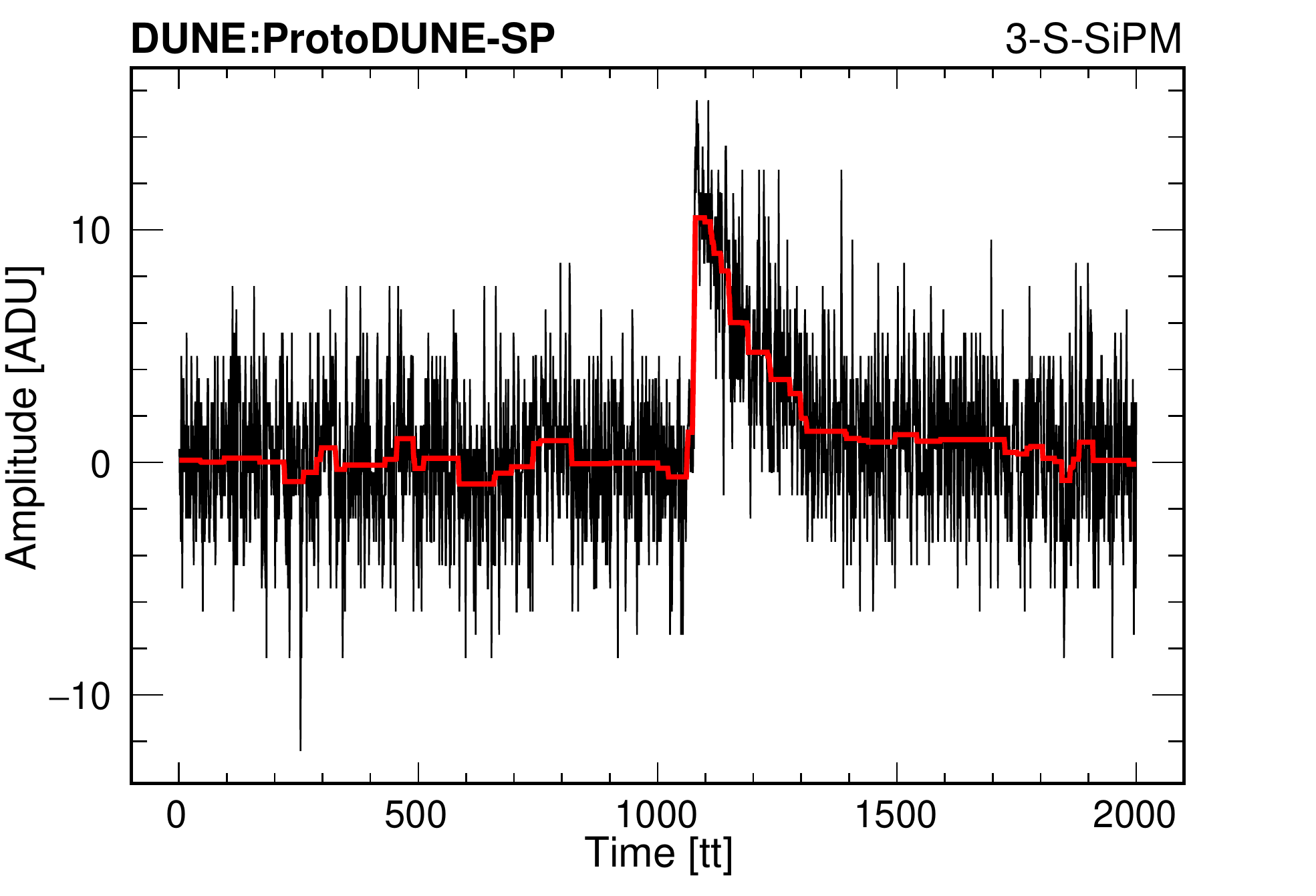}
\end{minipage}
\begin{minipage}[t]{0.5\textwidth}
\includegraphics[width=1.05\textwidth]{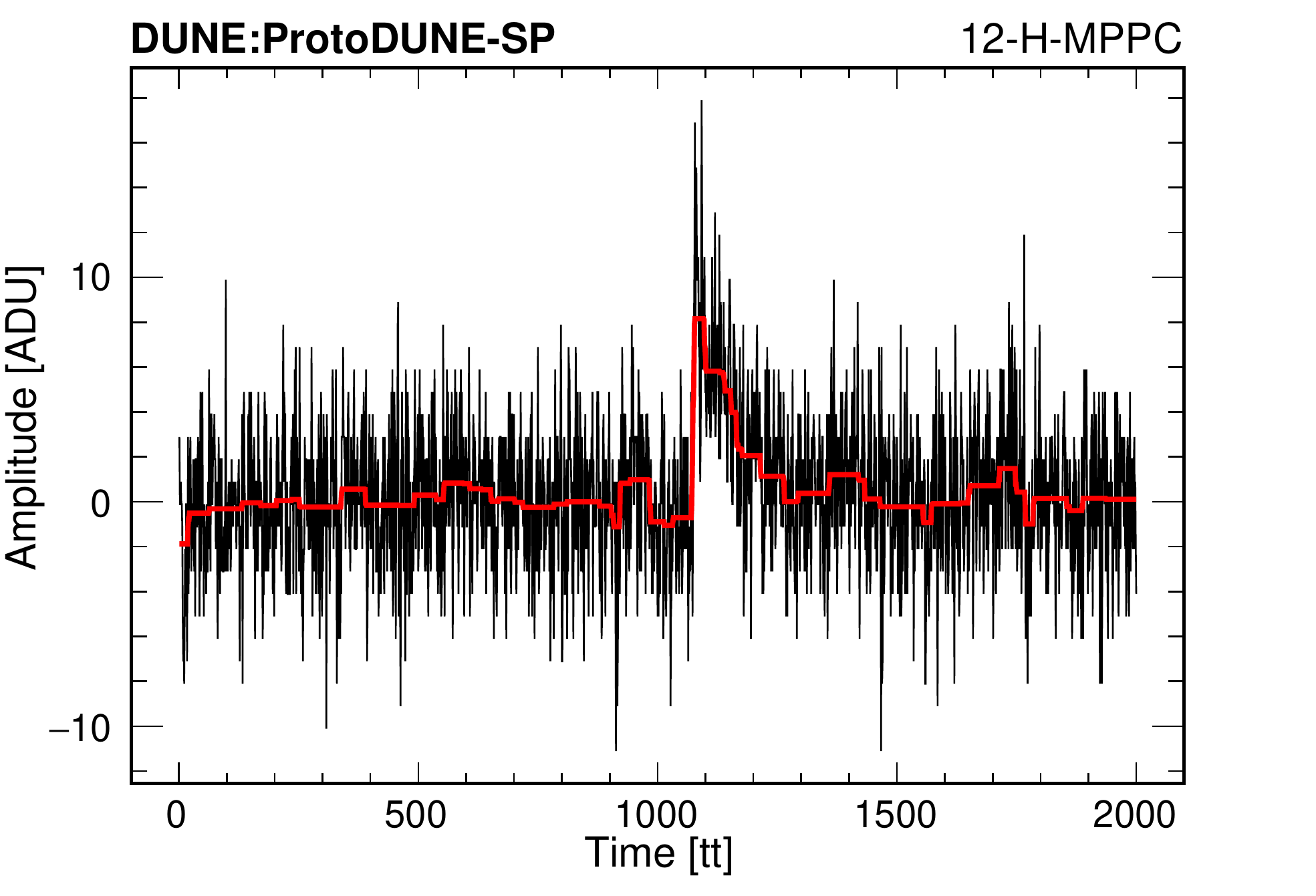}
\end{minipage}
\caption{Sample waveforms of single photon signal from a 3-S-SiPM channel (left) and  12-H-MPPC channel (right) (three and twelve sensors passively ganged in parallel, respectively). Noise filtered waveforms (red histogram) are superimposed, from noise-removal algorithms applied in data processing.}
\label{fig:protodune-sample-SensL-wvfm}
\end{figure}
Signal extraction and noise evaluation are performed for each recorded waveform. The waveform consists of 2000 samples of the photosensor output (in ADC units - ADU) at discrete, evenly spaced points in time (``time ticks'' - tt). The sampling interval is 6.67~ns and the duration of the waveform is 13.3 $\mu$s.  The trigger time, either from the global trigger or from the SSP internal trigger, is at a fixed time relative to the start of the waveform.
Typical recorded waveforms are shown in figure \ref{fig:protodune-sample-SensL-wvfm}. 

The mean of the pedestal distribution, determined from a pre-sample portion of the waveform before the trigger, gives the baseline value and the spread ($\sigma_N$) is an estimate of the noise in the recorded event.  After baseline subtraction, the charge of the signal (in ADU$~\times~$tt units, proportional to number of electrons or fC, where 1 ADU$~\times~$tt = 2750 electrons) is evaluated by integration of the portion of the waveform starting from the trigger time and extending over a 7~$\mu$s (1050 tt) time window.
De-noising algorithms that preserve the signal rise time and integral~\cite{condat:hal-00675043} are applied to more precisely evaluate the maximum amplitude in the same window corresponding to the photoelectron current of the signal (in ADU or $\mu$A). 

During detector assembly, the photosensors for each of the three- and twelve-unit channels were pre-selected based on minimal difference in their nominal breakdown voltage (at warm temperature) from that listed in data sheets.  After this selection, the spread in breakdown voltages among the sensors in the same 12-H-MPPC channel is typically $\Delta V^{\rm{max}}_{\rm{bd}}\le 300$~mV. All sensors in a channel are biased at the same common voltage $V_{\rm{B}}$, and the pre-selection thus enables the photosensors in the channels to all operate in relatively similar working conditions.

\begin{figure}[tbh]
\begin{minipage}[t]{0.5\textwidth}
\includegraphics[width=1.0\textwidth]{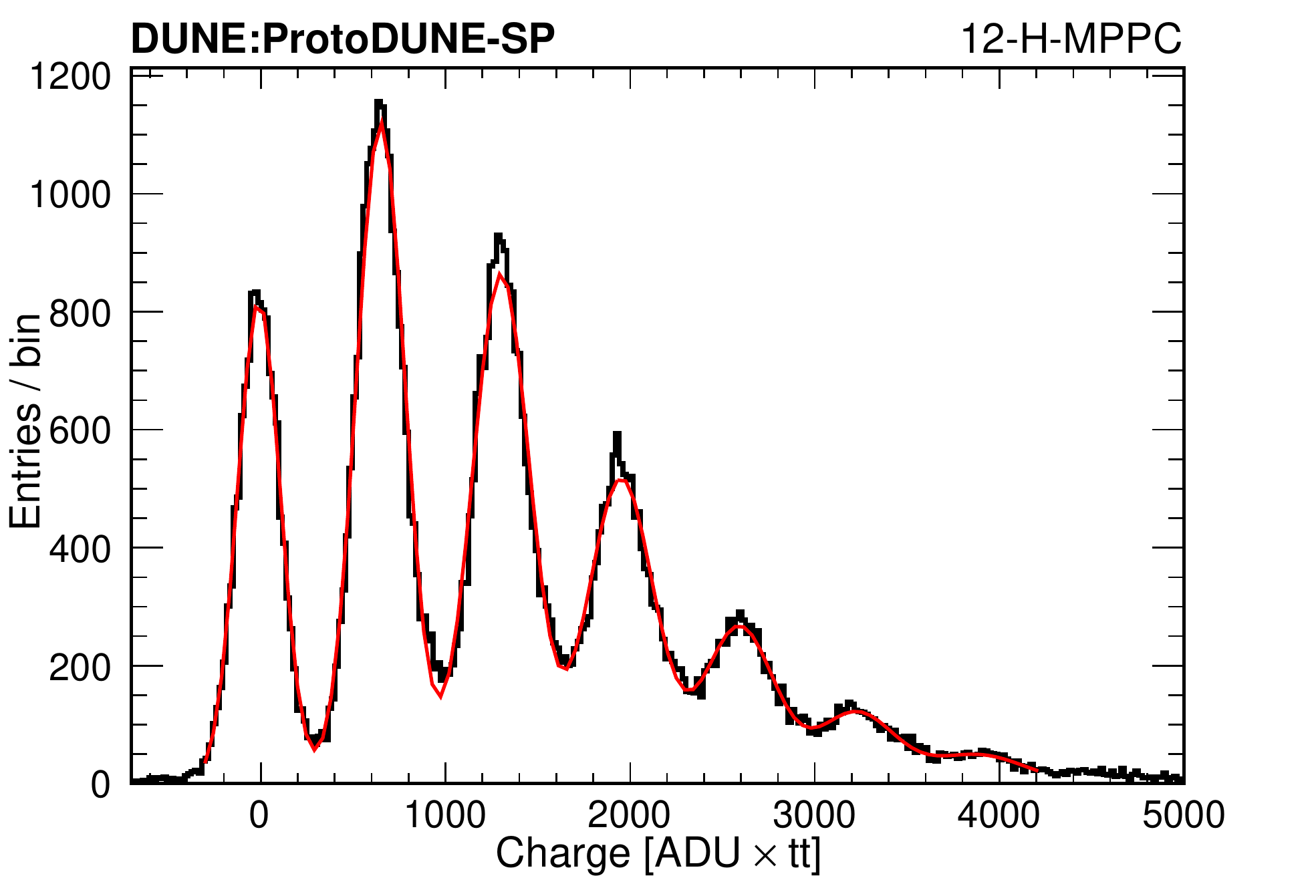}
\end{minipage}
\begin{minipage}[t]{0.5\textwidth}
\includegraphics[width=1.0\textwidth]{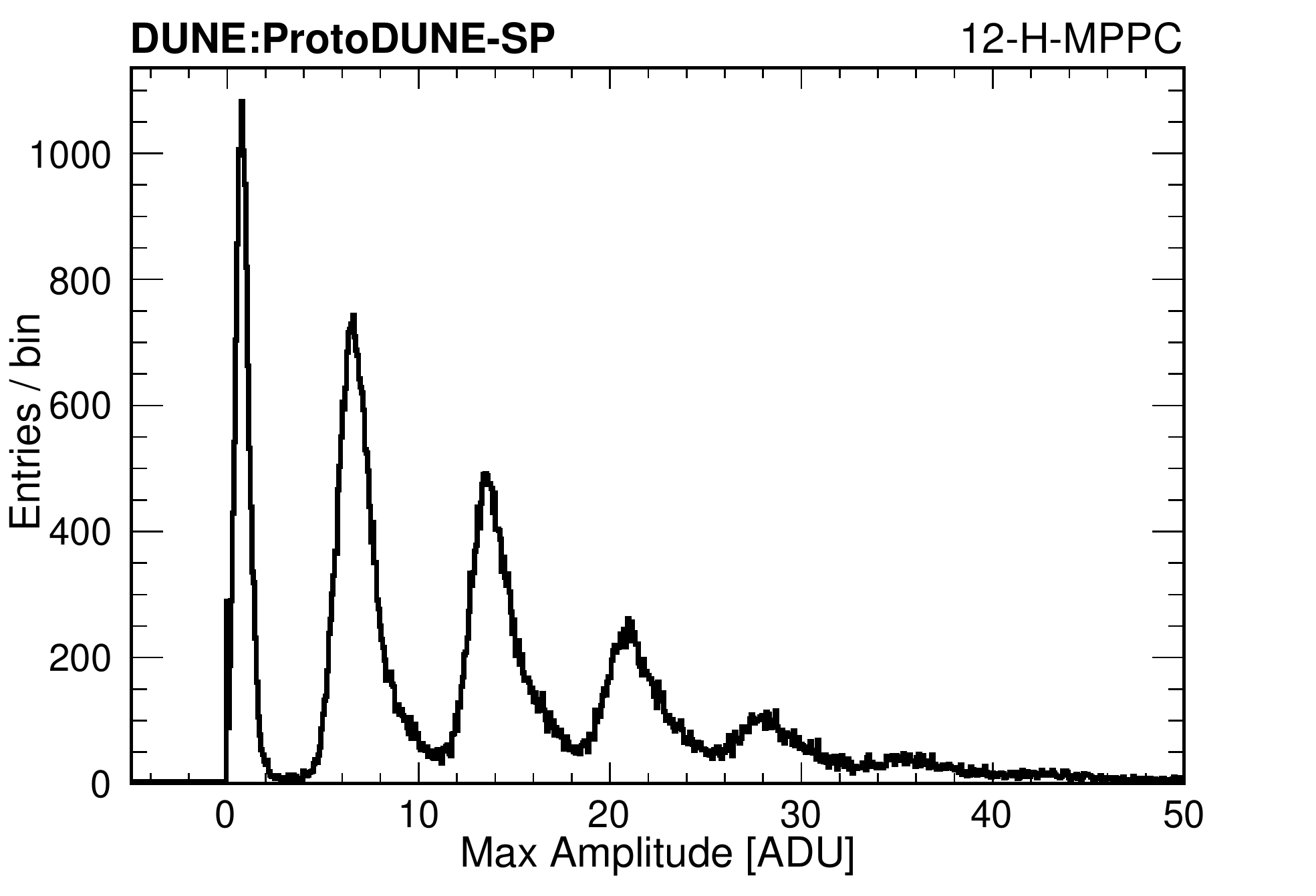}
\end{minipage}
\begin{minipage}[t]{0.5\textwidth}
\includegraphics[width=1.0\textwidth]{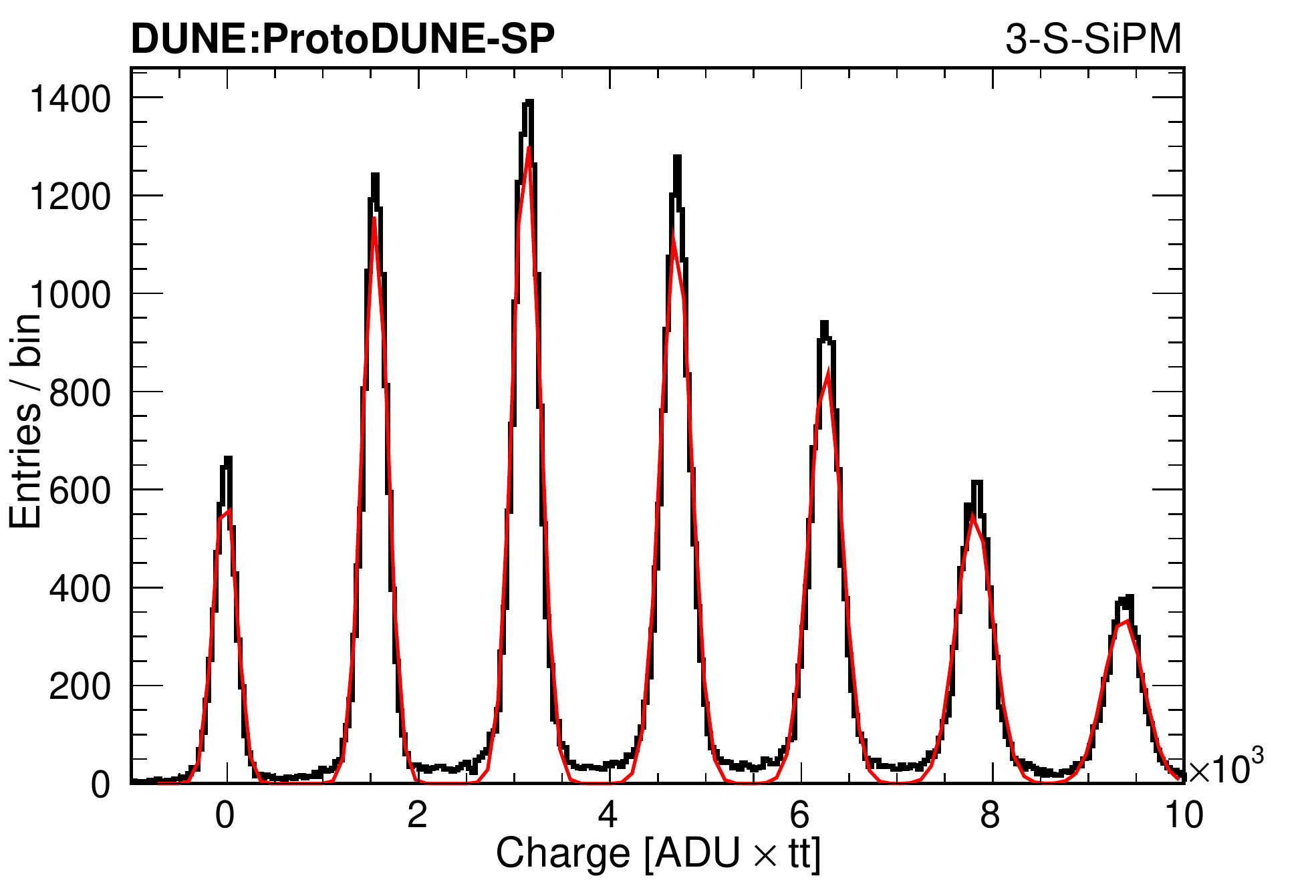}
\end{minipage}
\begin{minipage}[t]{0.5\textwidth}
\includegraphics[width=1.0\textwidth]{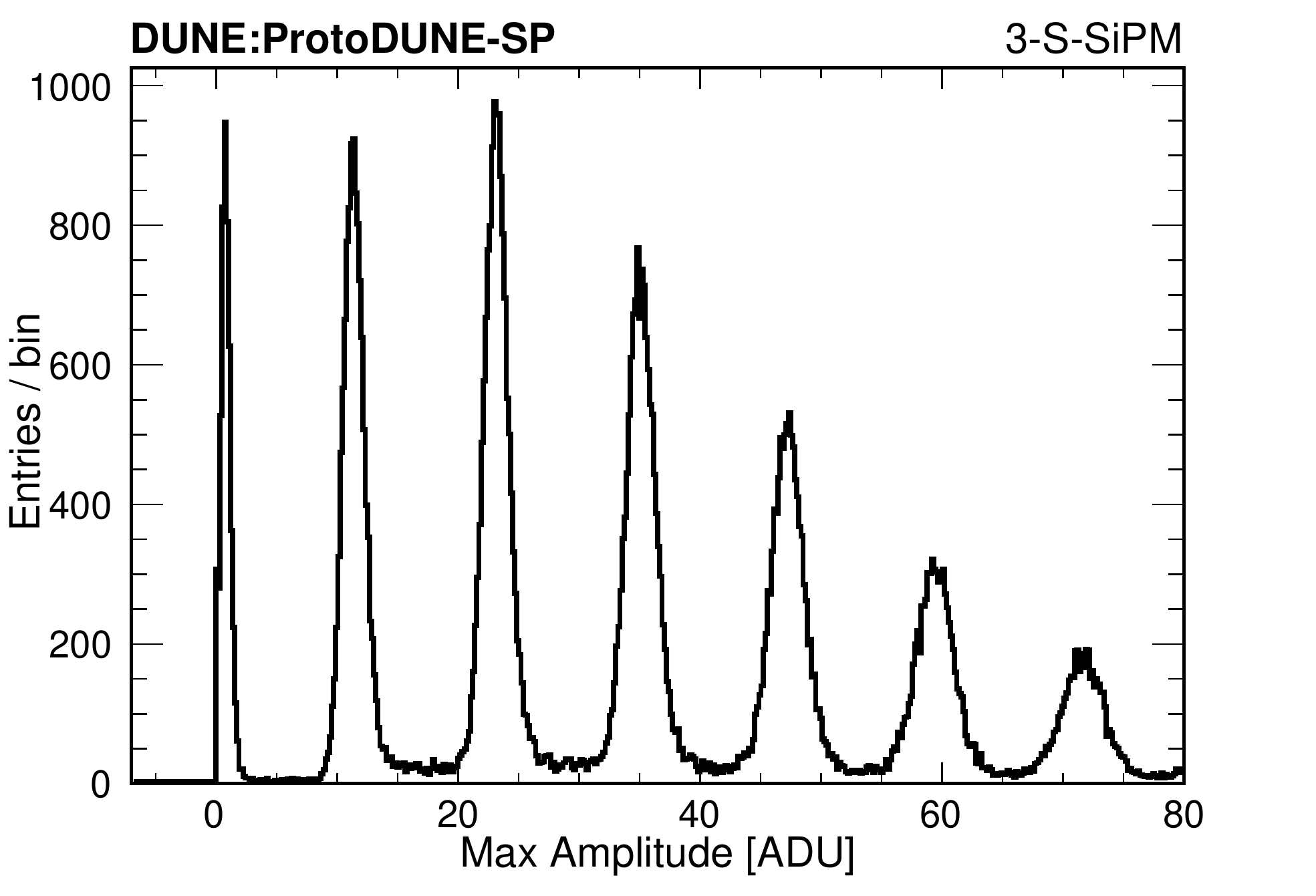}
\end{minipage}
\caption{Charge (left) and current  (right)  distribution for typical 12-H-MPPC channel ($V_{\rm{B}}=48~V$) (top)  and  3-S-SiPM channel ($V_{\rm{B}}=26~V$) (bottom) under low amplitude pulsed  LED illumination. A fit of the charge distributions with a multi-Gaussian function giving peak positions and widths is shown in red.
}
\label{fig:Q-and-I-MPPC-SensL}
\end{figure}

Typical charge (signal integral) and current (signal amplitude) distributions under pulsed LED illumination and with a nominal operating $V_{\rm{B}}$ setting are shown in figure \ref{fig:Q-and-I-MPPC-SensL} for a 12-H-MPPC channel (top row) and for a 3-S-SiPM channel (bottom row). 

The multi-peak structure corresponds to detection of 0, 1, 2, \textellipsis photoelectron-induced avalanches.
The clear peak separation confirms good sensitivity to single PE detection for both the three-sensor and the twelve-sensor channels. The spread around the peaks in the charge spectra (left panels) is partly due to over-voltage difference among sensors in the array. The asymmetric distributions around the peaks in the signal amplitude spectra (top right panel) may be due to secondary avalanches from afterpulses and crosstalk in the sensors more visible in the H-MPPC due to the faster recharge time.
The 1-PE charge directly measures the overall gain of the photosensor array at the applied bias voltage. The gain $g_i$ thus provides the charge per avalanche issued by the $i-$th channel.

\begin{figure}[tbh]
\begin{minipage}[t]{0.5\textwidth}
\includegraphics[width=1.0\textwidth]{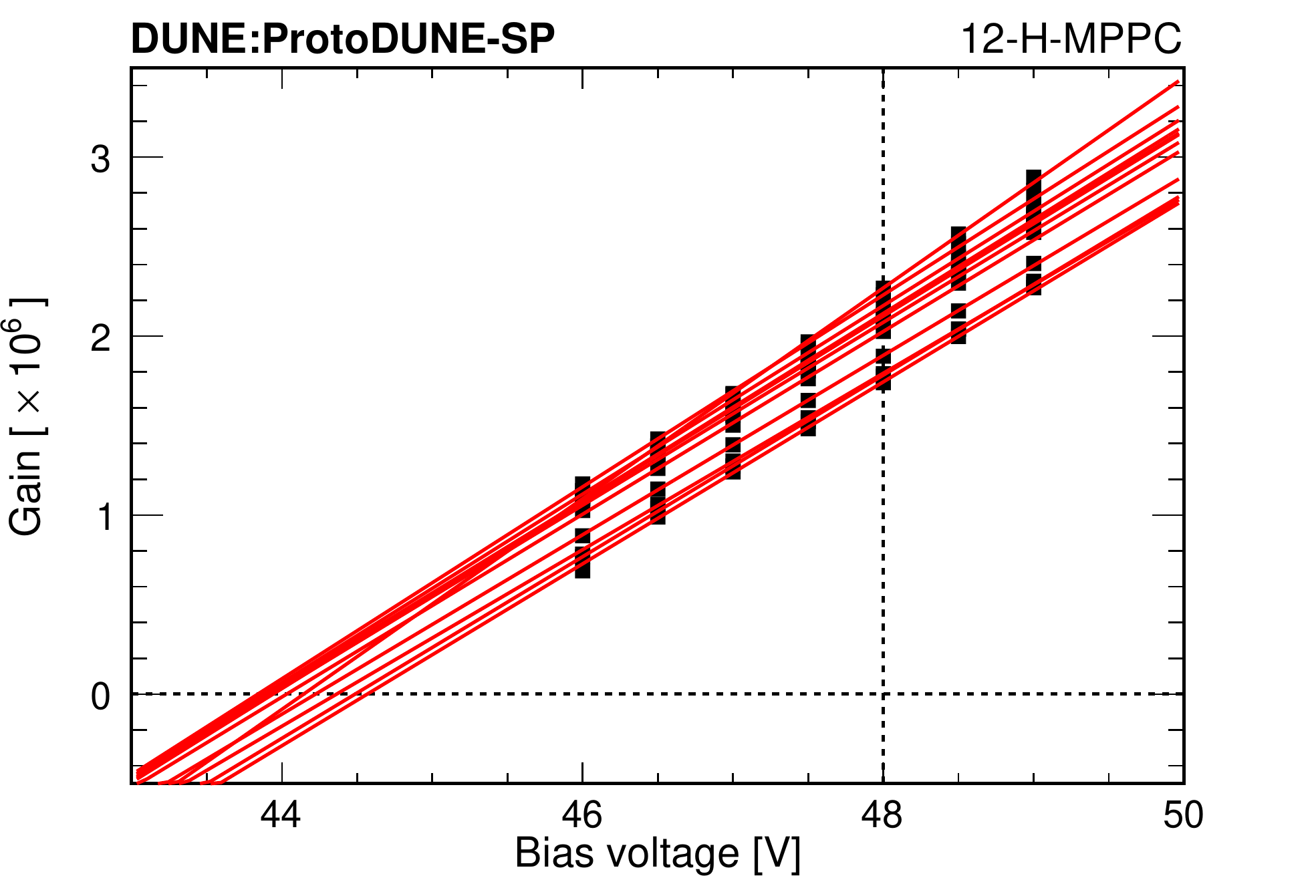}
\end{minipage}
\begin{minipage}[t]{0.5\textwidth}
\includegraphics[width=1.0\textwidth]{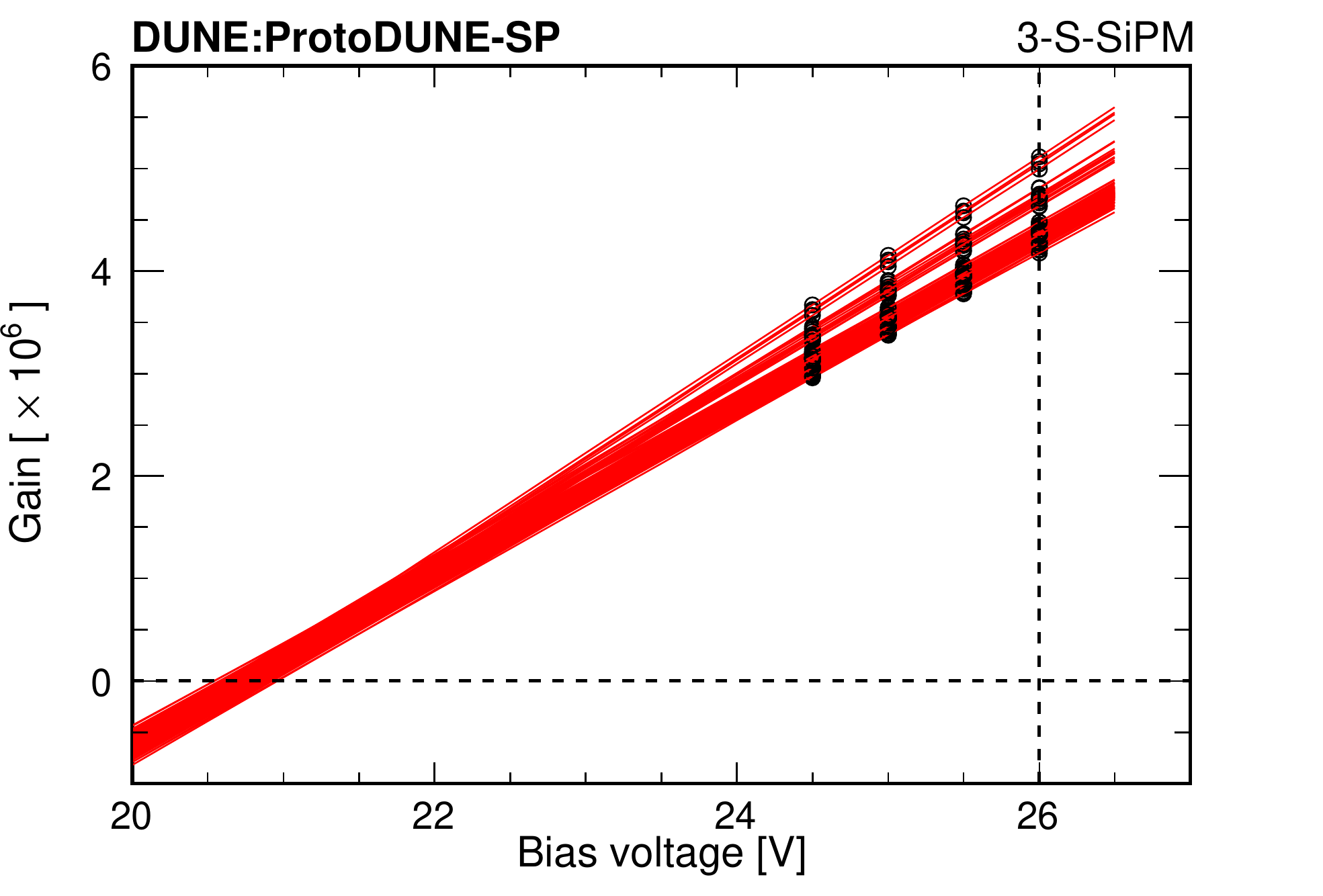}
\end{minipage}
\caption{Gain as a function of applied bias voltage for 12-H-MPPC channels (left), and for 3-S-SiPM channels (right). Linearity of individual channel response is shown by the linear fit  (red line) across the points at different bias voltage setting. The intercept of the fit line provides a direct evaluation of the breakdown voltage at LAr temperature for each 12-H-MPPC and 3-S-SiPM photosensor.}
\label{fig:gains-MPPCsSensLs}
\end{figure}

The gain as a function of bias voltage is shown in figure \ref{fig:gains-MPPCsSensLs} for some of the 12-H-MPPC channels and 3-S-SiPM channels. The gain response to varying $V_{\rm{B}}$ is very uniform channel by channel, as indicated by the slopes of the lines in figure \ref{fig:gains-MPPCsSensLs}. On the other hand, the intercept with the horizontal axis, that defines the actual breakdown voltage $V_{\rm{bd}}$ of the multi-sensor channel at LAr temperature, shows a relatively large spread, particularly for the 12-H-MPPC channels.

At the reference bias setting adopted for PDS operation ($V_{\rm{B}}=48$~V for the 12-H-MPPC, $V_{\rm{B}}=26$~V for the 3-S-SiPM - in either case within the range suggested by the different manufacturers) the two types of photosensors are operated at different over-voltages $V_{\rm{oV}}=(V_{\rm{B}}-V_{\rm{bd}})$ in the range (3.3 - 4.2)~V for the 12-H-MPPC channels and (5.0 - 5.5) V for the 3-S-SiPM channels. 
Gains are correspondingly higher, by a factor of $\sim 2$, for the 3-S-SiPM channels as indicated in figure \ref{fig:gains-MPPCsSensLs}.      




\subsubsection{Signal to noise in photosensors in passive ganging configurations}
\label{sec:StoN}
\begin{figure}[tbh]
\begin{minipage}[t]{0.33\textwidth}
\includegraphics[width=1.1\textwidth]{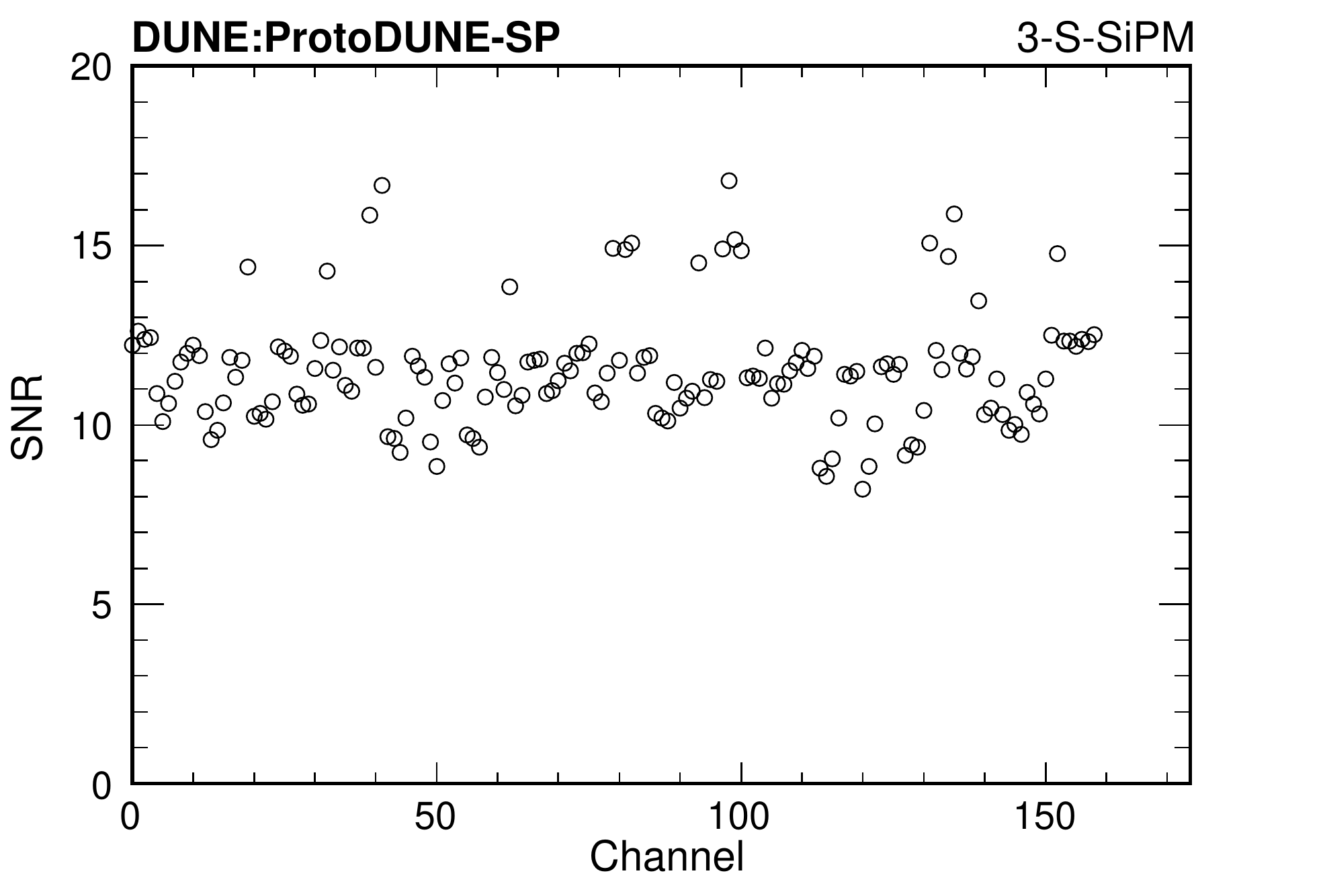}
\end{minipage}
\begin{minipage}[t]{0.33\textwidth}
\includegraphics[width=1.1\textwidth]{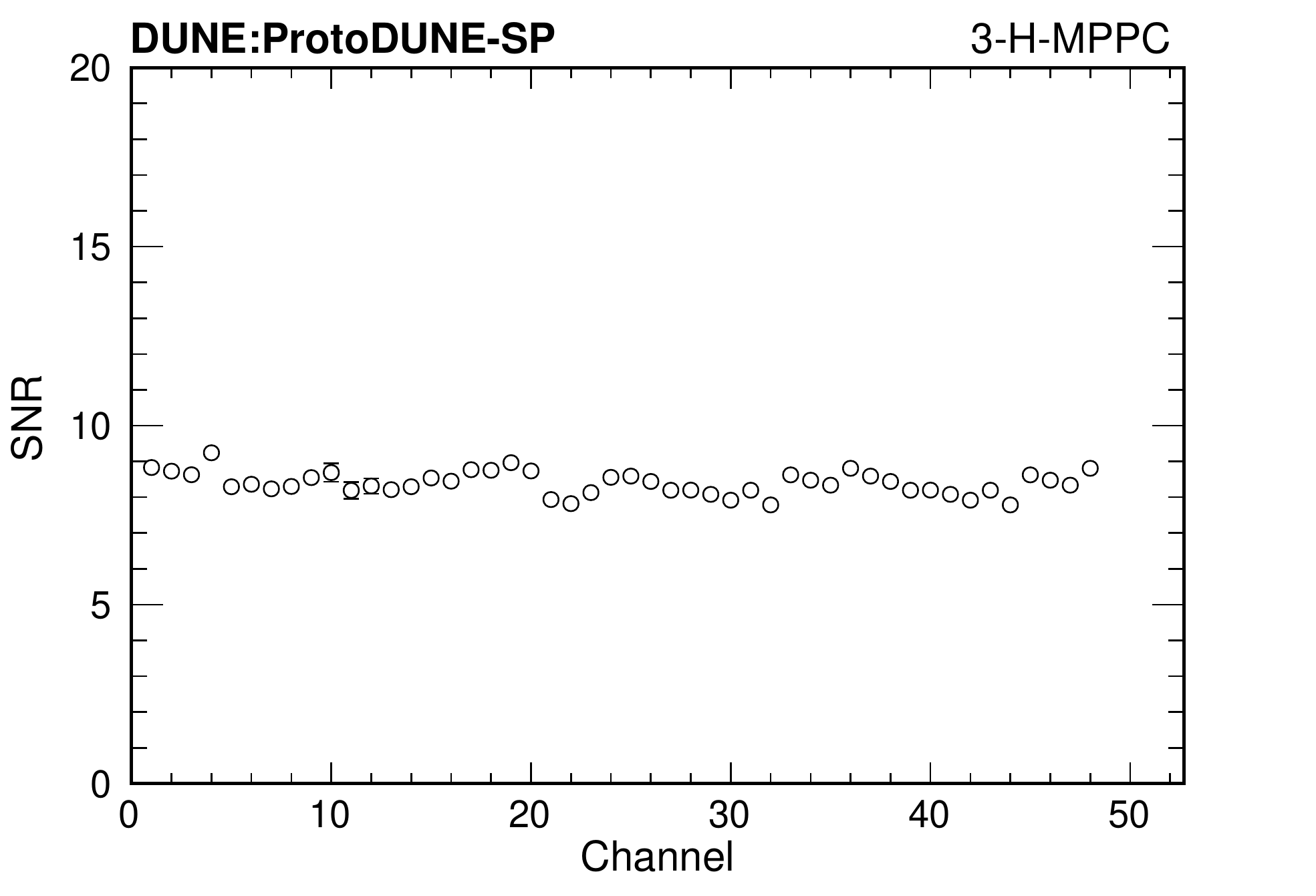}
\end{minipage}
\begin{minipage}[t]{0.33\textwidth}
\includegraphics[width=1.1\textwidth]{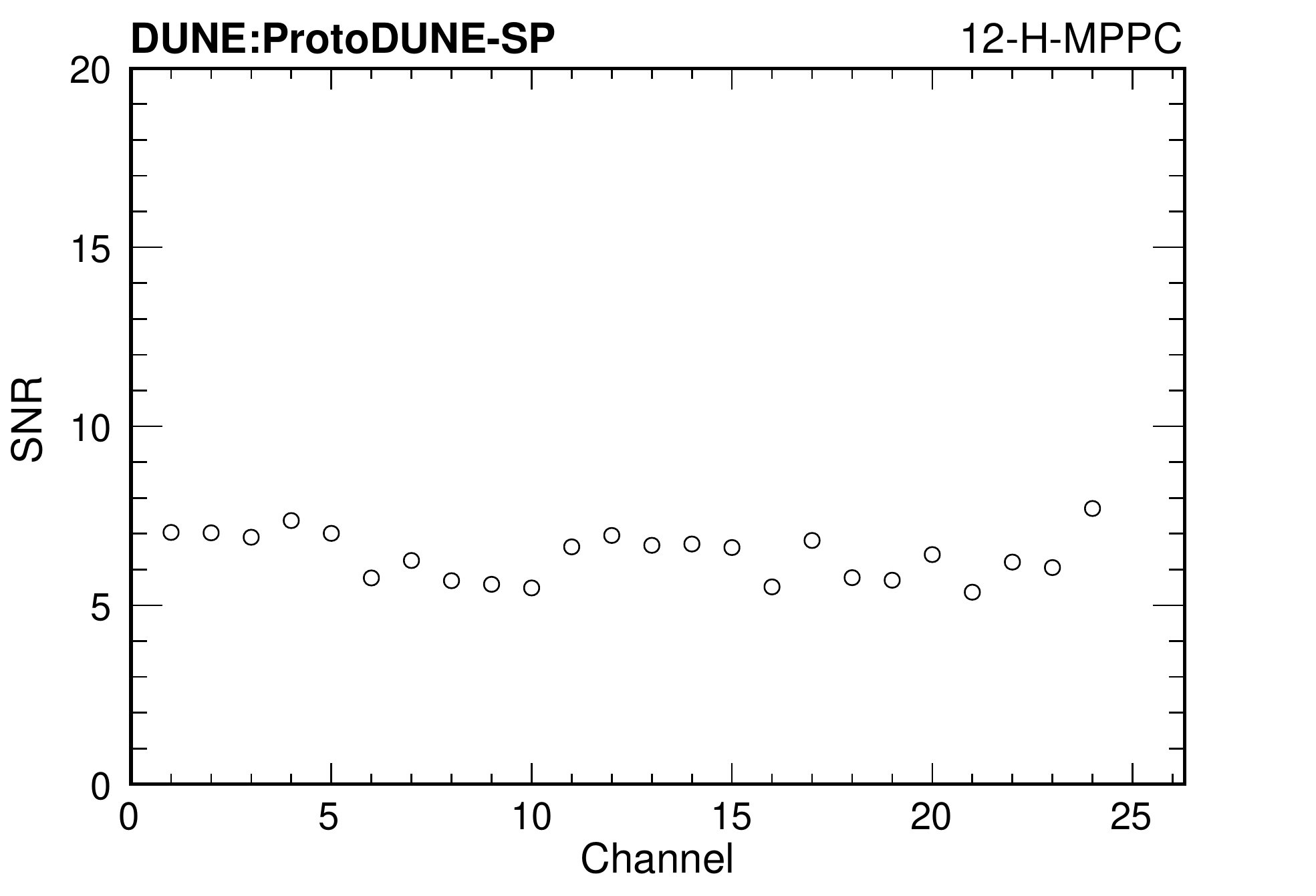}
\end{minipage}
\caption{Signal-to-Noise Ratio (SNR) for the 3-S-SiPM channels (left), for the  3-H-MPPC channels (center), and for the 12-H-MPPC channels (right).}
\label{fig:SNR-distribution}
\end{figure}

The signal-to-noise ratio (SNR) is a good performance metric for the characterization of the photosensor component of the PDS during normal operating conditions. The SNR of the individual channel (three or twelve photosensors in parallel) at the reference bias voltage setting is here defined as:
\begin{equation}
 {\rm{SNR}}~=~\frac{\mu_1}{\sigma_0}
 \label{eq:SNR}
\end{equation}
where, referring for example to figure \ref{fig:Q-and-I-MPPC-SensL} (left panels), the signal $\mu_1$ is the mean value from the Gaussian fit of the one-PE peak (the minimal detectable signal), and the noise is evaluated from the Gaussian spread, $\sigma_0$ of the zero-PE peak.  The SNR for all channels of the three types are shown in figure \ref{fig:SNR-distribution}. For the 12-H-MPPC channels of the ARAPUCA modules, the SNR values are around 6, while for the 3-S-SiPM channels of the double-shift and dip-coated bar modules the SNR is in the range 10 to 12. The signal-to-noise ratio, as defined in equation~\ref{eq:SNR}, is directly proportional to the gain and the higher SNR shown by the 3-S-SiPM channels is primarily due to their higher $V_{\rm{oV}}$ setting adopted for operation.


\subsubsection{Light calibration}
\label{sec:LightCalib}
Calibration of the light response is necessary to convert the charge signal from the photosensors into the corresponding number of photons detected. 
The detector calibration LED pulser is flashed synchronously with the data acquisition. Calibration runs at varying intensities of the flasher are needed in order to produce a suitable illumination in each PDS element. For each channel, the digitized waveform recorded in coincidence with the short LED pulse is baseline subtracted and the photosensor charge output is measured by waveform integration over a predefined time window (see section~\ref{sec:PDSDescription}). Typical charge distributions with multi-peak structure from a LED calibration run are shown in figure~\ref{fig:Q-and-I-MPPC-SensL} (left panels, top for a 12-H-MPPC channel and bottom for a 3-S-SiPM channel).  
Among the possible different calibration methods, the one adopted here relies on the statistical features of photon counting measurements under stable pulsed, low illumination conditions. 
The number of detected photons ($n$) per light flash follows the Poisson distribution with $\lambda$, the expected mean number of photons detected per flash, whose value is directly related to the probability of detecting 0-photons in that flash: 
\begin{equation}
P(n)=\frac{\lambda^ne^{-\lambda}}{n!} \qquad {\rm with} \qquad P(0)=e^{-\lambda} \qquad \to \qquad      \lambda=-\ln{P(0)}
\end{equation}
The probability $P(0)$ can be estimated by the relative frequency of detecting zero photoelectrons in many LED trials, and from this the mean number of photons detected per flash is inferred: 
\begin{equation}
     \lambda=-\ln{\left(\frac{N_0}{N_{\rm{Tot}}}\right)}
\label{Eq:lambda}
\end{equation}
where $N_0$ is the observed number of counts  under the zero-PE peak (1st peak in the charge distribution of figure \ref{fig:Q-and-I-MPPC-SensL} - left panels) and $N_{\rm{Tot}}$ is the number of LED flashes in the calibration run.

The mean number of photons per flash, $\lambda$, depends on the illumination level (LED flash amplitude). The illumination is maintained constant during the run and low enough to have sufficient probability of 0-photon detected.
In addition to this, the measured rate is corrected for the accidental background rate of environmental photons, which are not correlated to the LED flash, that may be detected in the trigger window.  The background rate is measured in the portion of the recorded waveforms before the LED trigger.

\begin{figure}[tbh]
\begin{minipage}[t]{0.5\textwidth}
\includegraphics[width=1.0\textwidth]{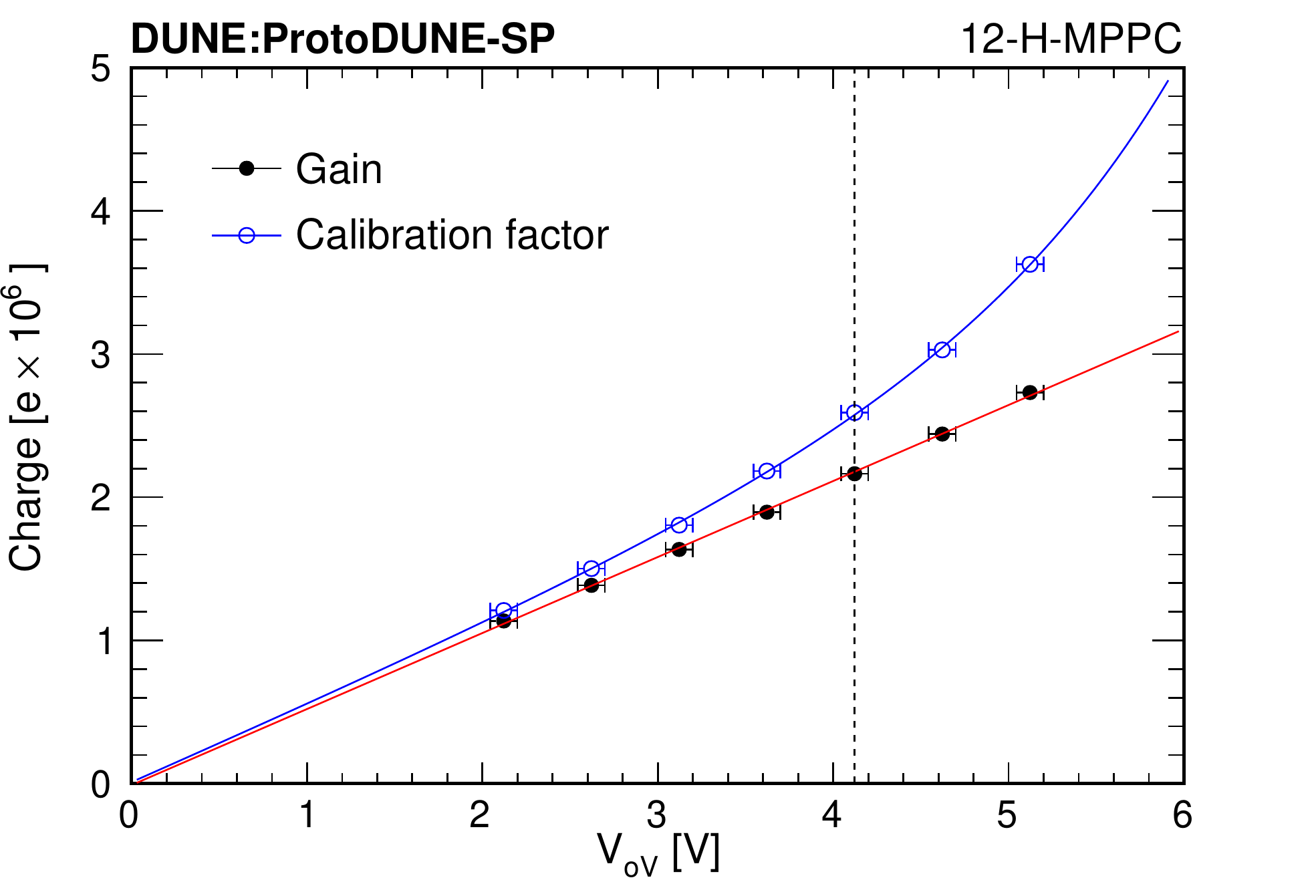}
\end{minipage}
\begin{minipage}[t]{0.5\textwidth}
\includegraphics[width=1.0\textwidth]{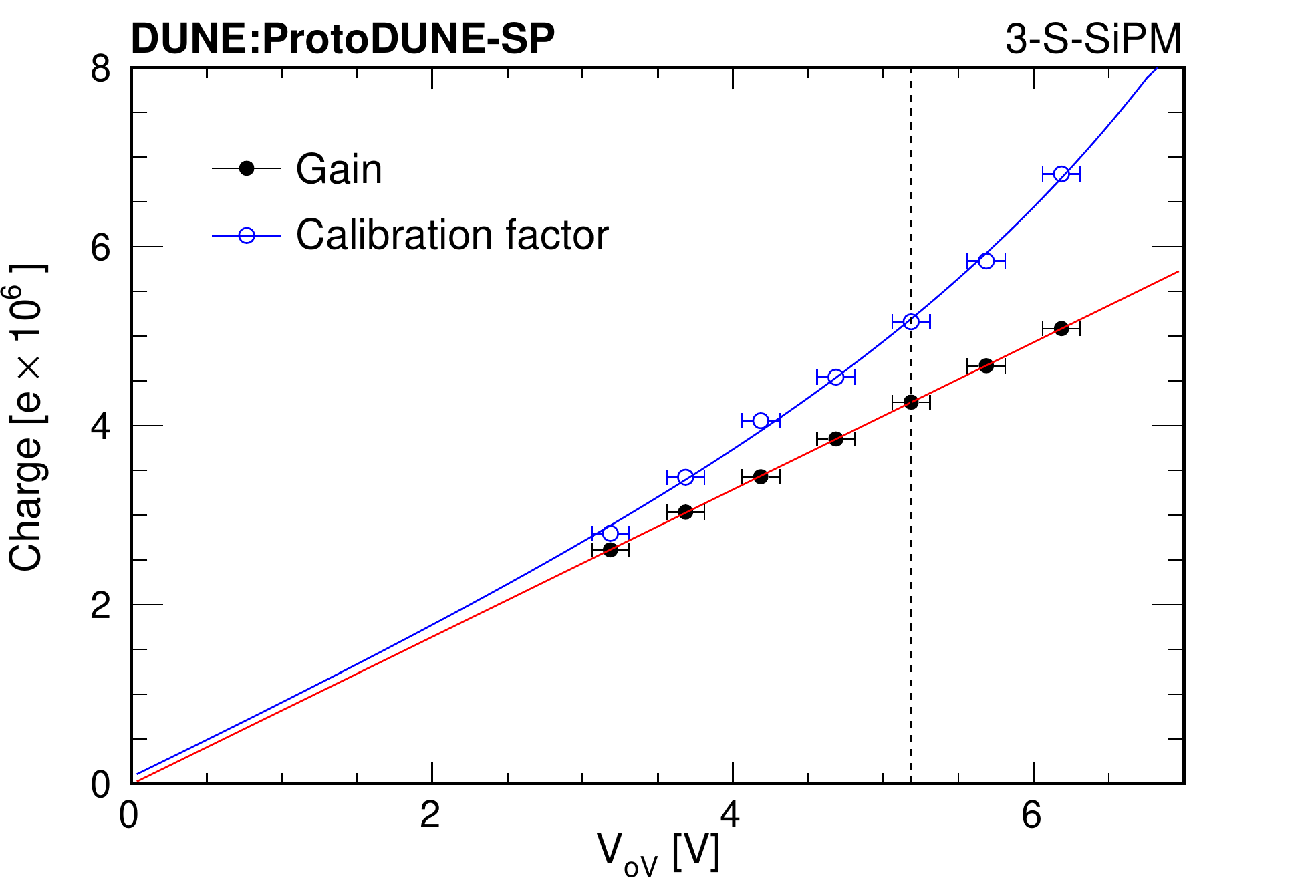}
\end{minipage}
\caption{Charge signal per detected photon (blue points) and charge signal per avalanche (black points) as a function of applied over-voltage $V_{\rm{oV}}$ for a typical 12-H-MPPC channel (left), and 3-S-SiPM channel (right). The difference is due to the correlated noise contribution, mainly from afterpulse and crosstalk in neighboring microcells of the photosensor. The vertical dotted line at the operation over-voltage set point indicates the gain $g_i$ and the calibration factor $c_i$  used in data analysis for the $i$-th channel shown in the figure.}
\label{fig:gain-and-calib}
\end{figure}
The photosensor response to $\lambda$ detected photons, estimated by equation~(\ref{Eq:lambda}), is the mean charge output per flash $\langle Q\rangle$ in the calibration run (average of the distribution shown in figure \ref{fig:Q-and-I-MPPC-SensL} - left panels). The calibration factor for the i-$th$ PDS channel is thus determined by the ratio $c_i={\langle Q\rangle}_i/{\lambda_i}$ and represents the output charge per photon detected by the individual photosensor channel.

The charge issued when an incident photon is detected is expected to be, on average, larger than the single-avalanche induced charge. The comparison of the charge per photon detected (calibration factor $c_i$ - blue line) and charge per avalanche (gain $g_i$ - red line) as a function of the applied over-voltage $V_{\rm{oV}}$ is shown in figure \ref{fig:gain-and-calib} for a typical 12-H-MPPC channel (left), and  3-S-SiPM channel (right). The difference is due to the correlated noise contribution to the signal formation in the photosensor. This is found to grow exponentially with increasing voltage.






\subsubsection{Afterpulses and crosstalk}
\label{sec:AP-CT}
A common feature of Si-photosensors is the generation of avalanche pulses subsequent to a primary event. The avalanche of a single microcell in a device has a finite probability of inducing an avalanche in neighboring microcells (optical crosstalk), or/and of re-triggering itself before the microcell is fully recovered (afterpulse). 
\begin{figure}[tbh]
\begin{minipage}[t]{0.33\textwidth}
\includegraphics[width=1.1\textwidth]{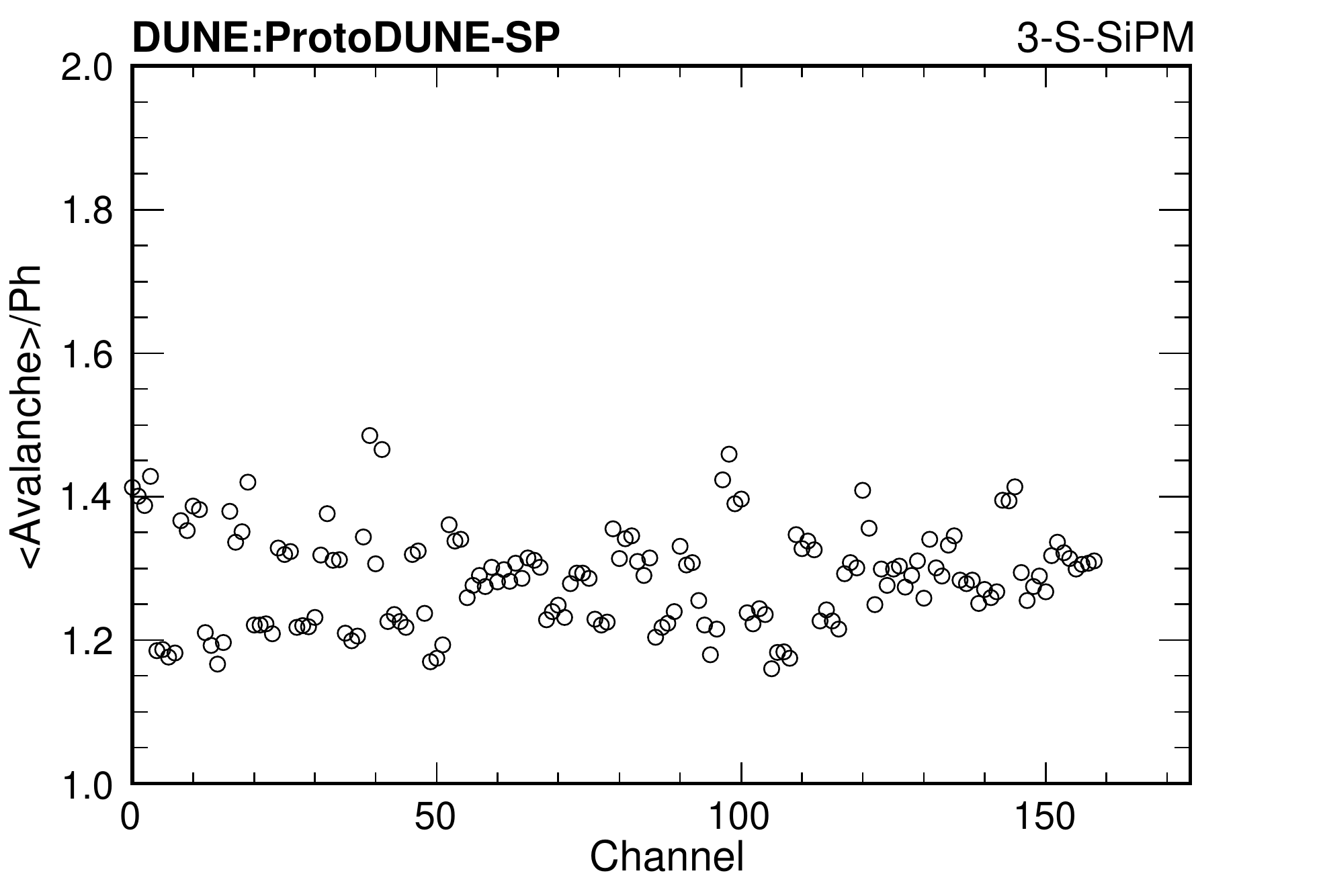}
\end{minipage}
\begin{minipage}[t]{0.33\textwidth}
\includegraphics[width=1.1\textwidth]{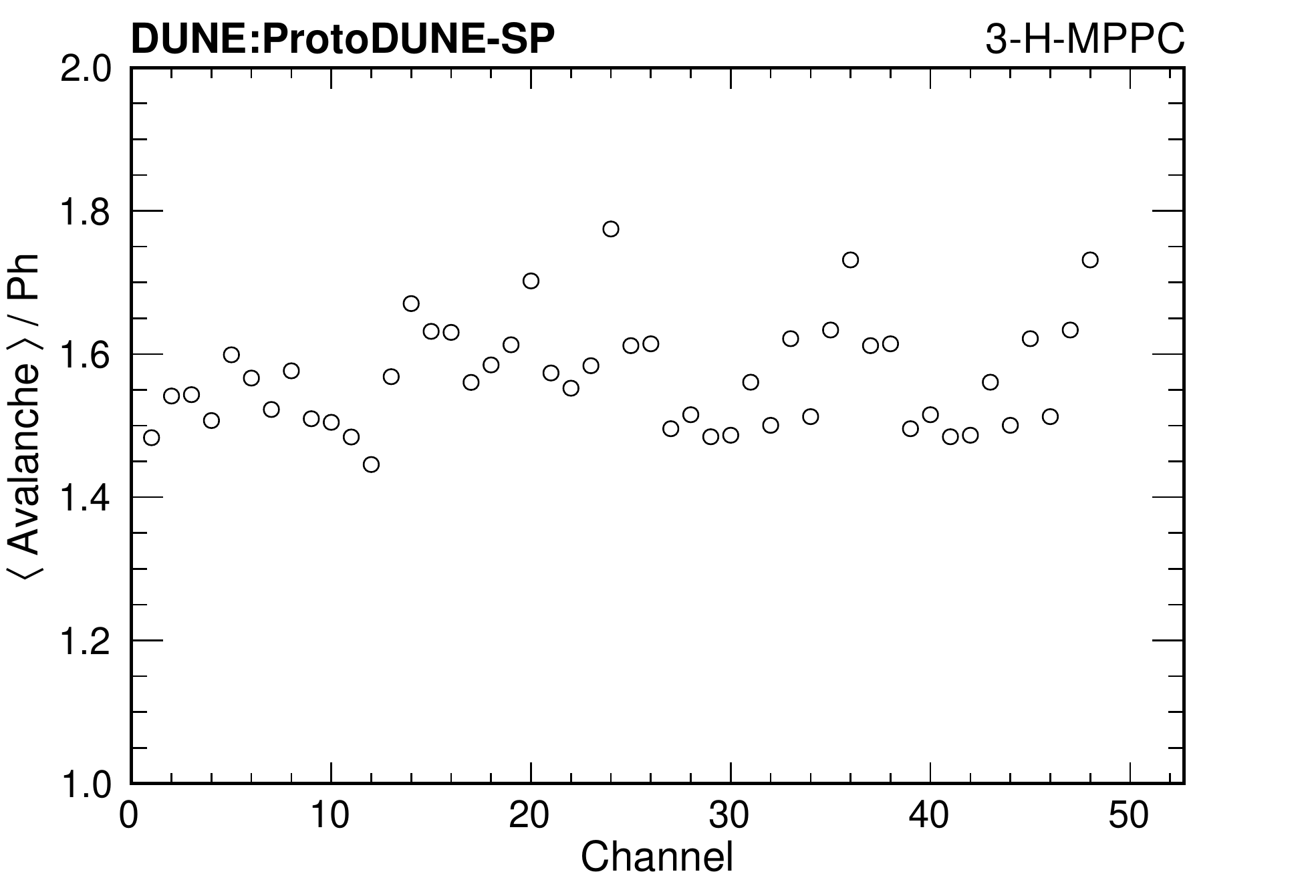}
\end{minipage}
\begin{minipage}[t]{0.33\textwidth}
\includegraphics[width=1.1\textwidth]{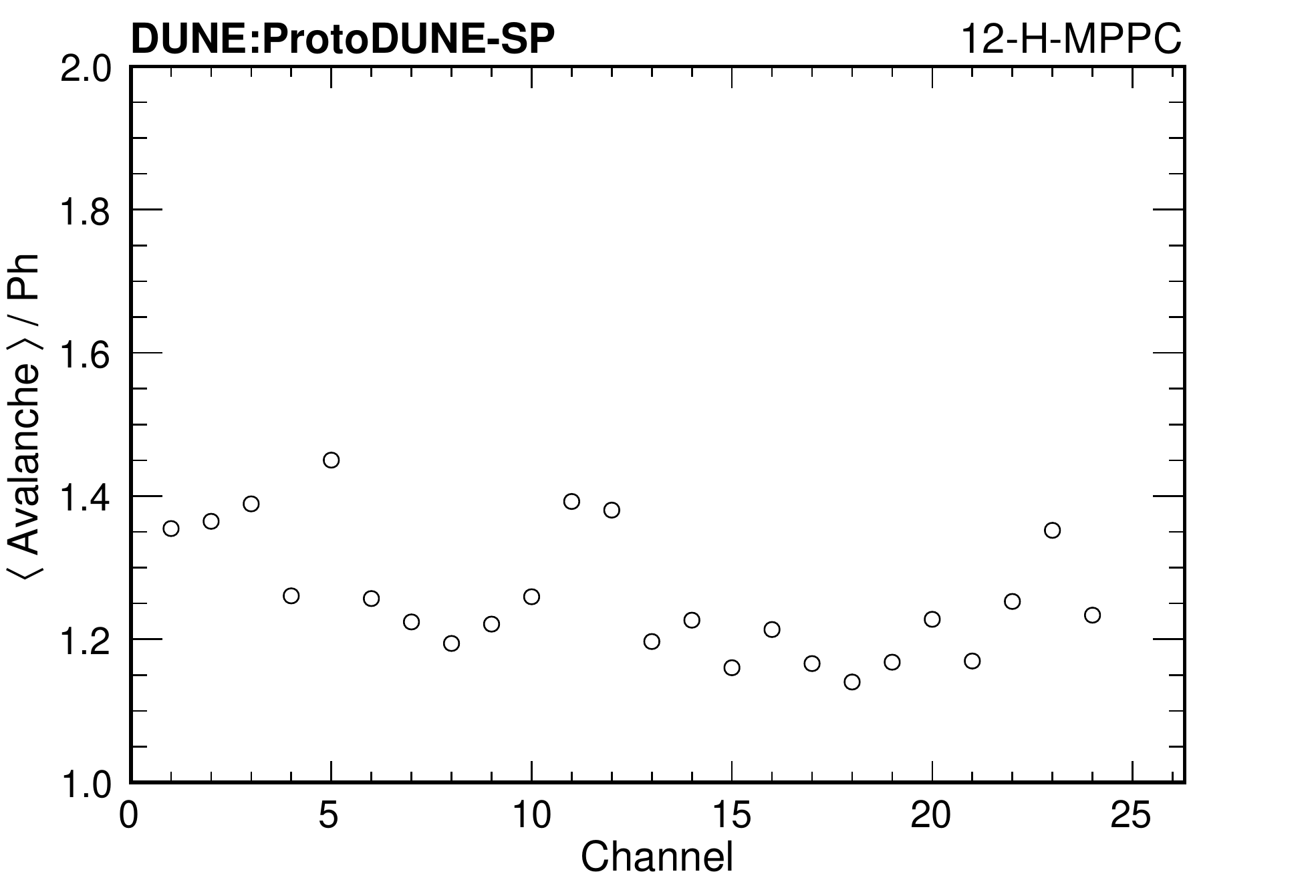}
\end{minipage}
\caption{Afterpulse and crosstalk contribution to the photosensor signal expressed by the average number of avalanches generated per detected photon for the 3-S-SiPM channels (left), for the  3-H-MPPC channels (center), and for the 12-H-MPPC channels (right).}
\label{fig:PHperAvalache}
\end{figure}

The rate of these secondary pulses increases at higher gain settings. The correlated noise due to these effects is a well-known limiting factor for a precise photon counting with silicon photosensors. 
The measurement of the charge per photon detected (the calibration factor $c_i$, defined in section~\ref{sec:LightCalib}) and the charge per avalanche (the gain $g_i$, defined in section~\ref{sec:SPESensitivity})
allows the calculation of the crosstalk and afterpulse probability for each photosensor by the ratio $c_i/g_i$ in units of [Ava/Ph], average number of avalanches per photon detected. 
This ratio is sensitive to the over-voltage on the photon detector and can be used to monitor for changes in the operating characteristics of the photosensor as a measure of the stability of the PD system.
Figure \ref{fig:PHperAvalache} shows the measured avalanche/photon value for each channel in the PDS. An average of $\sim$1.3~Ava/Ph is found for the 3-S-SiPM channels and the 12-H-MPPC, while a larger factor $\sim$1.6~Ava/Ph characterizes the 3-H-MPPC channels.

\subsubsection{Response stability over time}
\label{sec:stabTime}
\begin{figure}[tbh]
\begin{minipage}[t]{0.50\textwidth}
\includegraphics[width=1.0\textwidth]{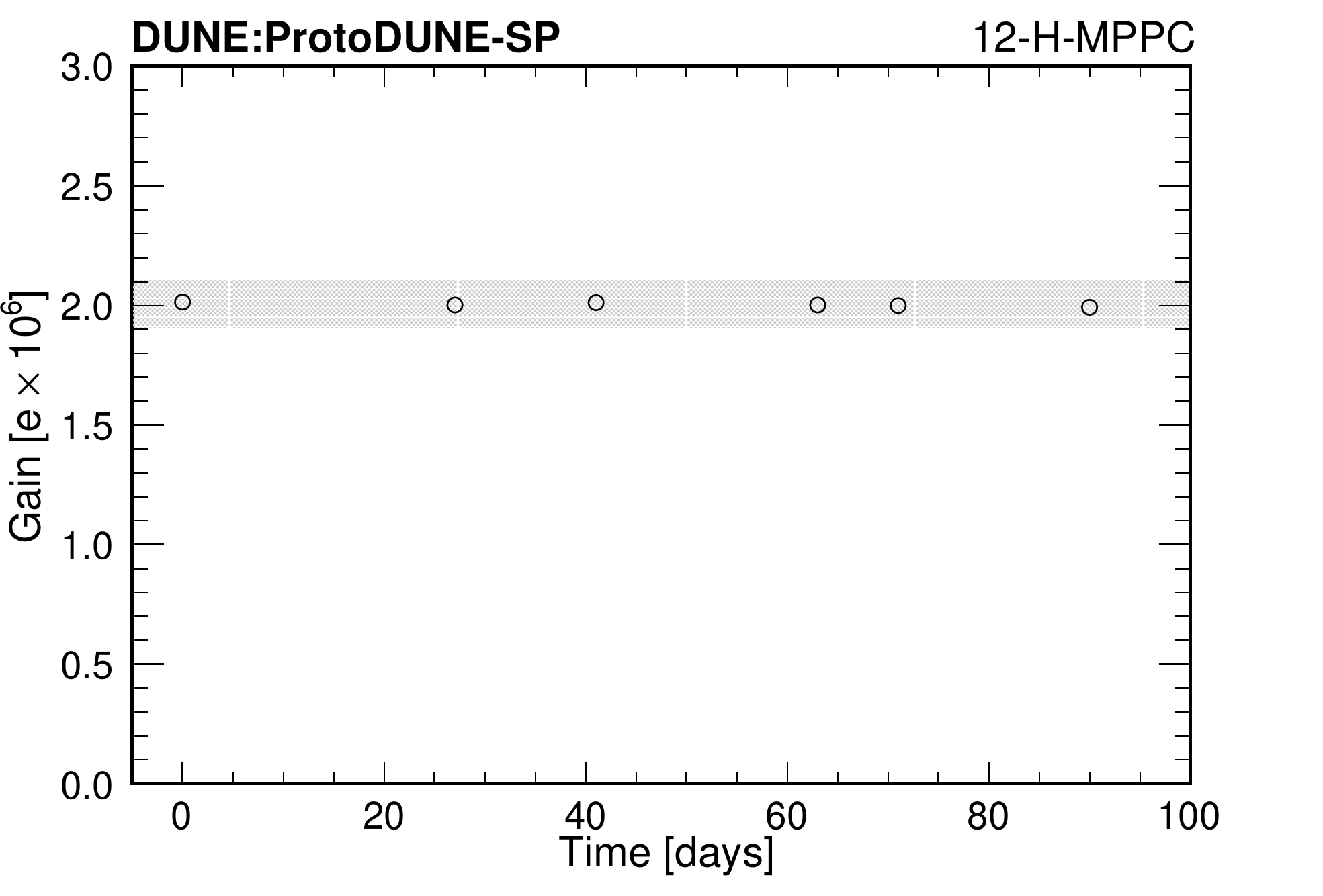}
\end{minipage}
\begin{minipage}[t]{0.50\textwidth}
\includegraphics[width=1.0\textwidth]{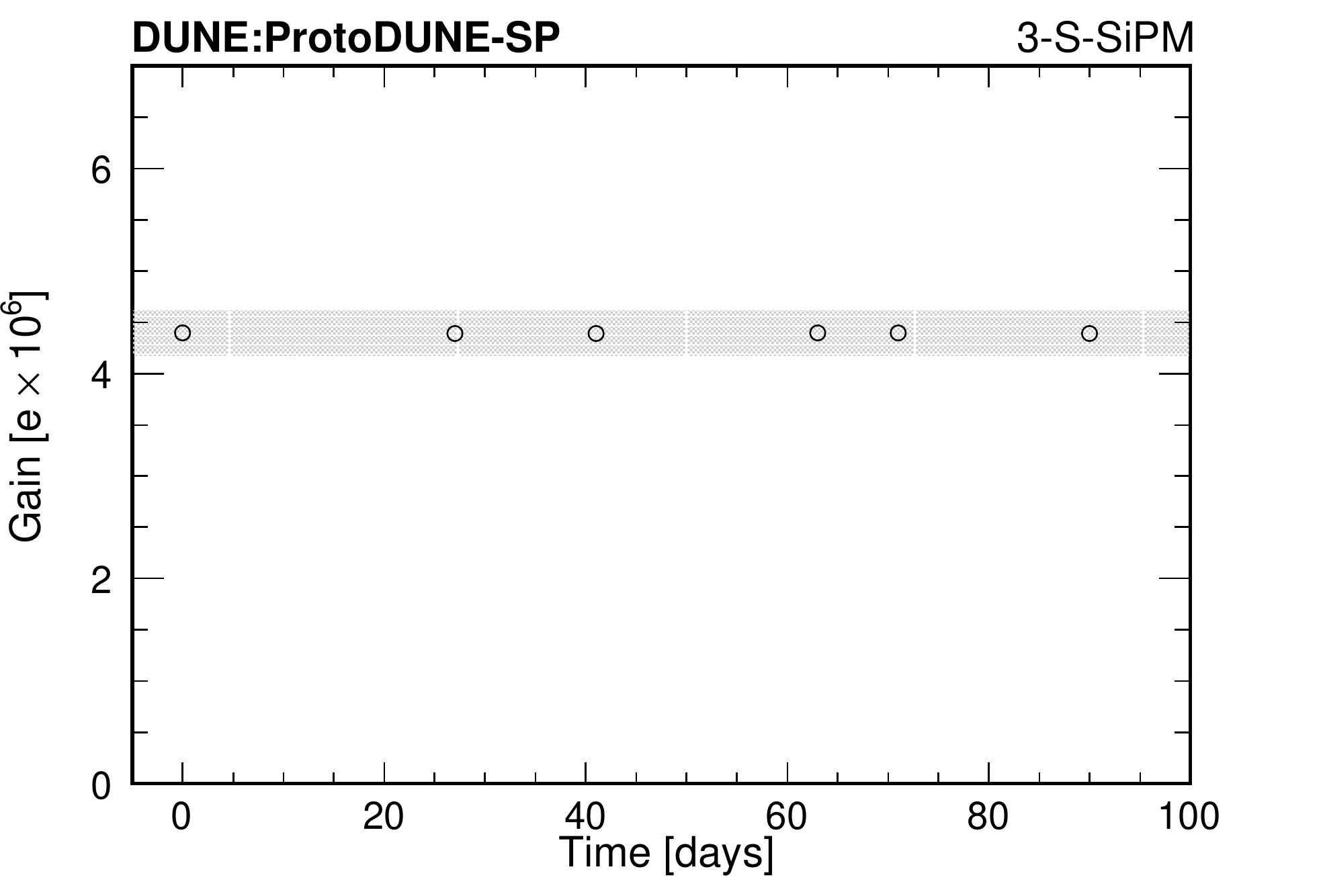}
\end{minipage}

\caption{Stability of the photo-sensor response over time: 
gain stability (charge signal per avalanche) for typical 12-H-MPPC channel (left) and 3-S-SiPM channel (right), from calibration runs performed over $\sim$ 100 days of operation. The shaded band corresponds to a $\pm5\%$ gain interval. Gain variations over time are contained well within the band. Statistical error bars are small, not visible inside the symbol}
\label{fig:sensorStability}
\end{figure}

The sensor gain, the calibration factor and the size of the afterpulse and crosstalk component of the signal can be used as a system monitor.  Any drift in these parameters is an indication of instability in the system.  
The calibration data taken at various times during operation provide measurements that indicate the system stability as a function of time.  Figure~\ref{fig:sensorStability} shows the value of the gain for typical 12-H-MPPC channels and 3-S-SiPM channels over the course of several months.  Within the uncertainties of the measurements, neither the gain nor the other parameters were found to be drifting over time for any of the sensors used in the ProtoDUNE-SP photon detector system.


\subsection{Photon detector performance}
\label{sec:PDPerf}
The PD modules (ARAPUCA and light-guide bars) are exposed to scintillation light from ionization events in the drift volume. A fraction of the emitted photons impinge upon the optical surface of any given PDS module, and a charge signal is issued, proportional to the number of photons detected by the photosensors of the module. The detection efficiency $\epsilon_D$ of a PDS module is defined here as the ratio of detected photons to impinging photons. Test-beam data from particles of known type, energy and incident direction in the LAr volume are used to determine $\epsilon_D$ and thus evaluate the performance of the different detection technologies implemented in the PDS. 

\begin{figure}[!htbp]
\centering
\includegraphics[width=0.75\textwidth]{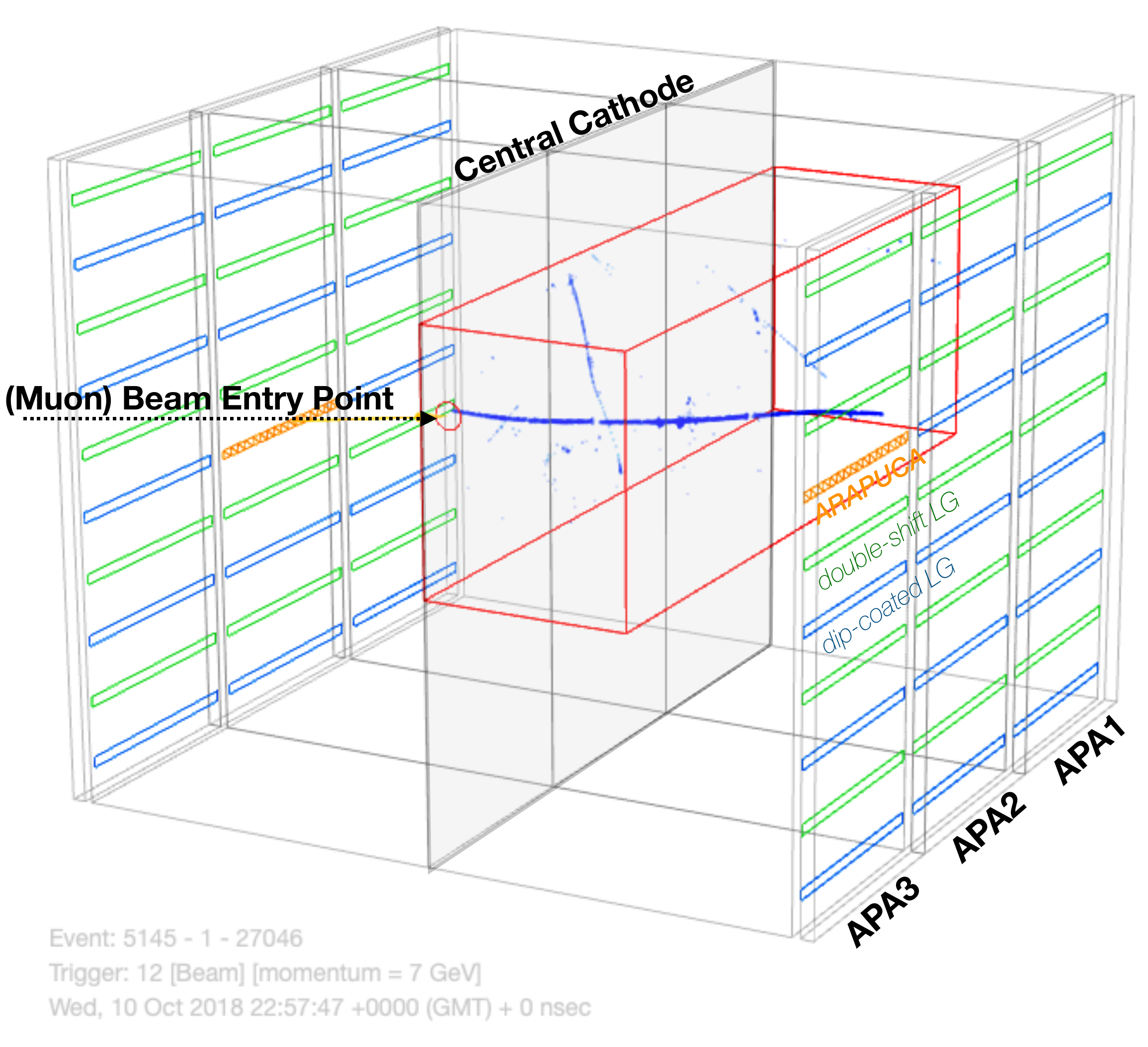}
\caption{3D event display made with the Wire-Cell BEE display~\cite{ref:bee_evd} showing data from a 7~GeV/$c$  beam muon crossing the whole TPC volume, fully reconstructed by the LArTPC. Only tracks inside a predefined sub-volume (red box) are shown. The beam muon track enters near the cathode and propagates along the beam direction about $10^{\circ}$ downward and $11^{\circ}$ toward the anode plane. Ten bars are located in each APA frame. The dip-coated bars are indicated in blue, the double-shift bars in green and the segmented ARAPUCA bar in orange.}
\label{fig:TPC-PDS-3D}
\end{figure}

For each beam event, the number of detected photons $N^{\rm{Det}}_j$ is evaluated from offline data reconstruction (baseline subtraction, waveform integration and charge-to-photon conversion) for each of the 29 
light-guide bars and for each of the 12 cells of the ARAPUCA module in the beam side 
of the PDS.  A Monte-Carlo simulation of test beam events is used for extracting the corresponding number of photons incident $N^{\rm{Inc}}_j$ on each PDS element (light-guide module or cell in the ARAPUCA module).  This simulation is performed with the LArSoft toolkit~\cite{Church:2013hea}, which has 
a detailed description of the geometry of the ProtoDUNE-SP detector, including a proper description of the materials and the positions of the TPC components surrounding the LAr volume (APA, CPA, FC, as shown in figure \ref{fig:TPC-PDS-3D}). 

Simulation of beam events is performed with standard {\textsc{Geant4}}/LArG4 generator within LArSoft. This accounts for the well known features of scintillation in liquid argon ensuing ionization processes.

{\it Photon emission}: The emission spectrum is a narrow band in the Vacuum-UV (VUV) wavelength range peaking around $\lambda = 128$~nm (FWHM$\simeq 6$~nm), exponentially distributed in time with two very different time components (fast $\sim 5$~ns and slow $\sim 1.3-1.4~\mu$s, with intensity ratio $0.3$ in case of minimum ionizing particles). Electric fields applied to the LAr medium  affect the intensity of scintillation emission. At 500~V/cm (ProtoDUNE-SP operation), a photon yield of $2.4\times 10^4$~photons/MeV for minimum ionizing particles is assumed in simulations, 60\% of the maximum yield measured at zero field. The relative uncertainty on the photon yield value is 8.5\%~\cite{Doke:2002oab}.  The photon yield dependence on increasing linear energy transfer, the rate of energy deposited by ionizing particles, is not included in the current simulation.

{\it Photon propagation}: LAr is transparent to its own scintillation light. However, during propagation through LAr, VUV photons may undergo Rayleigh scattering, absorption from residual photo-sensitive impurities diluted in LAr and reflections at the boundary surfaces that delimit the LAr volume. In the MC simulation, the Rayleigh scattering length, the reflectivities of materials for VUV photons, and the absorption length as a function of the impurity concentration are parameters that are fixed at their best estimates from existing data.  
Tracking each of the large number of VUV photons emitted in an event using {\textsc{Geant4}} is computationally expensive, so a pre-computed optical library is used to look up the probability that a photon produced at a particular location in the liquid argon volume is detected by a specific PDS channel.  In order to create the optical library, the liquid argon volume is segmented into small sub-volumes (voxels) of size $\sim 6~\times$ 6 $\times$ 6 cm$^3$.  For each voxel,  a large number (of order $5\times 10^5$) of VUV photons is sampled with an isotropic angular distribution.  All photons are tracked using {\textsc{Geant4}}, recording how many reach the sensitive area of each optical detector. In the simulation used to create the optical library, the Rayleigh scattering length for VUV photons in liquid argon is assumed to be 90~cm, according to the most recent experimental determination ~\cite{Babicz:2018gqv,Babicz:2020den}.
Due to the high level of purity during the beam run (Oxygen equivalent impurity concentration < 100 ppt), absorption by impurities is assumed to be negligible. Light reflection at VUV wavelength is  low for perfectly polished metal surfaces (20\% or less) and effectively null for any other material. The actual reflectance of the (extruded, non polished) Al profiles of the field cage surrounding the LAr drift volume is unknown and therefore it is set to zero in the current simulations. Once the library is created, ProtoDUNE-SP detector simulation jobs retrieve information from the library when the trajectory of a ionizing particle in the LAr volume is simulated by {\textsc{Geant4}}, converting the number of emitted photons from energy deposited in each voxel directly into the number of photons impinging upon the area of each PD module coming from this given voxel. The uncertainty on the rate at which photons arrive at the detector after photon transport is dominated by the uncertainty on the Rayleigh scattering length.  Neglecting reflections at the LAr volume boundaries is expected to be a subdominant effect. A relative uncertainty of 5\% is assigned to the number of photons incident on the detector surface by varying the Rayleigh length by 20\% around the nominal value in the simulation.  The uncertainty on the possible bias due to the photon library parameterization is not included in the total uncertainty. 
\begin{figure}[!htbp]
\centering
 \includegraphics[width=0.85\textwidth]{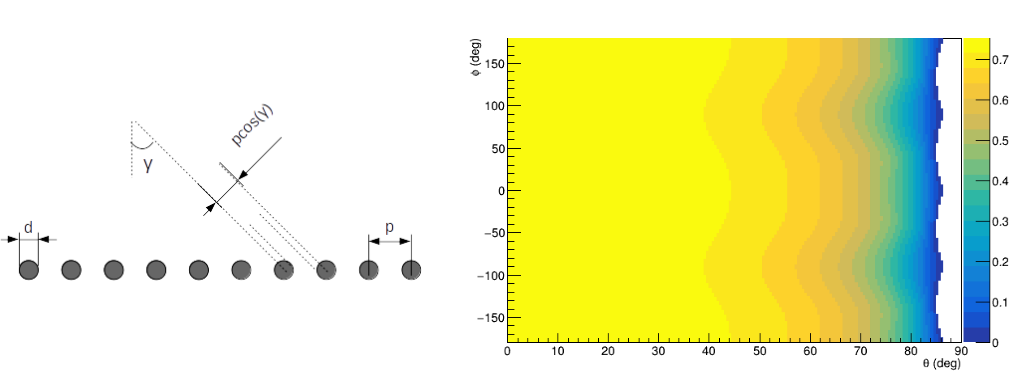}
\caption{Schematic diagram illustrating photons impinging on a TPC wire plane (left)  where the wire pitch $p$, the wire gauge $d$, and the incident photon angle ($\gamma$) are defined. On the right, the map of transmission -- color scale from 0 to 1 -- through the set of parallel planes (TPC wire planes and the mesh) as a function of the polar angles $\theta,~\phi$ of the
incident photon direction (the planes lie in the $(y,z)$ plane, $\theta=\gamma$ when $\phi=\pm\pi/2$).}
\label{fig:wireLayerModel}
\end{figure}

{\it Photon transmission at the anode plane}: The optical surfaces of the PD modules lie immediately behind the four wire planes of the TPC and a fifth (grounding) plane made by the woven metallic mesh stretched across the APA frame (see section~\ref{sec:det_tpc}).
A correction is applied to account for the light transmission through this series of parallel planes, which is not included in the detector simulation in LArSoft. The geometrical transparency of the mesh (percentage ratio of opening to total area, function of wire gauge and pitch) is 85\% The transparency is reduced to 75\% when the TPC wire planes above the mesh plane are also considered. This corresponds to the transmission upper value for orthogonal incident light. 
Transmission at any angle is then obtained based on a geometrical model\footnote{A simplified geometrical model is used in which VUV photons intercepting a wire of the mesh are absorbed (no reflection). The transmission coefficient shows a dependence on the polar angle of incidence ($\theta$) almost flat with T=0.75-0.7 for photons incoming with $\rm \theta < 45^{\circ}$ and then decreasing above that angle. Only a small modulation in the azimuthal angle $\phi$ is expected across the whole range due to the geometrical orientation of the wires and mesh planes ($\rm \phi=\pm 35.7
^{\circ},~0^{\circ},~90^{\circ}$) and the gauge per pitch ratios ($d/p$).}, as illustrated in figure \ref{fig:wireLayerModel}.
A stand-alone simplified MC simulation is then performed to evaluate the transmission of light from beam events.  Optical photon emission is sampled over straight trajectories crossing the LAr volume along the beam direction, nearly representing beam muon tracks, or sampled according to a spatial parametrization of electromagnetic showers, representing the longitudinal and transverse energy deposition from incident beam electrons. 
After photon propagation to the APAs, the angular distribution of incident photons on each PD module is folded with the transmission map to obtain the transmission coefficients for beam muons and beam electrons. These were found in the 65-71\% range, mildly depending on the position of the PD module. The relative uncertainty  on the transmission coefficients is evaluated to be 7\% (one-sided) to account for the simplified assumptions in the model (no reflection).
The transmission coefficients for each module so determined are then used to scale down the number of photons arriving at the APA from the {\textsc{Geant4}} MC simulation of the beam events into the actual number of photons $N^{\rm{Inc}}_j$ incident on the surface of the PD module behind the APA.

\subsubsection{Efficiency}
\label{Sec:Efficiency}
Runs with beam momentum settings from 2 to 7~GeV/$c$ are considered for the efficiency study. The muon and electron samples for the runs at different momenta are selected using the PID information from the beam instrumentation and the recorded light signals passing quality cuts are fully reconstructed (${\mathcal O}$(10k events/sample) for each run). Correspondingly, Monte Carlo runs were generated with muons or electrons entering the TPC volume from the beam-plug with the same momentum (nominal value and spread) and direction to reproduce the features of the H4-VLE beam line (see section \ref{sec:beamline}). The MC samples were generated with the same number of triggers as were collected in the corresponding data samples.
For each run the MC distribution of the number of photons in the event impinging upon the $j$-th PDS element (light-guide module or ARAPUCA cell) and the distribution from real data sample of photons detected by the same module/cell are extracted. 

\begin{figure}[tbh]
\begin{minipage}[t]{0.50\textwidth}
\includegraphics[width=1.1\textwidth]{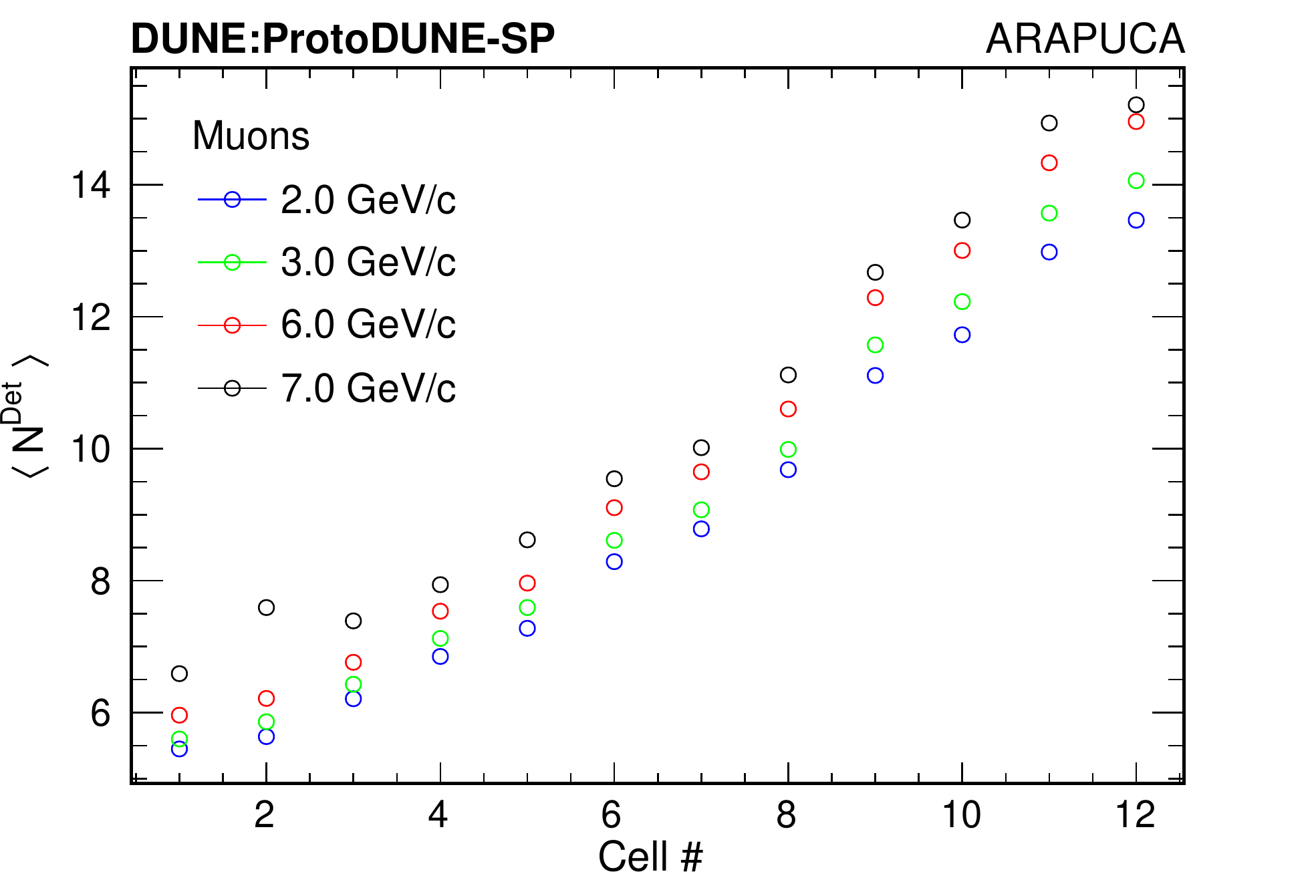}
\end{minipage}
\begin{minipage}[t]{0.50\textwidth}
\includegraphics[width=1.1\textwidth]{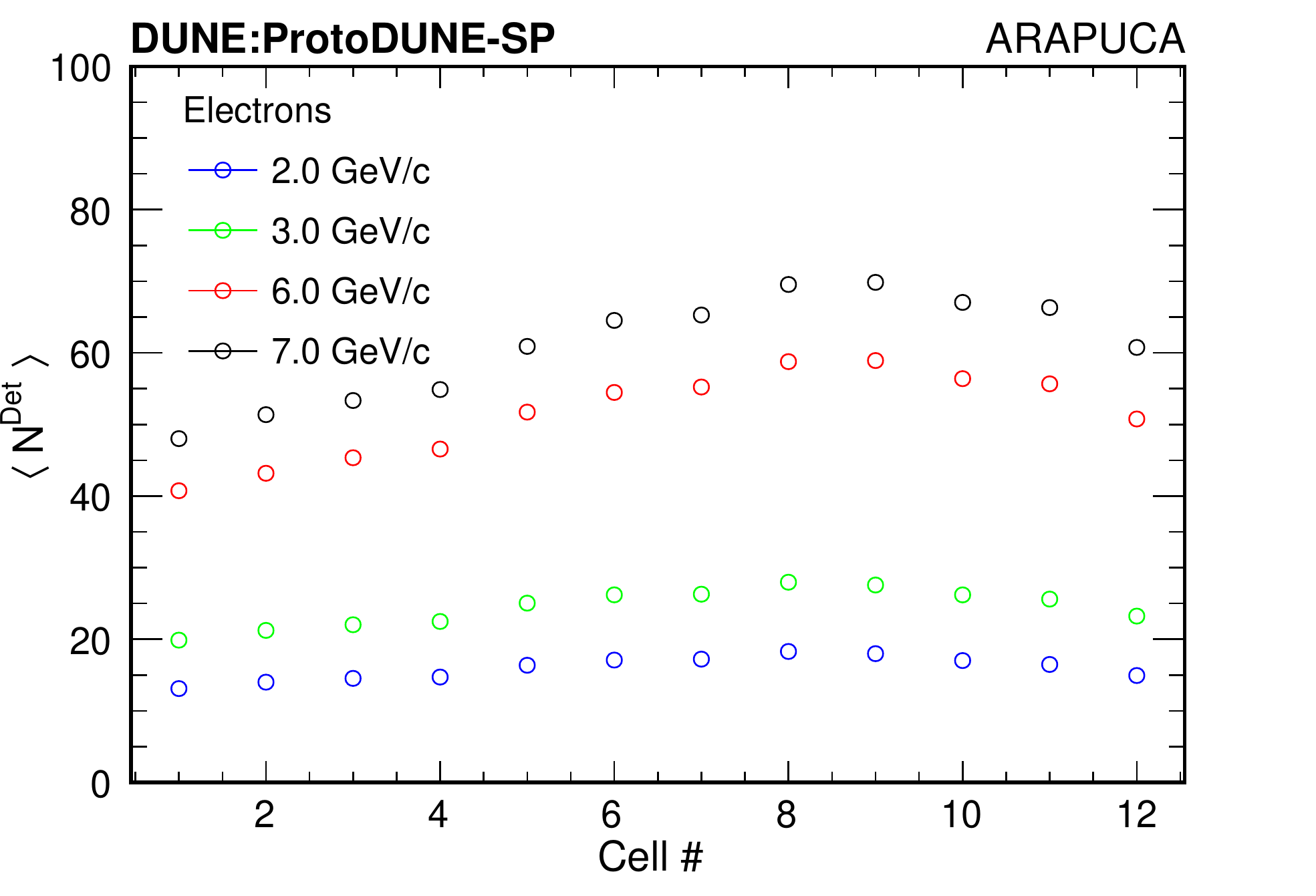}
\end{minipage}

\begin{minipage}[t]{0.50\textwidth}
\includegraphics[width=1.1\textwidth]{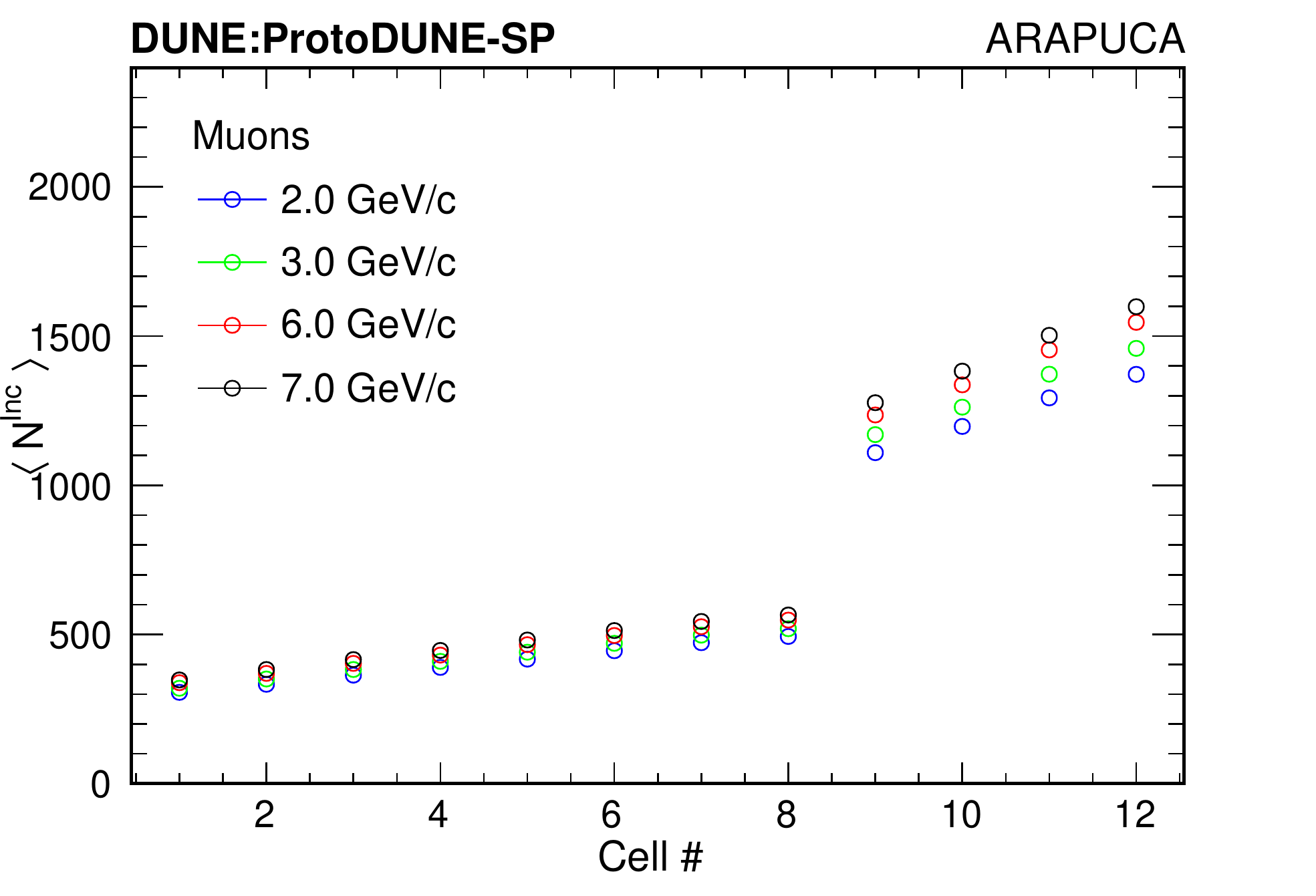}
\end{minipage}
\begin{minipage}[t]{0.50\textwidth}
\includegraphics[width=1.1\textwidth]{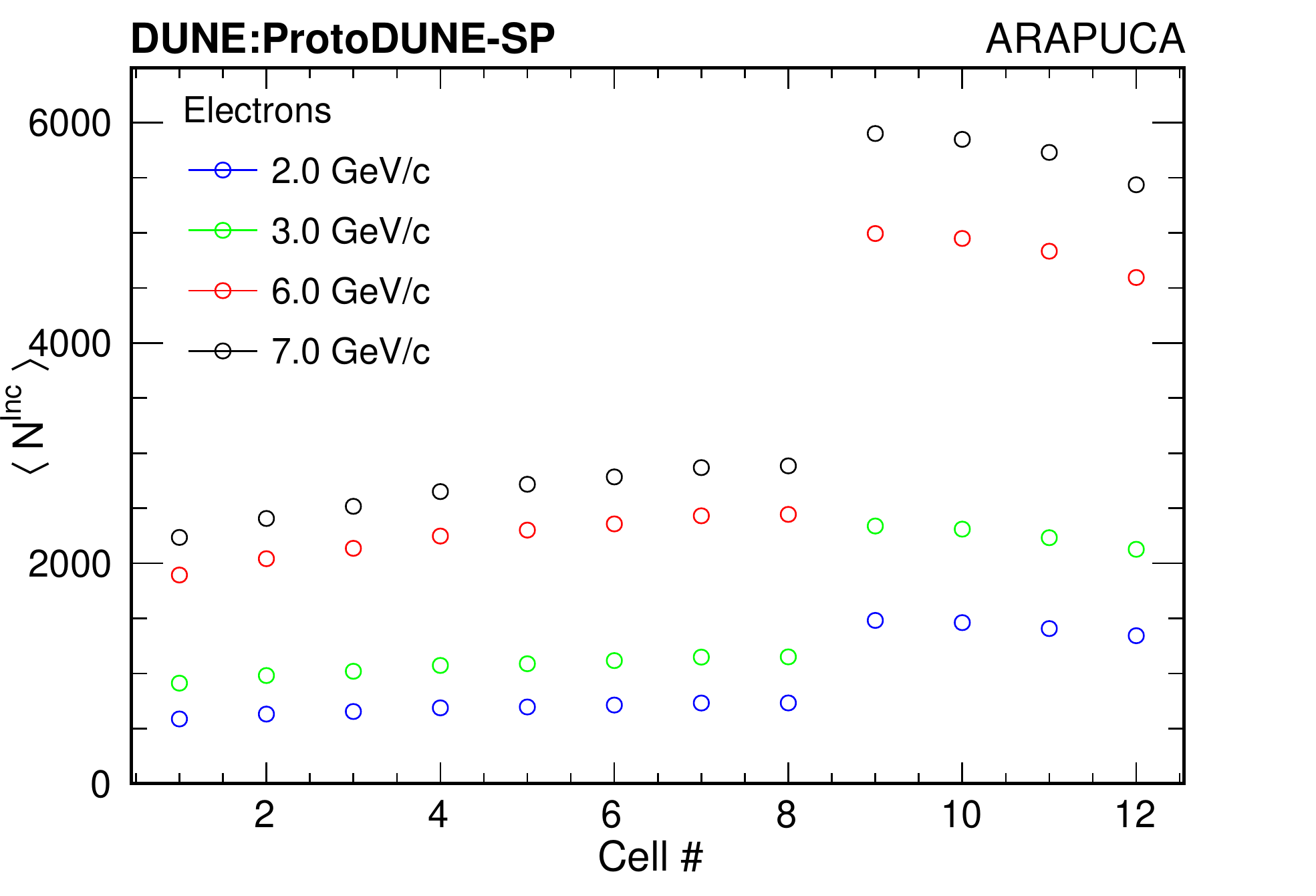}
\end{minipage}

\begin{minipage}[t]{0.50\textwidth}
\includegraphics[width=1.1\textwidth]{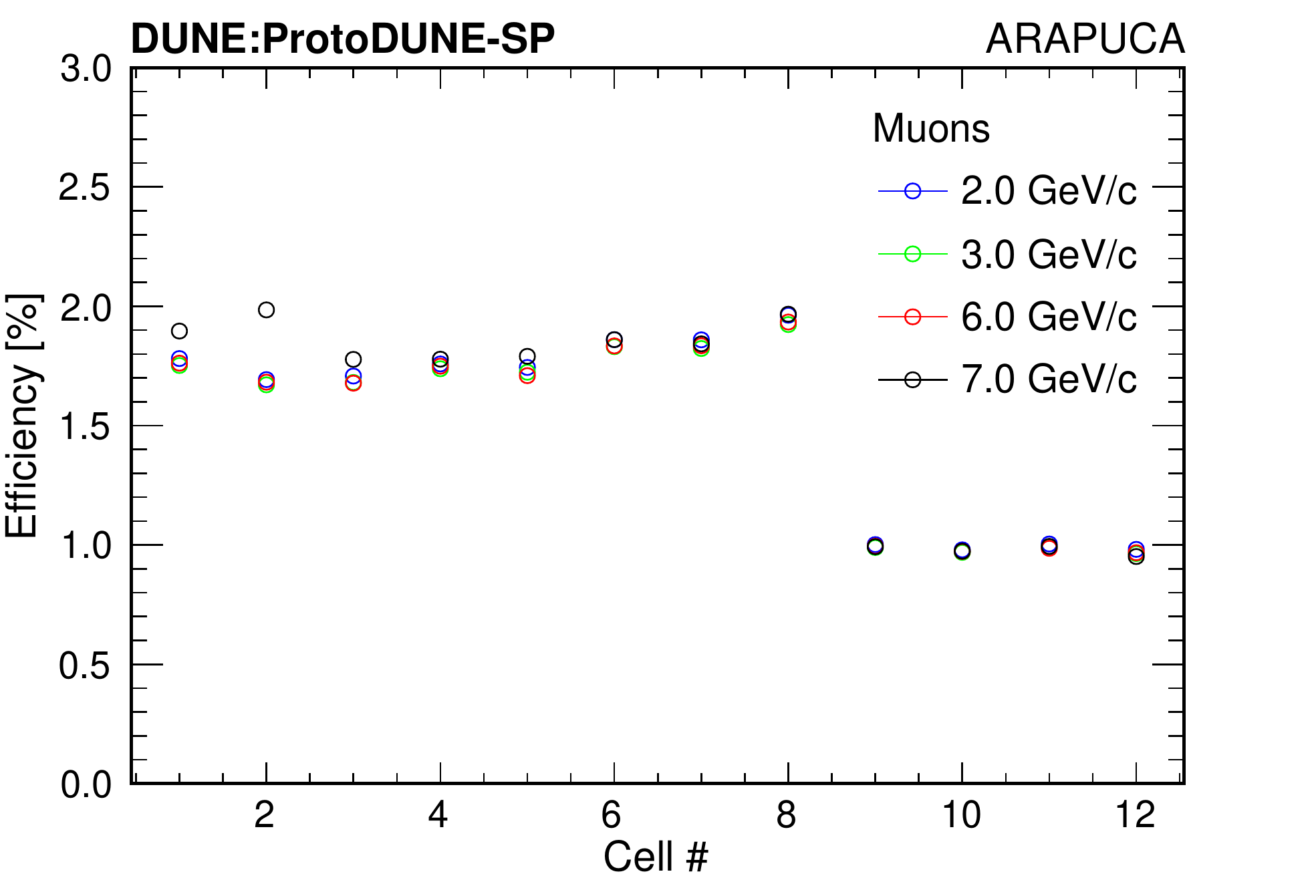}
\end{minipage}
\begin{minipage}[t]{0.50\textwidth}
\includegraphics[width=1.1\textwidth]{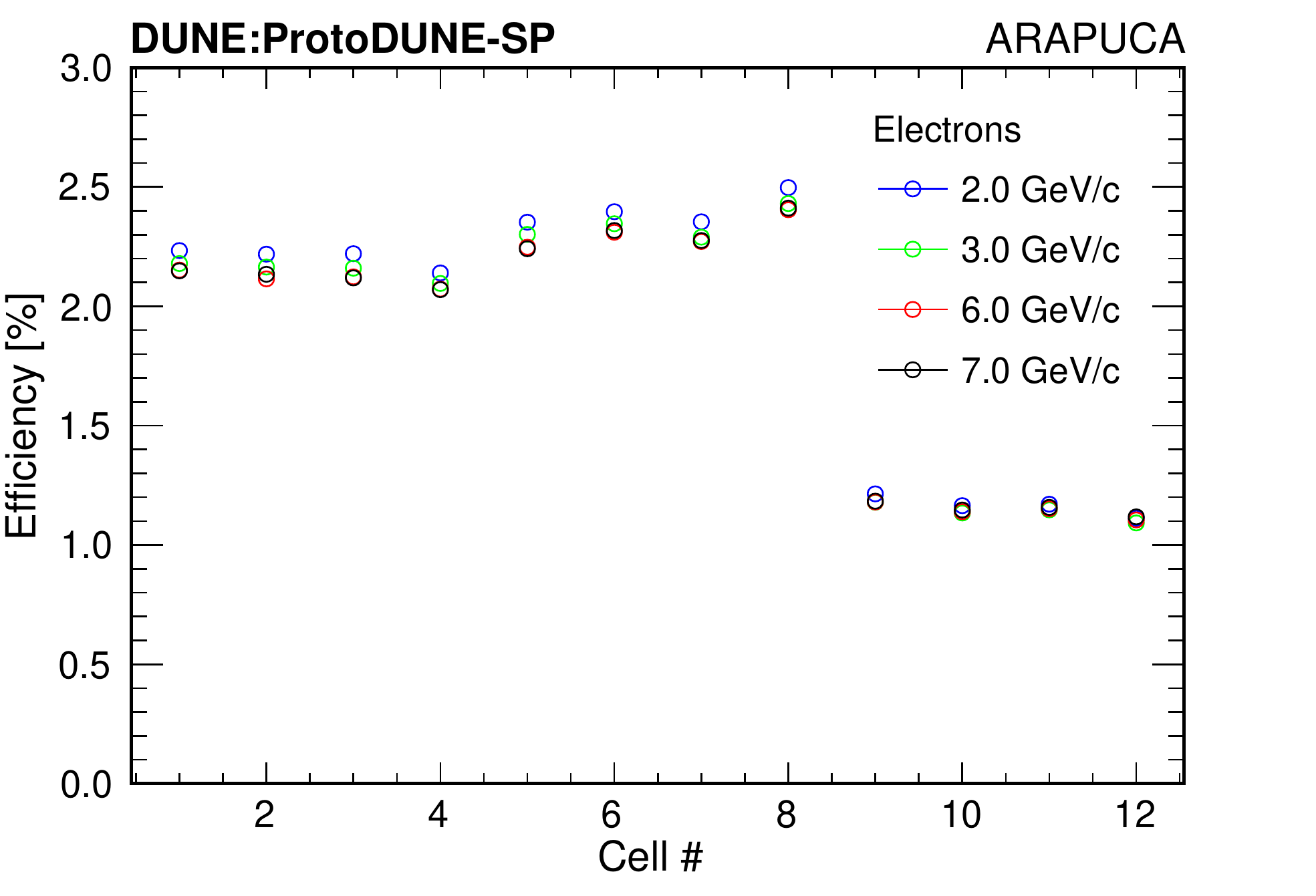}
\end{minipage}
\caption{ARAPUCA cell efficiency as determined from beam muons (left) and beam electrons (right).
Cells are of two types with the last four cells (channels $j=9, \ldots ,12$) have double size but equal number of photosensors than the first eight. Top: average number of detected photons with beams at different momenta. Center: average number of photons incident on the cell surface from MC simulation of electron and muon beams at corresponding momenta. Bottom: efficiency of the cell from the detected-to-incident ratio. Statistical error bars are small, not visible inside the symbols.}
\label{fig:PhL-PhD-eff}
\end{figure}

{\it Muon data}: for each of the 12 cells of the ARAPUCA module located in APA 3 of the PDS beam side, the mean value $\langle N^{\rm{Det}}_j\rangle$ of the detected photon distribution  from the muon data samples with beam momenta of 2, 3, 6 and 7 GeV/$c$ (open circles of assigned color) are displayed in figure \ref{fig:PhL-PhD-eff} (top-left), the mean values $\langle N^{\rm{Inc}}_j\rangle$ of the photons incident on the cell surface from the MC muon event samples is shown in the (center-left) panel. Statistical errors are small (few per-mille relative to the mean values, not visible inside the symbols), systematic uncertainties not shown in the figure are discussed later in this section. The detection efficiency $\epsilon_j=\langle N^{\rm{Det}}_j\rangle/\langle N^{\rm{Inc}}_j\rangle$ given by the ratio of the two mean values in the (bottom-left) panel, for each cell at all momenta.
Cells in the ARAPUCA module corresponding to channels $j=1,\ldots ,12$ are ordered along the $z$ axis
 with the upstream cell$\#1$ at the beam entry point ($z=0$) into the LArTPC volume. Muons at all incident momenta are energetic enough to cross the entire LAr volume and exit from the downstream side (see figure \ref{fig:TPC-PDS-3D}). The number of detected photons increases from cell to cell along $z$ due to the increasing visibility of the muon track from the cells deeper into the LAr volume. In every cell the number of detected photons is observed to increase with incident muon beam momentum (open circles of different color in figure \ref{fig:PhL-PhD-eff}) due to the increase in the energy loss along the track for more energetic muons. 
 Cells in the ARAPUCA module are of two types: the last four cells (channels $j=9, \ldots ,12$) have double size but equal number of photosensors than the first eight. The high step in the number of collected photons $N^{\rm{Inc}}_j$ at $j=9$ - figure \ref{fig:PhL-PhD-eff} (center) - reflects the double geometrical acceptance of these cells. A smaller step is observed in the detected photons $N^{\rm{Det}}_j$ (top). This is due to the halved photocathode coverage partly mitigated by the light trapping in the ARAPUCA cell.
 
\begin{figure}[tbh]
\begin{minipage}[t]{0.5\textwidth}
\includegraphics[width=1.1\textwidth]{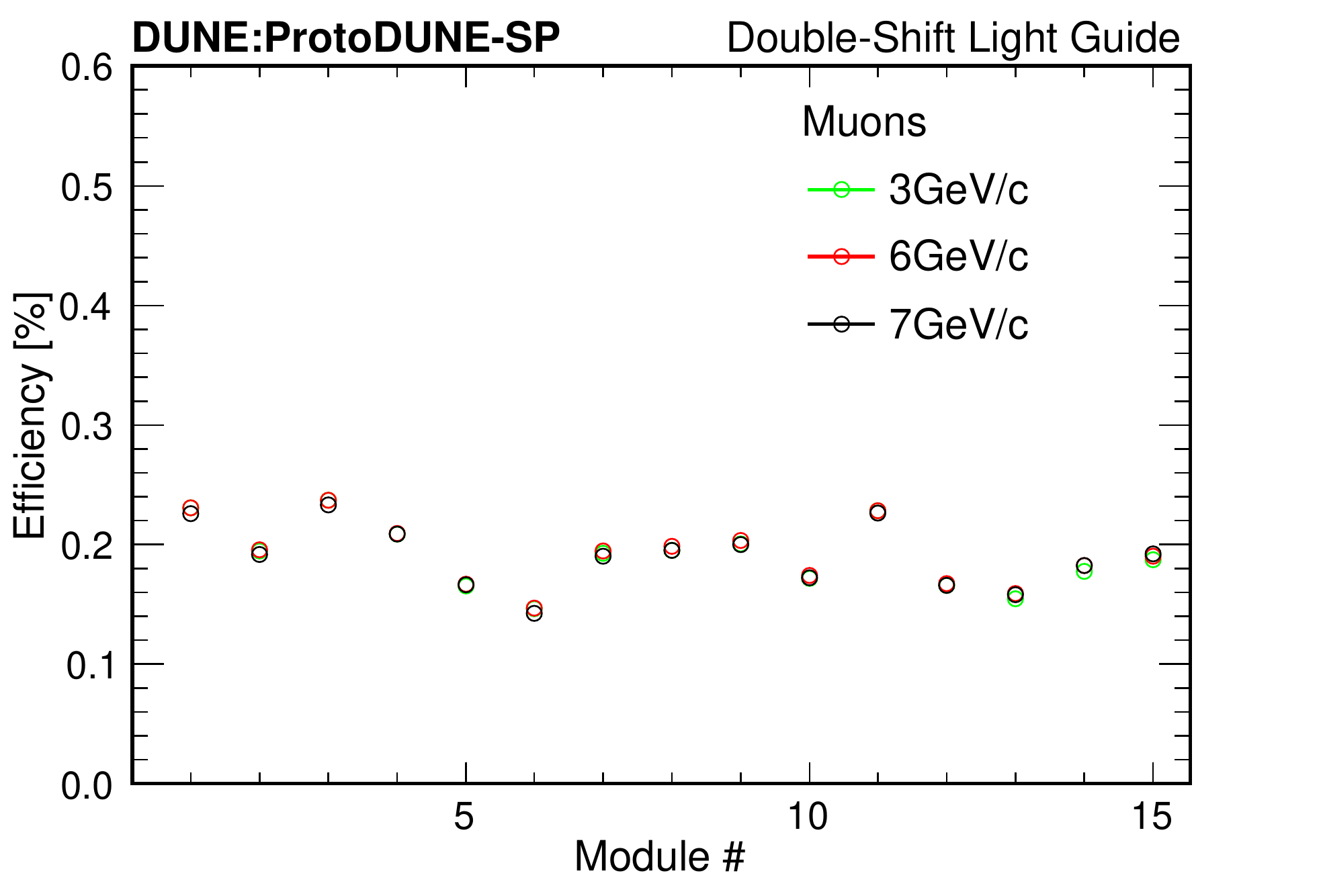}
\end{minipage}
\begin{minipage}[t]{0.5\textwidth}
\includegraphics[width=1.1\textwidth]{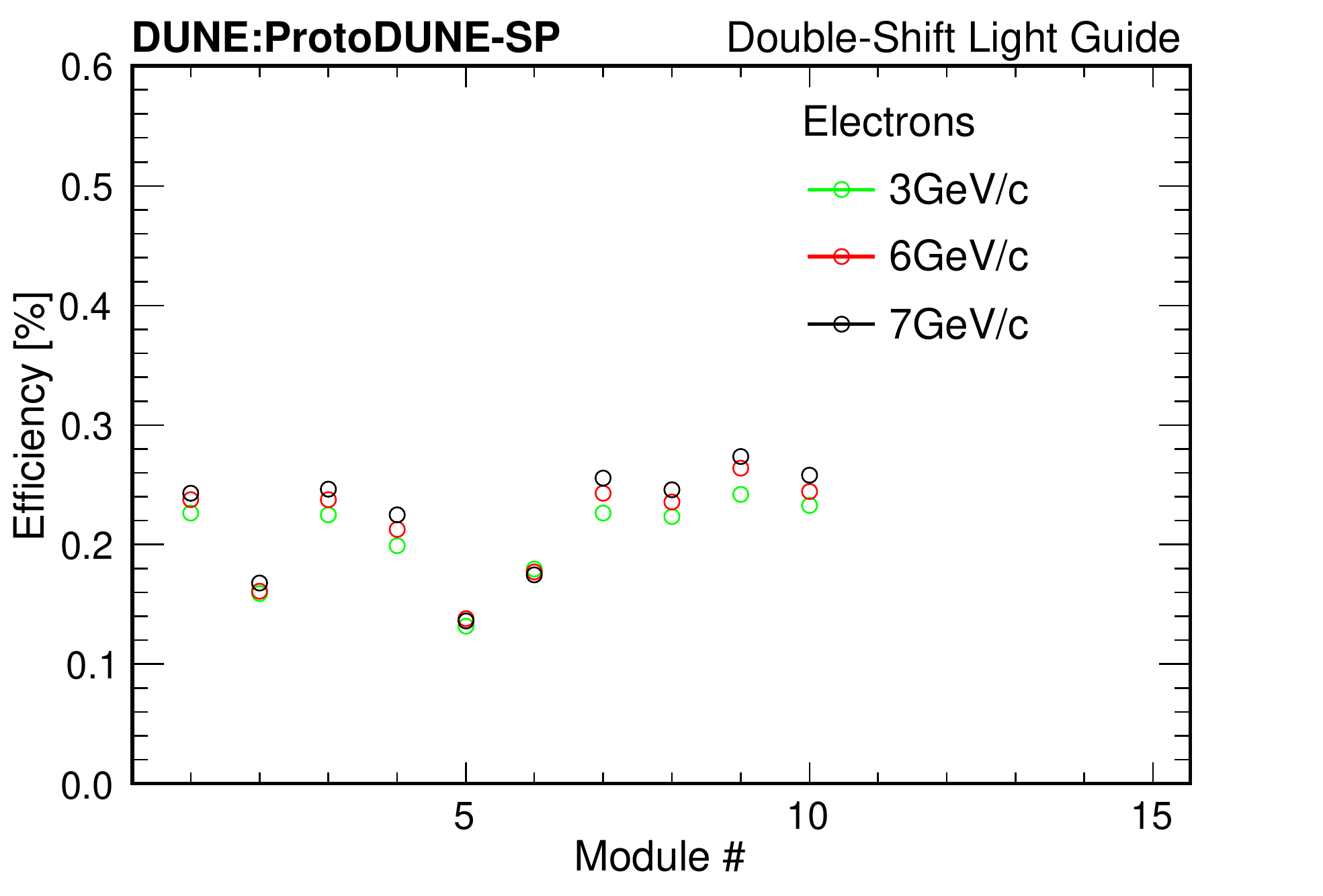}
\end{minipage}
\caption{Efficiency measurements of 15 double-shift light-guide modules (PDS beam side), as determined from beam muons (left) and beam electrons (right) data at different momenta (only modules $j=1,..,10$ in APA3, and 2 at the shorter distance from the shower and higher photon counting are displayed).}
\label{fig:eff-IU}
\end{figure}

\begin{figure}[tbh]
\begin{minipage}[t]{0.5\textwidth}
\includegraphics[width=1.1\textwidth]{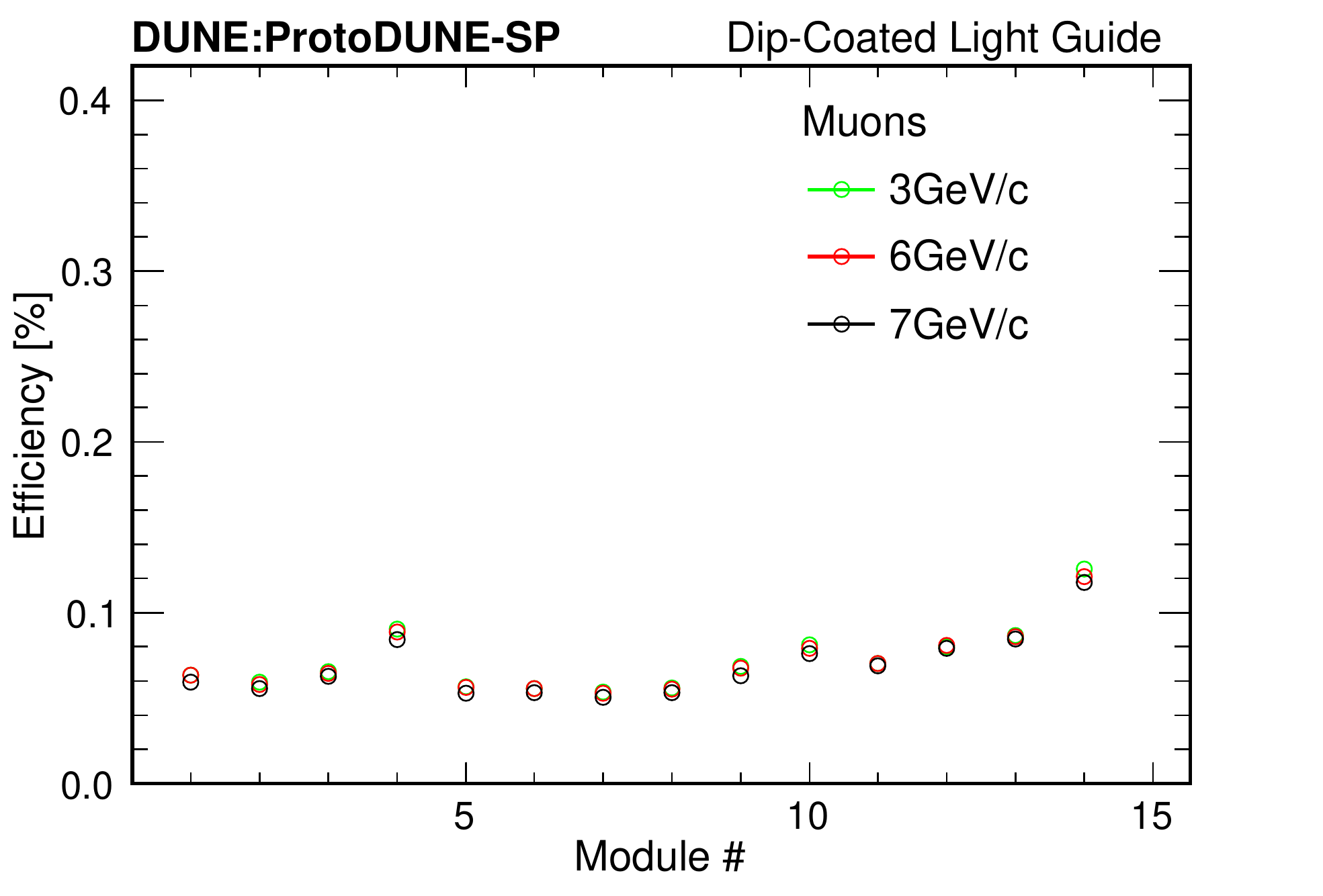}
\end{minipage}
\begin{minipage}[t]{0.5\textwidth}
\includegraphics[width=1.1\textwidth]{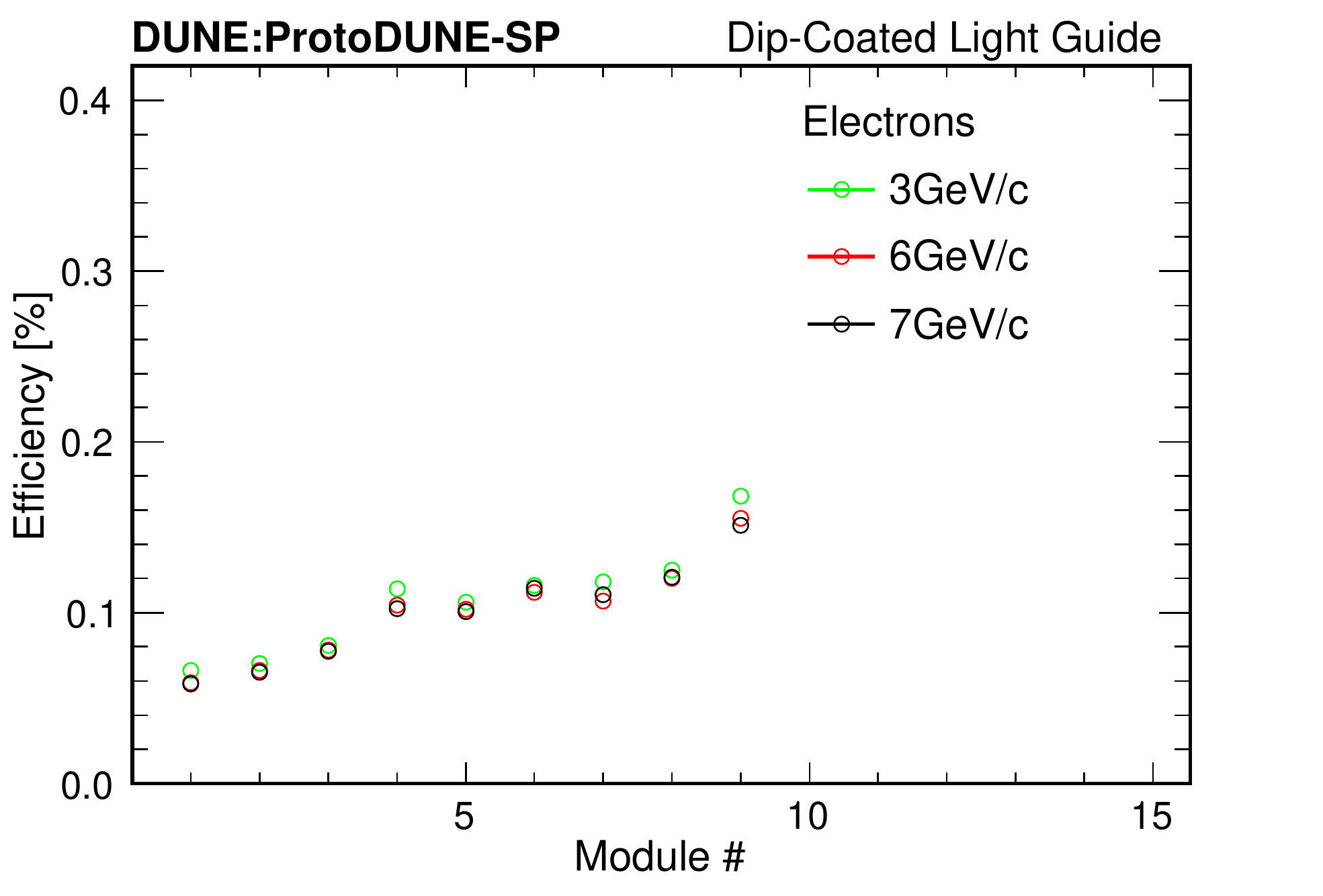}
\end{minipage}
\caption{Efficiency measurements of 14 dip-coated light-guide modules (PDS beam side), as determined from beam muons (left) and beam electrons (right) data at different momenta (only modules $j=1,..,9$ in APA3, and 2 at the shorter distance from the shower and higher photon counting are displayed).}
\label{fig:eff-MIT}
\end{figure}

 For each light-guide module in the PDS beam side the number of detected photons and  incident photons were evaluated in the same manner from the same beam muon samples (data and MC runs with beam momenta of 2, 3, 6 and 7 GeV/$c$). The efficiency from the ratio of the detected to incident photons is shown in figure \ref{fig:eff-IU} (left) for the 15 double-shift light-guide modules in APA3, 2 and 1 and in figure \ref{fig:eff-MIT} (left) for the 14 dip-coated light-guide modules. Statistical error bars are small, not visible inside the symbols.  The locations of APAs 1, 2, and 3 in the ProtoDUNE-SP detector are shown in figure~\ref{fig:TPC-PDS-3D}. The efficiency of light-guide modules is expected to be proportional to the number of suitably placed photosensors, and inversely correlated with the length of the optically active surface of a module with fixed width due to the attenuation of internally reflected optical photons. As a crude characterization, the smallness of the ratio of number of photosensors to optically active surface area of the light-guide modules relative to the ARAPUCA cells underlies the corresponding ratios of efficiencies. A factor of two can be gained by instrumenting both ends of the light guides, but improvement beyond that would require modification to the module design to enable effective deployment of additional photosensors. 

{\it Electron data}: electrons with beam momenta of 2, 3, 6 and 7 GeV/$c$ provide data samples for a second independent set of efficiency measurements.  Electrons deposit all their incident energy in showers localized in a limited portion of the LAr volume, unlike muons on long, throughgoing tracks. Electromagnetic showers develop in front of APA3, where the ARAPUCA module is positioned nearly at the height of the entering beam (see figure \ref{fig:3D-EM-showewr} with a 3D display of 7~GeV/$c$ beam electron event). Light detected in the ARAPUCA cells, the MC estimate of the light arriving on the cells' optical surfaces and the corresponding detector efficiency are shown in the right-panels of figure \ref{fig:PhL-PhD-eff} (top), (center) and (bottom) respectively. For any given beam energy (open circle colors in the plots), the distribution of the detected photons by the cells along the bar exhibits a shower-like longitudinal profile, an indication of the position  reconstruction capability of the segmented ARAPUCA module. The number of detected photons is also clearly correlated with the shower energy. The calorimetric energy reconstruction from scintillation light signals is discussed in section \ref{sec:calorimetry-light}.

The response of the light-guide modules (beam side) to beam electrons were also used for efficiency measurements with beam momenta of 3, 6 and 7 GeV/$c$, excluding electron data at 2~GeV/$c$ with the modules in APA1 at the farthest distance from the shower and low photon counting. Results are shown in figure \ref{fig:eff-IU} (right) for the ten double-shift light-guide modules in APA3 and APA2 and in figure \ref{fig:eff-MIT} (right) for the nine dip-coated light-guide modules. 

{\it Efficiency}: the photon detection efficiency was evaluated through 8 independent measurements using muon data and electron data at four different beam momenta, for each element of the PDS (12 cells in one ARAPUCA module, 15 double-shift light-guide modules and 14 dip-coated light-guide modules of the PDS beam side). 
By comparing the results, efficiency estimated from the electron data is found in all elements systematically higher than from the muon data, regardless of the energy of the particle [see figures \ref{fig:PhL-PhD-eff} (bottom), \ref{fig:eff-IU} and \ref{fig:eff-MIT}].
The systematic difference may be due to bias in the MC simulation at the photon emission stage (e.g., an unaccounted deviation in scintillation yield for GeV-scale electrons and muons with respect to minimum-ionizing particles) and at the propagation stage (e.g., a 
difference due to the computational method used to approximate the number of photons reaching the PD optical window from localized volumes (EM showers) and long tracks (muons)).
The mean value from all available measurements $\langle \epsilon_j\rangle$ for the $j$-th element is taken as the best estimate of the efficiency of that element, and the standard deviation $s_j$ that measures the dispersion around the mean is taken as an estimate of the systematic uncertainty on the efficiency. 
The statistical uncertainty, evaluated from the standard errors of the mean numbers of detected and incident photons in the data and MC samples of muons and electrons at different energies, is negligible. 
\begin{table}[!htb]
\centering
\caption{Efficiencies of the detector technologies in the ProtoDUNE-SP PD system: median value among detectors of the same type, determined from the average of independent measurements with beam muons and electrons at different energies. The error is from systematic uncertainty, with negligible statistical uncertainty. The number of detectors of different types examined correspond to the fraction of PDS elements in the beam side upstream APAs (\# 3 and 2), selected to determine the median efficiency reported in the efficiency column.}
\label{tab:median-effic-all}
\begin{tabular}{ c|c|c}
  {\bf N\textsuperscript{\underline{o}} of PDS elements examined}     &  {\bf Detector Type} &  {\bf Efficiency}   \\
\hline
      8    &  ARAPUCA cell        &   $\tilde \epsilon_{A} =(2.00 \pm 0.25)~\%$   \\
\hline
      4     &  ARAPUCA cell (double area)      &  $\tilde \epsilon_{A2} =(1.06 \pm 0.09)~\%$  \\
\hline
    10     &    Double-shift module            &  $\tilde \epsilon_{DS} =(0.21  \pm 0.03)~\%$     \\
\hline
  9     &    Dip-coated module            &    $\tilde \epsilon_{DC} =(0.08   \pm 0.02)~\%$     \\
\end{tabular} 
\end{table}
Comparing modules or cells of the same type, relative variations in efficiency are within $\pm6\%$ for the ARAPUCA cells, $\pm20\%$ for the double-shift modules and greater than $\pm25\%$ for the dip-coated modules. The median efficiency value with its statistical and systematic uncertainty from each group of detectors ($\tilde \epsilon_{A},~\tilde \epsilon_{A2},~\tilde \epsilon_{DS},~\tilde \epsilon_{DC}$) is selected to characterize the different technologies implemented in the ProtoDUNE-SP PD system. These value are reported in table \ref{tab:median-effic-all}. The overall relative uncertainty on the efficiency for all detector types is thus found to be $8.5\% \lesssim \sigma_\epsilon/\epsilon \lesssim 13.5\%$, as determined from the set of measurements described above. This appears compatible with the systematic error expected from uncertainty in the parameters of the photon emission and propagation used in the MC generation.  

\subsubsection{Comparisons of cosmic-ray muons to simulation}
The analysis of photon signals from cosmic-ray muons provides a check on the validity of the photon simulation used to determine efficiency.
Throughgoing cosmic-ray muons, or those which enter through the upstream face and exit through the downstream face of the cryostat, are isolated in the scintillation medium using CRT triggered events that have been matched to reconstructed tracks in the TPC. 
A trigger from this sub-detector involves a four-fold coincidence of the $x$- and $y$-measuring planes of the CRT over the span of 60~ns producing events that are very likely to contain a throughgoing cosmic-ray muon.
Two strategies were employed, both using an event definition described by the sum of all detected photons in the twelve cells of the non-beam side ARAPUCA module during a 13.33~$\mu$s externally tagged PDS trigger.
Before either analysis, the sum of photons per event collected at the detector surface in the simulation were scaled by a fixed amount in order to eliminate a systematic normalization difference with data.


\begin{figure}[tbh]
\centering
\begin{minipage}[t]{0.49\textwidth}
\centering    
\raisebox{-0.5\height}{\includegraphics[width=1.0\textwidth]{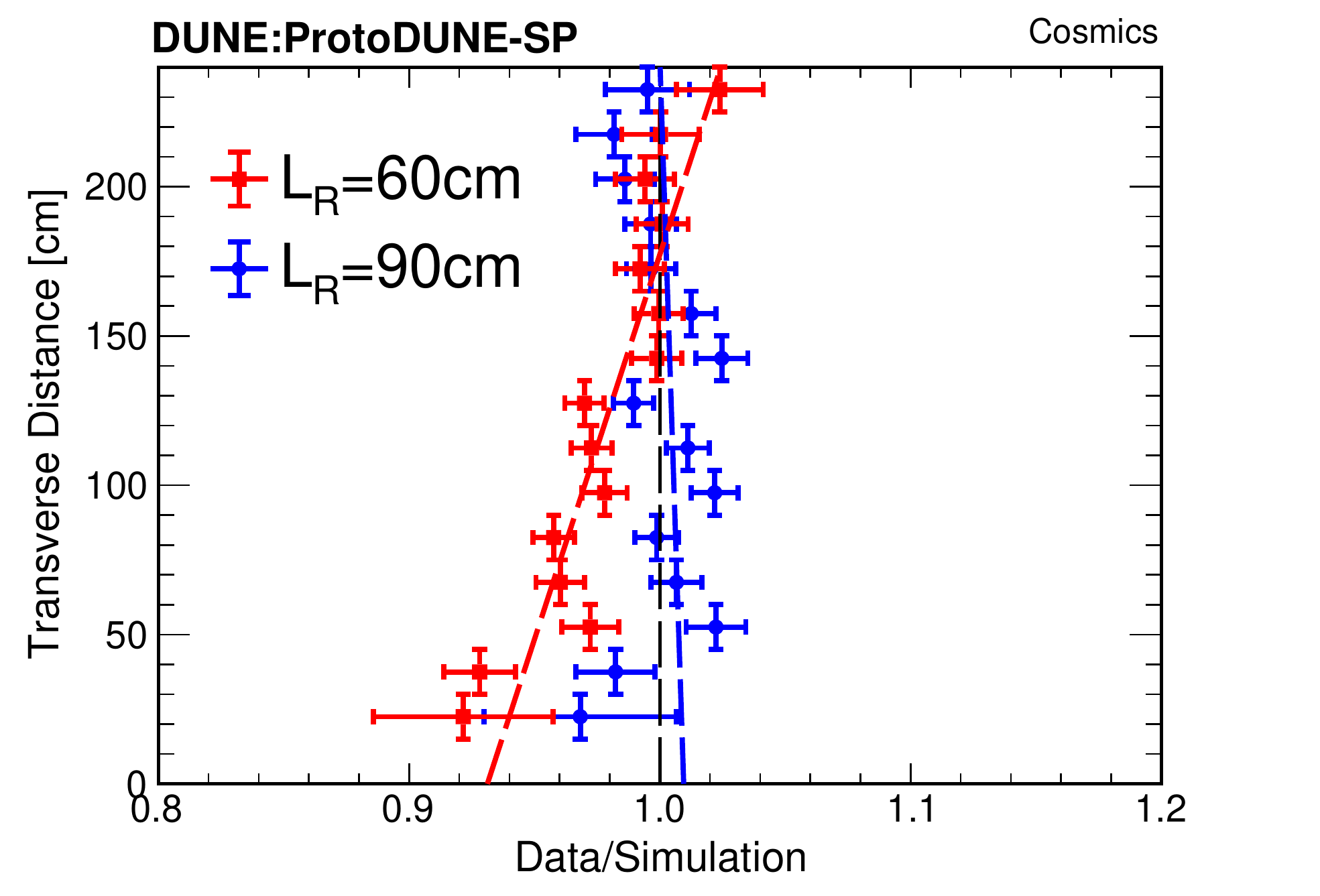}}
\end{minipage}
\begin{minipage}[t]{0.49\textwidth}
\centering    
\raisebox{-0.5\height}{\includegraphics[width=.85\textwidth]{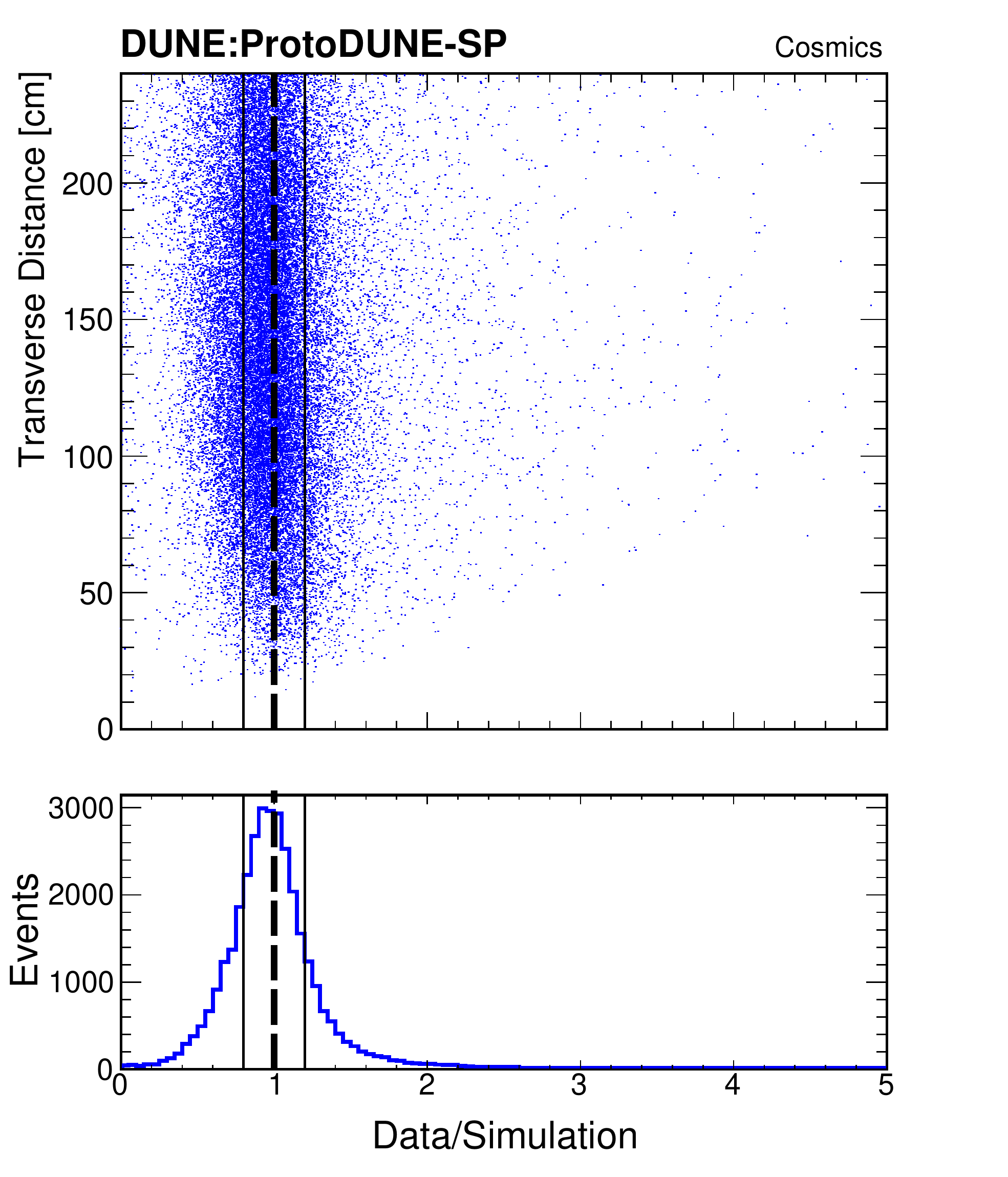}}
\end{minipage}
\caption{The left-hand panel shows the ratio of the observed to the predicted PE yields as a function of transverse distance, assuming two different values of the Rayleigh scattering length in the simulation. The data agree more with simulation that employs a Rayleigh scattering length (${\rm L}_{\rm R}$) of 90~cm as opposed to the 60~cm prediction.  On the right, an event-by-event comparison is shown for roughly 43,000 events taken over four months with a simulated Rayleigh scattering length of 90~cm. The dashed vertical lines in both plots represents where data and simulation agree. The solid vertical lines in the right-hand plot represent the bounds of the plot on the left. }
\label{fig:attmeas}
\end{figure}

The first analysis, a comparison of the average light response as a function of transverse distance, was employed to observe some of the bulk effects of the medium.
Here, the transverse distance is defined as the length of the segment between the reconstructed particle path and the center point of the ARAPUCA module as they occupy the same position along the $z$ axis.
Since the ARAPUCA module is in the central APA (APA6), this variable is exclusively a function of the particle position in $x$ and $y$ as it bisects the detector in the $z$ direction, exploiting the symmetry of the detector.
The results of this analysis, shown in the left plot of figure \ref{fig:attmeas}, suggest that the measured data favor simulation with Rayleigh scattering length hypothesis of 90~cm as opposed to a Rayleigh scattering length of 60~cm in the scintillation medium. 
A second analysis compares each data track to a simulated track generated with matching position and trajectory in a medium with Rayleigh scattering length set at 90~cm.
Results, which are shown in the right-hand plot of figure~\ref{fig:attmeas}, demonstrate excellent agreement with simulation as a function of the transverse distance from the ARAPUCA to the reconstructed track.
In comparisons of data to simulation, one standard deviation in the difference between the reconstructed simulated light and the measured reconstructed light is about 18.6\%. 
These two comparisons show that photons are successfully reconstructed in ProtoDUNE-SP and that the  simulation of the optical properties of scintillation medium is in a good agreement with the measurements. 




\subsubsection{Time resolution}
The timing performance of the PD system intrinsically depends upon a combination of factors, from the intrinsic time resolution of the photosensor, to the electronics response of the readout board and signal digitization, to the features of the light propagation, wave-length shifting and photon collection by the PD modules.
The overall timing performance is evaluated here in two different applications: time resolution of two consecutive light signals and time matching between light signal and TPC signal.  

\begin{figure}[!htbp]
\centering
\includegraphics[height=5cm,width=0.49\textwidth]{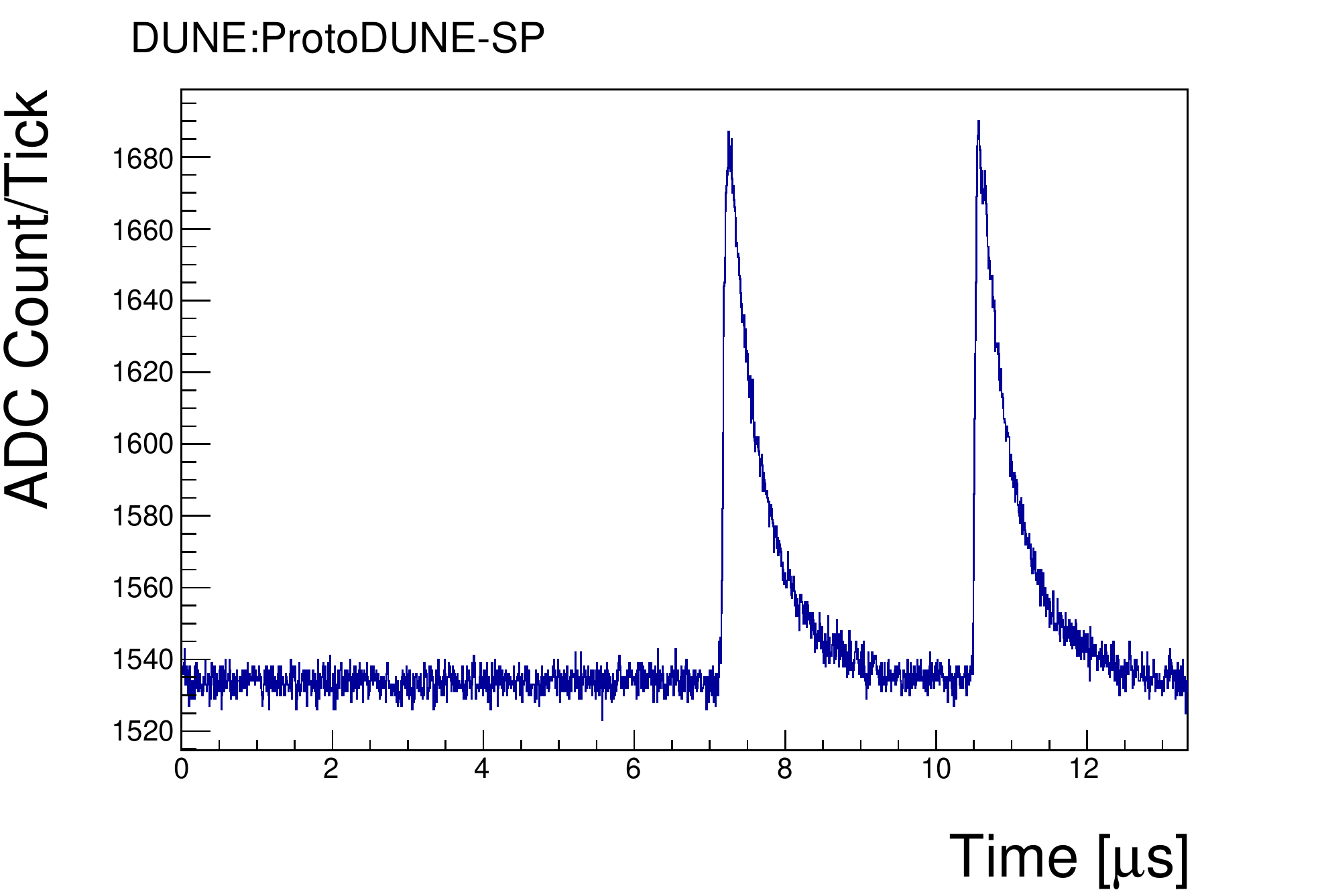}
\includegraphics[height=5cm,width=0.49\textwidth]{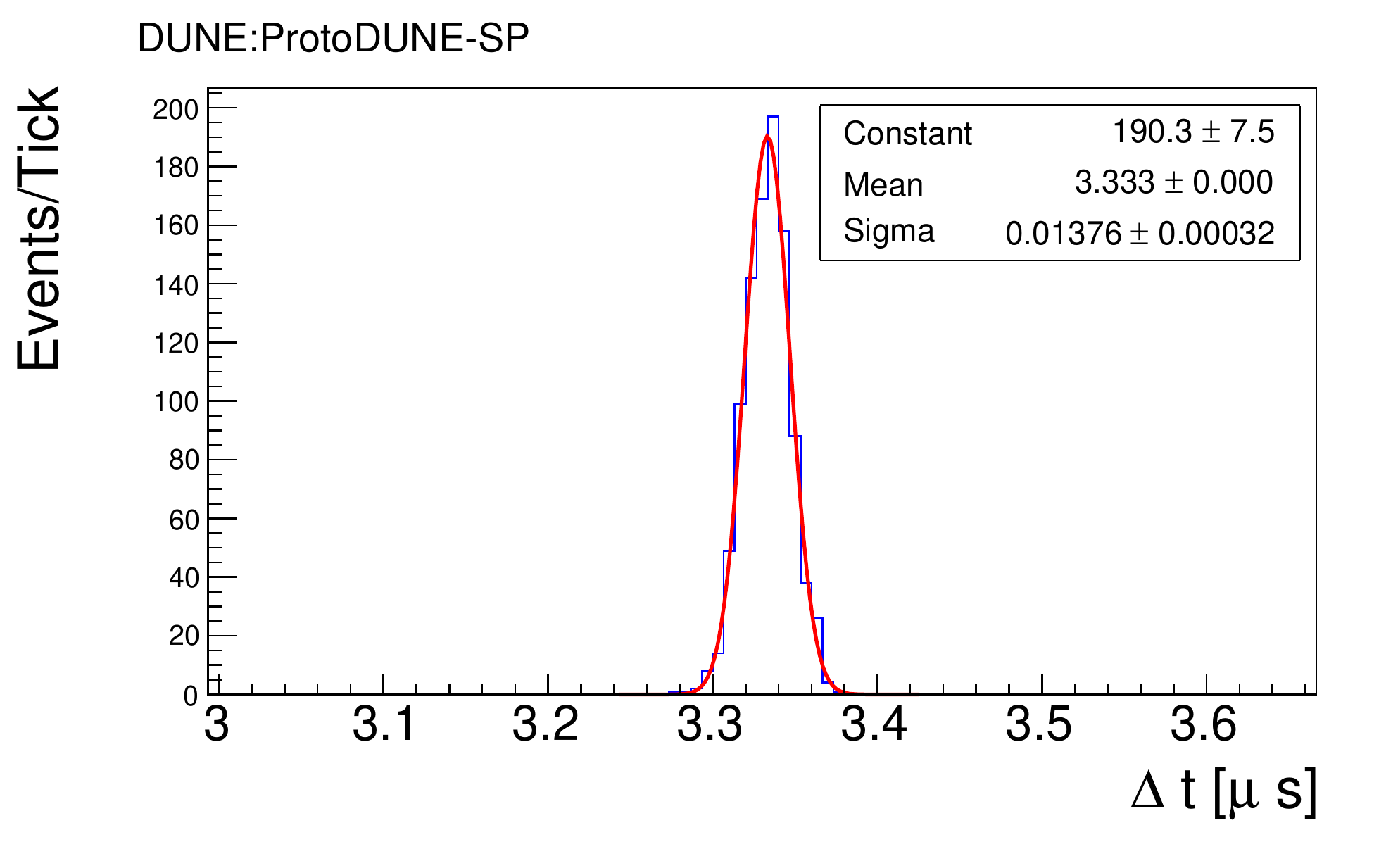}
\caption{PDS timing measurements: Double pulse light signal using the photon detector calibration system (left); Resolution in the time difference measurement between correlated light signals (right).}
\label{fig:calib_sys_time}
\end{figure}

Resolving successive light signals in time is of importance in physics reconstruction of correlated events, such as stopping muon with decay to Michel electron, or kaon decays, or nucleus de-excitation into gammas after neutrino interaction. Some of these correlations may be observed with light signals in LAr depending on the PD timing performance. To explore this, data were taken with an external trigger from the LCM (Light Calibration Module) producing two consecutive LED flashes with a fixed time difference among them and from the common trigger time.  The time difference between the pulses was set in the few $\mu$s range typical of the muon decay at rest, and much larger than the pulse width. In figure \ref{fig:calib_sys_time} (left) a recorded waveform  from the LCM trigger is shown with the two generated consecutive LED pulses as detected by an ARAPUCA cell/12-M-MPPC channel and digitized by the SSP readout board. The rise time of each of the two signals from the common trigger was measured in the events collected with the LCM trigger and the distribution of their time difference $\Delta t=(t_2-t_1)$ is shown in figure \ref{fig:calib_sys_time} (right). The time resolution to observe two separate light pulses is $\sigma_{\Delta t}\simeq$~14~ns. Time jitter in the LCM pulse formation is small (sub-ns range) and the dominant factor is from digitization (6.67~ns sampling period).  


\begin{figure}[!htbp]
\centering
\includegraphics[height=6cm,width=0.5\textwidth]{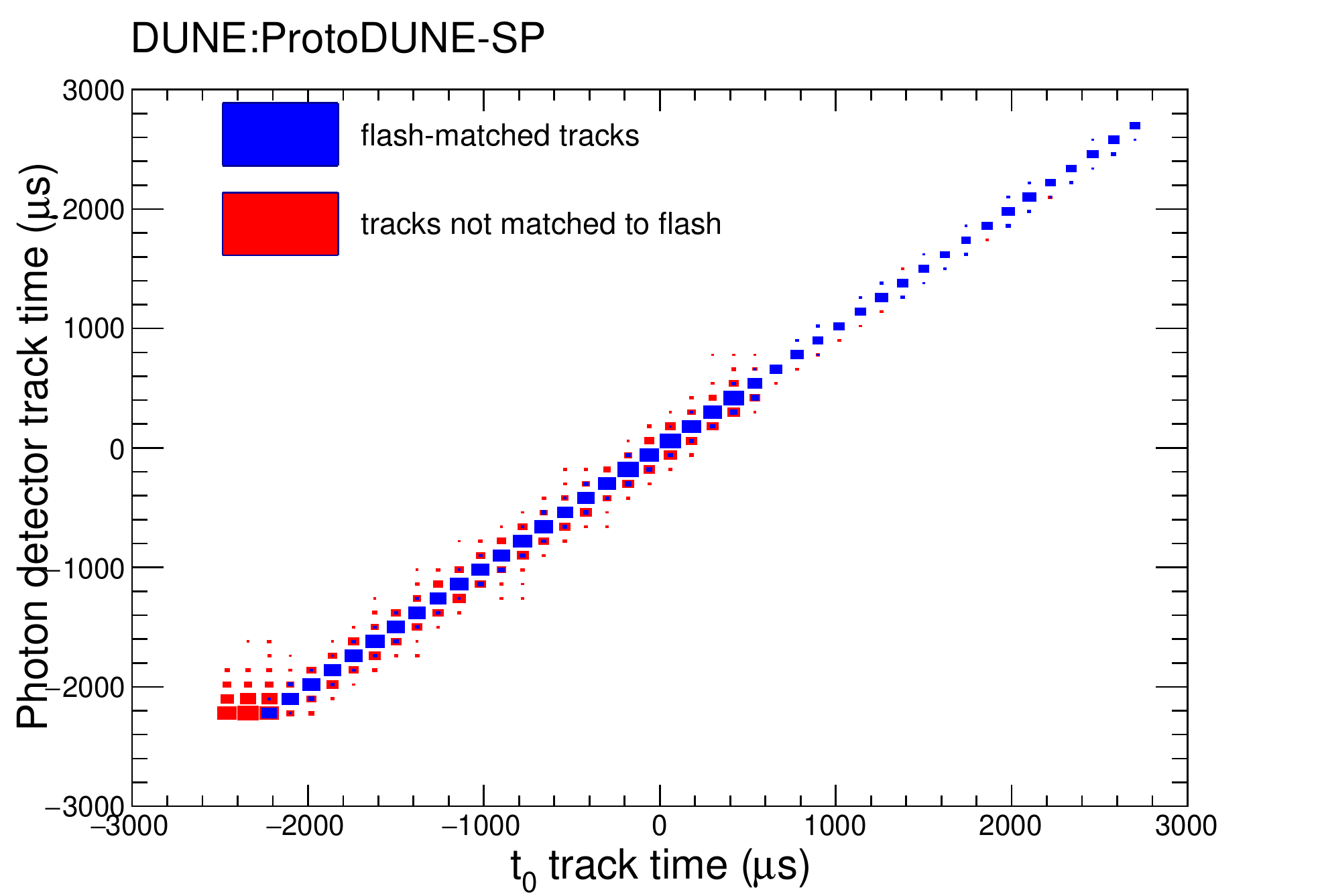}
\caption{PDS Timing Measurements: Correlation between the TPC track $t_0$ time and the PDS flash time. Non-matched tracks (red points) are mostly from CPA-crossing vertical muons (large negative $t_0$) whose flash was not recorded by the PDS at the opposite end of the drift distance.}
\label{fig:protodune-tpcpdstime}
\end{figure}
An efficient light flash-to-track matching in LArTPC is important for a correct event reconstruction, background rejection (especially for LArTPC's operated on the surface) and low-energy underground physics.   

An average of approximately 70 cosmic-ray tracks are observed overlaying each beam event during TPC readout window.
The cosmic-ray tracks arrive at random times relative to the beam trigger. For some of these tracks, such as those that cross the cathode plane or cross one of the anode planes, their actual time of entering the LArTPC volume ($t_0$ time) can be reconstructed offline from the TPC data by using the 3D track reconstruction algorithms and the geometrical features of these CPA or APA crossing tracks. With the ProtoDUNE-SP TPC at its nominal electric field of 500~V/cm, the full drift time is 2.2~ms. Additional 2.25~ms (0.55~ms) are also recorded  before (after) the drift time period, for a total of 5 ms TPC readout window per recorded event. Cathode or anode crossing muons have $t_0$ time distributed over the entire recorded window and reconstructed with a precision of about 20~$\mu$s.  

PDS detected light flashes corresponding to the $t_0$-determined TPC tracks are efficiently found in the packets received from the SSP and contained in the PD fragment of the event. Matching is performed in time inside the TPC readout range,
looking for the closest flash timestamp to the $t_0$ time of the track, within a given coincidence window. In the case of CPA/APA-crossing tracks the matching efficiency depends on the SSP discriminator threshold for the packet recording and to the width of the coincidence window.  
An example of flash-to-track matching is given in figure \ref{fig:protodune-tpcpdstime}, showing the bisector correlation of PDS flash time and TPC track $t_0$ time, for a 4500 APA/CPA crossing track sample. Tracks not matched to a flash are mostly the shorter CPA-crossing tracks whose flash was below threshold.





\section{TPC response}\label{sec:tpcresponse}
\begin{figure}[!htp]
  \centering
  \begin{subfigure}[b]{0.45\textwidth}
    \centering
    \includegraphics[width=\linewidth]{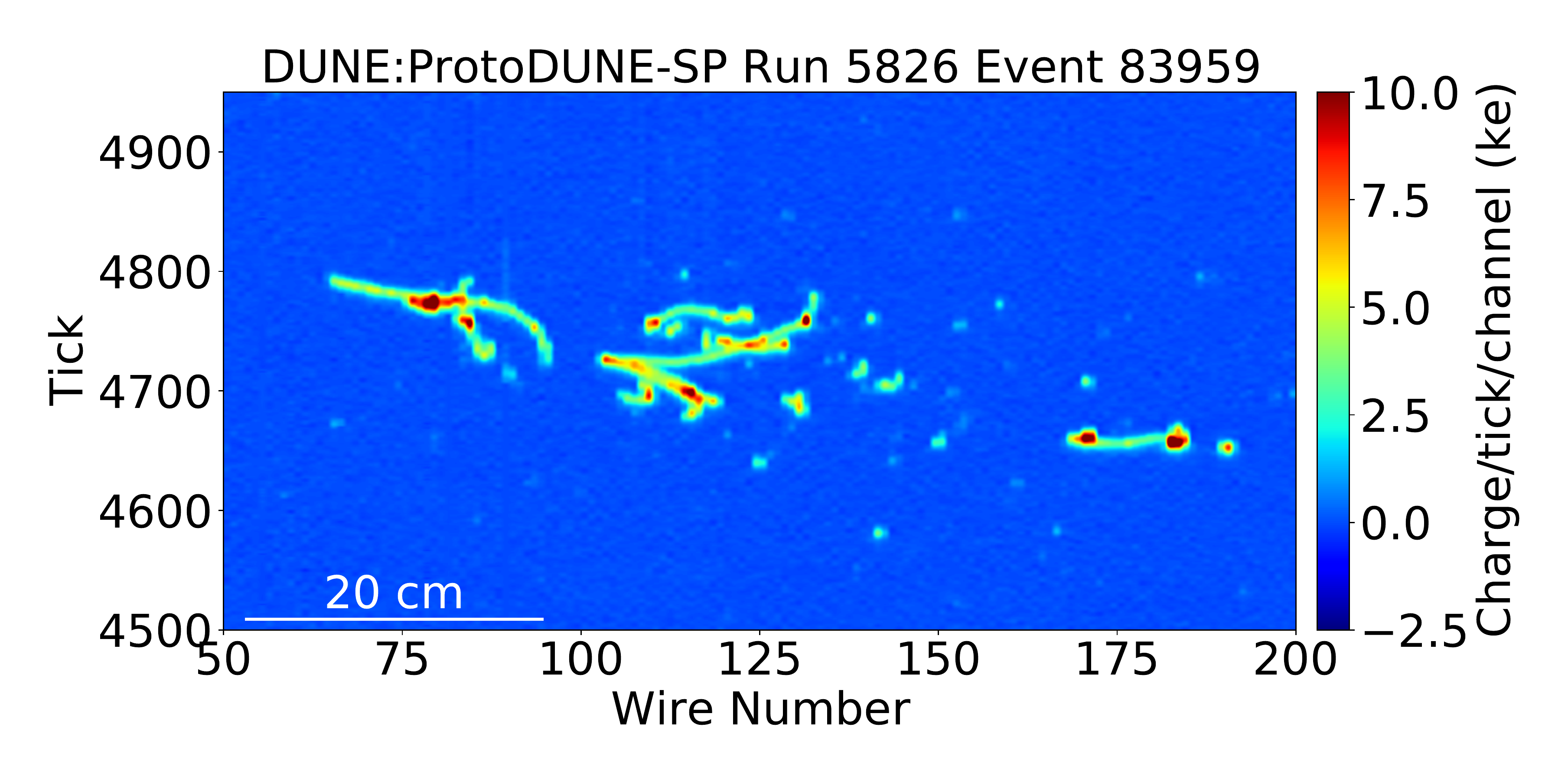}
    \caption{A 0.5~GeV/$c$ electron candidate.}
    \label{fig:0.5GeVEl}
  \end{subfigure}
  \hfill
  \begin{subfigure}[b]{0.45\textwidth}
    \centering
    \includegraphics[width=\linewidth]{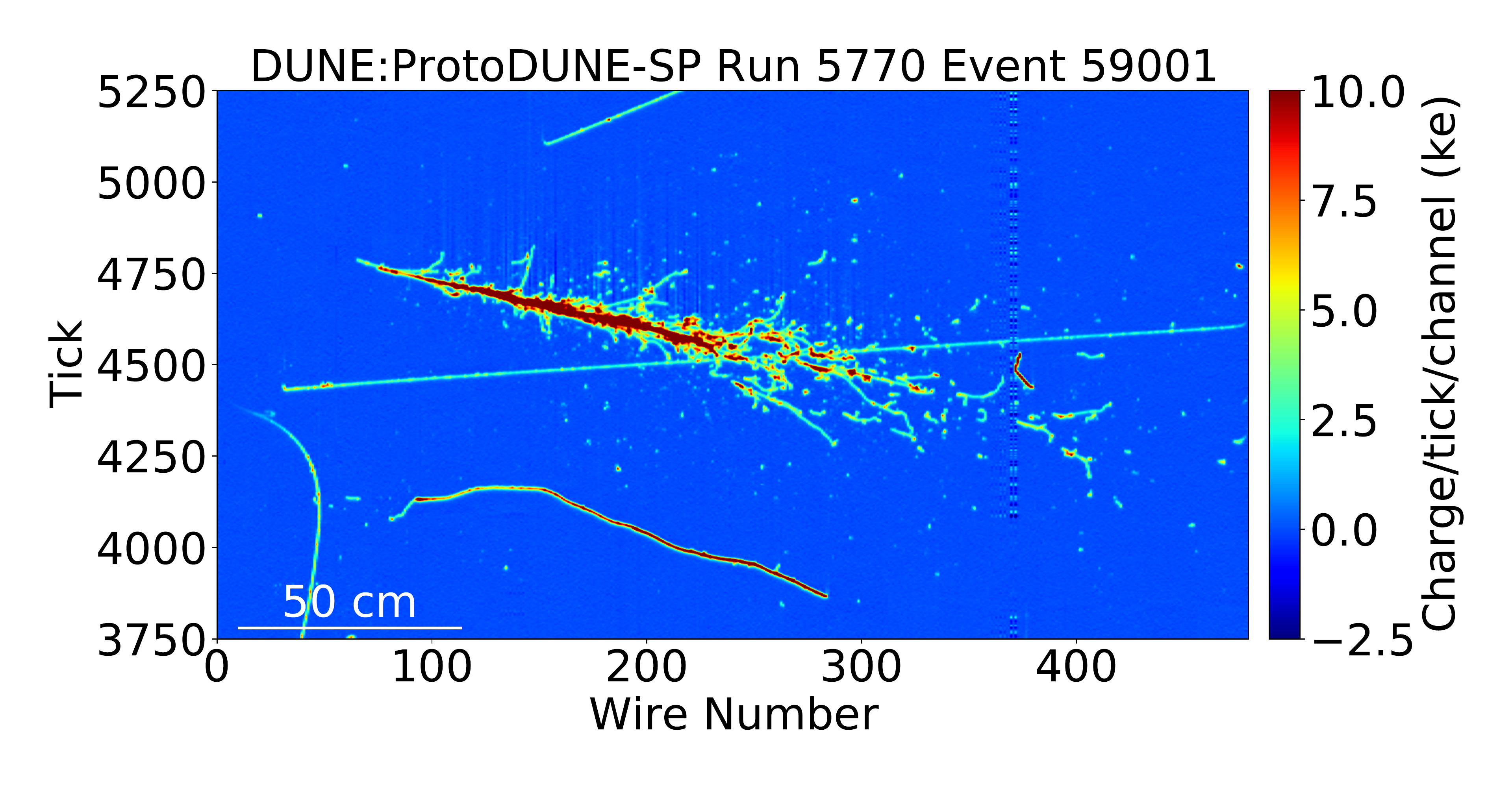}
    \caption{A 6~GeV/$c$ electron candidate.}
    \label{fig:6GeVEl}
  \end{subfigure}  
  \hfill
  \vspace{1mm}
  \begin{subfigure}[b]{0.45\textwidth}
    \centering
    \includegraphics[width=\linewidth]{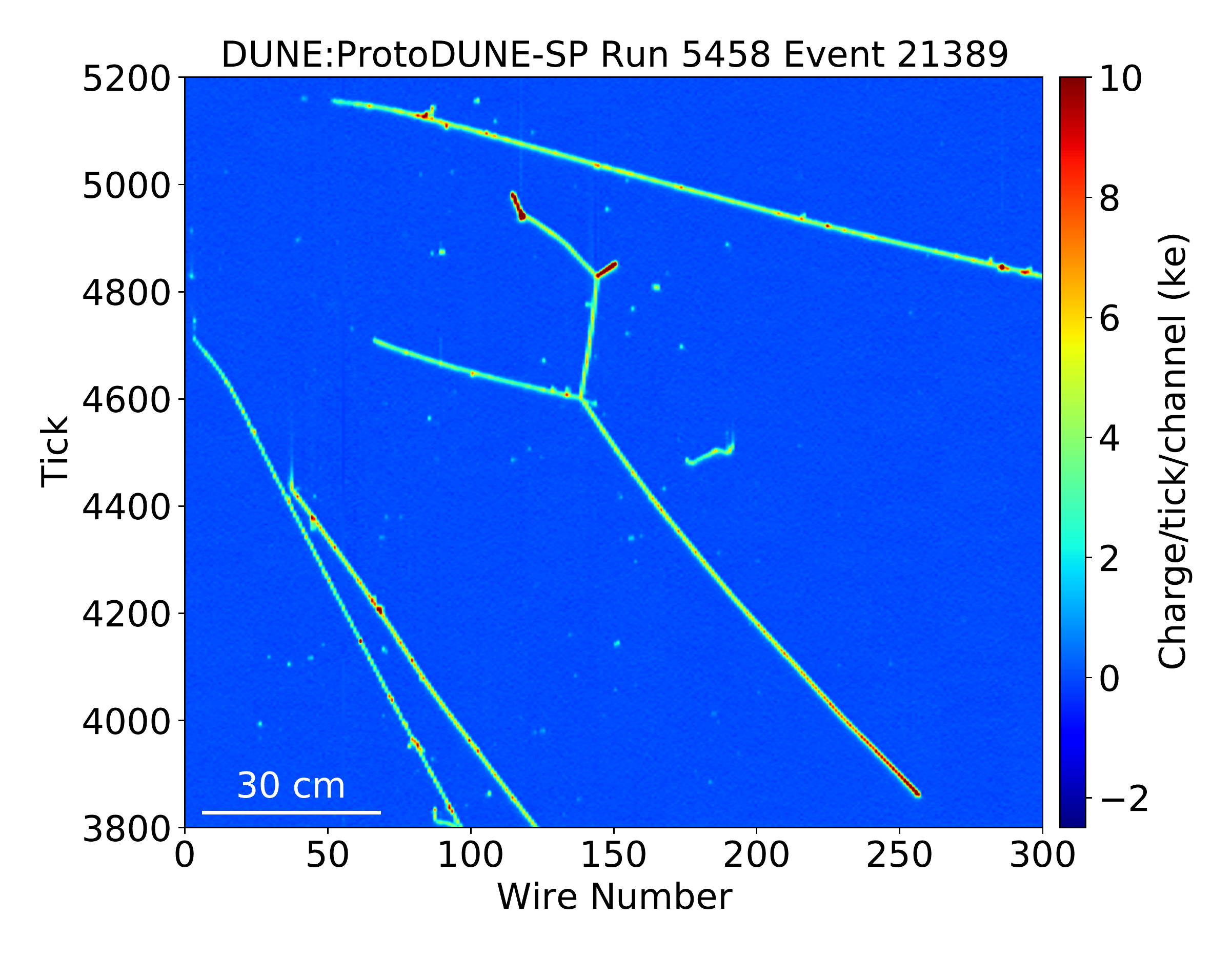}
    \caption{A 1~GeV/$c$ pion candidate.}
    \label{fig:1GeVPi}
  \end{subfigure}
  \hfill
  \begin{subfigure}[b]{0.45\textwidth}
    \centering
    \includegraphics[width=\linewidth]{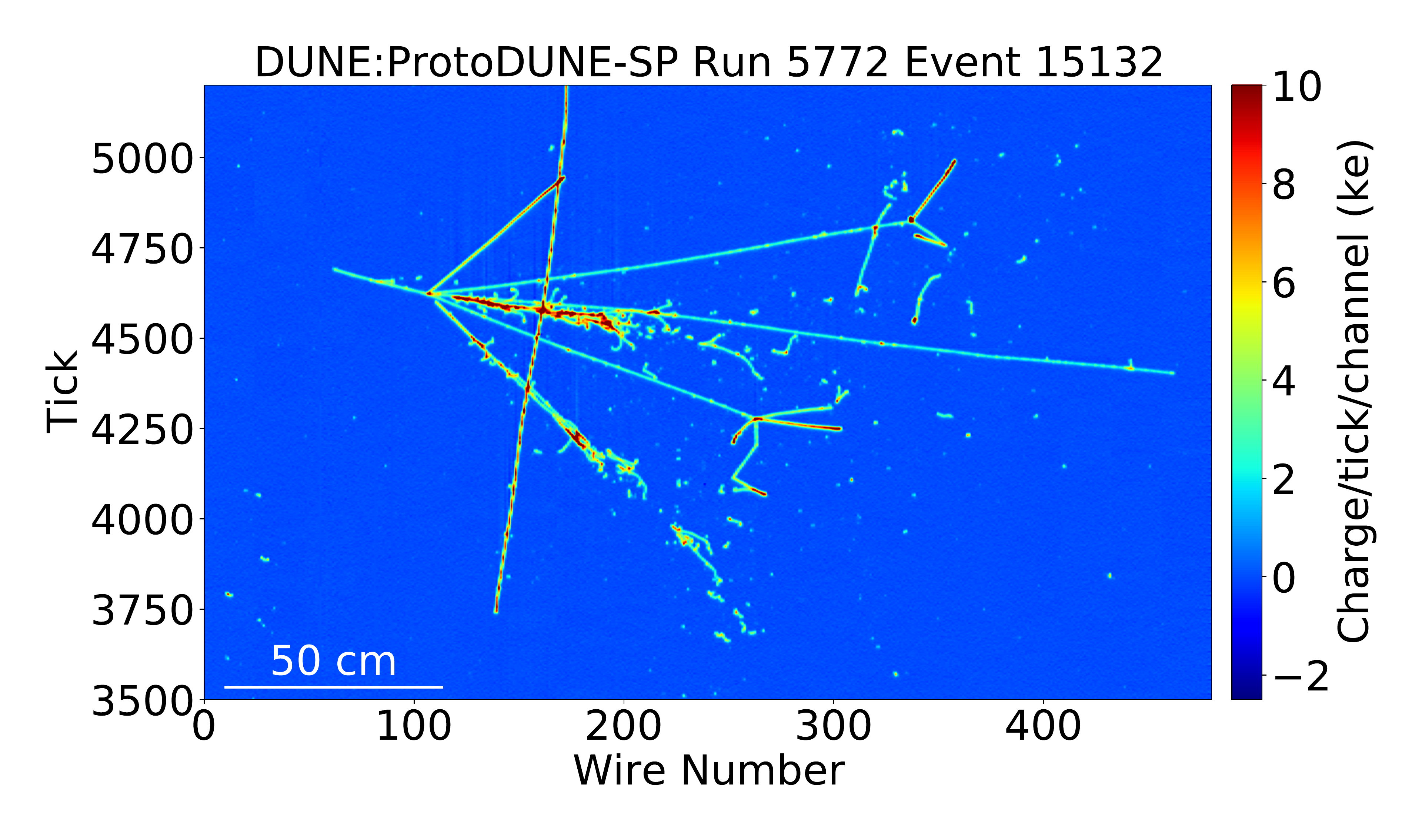}
    \caption{A 6~GeV/$c$ pion candidate.}
    \label{fig:6GeVPi}
  \end{subfigure}
  \hfill
  \vspace{1mm}
  \begin{subfigure}[b]{0.45\textwidth}
    \centering
    \includegraphics[width=\linewidth]{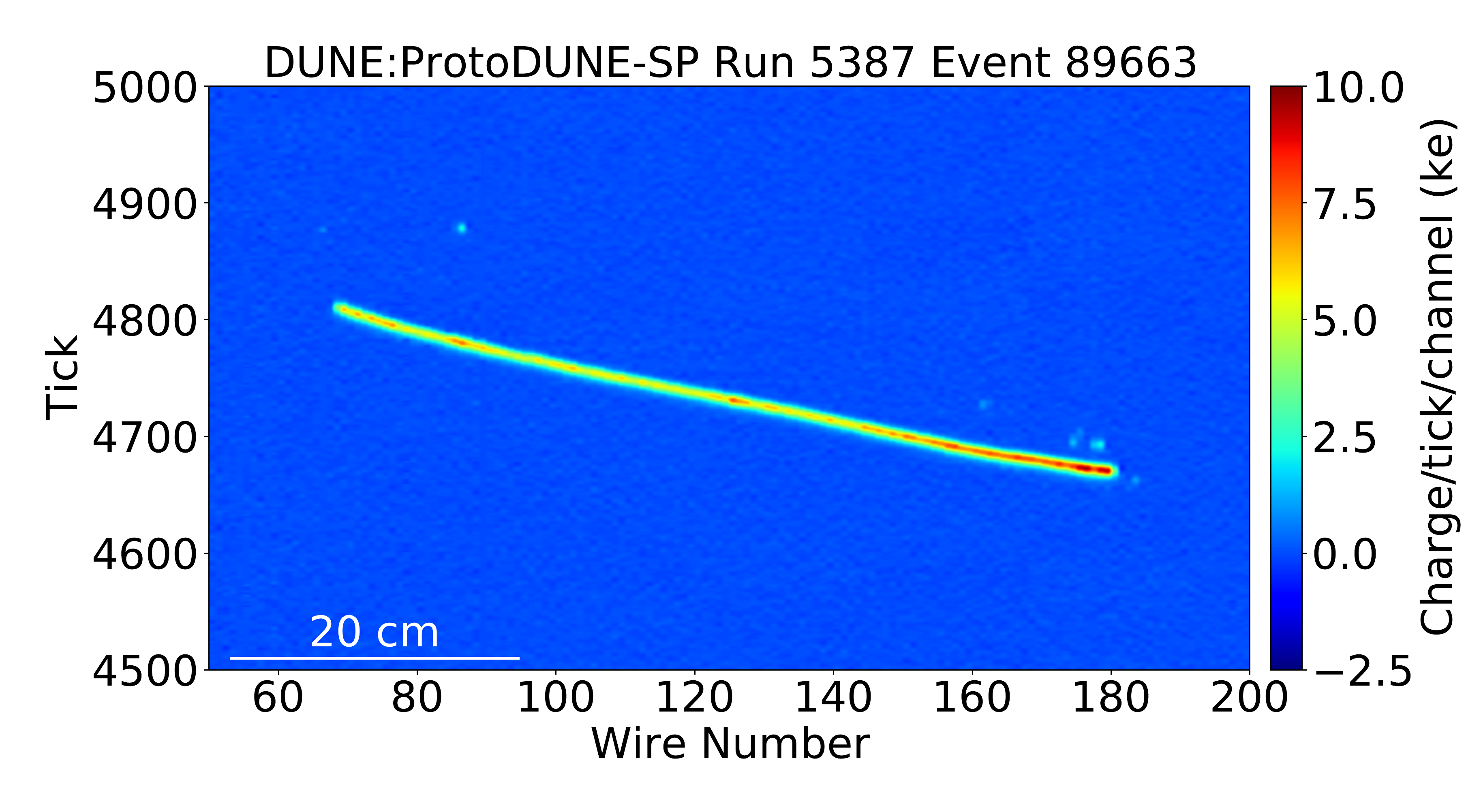}
    \caption{A 1~GeV/$c$ stopping proton candidate.}
    \label{fig:1GeVP}
  \end{subfigure}
  \hfill
  \begin{subfigure}[b]{0.45\textwidth}
    \centering
    \includegraphics[width=\linewidth]{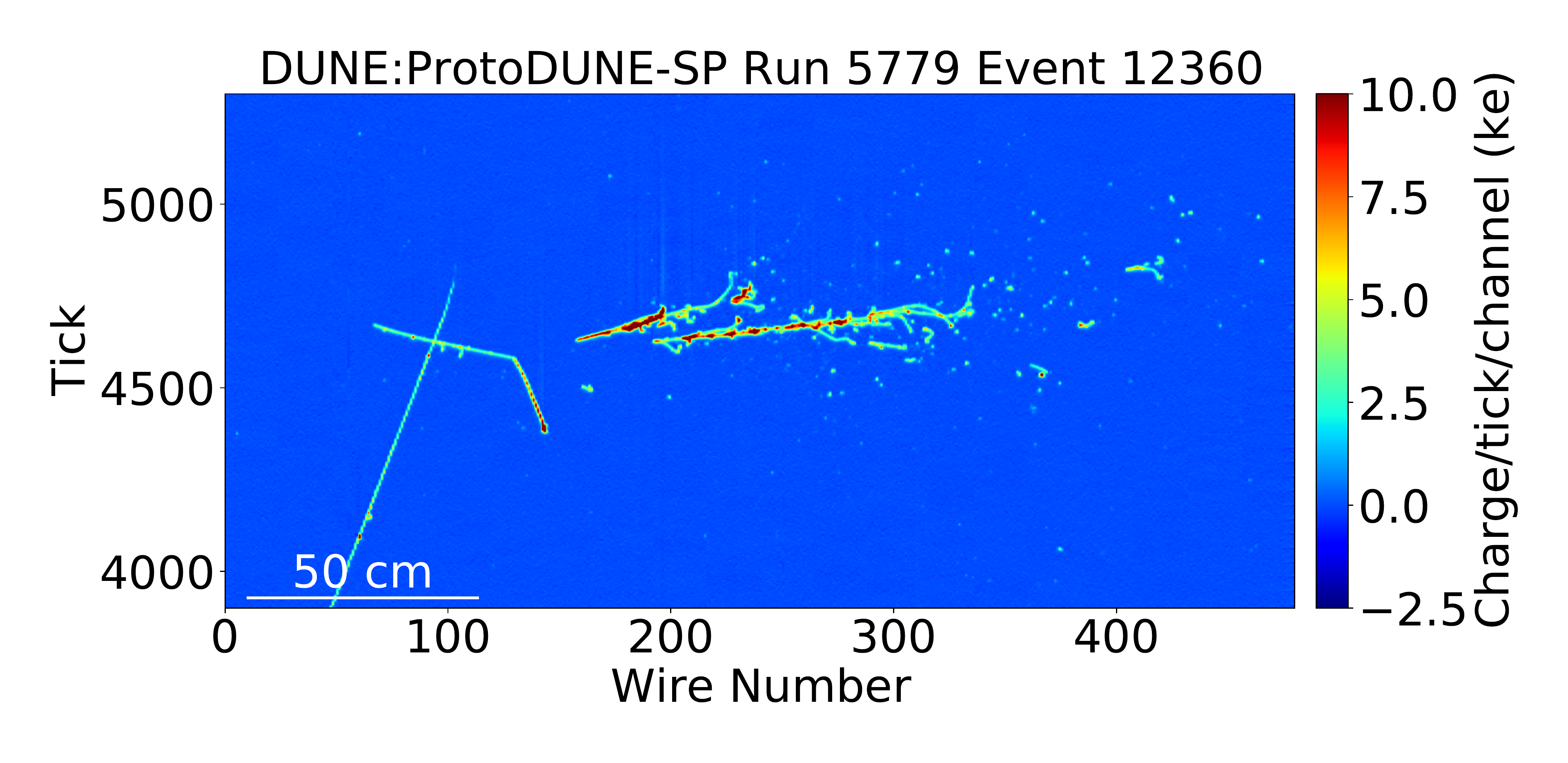}
    \caption{A 2~GeV/$c$ pion charge exchange candidate.}
    \label{fig:2GeVPi}
  \end{subfigure}
  \hfill
  \vspace{1mm}
  \begin{subfigure}[b]{0.45\textwidth}
    \centering
    \includegraphics[width=\linewidth]{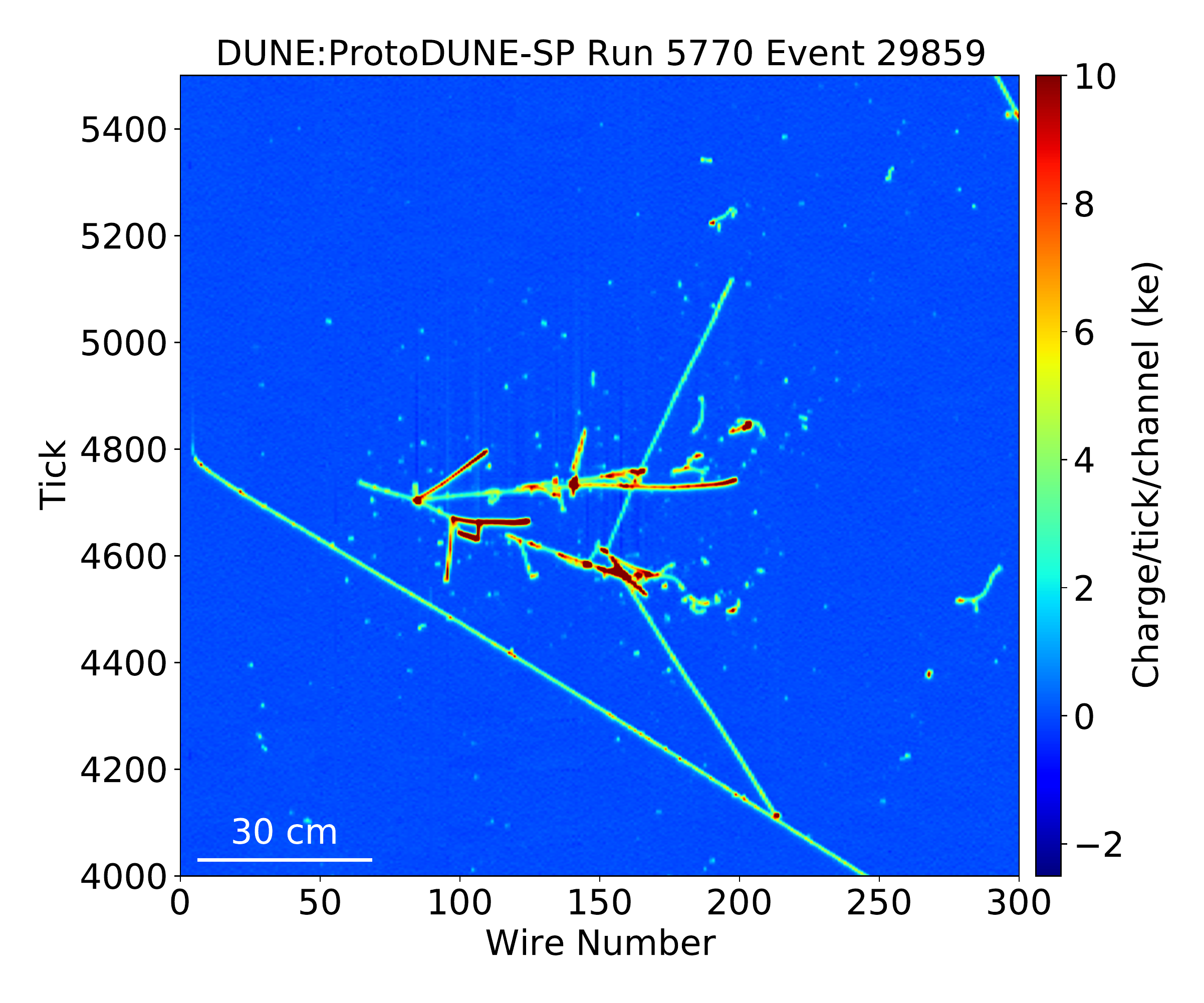}
    \caption{A 6~GeV/$c$ kaon candidate.}
    \label{fig:6GeVK}
  \end{subfigure}
  \hfill
  \begin{subfigure}[b]{0.45\textwidth}
    \centering
    \includegraphics[width=\linewidth]{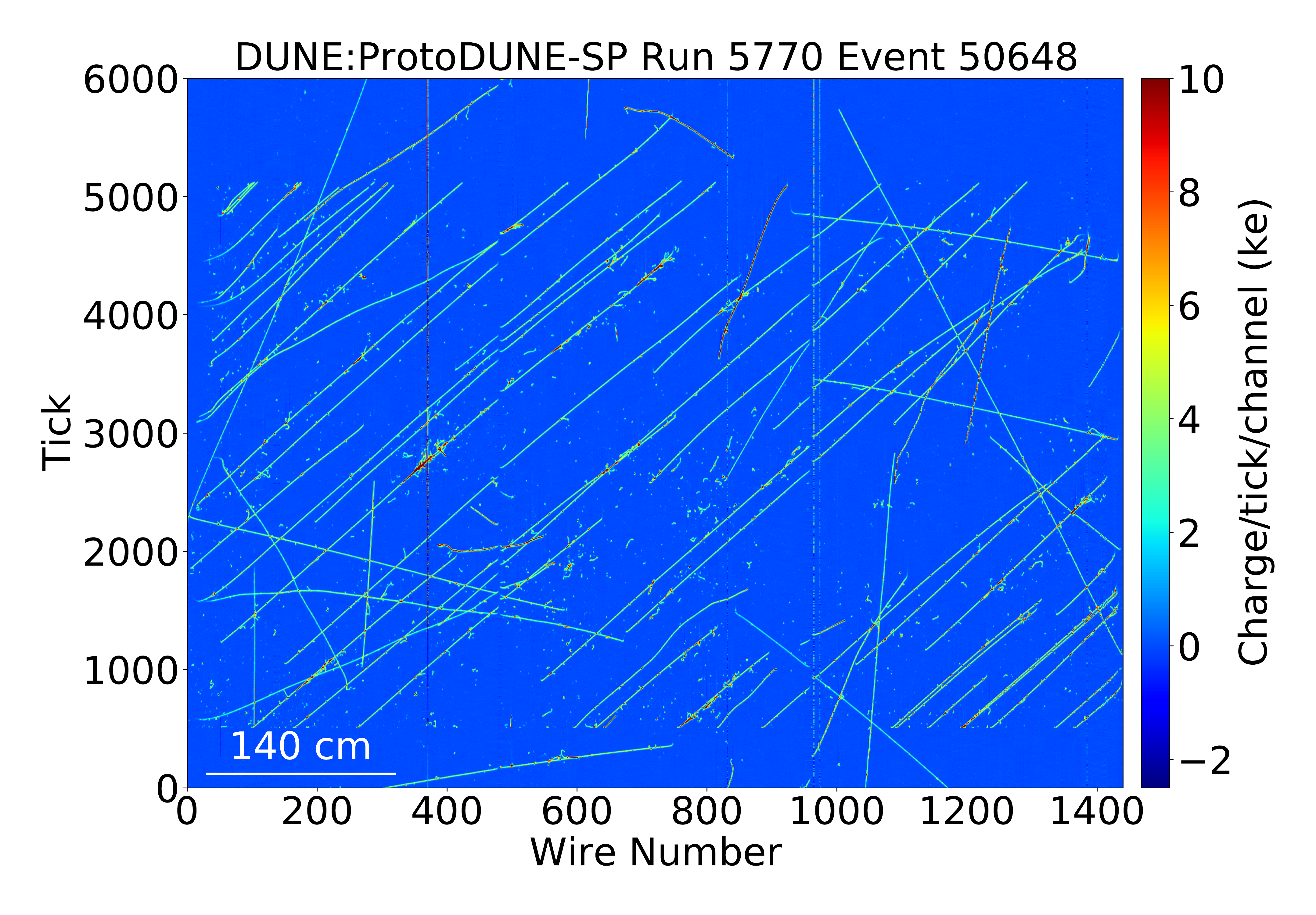}
    \caption{Large cosmic air shower candidate producing many parallel muons.}
    \label{fig:airshw}
  \end{subfigure}
  \hfill
  \caption{Various candidate events from ProtoDUNE-SP data, with beam particles entering from the left. The $x$ axis shows the wire number. The y axis shows the time tick in the unit of 0.5\,$\mu$s. The color scale represents the charge deposition. The beam particle starts approximately at wire number 90 and tick number 4800.}
  \label{fig:evd}
\end{figure}

The high-quality ProtoDUNE-SP data will be used to measure particle-argon interaction cross sections. Figure~\ref{fig:evd} shows various candidate events in the collection plane from ProtoDUNE-SP data after the noise mitigation described in section~\ref{tpcnoisesupp} and electronics gain calibration described in section~\ref{sec:gain_calib}. The color scale is in the unit of 1000 ionization electrons (ke). 

It is essential to understand the charge response of wires in a LArTPC for calorimetry. In order to measure the particle energy loss per unit length ($dE/dx$), it is important to correct for nonuniformities in the detector response and determine the energy scale to convert charge to energy. This section describes the procedure and results of the calibration of the charge and energy loss per unit length of the ProtoDUNE-SP LArTPC. Section~\ref{sec:SCE} discusses the calibration of space charge effects caused by the ion accumulation in the TPC. Section~\ref{sec:lifetime} describes the measurement of drift electron lifetime using TPC tracks and CRT information. Section~\ref{sec:muonCal} discusses the procedure to correct for remaining nonuniformities in the detector response and determine the energy scale. Section~\ref{sec:beamdedx} shows the calibrated energy loss per unit length of different beam particles, including 1 GeV/$c$ protons, muons, pions and electrons.

\subsection{Space charge effects in ProtoDUNE-SP}\label{sec:SCE}

As a detector located on the surface, ProtoDUNE-SP experiences a large flux of cosmic rays that results in a substantial amount of ionization produced in the detector per unit time.  Along with the production of ionization electrons, argon ions are also produced in the detector by these cosmic rays.  Because argon ions have drift velocities on the order of several millimeters per second at $\sim\SI{500}{V/cm}$ in liquid argon, $2\textup{--}4 \times 10^{5}$ times slower than ionization electrons at the same electric field, a considerable amount of positive space charge is expected in the detector as these argon ions build up on a timescale of roughly ten minutes. The convective flow of liquid argon in the detector has a similar velocity scale, and thus the space charge distribution depends on the velocity field of the liquid argon in the TPC.   Asymmetries in the fluid flow are expected to produce asymmetries in the space charge distribution.  The difficulty of predicting the exact fluid flow pattern necessitates a data-driven study. A steady flux of cosmic rays ensures that the positive ions are constantly replenished in the detector, leading to persistent distortions of the electric field in the TPC.

\begin{figure}[!htp]
  \centering
  \vspace{4mm}
  \includegraphics[width=.49\textwidth]{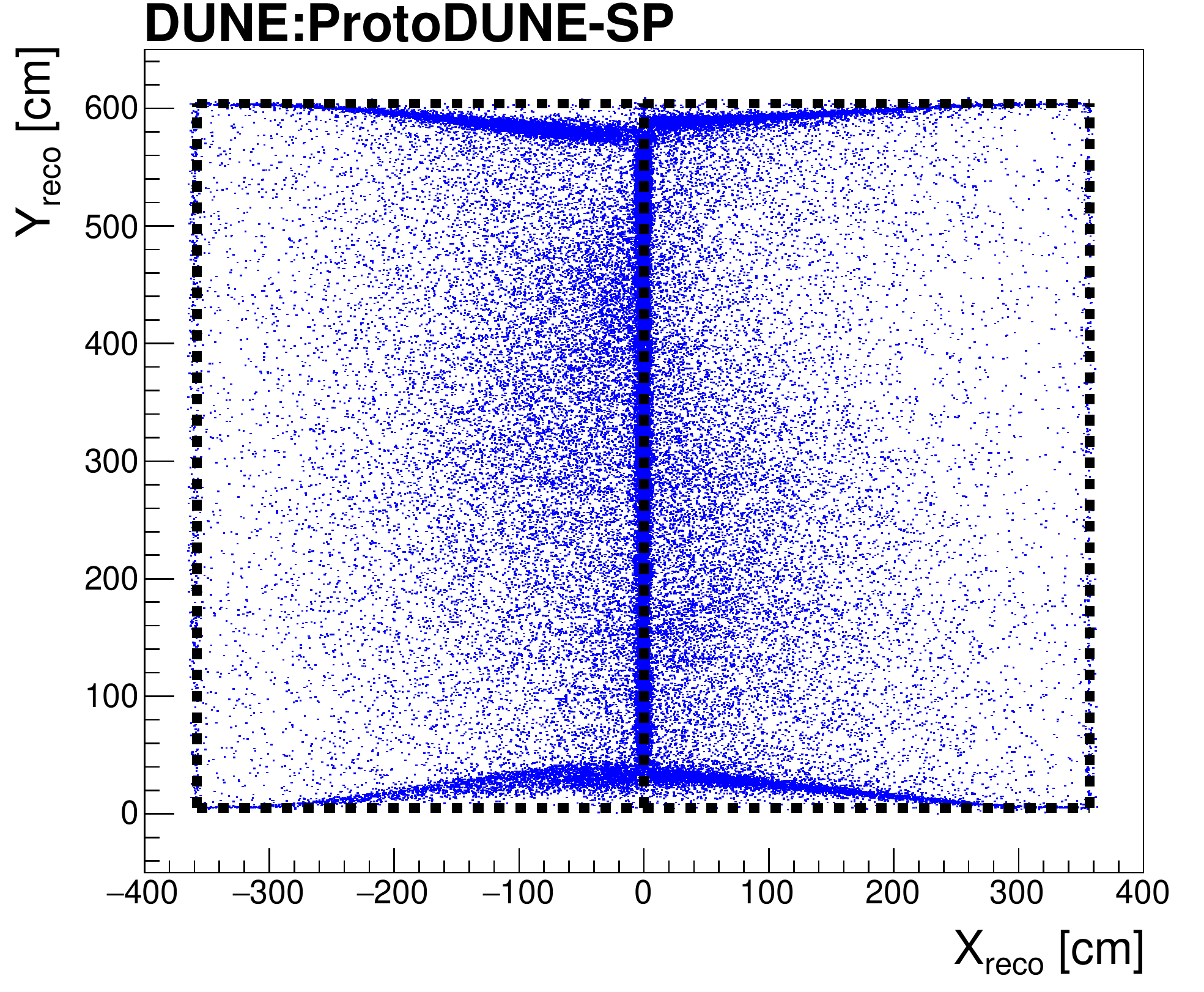}
  \includegraphics[width=.49\textwidth]{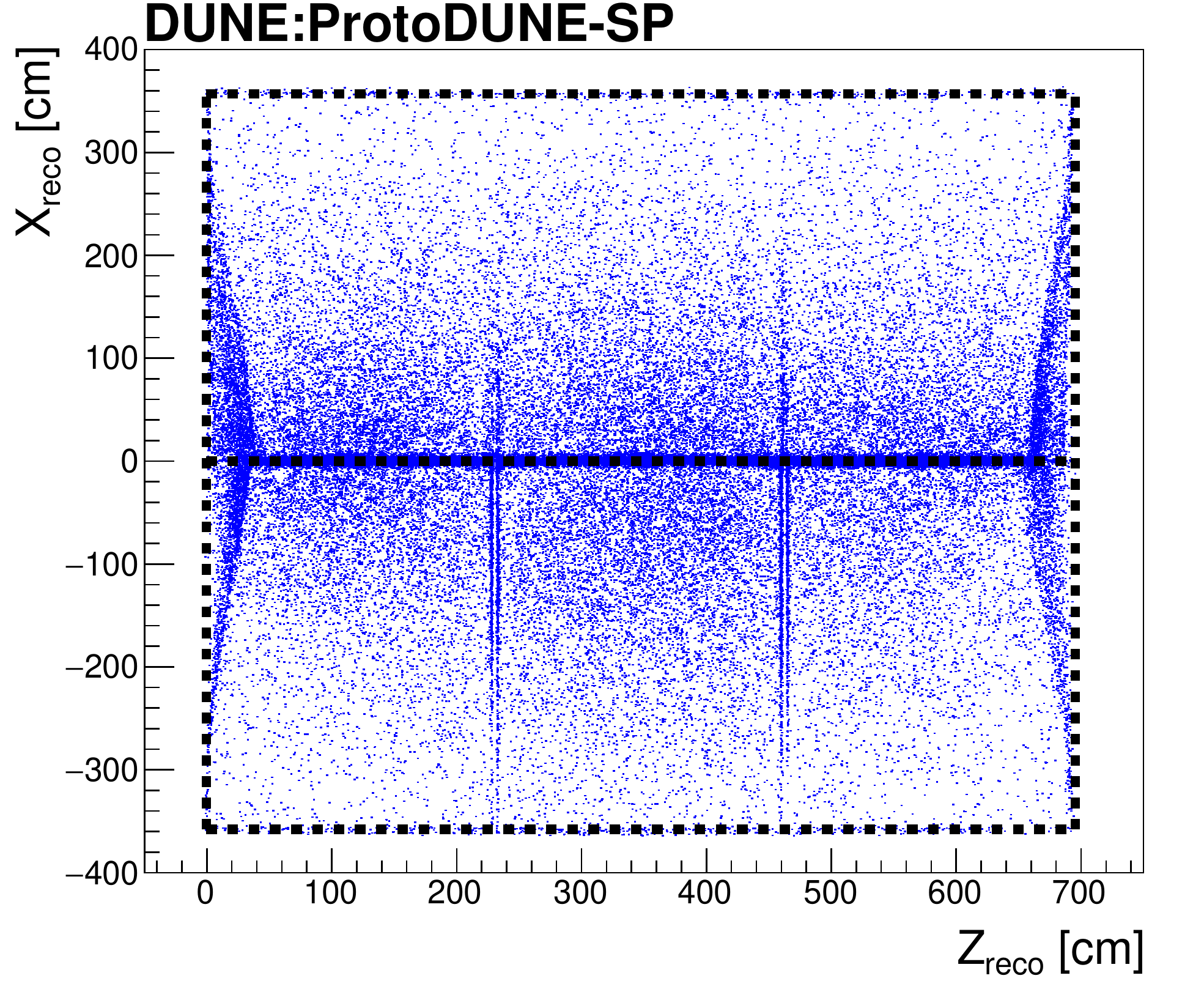} \\
  \caption{Projections of reconstructed and $t_{0}$-tagged cosmic-ray track end points in the $xy$ plane (left) and $zx$ plane (right) in ProtoDUNE-SP data; the selected tracks are $t_{0}$-tagged by requiring that they cross the cathode plane ($x=0$), as described in section~\ref{sec:reco:pandora}.  In the absence of space charge effects, the track end points should be reconstructed along the boundary of the TPC active volume (dashed lines).  The gaps between the APAs can also be observed in the $zx$ plane projection (vertical streaks in the middle of the image).}
  \label{fig:SCE_Projections_Data}
\end{figure}

These electric field distortions alter ionization electron drift paths in the detector, leading to ionization charge being reconstructed at different positions in the detector than where the charge originated from.  The electric field distortions also impact the amount of prompt electron-ion recombination experienced at the points of energy deposition in the detector.  Both of these effects can bias reconstructed particle energies and trajectories, and by modifying reconstructed $dE/dx$ along a particle track (or at the beginning of an electromagnetic shower), can lead to complications in particle identification in a LArTPC detector.  As a result, space charge effects should be carefully characterized at any large LArTPC detector operating at or near the surface, and calibrated out when reconstructing particle trajectories, energies and $dE/dx$.  The observation of transverse ionization charge migration during drift in early ProtoDUNE-SP data-taking, shown in figure~\ref{fig:SCE_Projections_Data}, is consistent with a large positive space charge density in the center of the TPC pulling ionization charge inward toward the middle of the detector during drift toward the anode planes.  This observation highlights the fact that significant space charge effects are present at ProtoDUNE-SP and need to be addressed when calibrating the detector.  In contrast to the case of ProtoDUNE-SP, space charge effects are expected to be negligible at the single-phase DUNE far detector due to the very low cosmic-ray rate deep underground.

\begin{figure}[!htp]
  \centering
  \includegraphics[width=.99\textwidth]{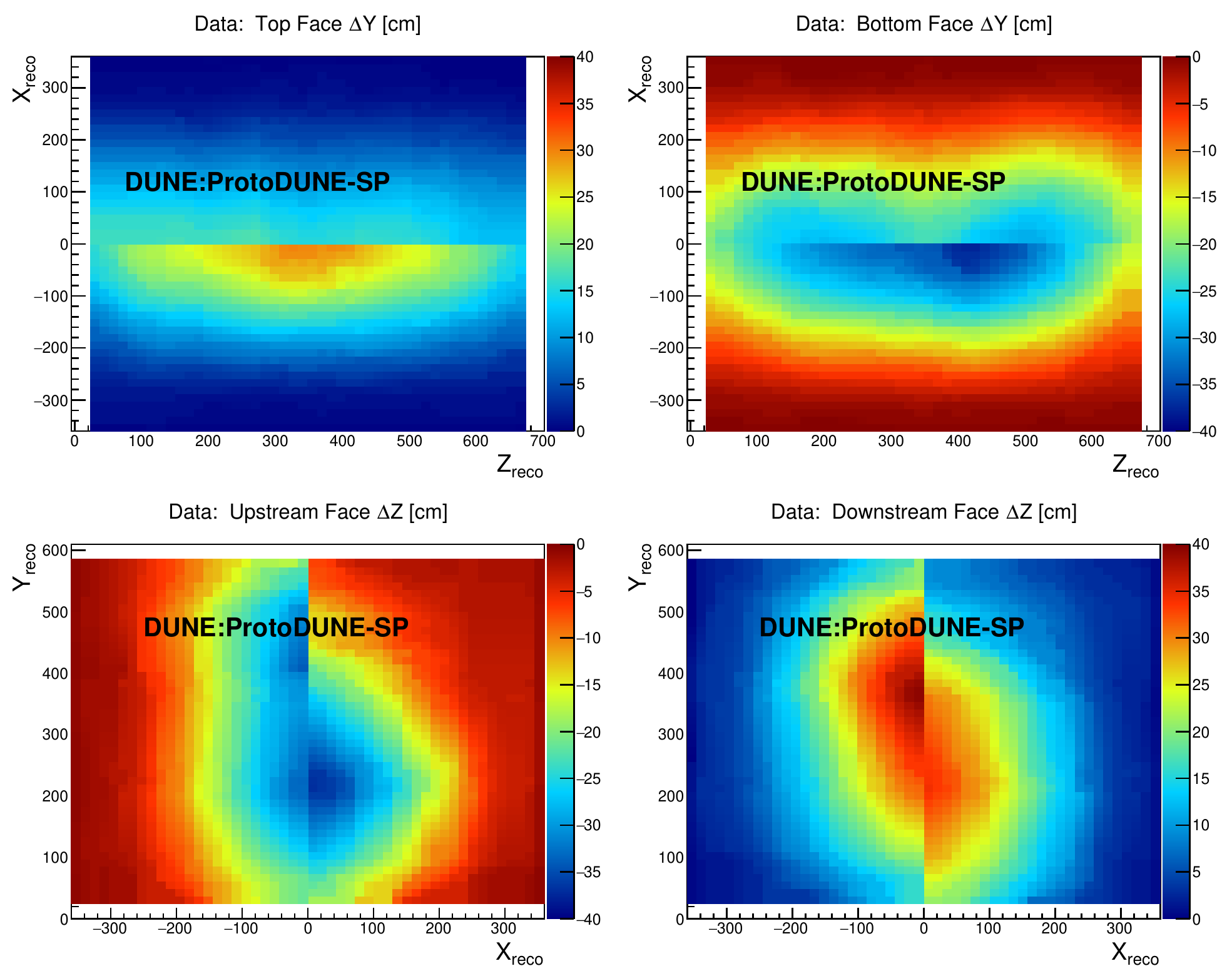} \\
  \caption{Spatial distortions normal to the top detector face (upper left), bottom detector face (upper right), upstream detector face (lower left), and downstream detector face (lower right) in ProtoDUNE-SP data.  The color axis represents the additive correction (in cm) one must apply to the start/end point of a track passing through the given detector face in order to correct its position to the true entry/exit point at the side of the detector.}
  \label{fig:SCE_FaceDistortions_Data}
\end{figure}

Figure~\ref{fig:SCE_FaceDistortions_Data} shows the magnitude of spatial offsets at four faces of the ProtoDUNE-SP TPC (top, bottom, front or upstream with respect to the beam direction, back or downstream with respect to the beam direction); specifically, spatial distortions in the direction normal to each detector face are shown.  These spatial distortions are estimated using the ends of reconstructed tracks that have been $t_{0}$-tagged in order to know their position in $x$, the ionization drift coordinate.  The transverse spatial distortion is determined by measuring the distance between the end of the track and the location of the detector face in the direction orthogonal to the detector face being considered.  As illustrated in figure~\ref{fig:SCE_FaceDistortions_Data}, spatial distortions as large as $\SI{40}{cm}$ are observed in ProtoDUNE-SP data, largest near the faces of the TPC and furthest from the anode planes; the latter observation can be understood as a result of charge originating further away from the anode experiencing space charge effects for a longer time, yielding a larger impact on drift path in the detector.

\begin{figure}[!htp]
  \centering
  \includegraphics[width=.99\textwidth]{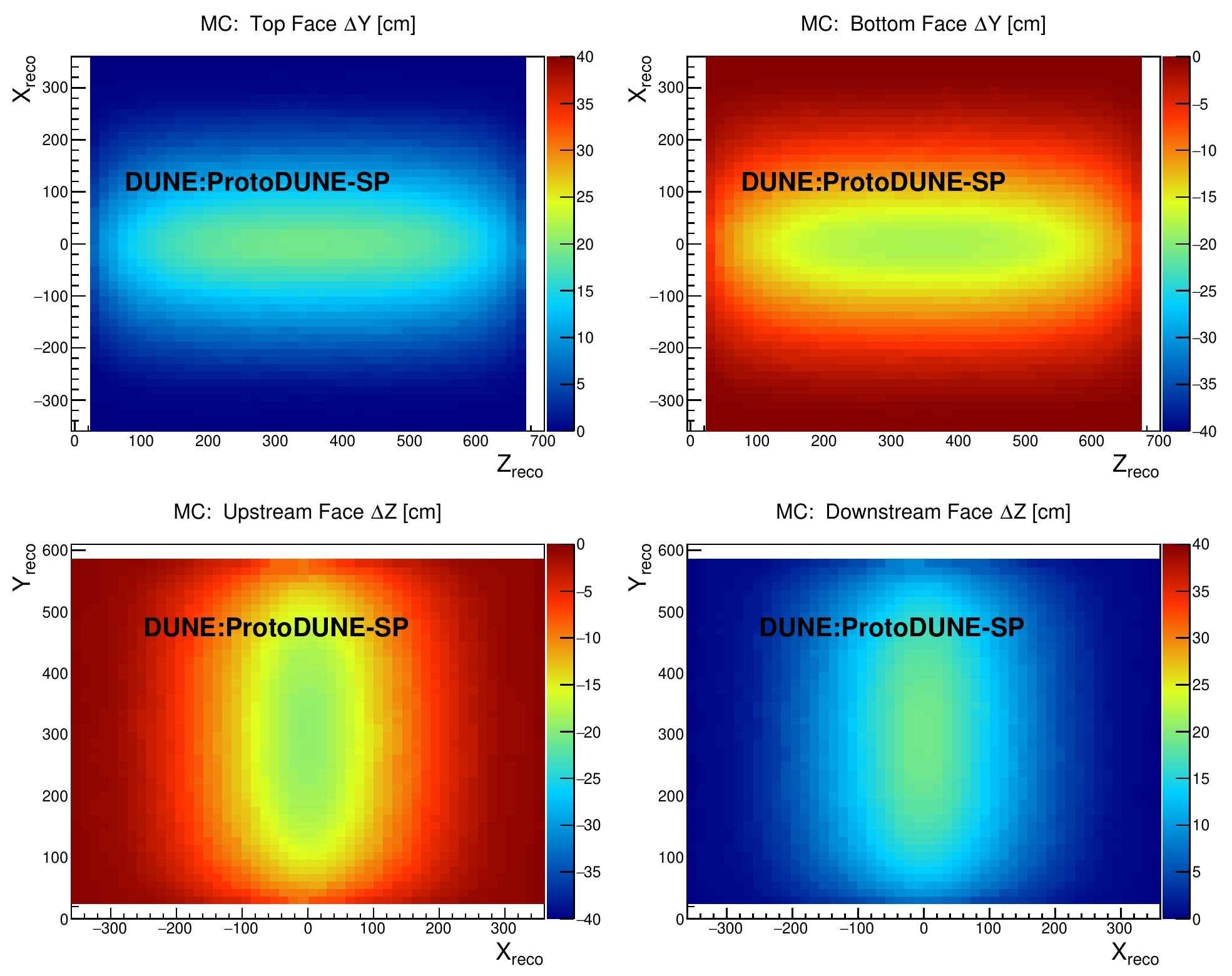} \\
  \caption{Spatial distortions normal to the top detector face (upper left), bottom detector face (upper right), upstream detector face (lower left), and downstream detector face (lower right) in the original ProtoDUNE-SP Monte Carlo simulation.  The color axis represents the additive correction (in cm) one must apply to the start/end point of a track passing through the given detector face in order to correct its position to the true entry/exit point at the side of the detector.}
  \label{fig:SCE_FaceDistortions_MC}
\end{figure}

A dedicated simulation of space charge effects was developed for ProtoDUNE-SP, using a software package originally developed for MicroBooNE~\cite{Mooney:2015kke,SCE_CosmicPubNote}. The analysis shown in figure~\ref{fig:SCE_FaceDistortions_Data} is for data; the same analysis applied to Monte Carlo simulated events is shown in figure~\ref{fig:SCE_FaceDistortions_MC}.

Comparing figures~\ref{fig:SCE_FaceDistortions_Data} and~\ref{fig:SCE_FaceDistortions_MC}, similar trends are observed in simulation as compared to data, building confidence that the spatial distortions observed in data are indeed a result of space charge effects.  However, there are also several differences:
\begin{itemize}
\item the magnitude of spatial distortions in data are generally larger than in simulation, by as much as a factor of two at some locations in the detector;
\item there is an asymmetry in the magnitude of the spatial distortions about the cathode ($x=0$), which is not present in simulation; and
\item the trends in the spatial distortion maps differ qualitatively from the simulation in certain parts of the detector, such as near the top of the upstream and downstream detector faces on the $x>0$ side of the cathode.
\end{itemize}
The first point above may be explained by a combination of potentially using an incorrect value for the argon ion drift velocity, which is not well known in liquid argon, and the possibility of liquid argon flow (not included in simulation) moving the argon ions around in the TPC in addition to their nominal drift in the applied electric field.  The second and third points are potentially explained by the effects of liquid argon flow alone.  These explanations are at present speculative; more detailed study is ongoing and will be reported in a future work.

Given the significant impact of space charge effects on particle trajectories, energies and $dE/dx$, as well as the inability for the dedicated space charge simulation to reproduce the observations seen in data, a data-driven simulation of space charge effects was produced for ProtoDUNE-SP using the results shown in figures~\ref{fig:SCE_FaceDistortions_Data} and \ref{fig:SCE_FaceDistortions_MC} as a starting point.  This data-driven map of space charge effects (both spatial distortions and electric field distortions, each with three components, for a total of six three-dimensional maps) can be ``inverted'' to remove effects of space charge in a calibration step for both events in actual data and those produced with this data-driven simulation.  The data-driven spatial distortion and electric field distortion maps are produced using the following steps:
\begin{itemize}
\item the two-dimensional transverse spatial offset maps at the four detector faces (top, bottom, upstream, and downstream) shown in figures~\ref{fig:SCE_FaceDistortions_Data} and \ref{fig:SCE_FaceDistortions_MC} are used to form a ``scale factor'' map at each detector face by taking the ratio of the data map to the Monte Carlo map on a pixel-by-pixel basis;
\item the scale factor maps are used to rescale the simulated three-dimensional spatial distortion map by linearly interpolating the scale factors between the top and bottom detector faces for spatial distortions in the $y$ direction, linearly interpolating the scale factors between the upstream and downstream detector faces for spatial distortions in the $z$ direction, and performing the average of the linear interpolations in these two directions for spatial distortions in the $x$ direction; the voxel-by-voxel scale factor obtained in this way is used as a multiplicative factor to rescale the spatial distortion magnitude in the corresponding three-dimensional voxel (a ``voxel'' here refers to a volumetric pixel);
\item the resulting data-driven spatial distortion maps are then inverted in order to obtain ``inverted spatial distortion maps'' that can be used to calibrate out spatial distortions in reconstructed data or Monte Carlo events via repositioning reconstructed ionization charge space points in three dimensions back to their point of original deposition (see more below); and
\item the gradient of the spatial distortion along the local drift direction, determined using the inverted spatial distortion maps, along with the known ionization electron drift velocity as a function of electric field, are used to obtain the electric field distortion maps (in three dimensions) using a method previously explored at MicroBooNE~\cite{Adams:2019qrr}.
\end{itemize}
The result of this procedure is a set of three-dimensional spatial distortion and electric field maps that are included in the ProtoDUNE-SP simulation used in the first results showcased in this work.

\begin{figure}[!htp]
  \centering
  \includegraphics[width=.45\textwidth]{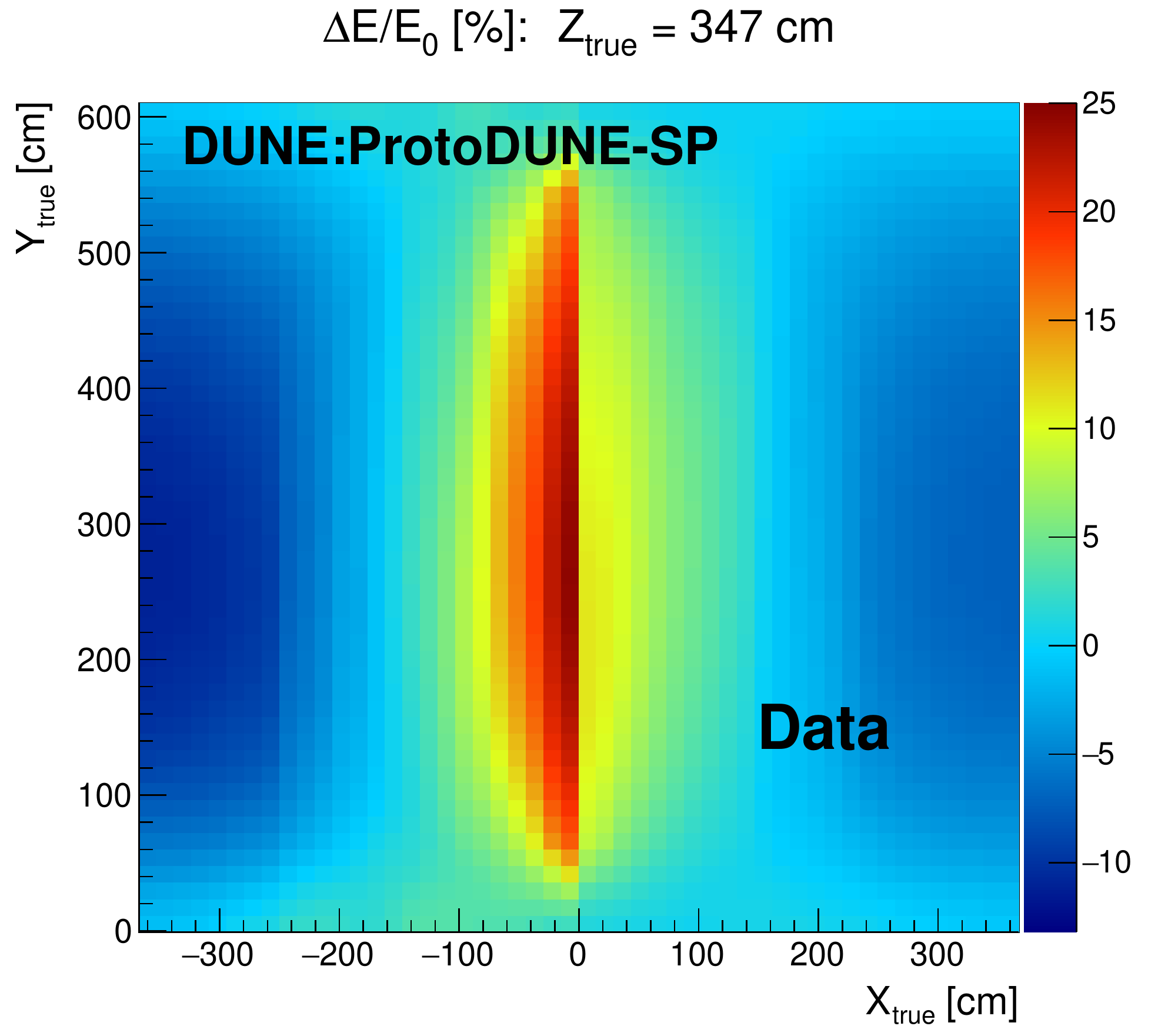}
  \hspace{4mm}
  \includegraphics[width=.45\textwidth]{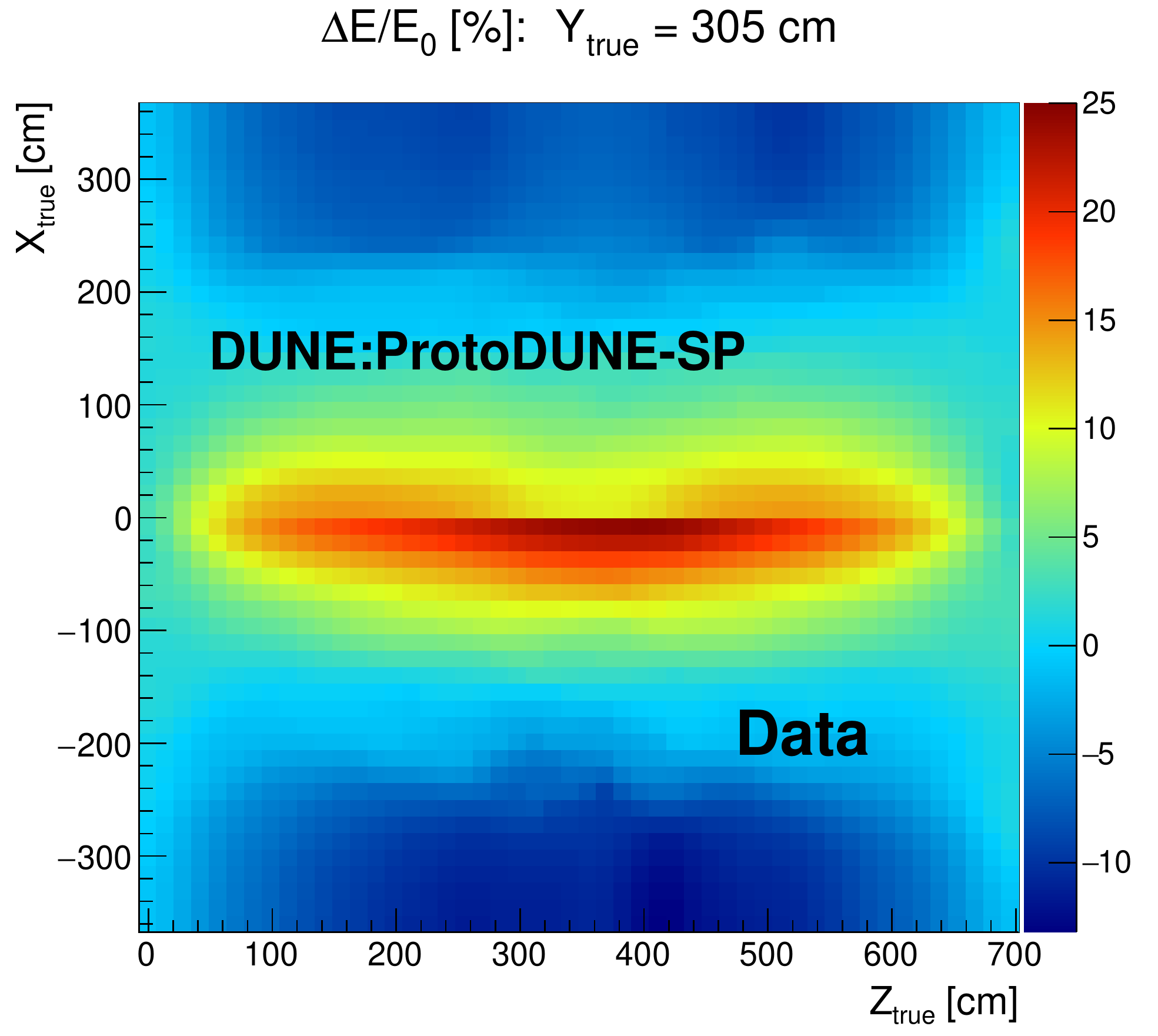} \\
  \includegraphics[width=.45\textwidth]{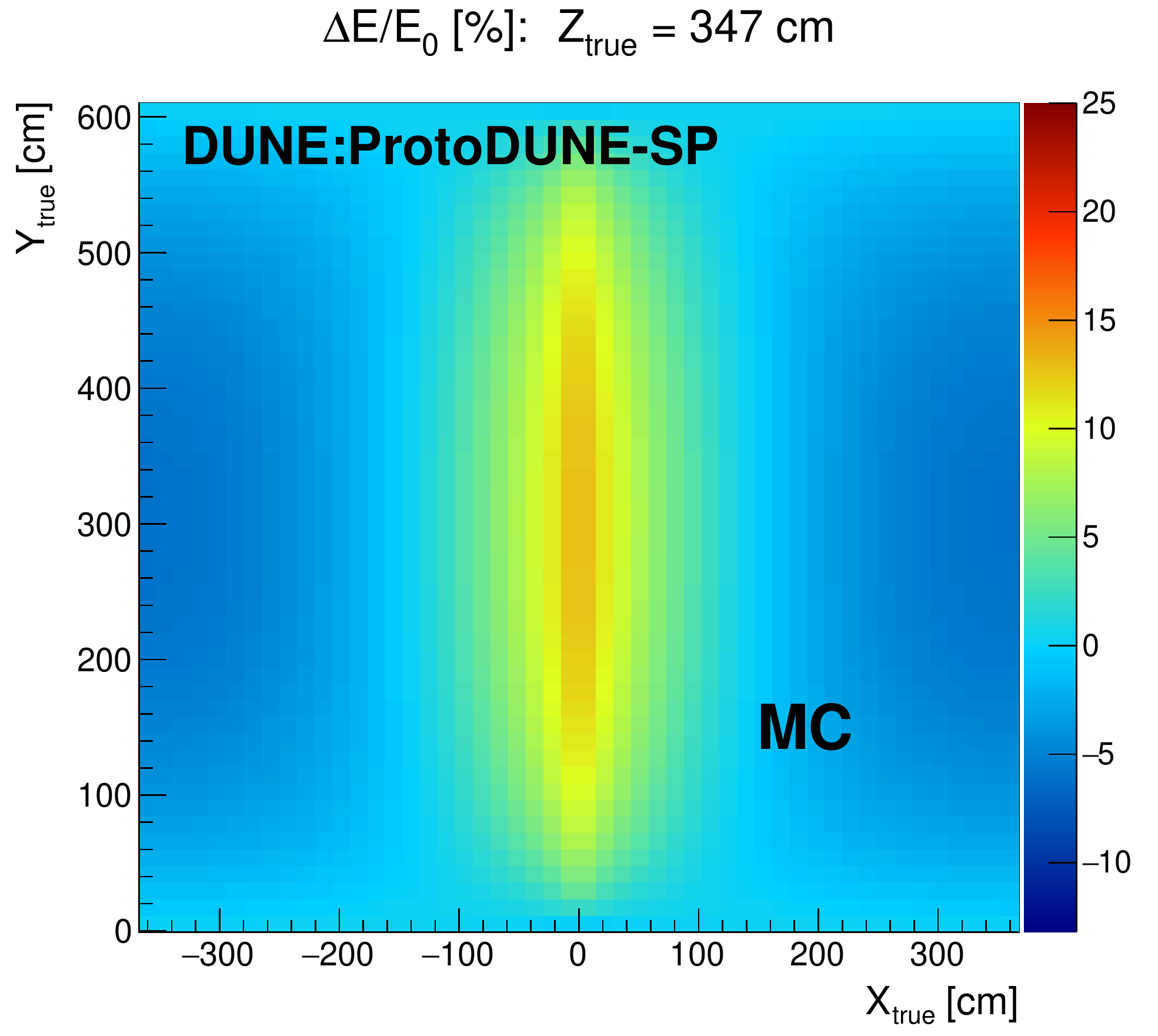}
  \hspace{4mm}
  \includegraphics[width=.45\textwidth]{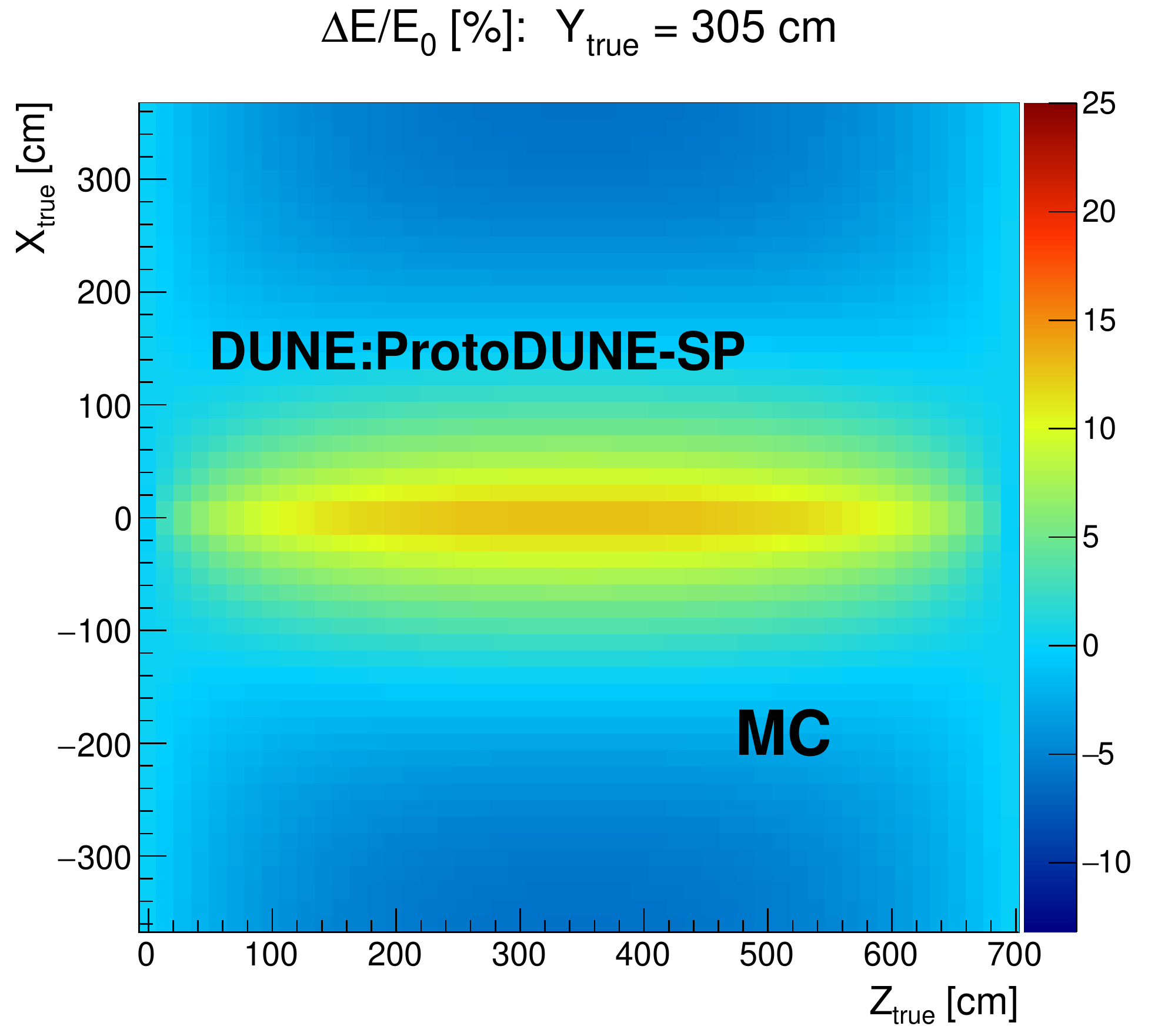} \\
  \vspace{-4mm}
  \caption{Two-dimensional slices of the three-dimensional electric field magnitude distortion map in both ProtoDUNE-SP data (top row) and the original ProtoDUNE-SP Monte Carlo simulation (bottom row); the local electric field distortion magnitude is shown as a percentage of the nominal drift electric field magnitude.  Shown are slices in the $z$ direction (left column) and $y$ direction (right column), looking at the center of the detector in both slices.}
  \label{fig:SCE_ElecField}
\end{figure}

The electric field magnitude variations in a couple of slices of the ProtoDUNE-SP TPC are shown in figure~\ref{fig:SCE_ElecField}, comparing to the prediction of the original space charge effect simulation.  It is observed in figure~\ref{fig:SCE_ElecField} that the electric field magnitude variations in data are as large as 25\% with respect to the nominal drift electric field; an approximate estimate of the systematic uncertainty on this number is 5\% with respect to the nominal drift electric field magnitude (20\% relative uncertainty with respect to the full systematic effect), driven by the uncertainty in extrapolating spatial offsets from the detector faces into the center of the detector.  The simulation utilizes the data-driven spatial distortion maps to modify the reconstructed position of ionization charge to better represent data events, while the data-driven electric field distortion maps are used to improve the prediction of charge yield after electron-ion recombination, as this effect is dependent on the local electric field magnitude.

Additionally, a calibration of particle $dE/dx$ was developed for both tracks and the track-like segments at the beginnings of electromagnetic showers measured by ProtoDUNE-SP.  In this calibration, the spatial distortion map is used to correct for spatial squeezing/stretching of charge, impacting $dx$, and the electric field distortion map is used to correct for the electric-field-dependence of electron-ion recombination in the liquid argon, impacting $dE$.  The performance of this first attempt at a calibration of space charge effects at ProtoDUNE-SP is demonstrated in several of the results presented in this work in sections~\ref{sec:muonCal} and ~\ref{sec:beamdedx}.

The time dependence of space charge effects has been observed to be relatively small during the period of time that ProtoDUNE-SP was taking beam data, on the order of 5\% of the total spatial distortion magnitude.  A detailed study of the time dependence of space charge effects at ProtoDUNE-SP will be presented in a future work.
\subsection{Drift electron lifetime}
\label{sec:lifetime}

The liquid argon of the ProtoDUNE-SP detector contains impurities, such as water and oxygen, that can capture the ionized electrons as they drift towards the APA. Although the negative ions formed by the attached electrons still drift to the APA, they drift much more slowly than the unattached electrons and contribute negligibly to the signals measured on the APAs. The charge measured by the APAs then becomes reduced due to the impurities capturing the electrons, lowering and biasing the amount of charge measured on the wire planes.
This effect is modeled as an exponential decay as a function of time:
 \begin{equation}\label{eqn:elifetime}
    Q(t)=Q_0\exp\left(-(t_{\rm{hit}}-t_0)/\tau\right),
 \end{equation}
where $Q(t)$ is the charge measured on a wire, $Q_0$ is
the initial charge created by the ionization of the argon accounting for recombination, $t_0$ is the time at which the ionization took place, $t_{\rm{hit}}$ is the time the drifting charge arrived the APA, and $\tau$ is the drift electron lifetime.
 A larger value of $\tau$ corresponds to higher liquid argon purity, as fewer drifting electrons will attach to impurities as they drift to the APA.

Purity monitors located inside the cryostat, but outside the field cage, measure the drift electron lifetimes for the argon inside their drift volumes, and thus are not expected to measure exactly the drift electron lifetime in the TPC.  Furtheremore, the electric field strength in the purity monitors is lower than the electric field strength inside the TPC.  Since the rate of drift electron attachment to impurities depends on the electric field strength, the measured lifetimes in the purity monitors are expected to further differ from that in the TPC. In situ measurements of the drift electron lifetime from signals in the TPC therefore are needed in order to calibrate the results of charge-based analyses.

The drift electron lifetime inside the TPC is measured by fitting the $dQ/dx$ of collection plane hits from cosmic-ray tracks as a function of drift times. A sample of cosmic-ray tracks that pass through the front and back faces of the TPC and the CRT are selected. CRT data are used to calibrate track positions and to provide timestamps for TPC tracks.

The electron lifetime measurement starts by matching a CRT track to a TPC track using the positions of X and Y from both tracks. The timestamp from the CRT hits serve as the $t_0$ for the TPC track. The $dQ/dx$ of a hit is defined to be the hit charge ($Q$) obtained from the area of a Gaussian fit to the deconvolved signal divided by the step length from the previous collection plane hit to the current collection plane hit.

\begin{figure}[!ht]
    \centering
    \includegraphics[width=.8\textwidth]{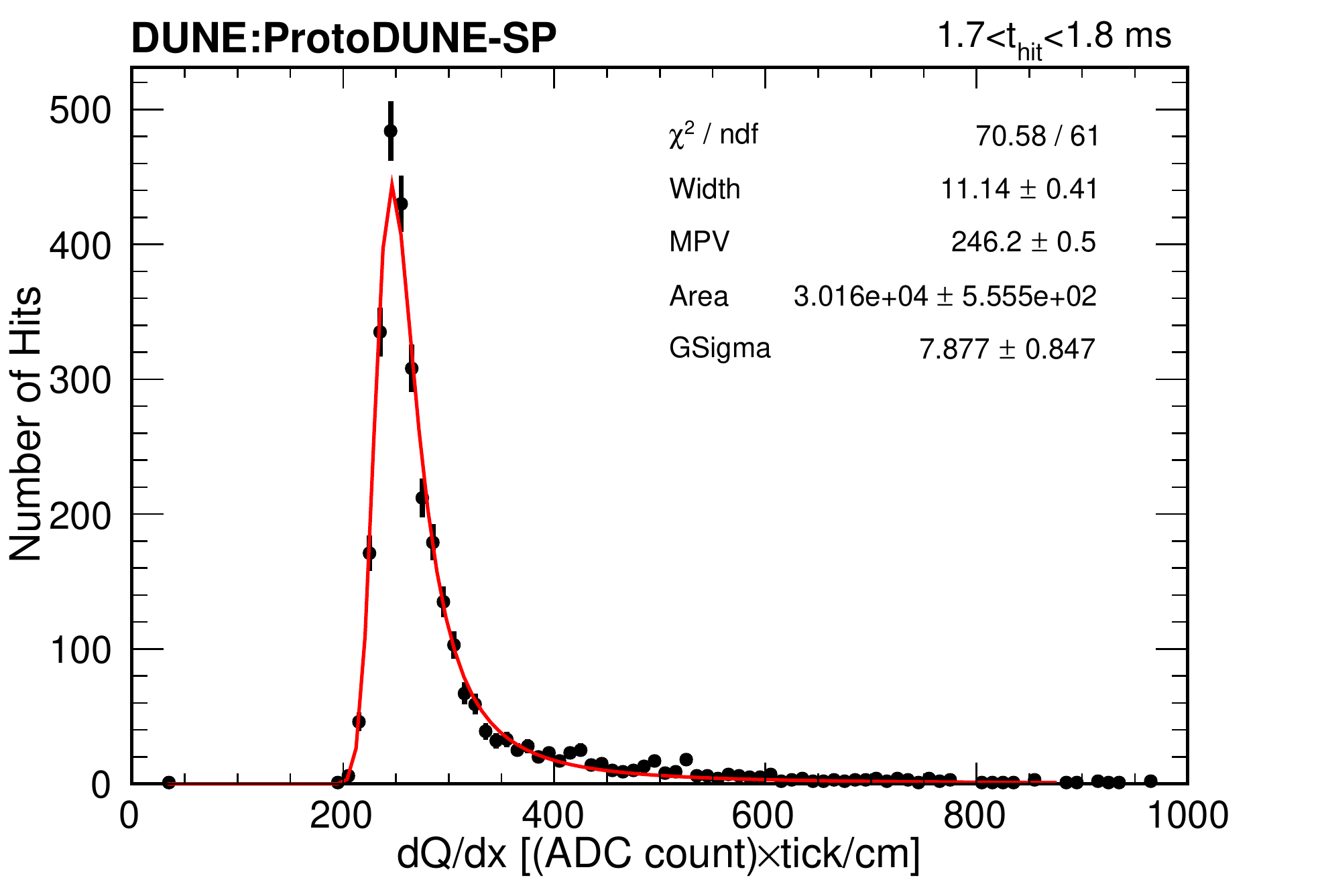}
    \caption{Distribution of $dQ/dx$ for a slice of 100~$\mu$s for a lower purity run in early November 2018.}
    \label{fig:lan}
\end{figure}

To avoid the space charge effect distortions on the location of a TPC hit, the CRT track is used to determine the hit's position in X and Y as a function of Z, which is found through which collection plane wire the hit occurred. The track's position is then fed into the electric field calibration map from figure~\ref{fig:SCE_ElecField}. This field map corrects the $dQ/dx$ by a scale factor based on electric field deviations from the space charge effect. The electric field distortions from the space charge effect were observed to cause at most a 1.75\% difference in $dQ/dx$ along the drift distance. These calibrated measurements of $dQ_{\rm{calibrated}}/dx$ are fitted to a Landau function convolved with a Gaussian to find the most probable value (MPV) for 100~$\mu$s bin in time. An example of such a fit is shown in figure~\ref{fig:lan}. The function that describes the drift electron lifetime then can be quantified by evaluating $dQ_{\rm{MPV}}/dx$ as such: 
  \begin{equation}
    \frac{dQ(t)_{\rm{MPV}}}{dx}=\frac{dQ_{0,\rm{MPV}}}{dx}\exp(-(t_{\rm{hit}}-t_{\rm{CRT}})/\tau)
    \label{eqn:ecalib}
    \end{equation} 
where adjustments in the timing are made based on the timestamp provided by the CRT.

\begin{figure}[!ht]
    \centering
    \includegraphics[width=0.8\textwidth]{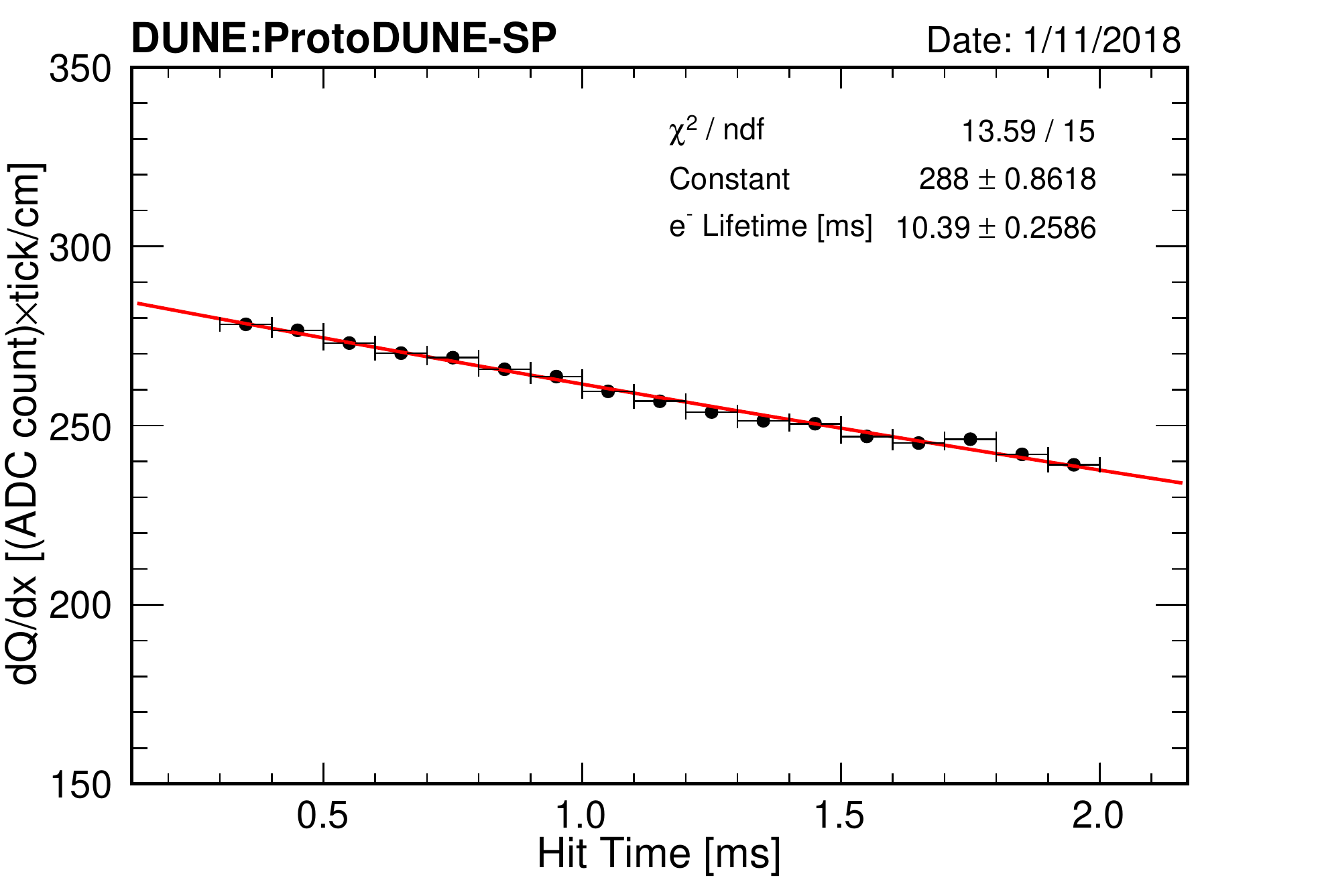}
     \includegraphics[width=0.8\textwidth]{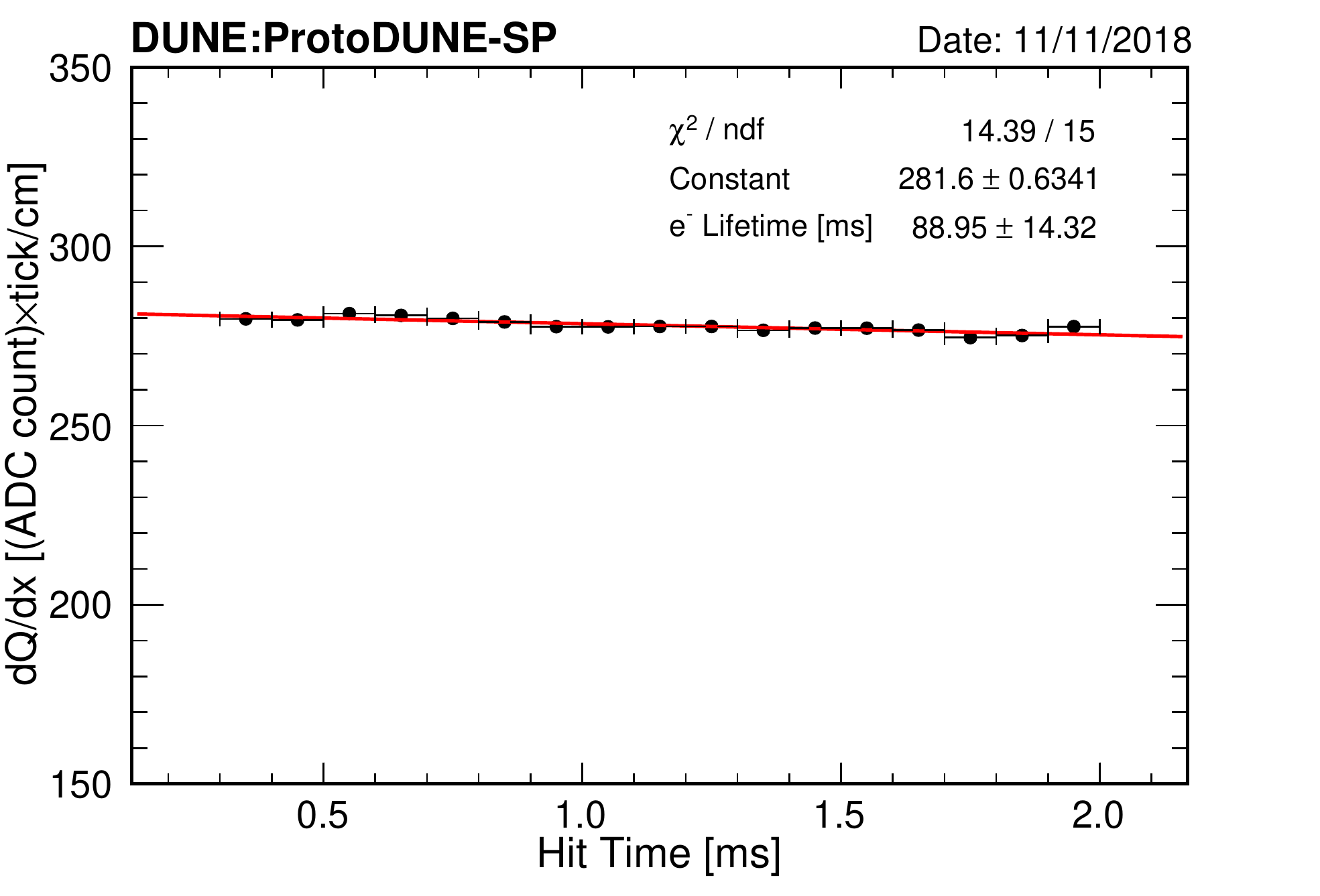}
    \caption{Plot the MPV of the $dQ/dx$ distribution as a function of the hit time, fit to an exponential decay function on November 1st, 2018 during a period of lower purity (top) and on November 11th, 2018 during a period of higher purity (bottom). Only statistical errors are included.}
    \label{fig:elifetime}
\end{figure}

 Data was taken using the CRT once it became operational on November 1st, 2018 and runs were taken with the CRT during beam data-taking from November 1st, 2018 to November 11th, 2018, the last day of beam. Fits to the MPV of the $dQ/dx$ distributions as functions of hit time are shown starting the first day in November and for the final day of beam data-taking in figure~\ref{fig:elifetime}. During the end of October 2018, the pumps that circulate and purify the liquid argon were not operating due to an external electrical issue. The pumps resumed recirculating and purifying on November 1st and typically take approximately a week to return the TPC back to its previous state of liquid argon purity. Because of this circumstance, these drift electron lifetime measurements show the structure of charge attenuation for a lower purity run and a higher purity run during purity recovery, respectively. 

 While the drift electron lifetime can represent the purity, another useful metric is the ratio of $dQ/dx$ at the anode and cathode or $\frac{Q_c}{Q_a}$. This is calculated as follows:
     \begin{equation}\label{eqn:qaqc}
    \frac{Q_c}{Q_a}=e^{\frac{-t_{\rm{full\, drift}}}{\tau}} 
  \end{equation}
where $t_{\rm{full\, drift}}$ is the time it takes to drift from the cathode to the anode, which was measured to be 2.3~ms.
 Runs were taken with the CRT operating for the last week of beam data-taking. The $Q_c/Q_a$ was measured for each day possible and occurred during a rise in purity in the detector.

The systematic uncertainties determined for this measurement are from the uncertainty in the SCE calibration and impacts diffusion have on the hits as a function of drift time. The SCE effect uncertainty is estimated by evaluating the electron lifetime using a different SCE calibration map measured using cosmic muons that cross the cathode and both anodes of the TPC.  The difference between the lifetime value obtained using this alternate map and the lifetime value obtained using the SCE calibration discussed in section~\ref{sec:SCE} is defined to be 1$\sigma$ of the SCE systematic uncertainty. The uncertainty due to diffusion is estimated by turning diffusion off in the Monte Carlo simulation of the ProtoDUNE-SP detector, with a lifetime set at 35~ms.  The difference in the electron lifetime values extracted with and without diffusion in the Monte Carlo simulation is taken to be the 1$\sigma$ variation due to the diffusion systematic, and is denoted $\sigma_{\rm{diff}}$. The fractional change in the lifetime due to diffusion is $\frac{\sigma_{\rm{diff}}}{\tau}=0.143$, which corresponds to a difference in $Q_c/Q_a$ by 0.7\%, assuming an electron lifetime of 35~ms. This value of the fractional uncertainty in the lifetime is assumed for all dates investigated.

\begin{figure}[!ht]
    \centering
    \includegraphics[width=0.8\textwidth]{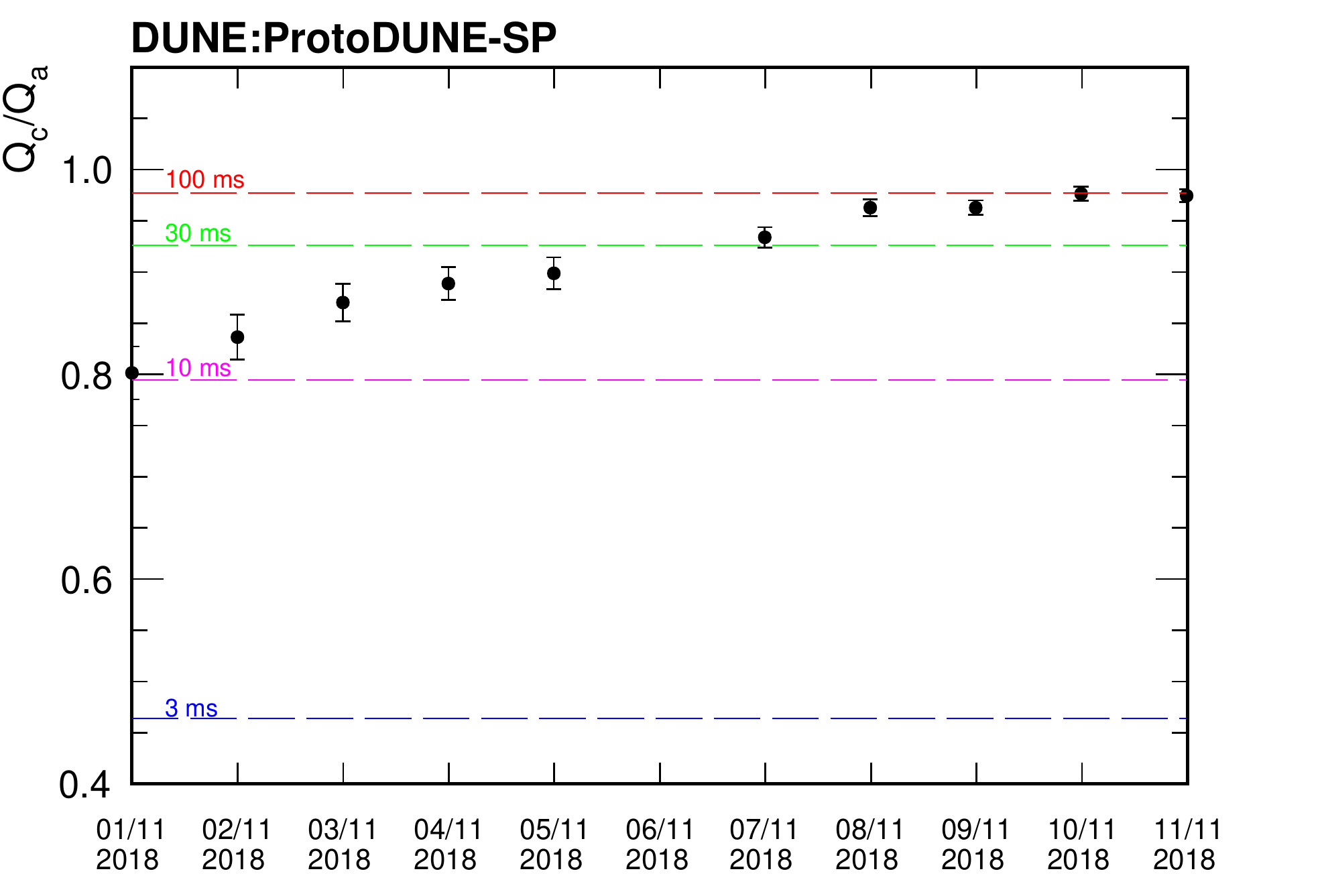}
    \caption{Plot of $Q_c/Q_a$ during November 2018 as measured with TPC tracks calibrated with CRT data. Due to the run plan on November 6th, 2018, there was not enough data to make a precision measurement of $Q_c/Q_a$ on that day. Error bars include the statistical uncertainty, the uncertainty from calibrating the SCE, and the uncertainty from diffusion's impact on the measurement.}
    \label{fig:purtime}
\end{figure}

  Towards the beginning of data-taking $Q_c/Q_a$ was measured as low as 0.801$\pm$0.026, which is equivalent to an electron lifetime of 10.4$\pm$1.5~ms, as seen in figure~\ref{fig:purtime}. Toward the end of data-taking, higher purity was achieved with a $Q_c/Q_a$ of 0.9745$\pm$0.0063 on November 11th of 2018, a value close to a 100~ms drift electron lifetime.

These electron lifetimes also approximate the amount of impurity in the detector as expressed in units of oxygen equivalent concentration. Measurements indicate that at approximately 500~V/cm, the impurity can be described as follows:    \begin{equation}\label{eqn:impurity}
   N_{{\rm{O}}_2}=\frac{1}{k_a\tau}=\frac{300 \, {\rm{ppt}} \, {\rm{ms}}}{\tau}
  \end{equation}
  with $N_{{\rm{O}}_2}$ being the concentration equivalent if all impurities were from oxygen and $k_a$ being the attachment constant for oxygen~\cite{bettini1991study}. Considering the inverse relationship between the drift electron lifetime and the amount of oxygen equivalent impurity, the estimate predicts the impurity never went above 40~ppt equivalent of oxygen in the week of data-taking. At the end of beam data-taking on November 11th, 2018, the impurity in the detector can be estimated to be approximately 3.4$\pm$0.7~ppt oxygen equivalent.

\subsection{Calibration based on cosmic-ray muons}\label{sec:muonCal}
The goal of detector calibration is to convert the measured charge in units of ADC counts to energy in units of MeV, which provides important information for particle identification and energy measurements. In order to get reliable calorimetric information, 
a two-step calibration procedure is employed following the same method developed by the MicroBooNE collaboration~\cite{Adams:2019ssg}. In the first step, the detector response is equalized using throughgoing cosmic-ray muons. In the second step, the absolute energy scale is determined using stopping cosmic-ray muons. In both steps muons that cross the cathode are used because their $t_0$ can be reconstructed (section~\ref{sec:reco:pandora}). The two steps are described in the following sections, using the results for Run 5770 that was taken on Nov. 3, 2018 as an example.
\subsubsection{Charge calibration} 
\label{sec:muonchargecalibration}
\begin{figure}[!ht]
  \begin{subfigure}[b]{0.49\textwidth}
    \centering
    \includegraphics[width=\textwidth]{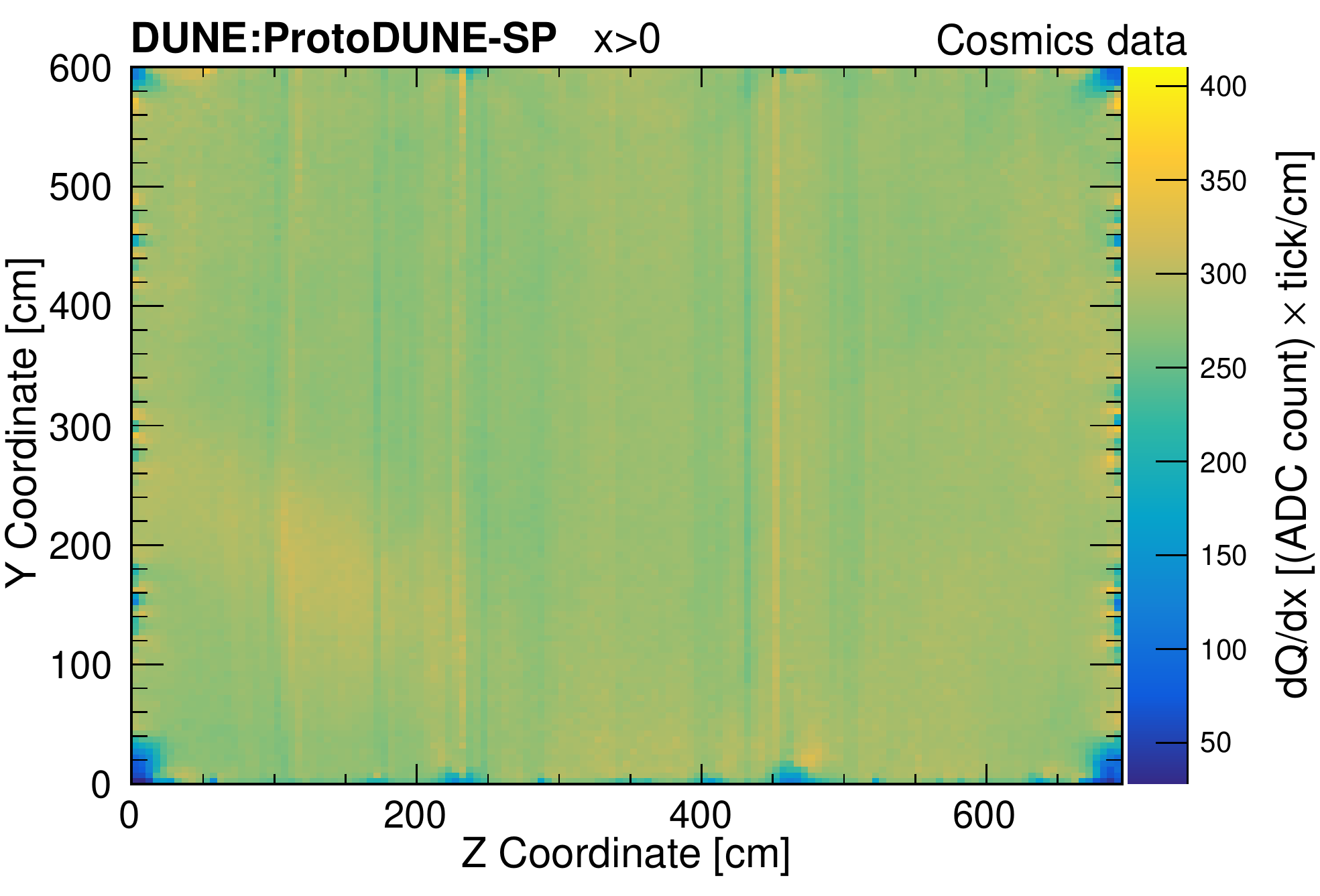}
    \caption{$dQ/dx$ distribution in the $yz$ plane, $x>0$}
    \label{fig:run5770_plane2_beam_left}
  \end{subfigure}
  \hfill
  \begin{subfigure}[b]{0.49\textwidth}
    \centering
    \includegraphics[width=\textwidth]{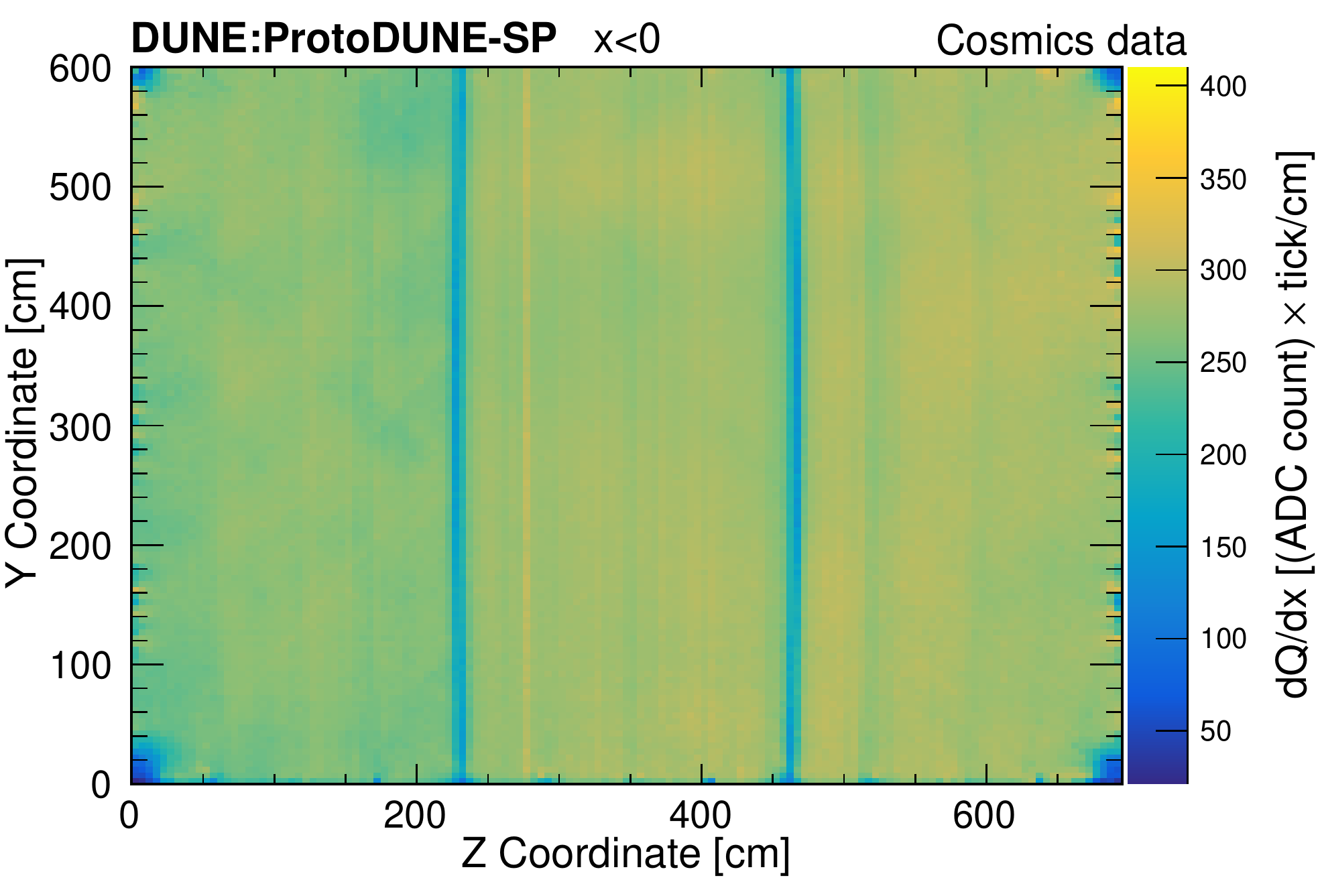}
    \caption{$dQ/dx$ distribution in the $yz$ plane, $x<0$}
    \label{fig:run5770_plane2_beam_right}
  \end{subfigure}
  \hfill
  \caption{$dQ/dx$ distributions for ProtoDUNE-SP Run~5770 in the $yz$ plane, for $x > 0$~\subref{fig:run5770_plane2_beam_left}, $x < 0$~\subref{fig:run5770_plane2_beam_right}, using cosmic-ray cathode-crossing muons. A sample of 99689 throughgoing cosmic ray muon tracks was used in making the plots, which constitutes 23.2\% of the total number of cathode-crossing tracks in Run~5770. }
  \label{fig:dQdx_crossingmuons}
\end{figure}

\begin{figure}[!ht]
  \begin{subfigure}[b]{0.49\textwidth}
    \centering
    \includegraphics[width=\textwidth]{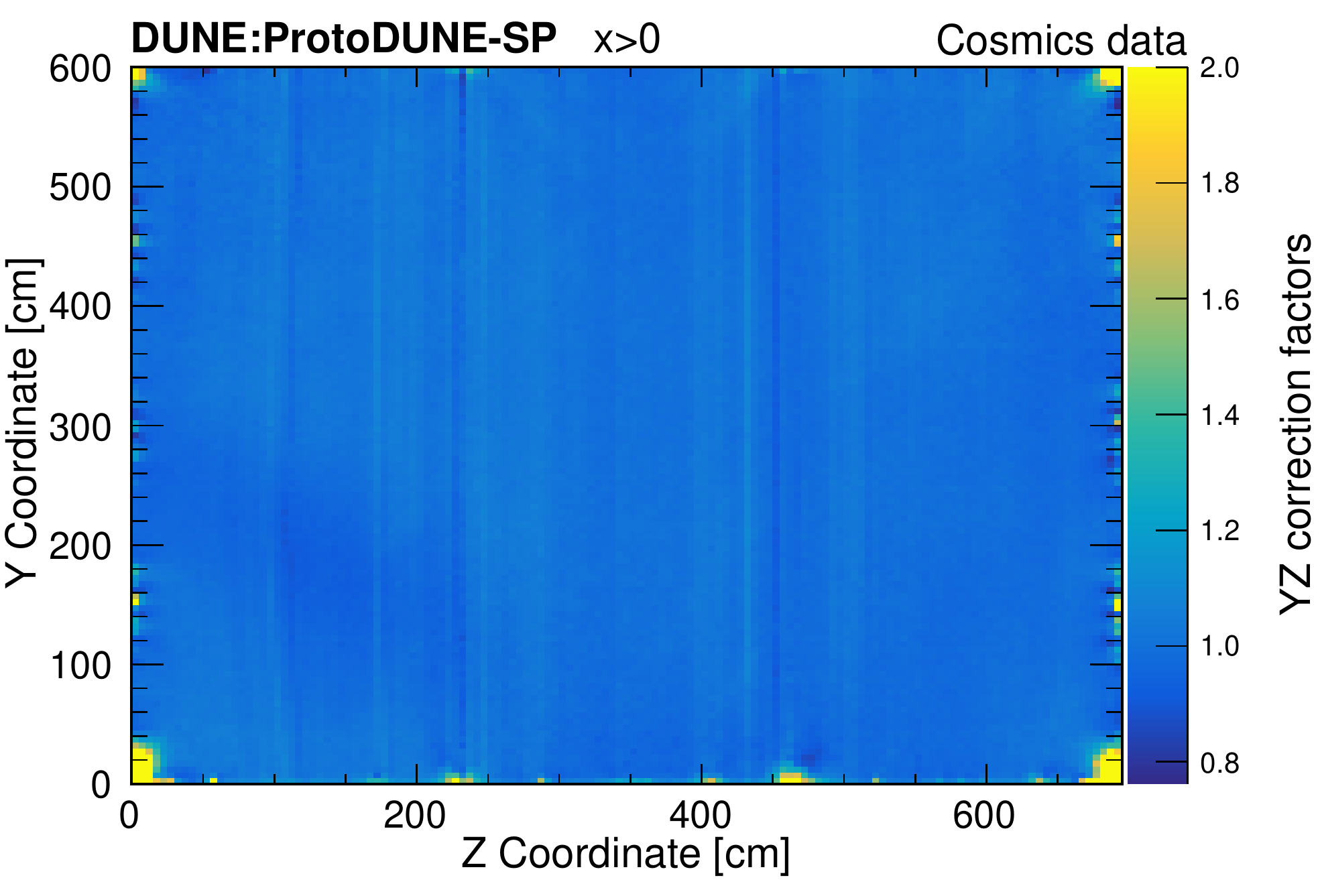}
    \caption{YZ correction factors, $x>0$}
    \label{fig:run5770_beamleft_corr}
  \end{subfigure}
  \hfill
  \begin{subfigure}[b]{0.49\textwidth}
    \centering
    \includegraphics[width=\textwidth]{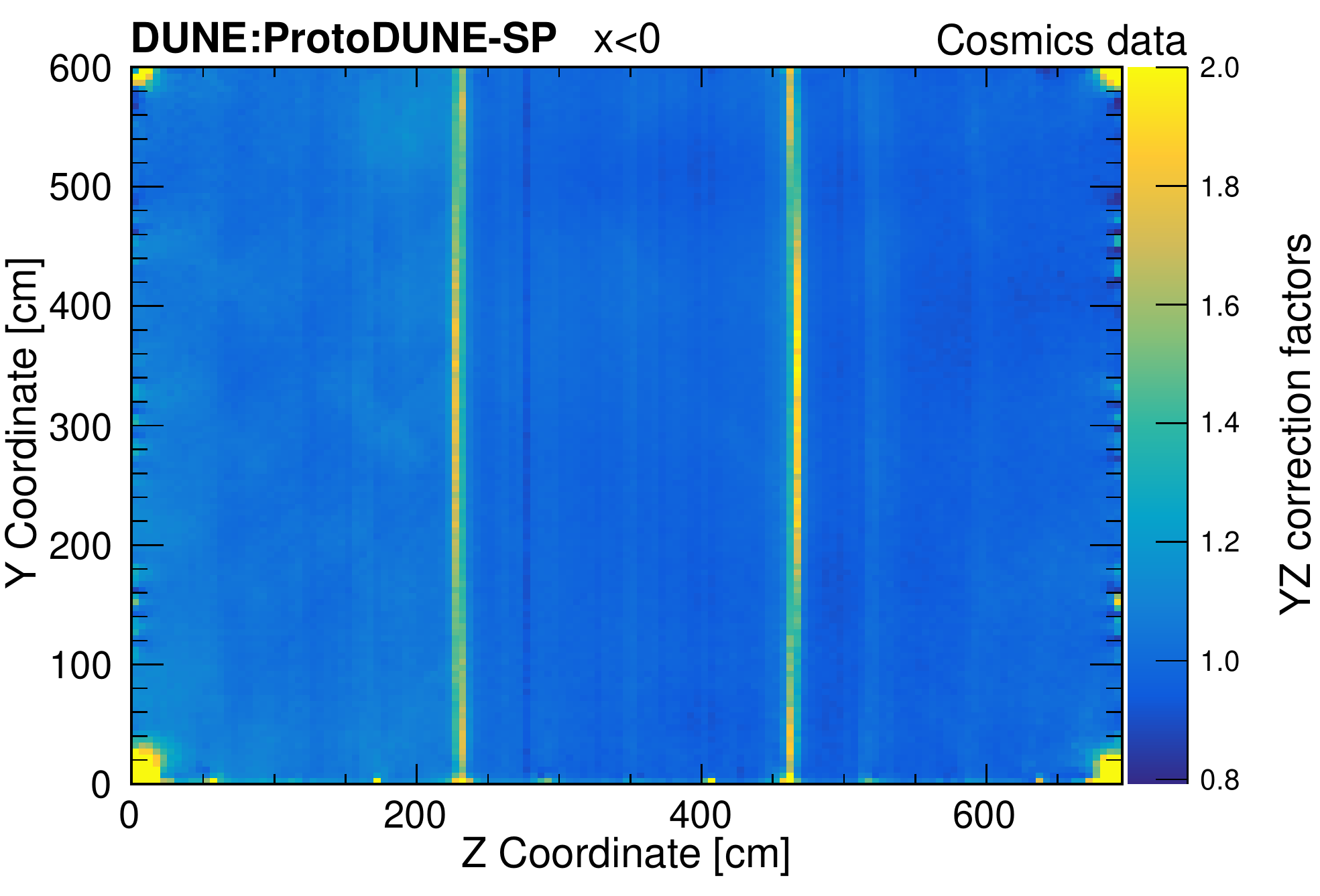}
    \caption{YZ correction factors, $x<0$}
    \label{fig:run5770_beamright_corr}
  \end{subfigure}
  \hfill
  \caption{YZ correction factors for ProtoDUNE-SP Run~5770 in the $yz$ plane, for $x > 0$~\subref{fig:run5770_beamleft_corr}, $x < 0$~\subref{fig:run5770_beamright_corr}, using cosmic-ray cathode-crossing muons.}
  \label{fig:yz_correction_factors}
\end{figure}

\begin{figure}[!ht]
  \begin{subfigure}[b]{0.49\textwidth}
    \centering
    \includegraphics[width=\textwidth]{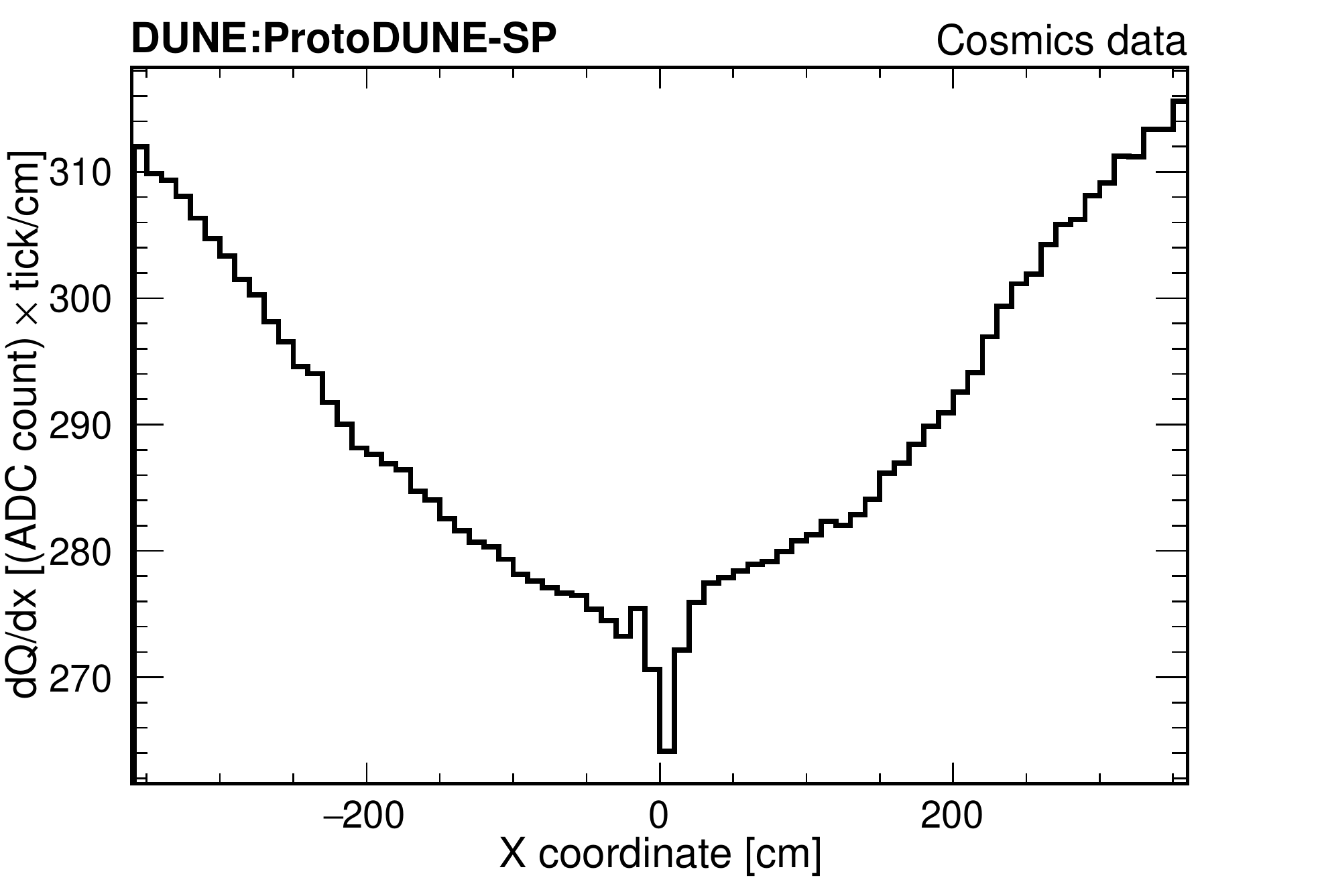}
    \caption{$dQ/dx$ vs drift dimension $x$}
    \label{fig:run5770_dqdx_x}
  \end{subfigure}
\hfill
   \begin{subfigure}[b]{0.49\textwidth}
    \centering
    \includegraphics[width=\textwidth]{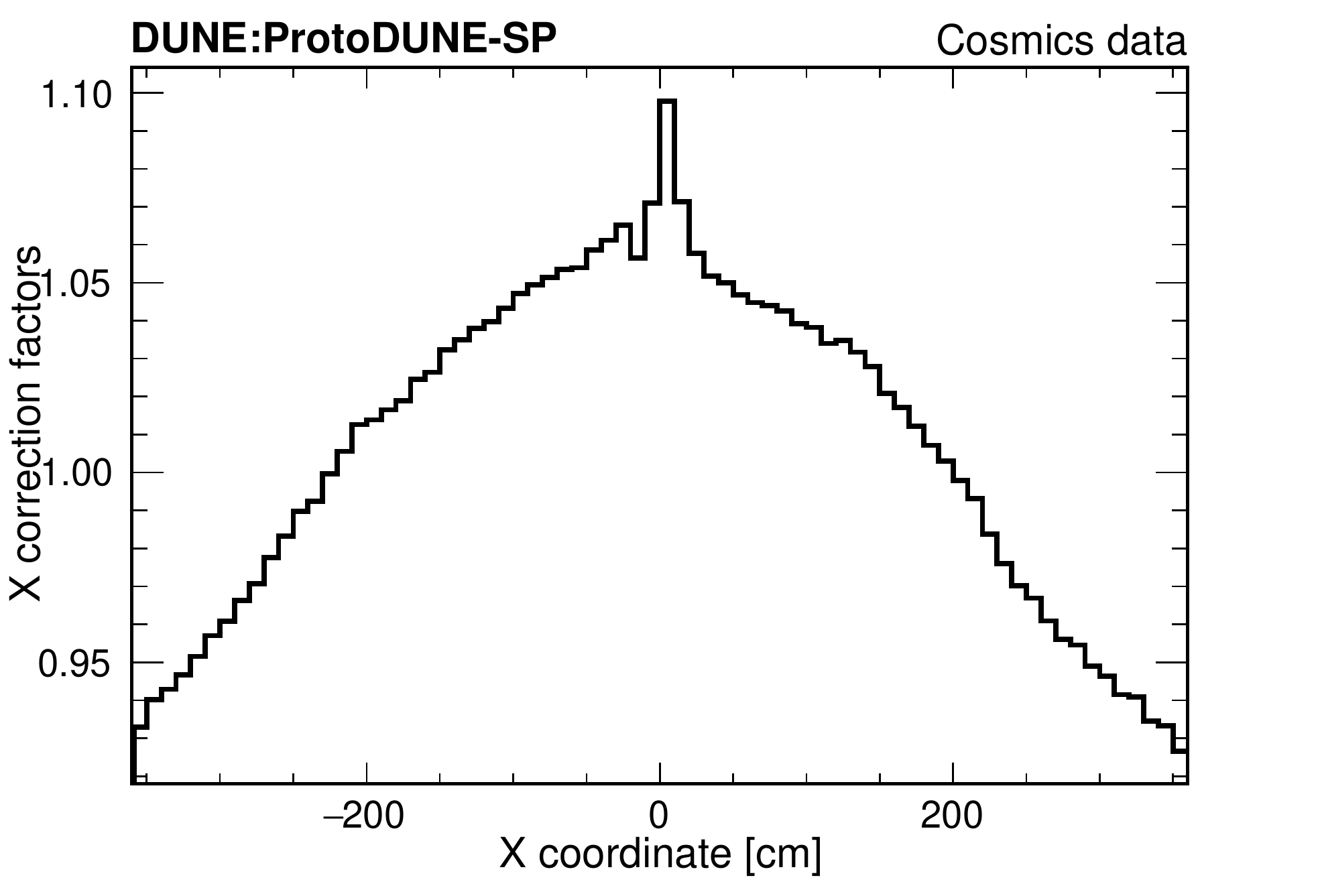}
    \caption{X correction factors vs drift dimension $x$}
    \label{fig:run5770_dqdx_x_corr}
  \end{subfigure}
  \caption{$dQ/dx$ distribution and X correction factors as a function of drift dimension $x$ for ProtoDUNE-SP Run~5770 ~\subref{fig:run5770_dqdx_x} $dQ/dx$ distribution and ~\subref{fig:run5770_dqdx_x_corr} X correction factors}
  \label{fig:x_correction_factors}
\end{figure}

The charge deposition per unit length $(dQ$/$dx)$ in a LArTPC is affected by a number of factors including electronics gain variations, space charge effects,  attenuation (due to electronegative impurities like O$_{2}$ and H$_{2}$O), diffusion, and other effects. Some effects are calibrated out using measurements described in previous sections such as electronics gains (section~\ref{sec:gain_calib}) and space charge effects (section~\ref{sec:SCE}). Calibrating the electron lifetime in situ via cathode-crossing muons is complicated by the very complex space charge effects. Using the CRT allowed a much more precise calibration, but that system was not operable until late in the run and therefore those more precise lifetime measurements were not available for runs taken before then. The effect of diffusion has not been measured yet. In the equalization step, cathode-crossing cosmic-ray muons are used to calibrate the residual nonuniformity in the $dQ/dx$ values throughout the TPC after the gain and space charge effect calibrations. The following requirements are applied for track selection:
\begin{itemize}
\item \textbf{Fiducial volume requirements:} The fiducial volume FV1 is defined as a rectangular prism shaped as follows: the boundary from the anode planes is 10 cm, the boundary from the upstream and downstream ends is 40 cm, and the boundary from the top and bottom of the TPC is 40 cm. In order for a track to be selected, its start point and its end point must be outside FV1. 

\item \textbf{Angular requirements:} The reconstruction capability of LArTPCs is limited for tracks that are parallel to the APA wires or contained in a plane containing a wire and the electric field direction. Figure~\ref{fig:dQdx_angular_distribution} shows the $dQ/dx$ distribution as a function of $\theta_{xz}$ and $\theta_{yz}$, the two angles are illustrated in figure~\ref{fig:thetaxzyz_definition}. To get a sample of well reconstructed tracks, tracks with 65$^{\circ}$ $<$ $|\theta_{xz}|<$ 115$^{\circ}$ or 70$^{\circ}$ $<|\theta_{yz}|<$ 110$^{\circ}$ are removed, as indicated by the dashed lines in figure~\ref{fig:dQdx_angular_distribution}.
  
  \begin{figure}[!ht]
     \begin{subfigure}[b]{0.49\textwidth}
      \centering
      \includegraphics[width=\textwidth]{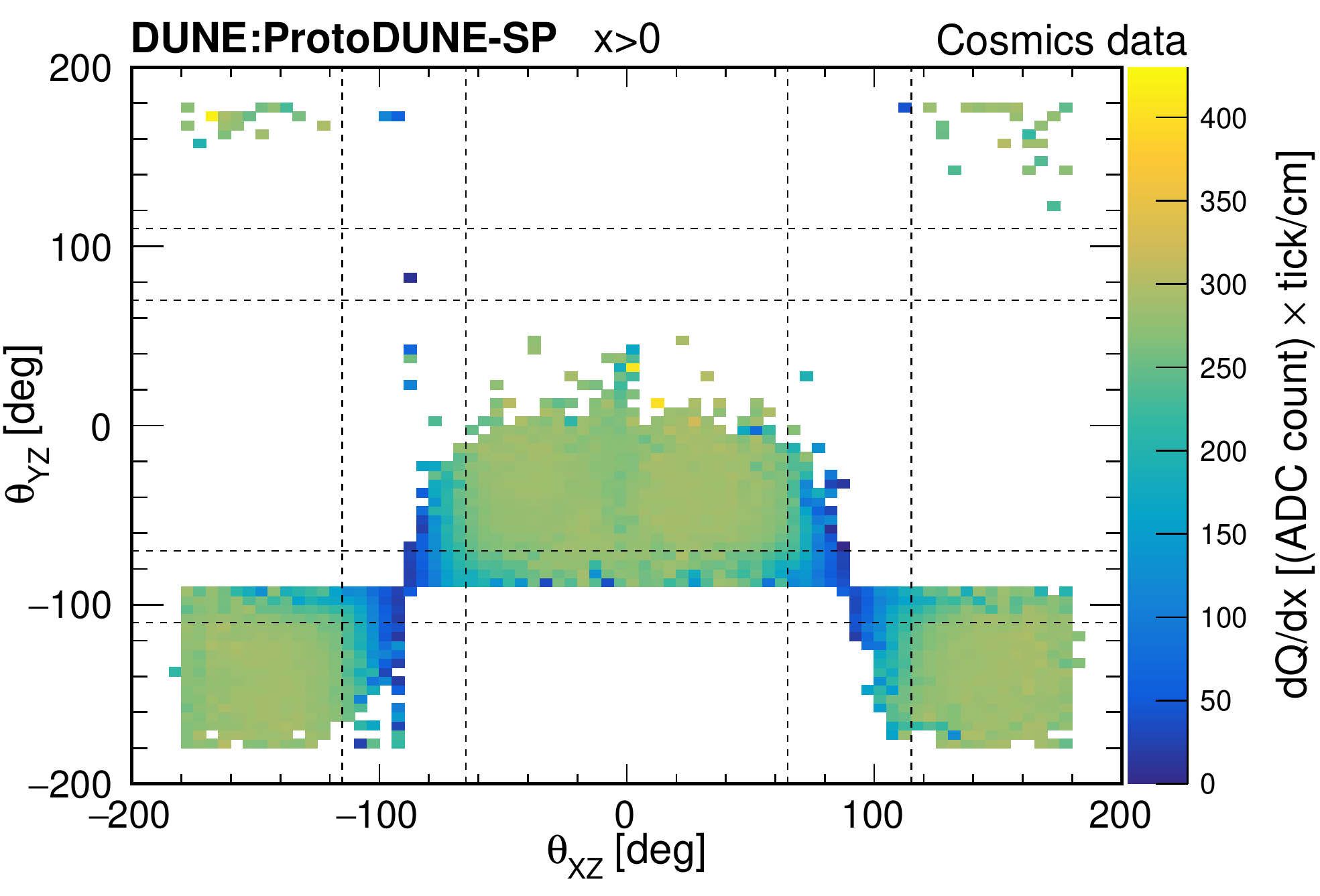}
      \caption{$x>0$}
      \label{fig:run5770_thetaxzyz_beam_left}
    \end{subfigure}
    \hfill
    \begin{subfigure}[b]{0.49\textwidth}
      \centering
      \includegraphics[width=\textwidth]{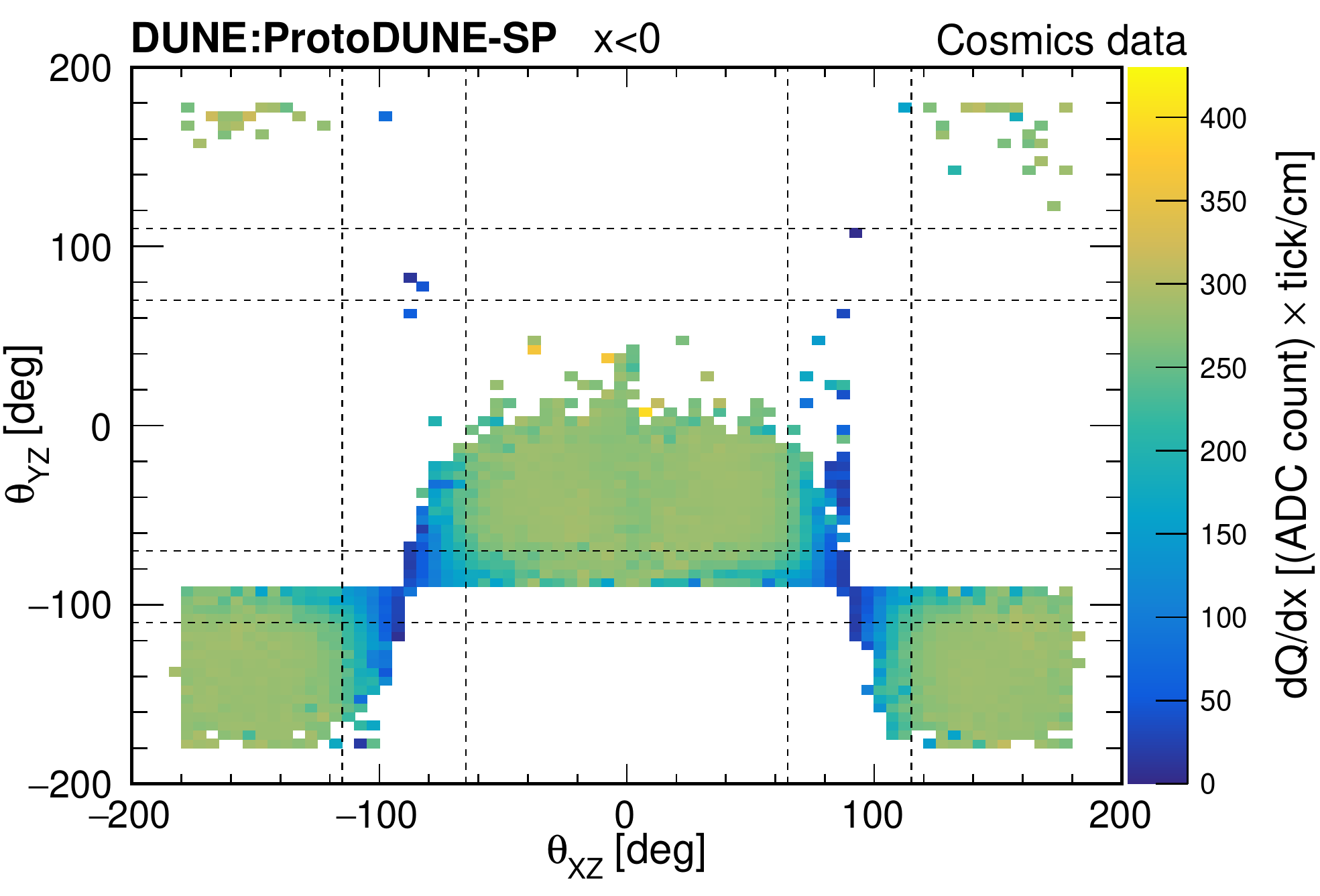}
      \caption{$x<0$}
      \label{fig:run5770_thetaxzyz_beam_right}
    \end{subfigure}
    \hfill
    \caption{Average $dQ/dx$ distributions for ProtoDUNE-SP Run~5770 as functions of $\theta_{xz}$ and $\theta_{yz}$ in the collection plane. The color scale represents average $dQ/dx$ for a track. The regions inside the dashed lines show the track incident angles excluded for the collection plane. 106764 throughgoing cosmic ray muon tracks were used in making the plots, which constitutes 24.8\% of the total number of cathode-crossing tracks in Run~5770. }
    \label{fig:dQdx_angular_distribution}
  \end{figure}

\end{itemize}
Tracks passing the above selection criteria are used for $dQ/dx$ calibration. 
Corrections are obtained in the $yz$ plane and as a function of the drift distance. 
\begin{itemize}
\item \textbf{YZ correction factors:} $dQ/dx$ values in the $yz$ plane are affected by many factors including non-uniform wire response caused by nearby dead channels or disconnected wires, detector features such as the electron diverters and the wire support combs, and transverse diffusion. Figures~\ref{fig:run5770_plane2_beam_left} and~\ref{fig:run5770_plane2_beam_right} show the $dQ/dx$ distribution in the $yz$ plane separately for the $x>0$ drift volume and the $x<0$ drift volume.  The vertical stripes in the $x<0$ plot show places where charge has been collected or distorted by the grounded electron diverters.  The first APA on the left in figure~\ref{fig:run5770_plane2_beam_right} has a lower-than average $dQ/dx$ because of the partially-charged disconnected G plane.  The wire support combs also distort the $dQ/dx$ averages~\cite{Abi:2020loh}, but only by around 5\%, and only in very localized positions that are narrower than the bin sizes in the figure. To correct for these non-uniformities we divide the $yz$ plane in the two ProtoDUNE-SP drift volumes into a number of $5\times5$~cm$^2$ bins. Considering the $dQ/dx$ values of all the hits lying in a particular bin, the median $dQ/dx$ value is calculated and denoted $(dQ/dx)^{\rm{local}}_{\rm{YZ}}$. Further, the median $dQ/dx$ value is calculated considering the hits throughout a drift volume, which is denoted $(dQ/dx)^{\rm{global}}_{\rm{YZ}}$. The YZ correction factor is then defined as 
  \begin{equation}
    C(y,z)=\frac{(dQ/dx)^{\rm{global}}_{\rm{YZ}}}{{(dQ/dx)^{\rm{}local}_{\rm{YZ}}}}.
  \end{equation}
Figures~\ref{fig:run5770_beamleft_corr} and~\ref{fig:run5770_beamright_corr} show the YZ correction factors for ProtoDUNE-SP Run ~5770.

\item \textbf{X correction factors:} The $dQ/dx$ values along the drift direction are affected by factors such as attenuation due to electronegative impurities and longitudinal diffusion. Figure~\ref{fig:run5770_dqdx_x} shows the $dQ/dx$ distribution as a function of $x$. The total drift volume is divided into 5\,cm bins in the $x$ coordinate. The $dQ/dx$ values are first corrected using YZ correction factors based on the $y$ and $z$ coordinates of the hit. After the YZ correction, the median $dQ/dx$ value $(dQ/dx)^{\rm{local}}_{\rm{X}}$ is calculated for each bin. The median $dQ/dx$ value for the whole TPC is denoted $(dQ/dx)^{\rm{global}}_{\rm{X}}$. The X correction factor is defined to be
  \begin{equation}
    C(x)=\frac{(dQ/dx)^{\rm{global}}_{\rm{X}}}{(dQ/dx)^{\rm{local}}_{\rm{X}}}.
  \end{equation}
Figure~\ref{fig:run5770_dqdx_x_corr} shows the X correction factors for ProtoDUNE-SP Run ~5770.
  The $dQ/dx$ value is then normalized to the average value at the two anodes by defining the normalization factor
  \begin{equation}
    N_Q=\frac{(dQ/dx)^{\rm{anode}}}{(dQ/dx)^{\rm{global}}}.
  \end{equation}

  Finally, the corrected $dQ/dx$ value is given by,
  \begin{equation}
    (dQ/dx)_{\rm{corrected}}=N_QC(y,z)C(x)(dQ/dx)_{\rm{reconstructed}}
  \end{equation}
\end{itemize}

\begin{figure}[!ht]
  \begin{subfigure}[b]{\textwidth}
    \centering
    \includegraphics[width=0.49\textwidth]{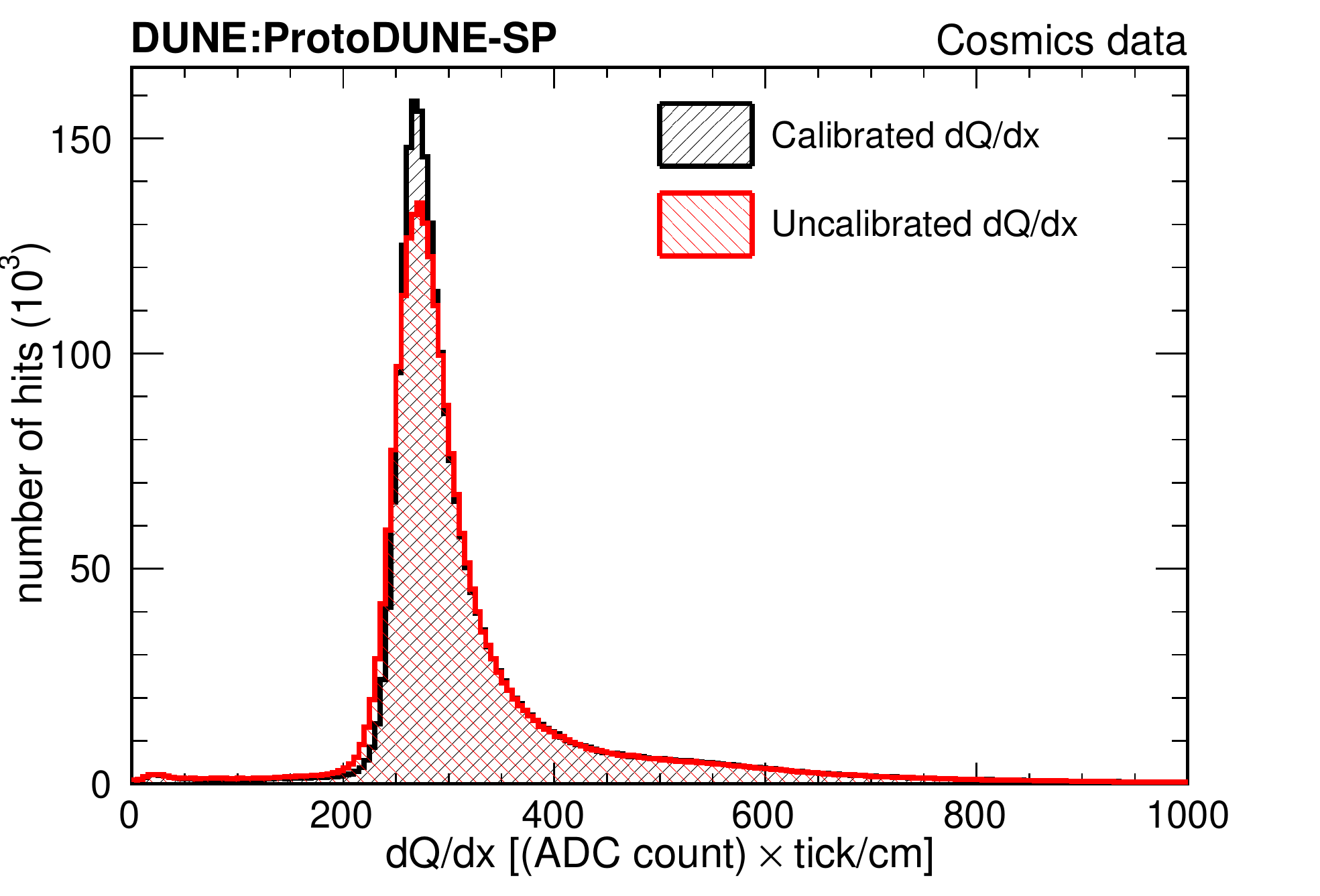}
    \label{fig:run5770_before_after}
  \end{subfigure}
  \caption{$dQ/dx$ distribution for throughgoing cosmic-ray muons before and after calibration}
  \label{fig:dQdx_beforeafter}
\end{figure}
Figure~\ref{fig:dQdx_beforeafter} shows the $dQ/dx$ distribution for throughgoing cosmic-ray muons before and after charge calibration. Once the detector response is equalized, a sample of stopping cosmic-ray muons are selected to determine the absolute energy scale.

\subsubsection{Energy scale calibration}\label{sec:Ecali}
The conversion between ADC counts and the number of electrons is primarily determined by the electronics response, including both the gain and the shaping time, and the field response. Even though the electronics gain is measured by the charge injection system (section~\ref{sec:gain_calib}) and the field response is calculated with Garfield (section~\ref{signalproc}), the estimated uncertainty on the measurement and calculation is at least a few percent. On the other hand, the energy loss per unit length for a minimum ionizing particle is known to better than 1\%. Therefore, a sample of stopping cosmic-ray muons is selected to determine the absolute energy scale for both data and MC. 
The following cuts are used to select the stopping muon sample:
\begin{itemize}   
\item \textbf{Fiducial volume cuts:} Cathode-crossing tracks which start outside FV1 and end inside a smaller volume FV2 are used. The FV2 volume is a rectangular prism inside FV1 shaped as follows: the boundary from the anode planes is 30 cm, the boundary from the upstream and downstream ends is 50 cm, and the boundary from the top and bottom of the TPC is 50 cm.
  \item \textbf{Angular cuts:} Tracks with 65$^{\circ}$$<|\theta_{xz}|<$115$^{\circ}$ and tracks with 70$^{\circ}<|\theta_{yz}|<$110$^{\circ}$ are removed.
\item \textbf{Removing broken tracks:} Some muons are reconstructed as two or more tracks, which mimic a stopping muon. If the end points of the two tracks are within 30 cm and the angle between them is less than 14$^{\circ}$, both tracks are removed. Additionally, any track which starts or stops within 5 cm of an APA boundary are removed.
  
\item \textbf{Removing tracks with early and late hits}: Tracks that are cut off by the 6000-tick TPC readout window boundaries may mimic a stopping muon.  If any hit associated with a track has a peak time less than 250 ticks or greater than 5900 ticks, the track is removed.
\item \textbf{Removing tracks with Michel hits attached:} The presence of an Michel electron can confuse the reconstruction of the muon end point so muons that decay into Michel electrons are removed. The Michel activities are identified by looking for isolated hits close to muon end point. The number of hits within $\pm$5 wires and $\pm$50 ticks from the last hit of the muon track and not belonging to the muon track or any other track longer than 100\,cm is counted. If the count is greater than 0, such tracks are removed.
\end{itemize}

\begin{figure}[!ht]
  \begin{subfigure}[b]{0.49\textwidth}
    \centering
    \includegraphics[width=\textwidth]{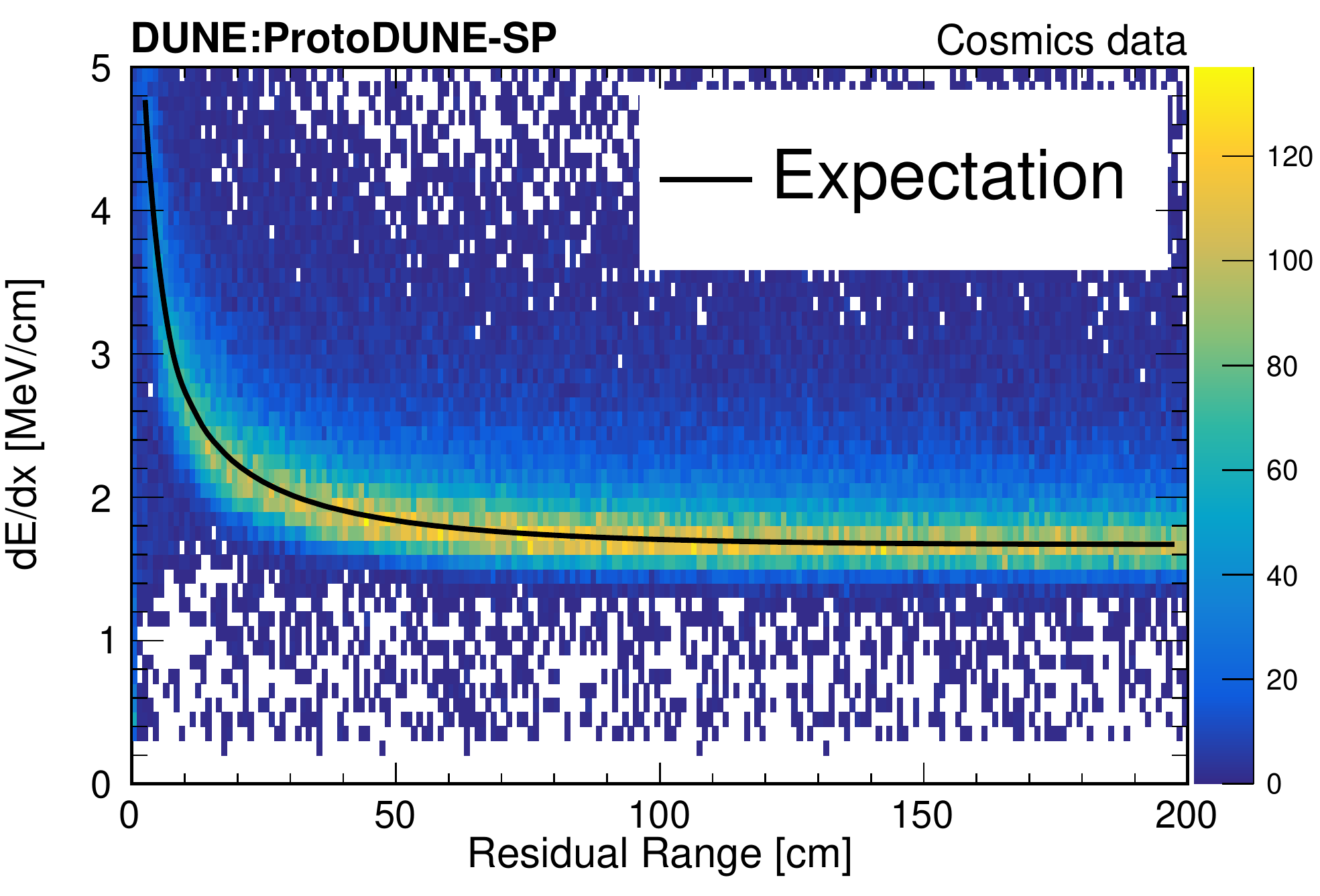}
    \caption{Data}
    \label{fig:muon_dedx_resrange_run5770}
  \end{subfigure}
  \hfill
  \begin{subfigure}[b]{0.49\textwidth}
    \centering
    \includegraphics[width=\textwidth]{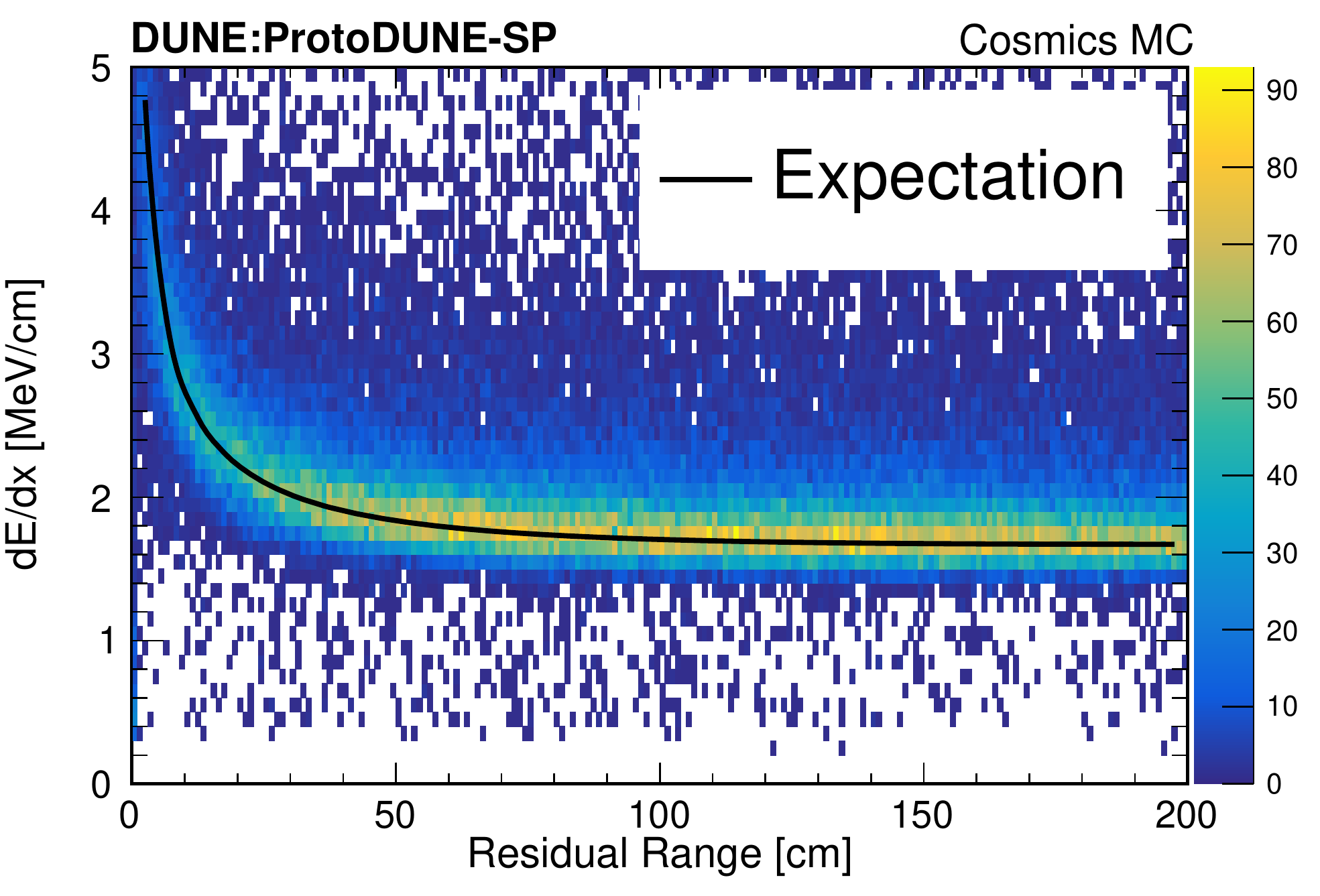}
    \caption{MC}
    \label{fig:muon_dedx_resrange_sce}
  \end{subfigure}
  \hfill
  \begin{subfigure}[b]{\textwidth}
    \centering
    \includegraphics[width=0.49\textwidth]{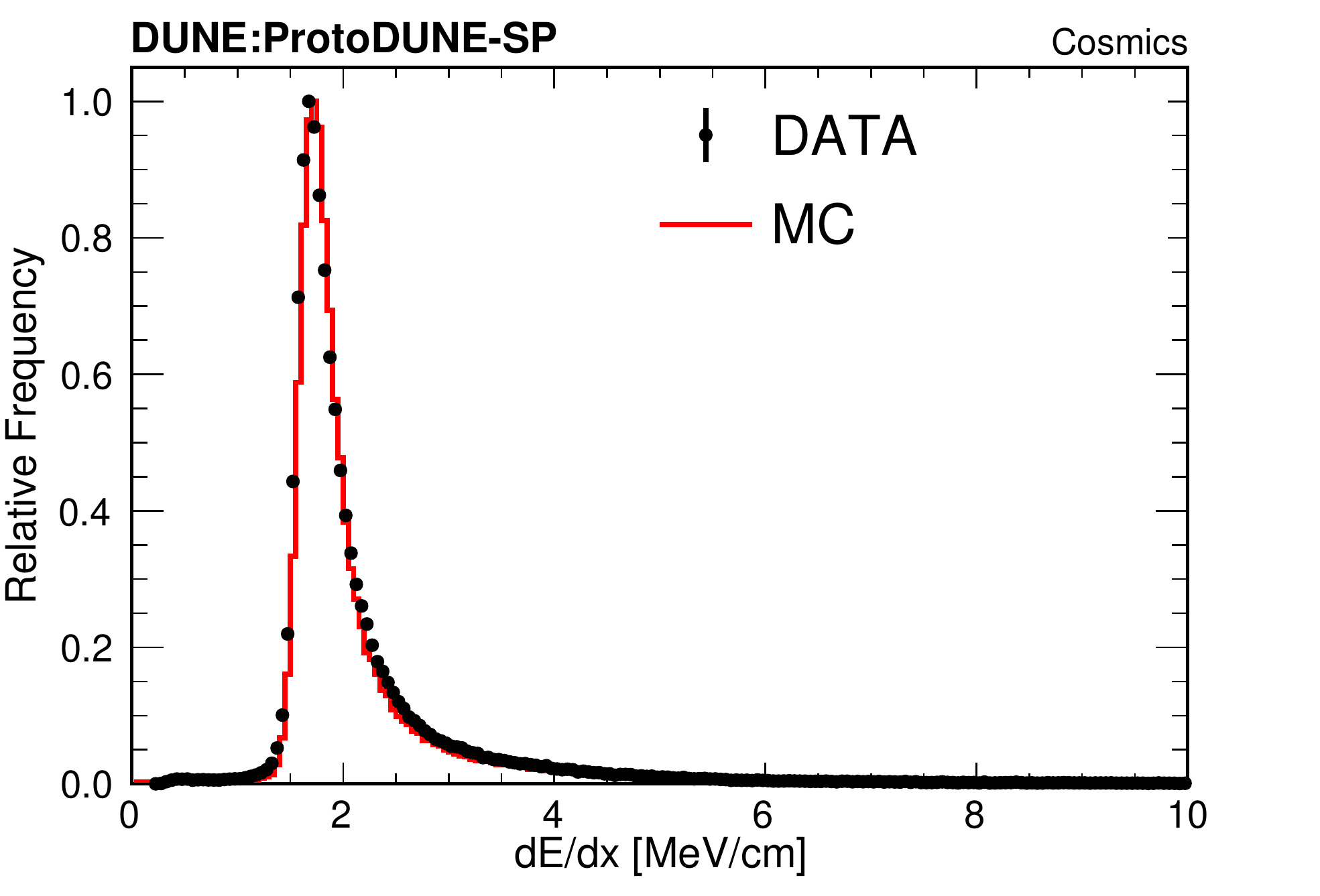}
    \caption{$dE/dx$ comparison}
    \label{fig:muon_comparison_dedx_sce_data}
  \end{subfigure}
  \caption{Stopping muon $dE/dx$ distributions for the ProtoDUNE-SP cosmic-ray data and MC. The black curves in~\subref{fig:muon_dedx_resrange_run5770} and~\subref{fig:muon_dedx_resrange_sce} are the predicted most probable values (using the Landau-Vavilov function) of $dE/dx$ versus residual range and~\subref{fig:muon_comparison_dedx_sce_data} is the $dE/dx$ distribution for the stopping muon sample. The histograms in ~\protect\subref{fig:muon_comparison_dedx_sce_data} are normalized such that the maximum frequency is one.}
  \label{fig:pandora_protodune_mu}
\end{figure}

After applying the above selection cuts, a highly pure sample of stopping muons remains. Defining the purity as the number of true stopping muons divided by the total number of candidate stopping muons in our sample, a purity of 99.74\% is achieved based on Monte Carlo study. The $dQ/dx$ values are corrected as described in the previous section. The most probable $dE/dx$ value as a function of residual range for stopping muon tracks in LAr is accurately predicted by Landau-Vavilov theory~\cite{PhysRevD.98.030001}.  
From the calibrated $dQ/dx$ values (in ADC/cm) along the muon track in its MIP region (120 to 200\,cm from stopping point), the $dE/dx$ (in MeV/cm) values are fitted using the modified Box model~\cite{Acciarri:2013met} function to correct for the recombination effect with the charge calibration constant $C_{\rm{cal}}$ (ADC$\times$tick/cm $\rightarrow$ e/cm) as a free parameter in the $\chi^{2}$ minimization. $C_{\rm{cal}}$ is effectively a scaling factor that accounts for the electronics gain, ADC conversion and other residual effects that are not explicitly calibrated out. The energy loss from the stopping muon sample and a comparison with the theoretical prediction in figure~\ref{fig:pandora_protodune_mu} show the result of the calibration procedure for ProtoDUNE-SP Run~5770 and the corresponding Monte Carlo sample.
\begin{equation}
  \label{eqn:de_dx}
  \left(\frac{dE}{dx}\right)_{\rm{calibrated}}=\left(\exp\left(\frac{(\frac{dQ}{dx})_{\rm{calibrated}}}{C_{\rm{cal}}}\frac{\beta^{\prime} W_{\rm{ion}}}{\rho\mathscr{E}}\right)-\alpha\right)\left(\frac{\rho\mathscr{E}}{\beta^{\prime}}\right),
\end{equation}
where
\begin{conditions*}
C_{\rm{cal}} & Calibration constant used to convert ADC values to number of electrons,\\
W_{\rm{ion}} & $23.6\times10^{-6}$ MeV/electron (the work function of argon), \\
\mathscr{E} & $E$~field based on the measured space charge map, \\
\rho & 1.38 g/cm\textsuperscript{3} (liquid argon density at a pressure of 124.106 kPa), \\ 
\alpha & 0.93, and \\
\beta\textsuperscript{$\prime$} & 0.212 (kV/cm)(g/cm\textsuperscript2)/MeV. 
\end{conditions*}
$\alpha$ and $\beta^\prime$ are the Modified Box model parameters which were measured by the ArgoNeuT experiment at an electric field strength of 0.481 kV/cm~\cite{Acciarri:2013met}.

The calibration constant $C_{\rm{cal}}$ is normalized so that the unit (``ADC$\times$tick'') corresponds to 200 electrons. In the case where the detector response is perfectly modeled (e.g. in the simulation), the calibration constant $C_{\rm{cal}}$ should be exactly 1/200 = 5$\times10^{-3}$ ADC$\times$tick/e. The calibration constants derived for the collection plane by fitting the stopping muon samples to the predicted $dE/dx$ curve are shown in table~\ref{tab:cal_cons}. The uncertainties are statistical only. The difference between data and MC calibration constants is caused by the uncertainties on the gain measurement and the simulation of detector response. 
\begin{table}[ht]
\centering
\caption{Calibration constants for the collection plane in MC and data.} 
\begin{tabular}[t]{lcc} 
\hline
&Data &MC \\
\hline
Fitted value of $C_{\rm{cal}}$  &(5.4 $\pm$ 0.1) $\times10^{-3}$~ADC$\times$tick/e &(5.03 $\pm$ 0.01) $\times10^{-3}$~ADC$\times$tick/e\\
\hline
\end{tabular}
\label{tab:cal_cons}
\end{table}

\subsection{Calorimetric energy reconstruction and particle identification}\label{sec:beamdedx}
The calibration constants derived from cosmic-ray muons are applied to beam particles. The following sections discuss the $dE/dx$ distributions for beam muons and pions (section~\ref{sec:PionMuon}), beam protons (section~\ref{sec:Proton}) and beam electrons (section~\ref{sec:electron}). The identification of MIP particles (muons) and non-MIP particles (protons) is discussed in section~\ref{sec:Proton}.
\subsubsection{Identification and calorimetric energy reconstruction of 1 \texorpdfstring{GeV/$c$}{GeV/c} beam pions and muons}\label{sec:PionMuon}
Measurements of charged pion interactions with argon nuclei are an important physics goal of the ProtoDUNE-SP experiment. Accurate measurement of these interactions allows a more precise understanding of neutrino interactions producing final-state pions, a key study channel of the DUNE experiment. Reconstructing these final state pions and any particles produced by their secondary interactions is important for estimations of neutrino energy. This section describes the results obtained from ProtoDUNE-SP Run 5387 taken on Oct. 18, 2018 corresponding to approximately 12 hours of exposure to a 1 GeV/$c$ beam. The electron drift lifetime of this run was approximately 14 ms as measured by the purity monitor.

\paragraph{Selection of events with reconstructed beam pions:}
The beam line instrumentation as described in section~\ref{sec:beamline} can be used to find events where the beam line has delivered a pion to the TPC. For the 1 GeV beam energy runs considered here, the beam line PID conditions for pions and muons can be found in table~\ref{tab:PID}. At a beam momentum of 1~GeV/$c$ the measured TOF of pions and muons is indistinguishable so the PID will select both pions and muons.
Events are first removed for periods of unstable HV or one or more inactive TPC readout boards. The Pandora pattern recognition framework, described in section~\ref{sec:reco:pandora} is used to reconstruct the particle trajectories in the TPC as well as identify a reconstructed particle as a likely candidate for the beam line track.
To remove events where a track has been incorrectly identified as the primary beam particle, quality cuts are placed on the distance and angle between the end of the beam line particle's reconstructed trajectory and the start of the assigned reconstructed TPC beam particle track. These cuts remove backgrounds that are not aligned with the measured beam line track such as cosmic rays and secondary particles produced by the beam particle interacting upstream of the TPC. The cuts used for data are as follows:
\begin{itemize}
\item \textbf{Beam quality cuts: angle}\\
The cosine of the angle between the beam line track and reconstructed TPC track is required to be $>0.93$.
\item \textbf{Beam quality cuts: position }\\
$\circ$~0\,cm~$\leq (X^{\rm{Start}}_{\rm{TPC}}-X^{\rm{End}}_{\rm{Beam}}) \leq$~10\,cm,\\
$\circ$~-5\,cm~$\leq (Y^{\rm{Start}}_{\rm{TPC}}-Y^{\rm{End}}_{\rm{Beam}}) \leq$~10\,cm,\\
$\circ$~30\,cm~$\leq (Z^{\rm{Start}}_{\rm{TPC}}-Z^{\rm{End}}_{\rm{Beam}}) \leq$~35\,cm,\\
where $X^{\rm{Start}}_{\rm{TPC}}$ is the start position in X of the reconstructed candidate beam track in the TPC before SCE corrections. $X^{\rm{End}}_{\rm{Beam}}$ is the projected X location of the intersection of the track from the beam line instrumentation with the TPC. The cuts are not centered around zero due the SCE shifting the start points of the reconstructed TPC tracks. 
\end{itemize}
A replica selection is also placed on beam MC events where the true simulated beam particle is either a pion or muon. In MC the truth information of the beam particle is used in place of the beam line information.  The cuts used for MC are as follows:
\begin{itemize}
\item \textbf{Beam quality cuts: angle}\\
The cosine of the angle between the true beam particle  and reconstructed TPC track is required to be $>0.93$.
\item \textbf{Beam quality cuts: position }\\
$\circ$~-3\,cm~$\leq (X^{\rm{Start}}_{\rm{TPC}}-X^{\rm{End}}_{\rm{Beam}}) \leq$~7\,cm,\\
$\circ$~-8\,cm~$\leq (Y^{\rm{Start}}_{\rm{TPC}}-Y^{\rm{End}}_{\rm{Beam}}) \leq$~7\,cm,\\
$\circ$~27.5\,cm~$\leq (Z^{\rm{Start}}_{\rm{TPC}}-Z^{\rm{End}}_{\rm{Beam}}) \leq$~32.5\,cm.\\
Here $X^{\rm{End}}_{\rm{Beam}}$ is the projected X location of the true beam particle's intersection with the TPC and $X^{\rm{Start}}_{\rm{TPC}}$ is same as for data. 
\end{itemize}
\paragraph{Selection of stopping muons:}
\begin{figure}[!ht]
\centering
\includegraphics[width=0.7\textwidth]{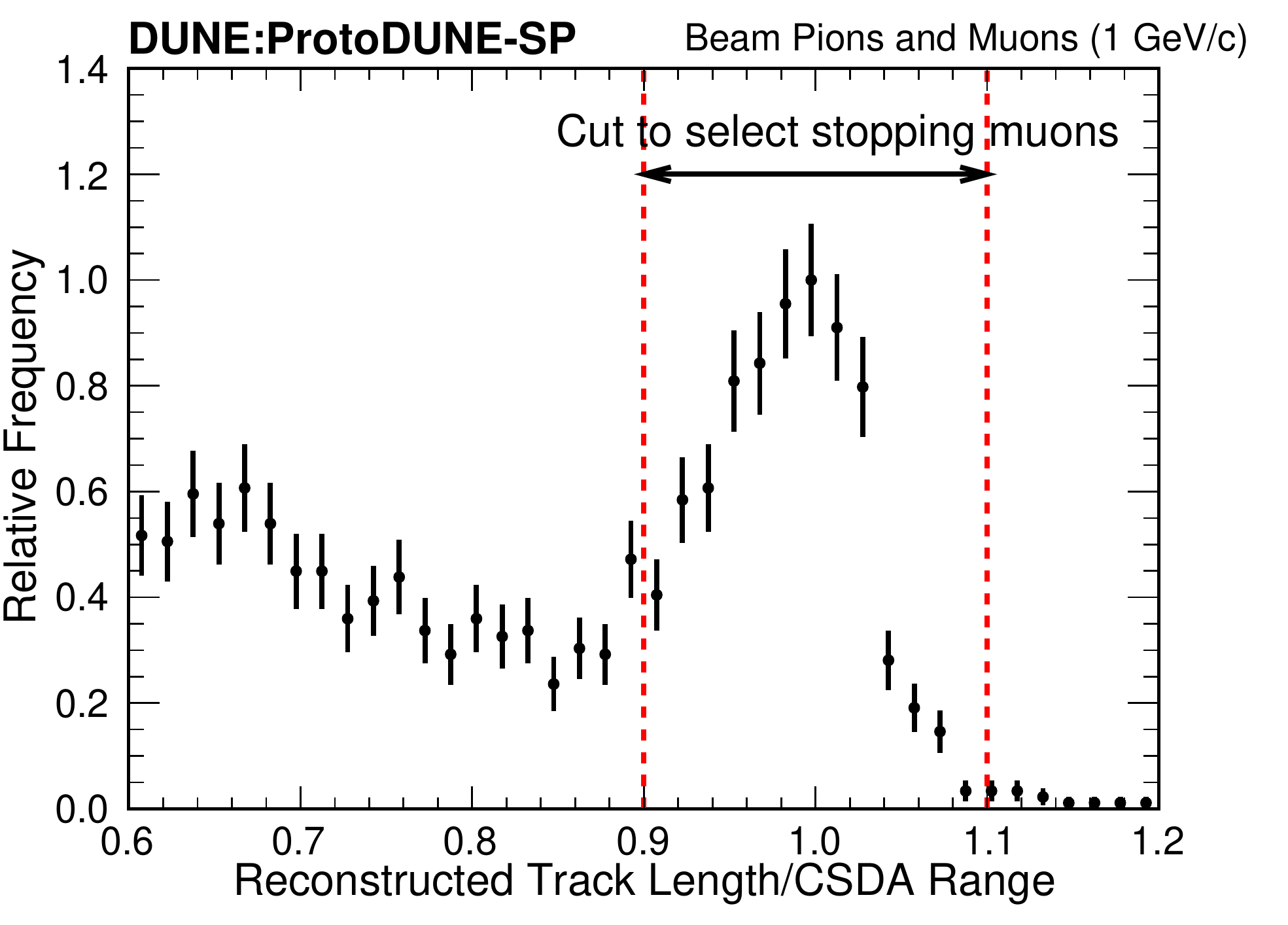}
\caption{The ratio of track length (SCE corrected) and continuous-slowing-down-approximation (CSDA) length for data candidates under the assumption the particle is a muon. The cut to select the stopping muons is indicated. The histogram is normalized such that the maximum frequency is one.}
\label{fig:muon_cut}
\end{figure}

The track lengths of the selected beam particles can be used to separate the beam pions and muons. As 1~GeV/$c$ pions have an expected interaction length in argon of \textasciitilde 1~m the majority of pions will interact before coming to a stop via ionization energy loss.
Whereas 1\,GeV/$c$ muons will  stop predominantly via ionization energy losses with an expected path length in argon of \textasciitilde 4\,m. The measured beam momentum is used to approximate the stopping range of each particle under the assumption it is a muon using the continuous-slowing-down-approximation (CSDA) range~\cite{csdamu}. The distribution of the beam tracks' reconstructed lengths divided by their calculated CSDA ranges is shown in figure~\ref{fig:muon_cut}. Requiring the ratio
\begin{equation}
\label{eqn:muon_CSDA}
0.9 < {\rm \frac{Track~Length}{CSDA~Stopping~Range}} < 1.1,
\end{equation}
selects a subsample of stopping particles, predominantly muons.
A replica selection cut is placed on the MC sample using the the simulated momentum for the true beam particle in each event as input to the ratio calculation.
\paragraph{Calorimetric energy information of beam pions}

\begin{figure}[!ht]
\begin{subfigure}[b]{0.5\textwidth}

\centering
\includegraphics[width=\textwidth]{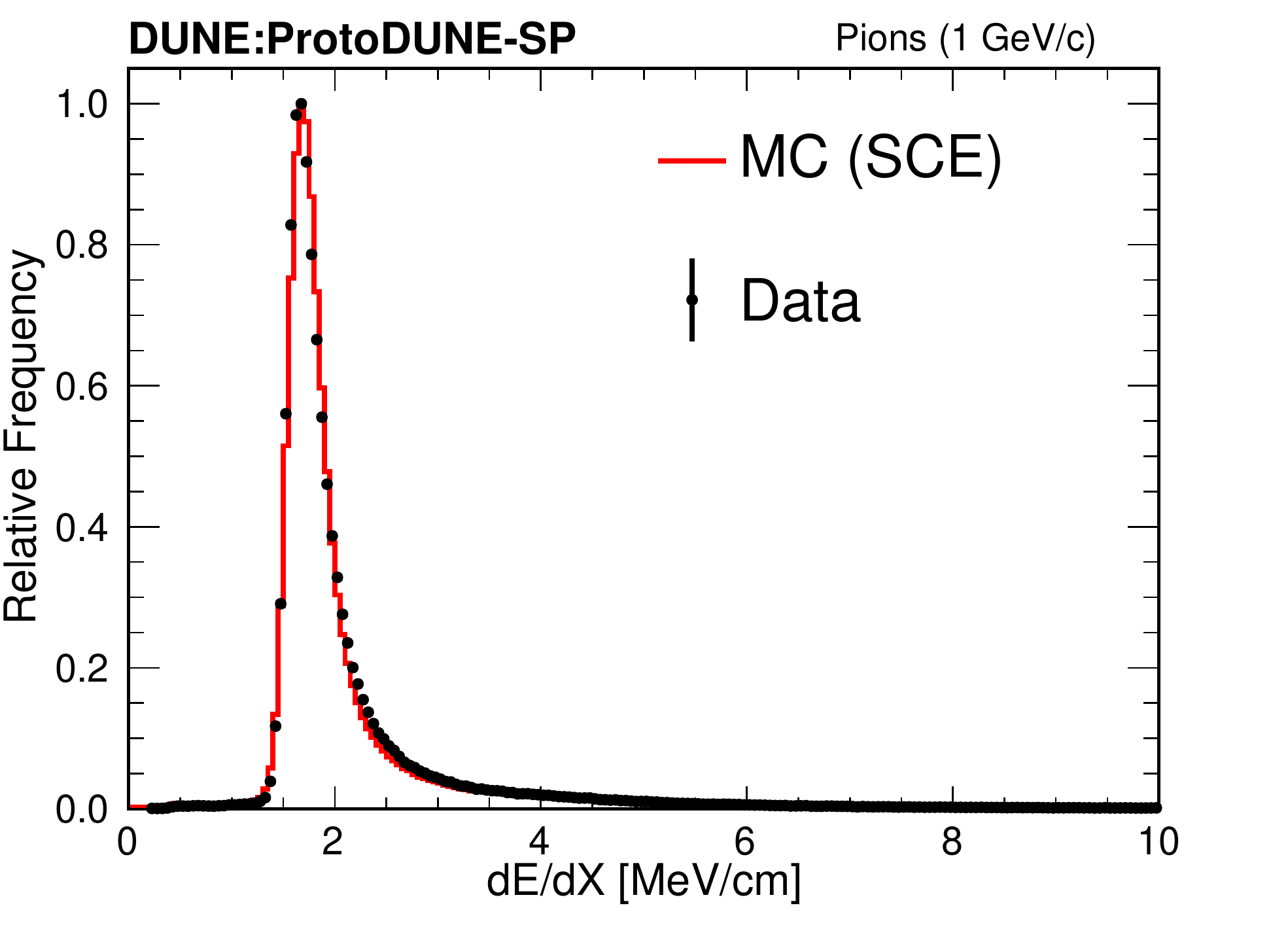}
\caption{}
\label{fig:pion_dedx1d}
\end{subfigure}
\hfill
\begin{subfigure}[b]{0.5\textwidth}
\centering
\includegraphics[width=\textwidth]{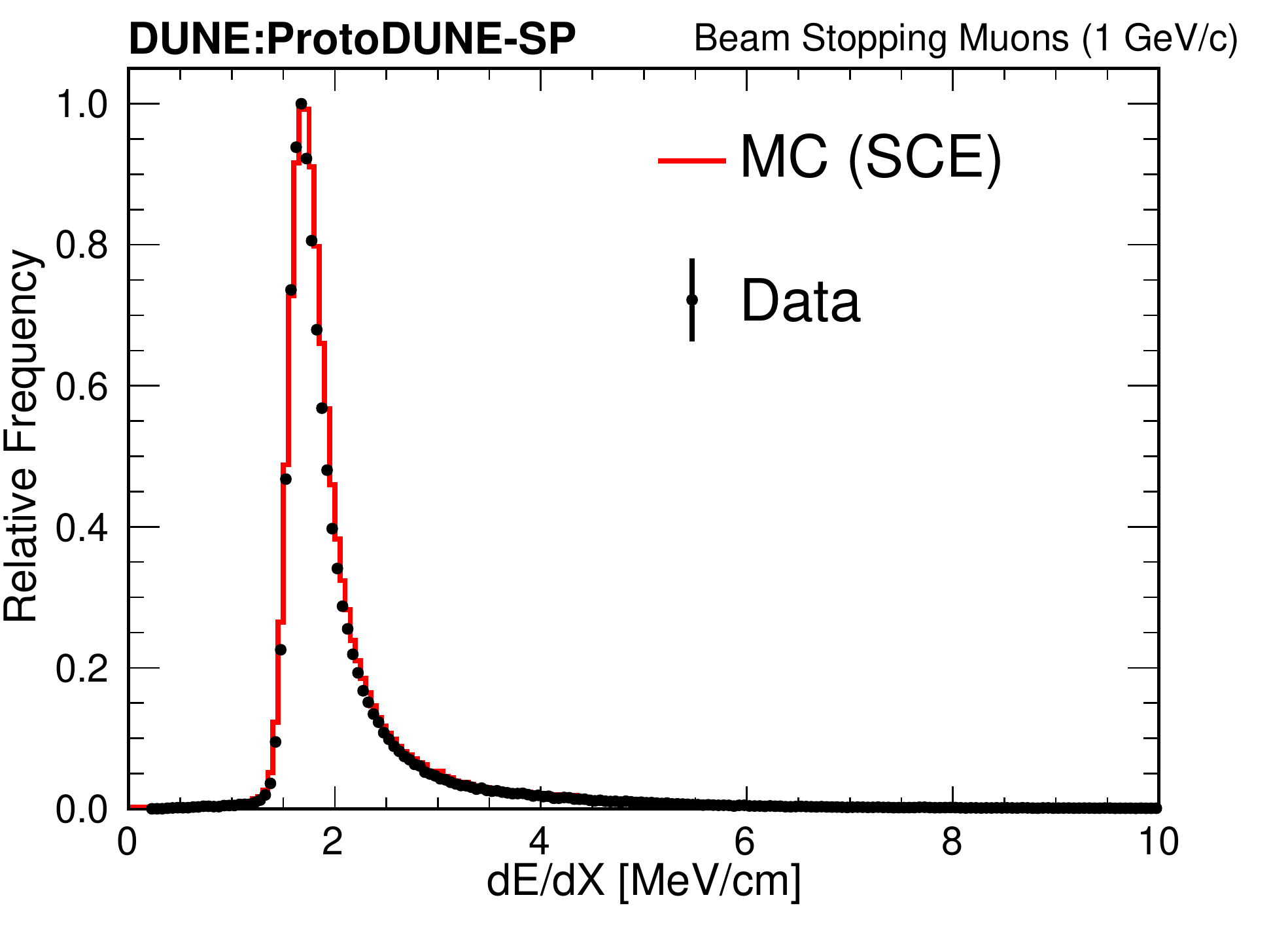}
 \caption{}
\label{fig:beammuon_dedx1d}

 \end{subfigure}
 \caption{
Pion~\subref{fig:pion_dedx1d} and stopping muon~\subref{fig:beammuon_dedx1d} $dE/dx$ distributions for the ProtoDUNE-SP beam data after applying the calibration derived using cosmic-ray muons.
The distribution for replica selections on a beam Monte Carlo sample with space charge effect simulated is also shown. The histograms are normalized such that the maximum frequency is one.
 }
\end{figure}

The charge signal calibration produced from stopping cosmic-ray muons collected during the same run period described in section \ref{sec:muonCal} is applied to the selected beam tracks to calculate the true deposited energy as a function of distance ($dE/dx$) along their trajectory. Figure~\ref{fig:pion_dedx1d} shows the distribution of $dE/dx$ values of all hits on the collection plane from the selected beam pion candidates. That is, all candidates that are not selected by the stopping muon cut~(eq.~\ref{eqn:muon_CSDA}) described above. The selected particles are MIPs. The MPVs of the calibrated data and MC samples agree to better than 1\%.

\paragraph{Calorimetric energy information of beam stopping muons}
\begin{figure}[!ht]
  \begin{subfigure}[b]{0.5\textwidth}
    \centering
    \includegraphics[width=\textwidth]{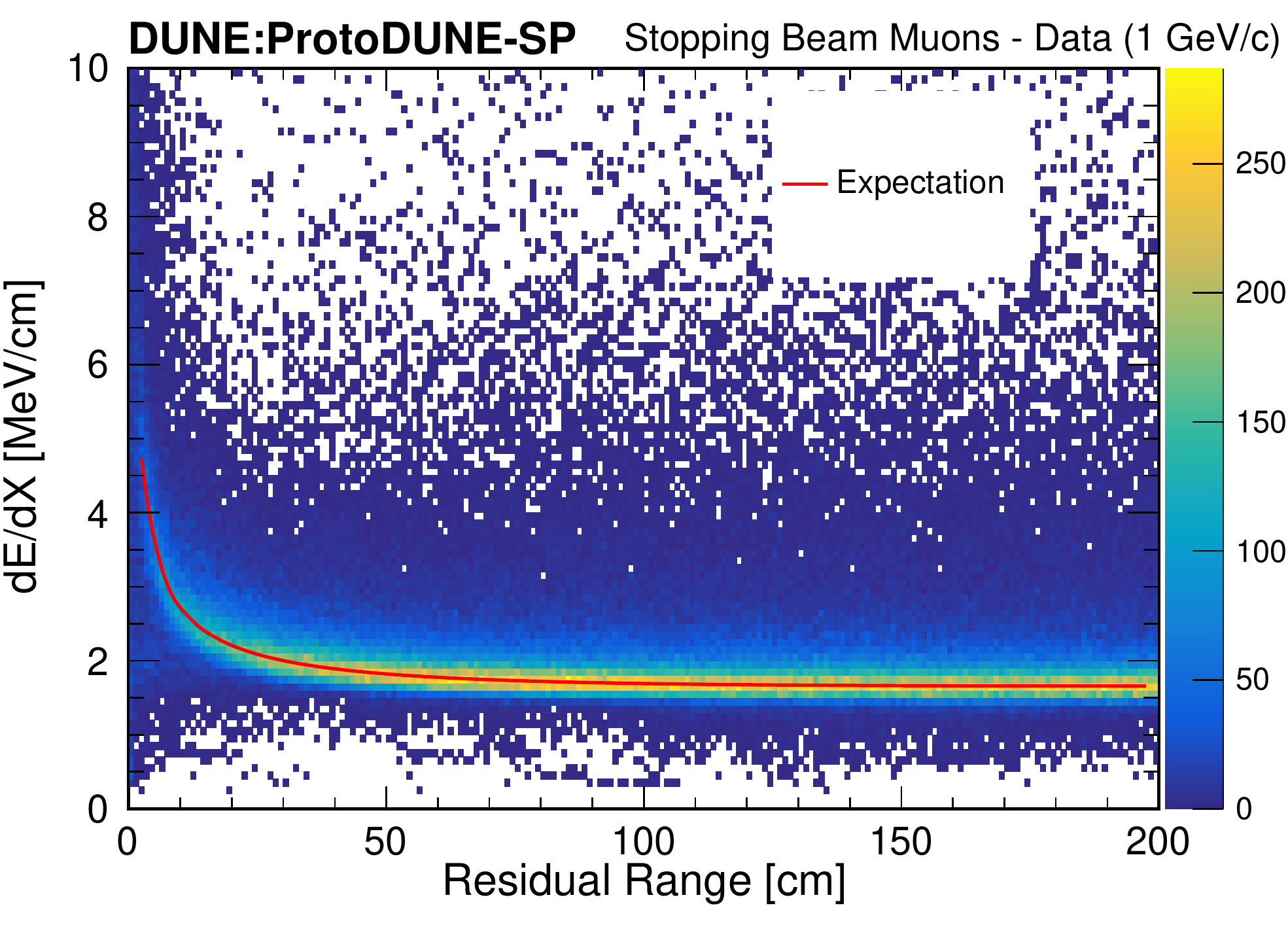}
    \caption{}
    \label{fig:beammuon_dedx_data}
  \end{subfigure}
  \hfill
  \begin{subfigure}[b]{0.5\textwidth}
    \centering
    \includegraphics[width=\textwidth]{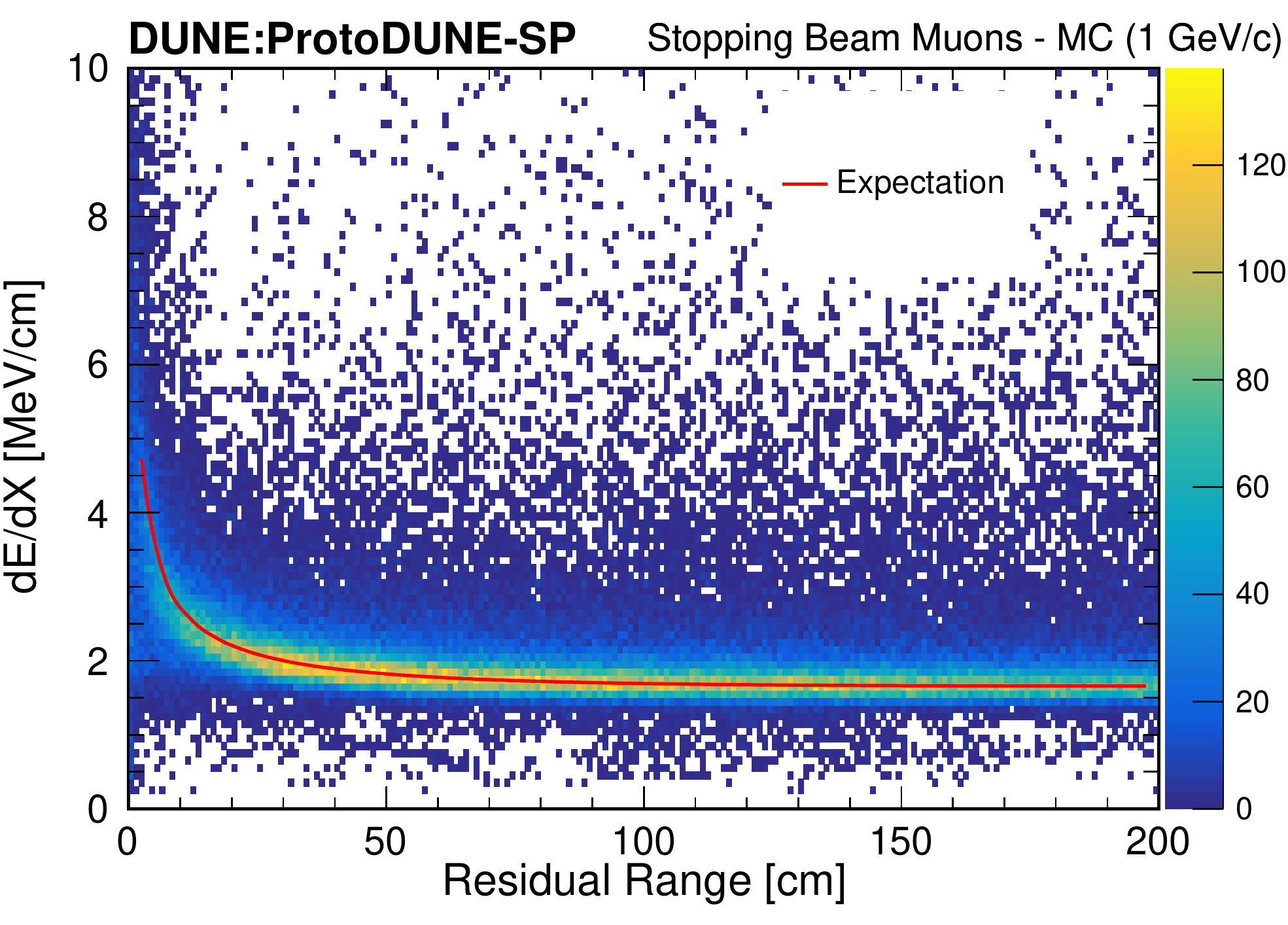}
    \caption{}
    \label{fig:beammuon_dedx_mc}
  \end{subfigure}
  \caption{$dE/dx$ vs residual range for selected stopping muons in the 1 GeV/$c$ beam after applying the calibration derived using cosmic-ray muons, in data~\subref{fig:beammuon_dedx_data} and MC~\subref{fig:beammuon_dedx_mc}. The expected most probable value of $dE/dx$ is plotted as a function of residual range for both.}
  \label{fig:muon_dedx_mc}
\end{figure}
Figure~\ref{fig:beammuon_dedx1d} shows the distribution of calibrated $dE/dx$ values of all hits on the collection plane from the beam muon candidates selected by the stopping particle cut~(eq.~\ref{eqn:muon_CSDA}). As with the pions, the MPVs of the data and MC samples agree to better than 1\%. The distribution of the $dE/dx$ vs residual range for the selected stopping muons is shown for data in figure~\ref{fig:beammuon_dedx_data} and for MC in figure~\ref{fig:beammuon_dedx_mc}. A clear Bragg peak is seen at low residual range, as expected. The measured distribution displays good agreement with the theoretical MPV curve for a stopping muon in argon, in both the minimum ionization region and Bragg peak region of residual range. 

\subsubsection{Identification and calorimetric energy reconstruction of 1 \texorpdfstring{GeV/$c$}{GeV/c} beam protons}\label{sec:Proton}


To understand the detector response to protons interacting in a LArTPC, an analysis procedure, including the selection of beam protons, detector calibration and calorimetric analysis, has been developed. This section describes the results obtained from ProtoDUNE-SP Run 5387.

\begin{figure}[!ht]
\centering
\includegraphics[width=0.7\textwidth]{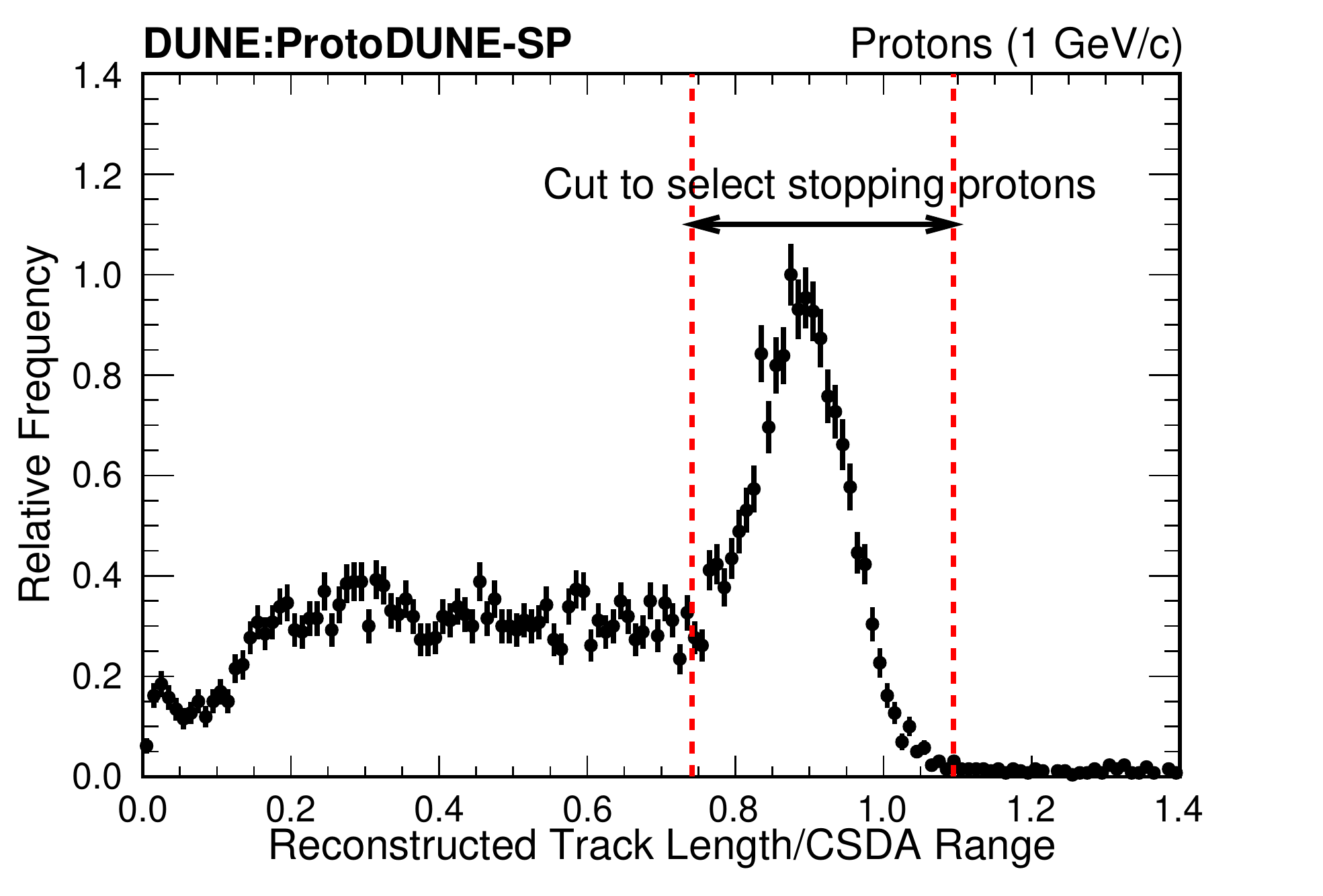}
\caption{The distribution of the reconstructed proton track length divided by the associated CSDA range. The histogram is normalized such that the maximum frequency is one. The incident beam momentum is 1~GeV/$c$. The cut to select the stopping protons is indicated. }
\label{fig:proton_normtrklen}
\end{figure}

Stopping protons are used for the detector characterization in terms of calorimetry and particle identification. Protons are selected using the same beam-TPC matching criteria described in section~\ref{sec:PionMuon}. For the 1 GeV/$c$ beam momentum runs considered here, the beam line PID conditions for protons can be found in table~\ref{tab:PID}.
The measured beam momentum is used to approximate the stopping range under the assumption it is a proton using the CSDA range.
Figure~\ref{fig:proton_normtrklen} shows the distribution of the reconstructed proton track lengths, divided by their expected CSDA ranges. The distribution peaks at 0.88, which is dominated by the stopping protons. The peak position is less than one because of the energy loss upstream and the SCE. The tail on the left of the distribution is due to the interacting protons, since their drift distances inside the LAr are shorter than those of the stopping protons. The ratio cut, 0.74 $\leq$ (reconstructed proton track length/CSDA range) $\leq$ 1.09, is used to select the stopping protons. 


After the proton event selection, the data-driven corrections, described in section~\ref{sec:SCE}, are applied to correct for both the spatial and the $E$-field distortions due to the SCE. After these corrections, the $dQ/dx$ values (ADC/cm) of the stopping protons are converted to the corresponding $dE/dx$ values (MeV/cm) using the calibration constants described in section~\ref{sec:muonCal}. The same analysis procedure is applied to a Monte Carlo sample. 

\begin{figure}[!ht]
  \begin{subfigure}[b]{0.5\textwidth}
    \centering
    \includegraphics[width=\textwidth]{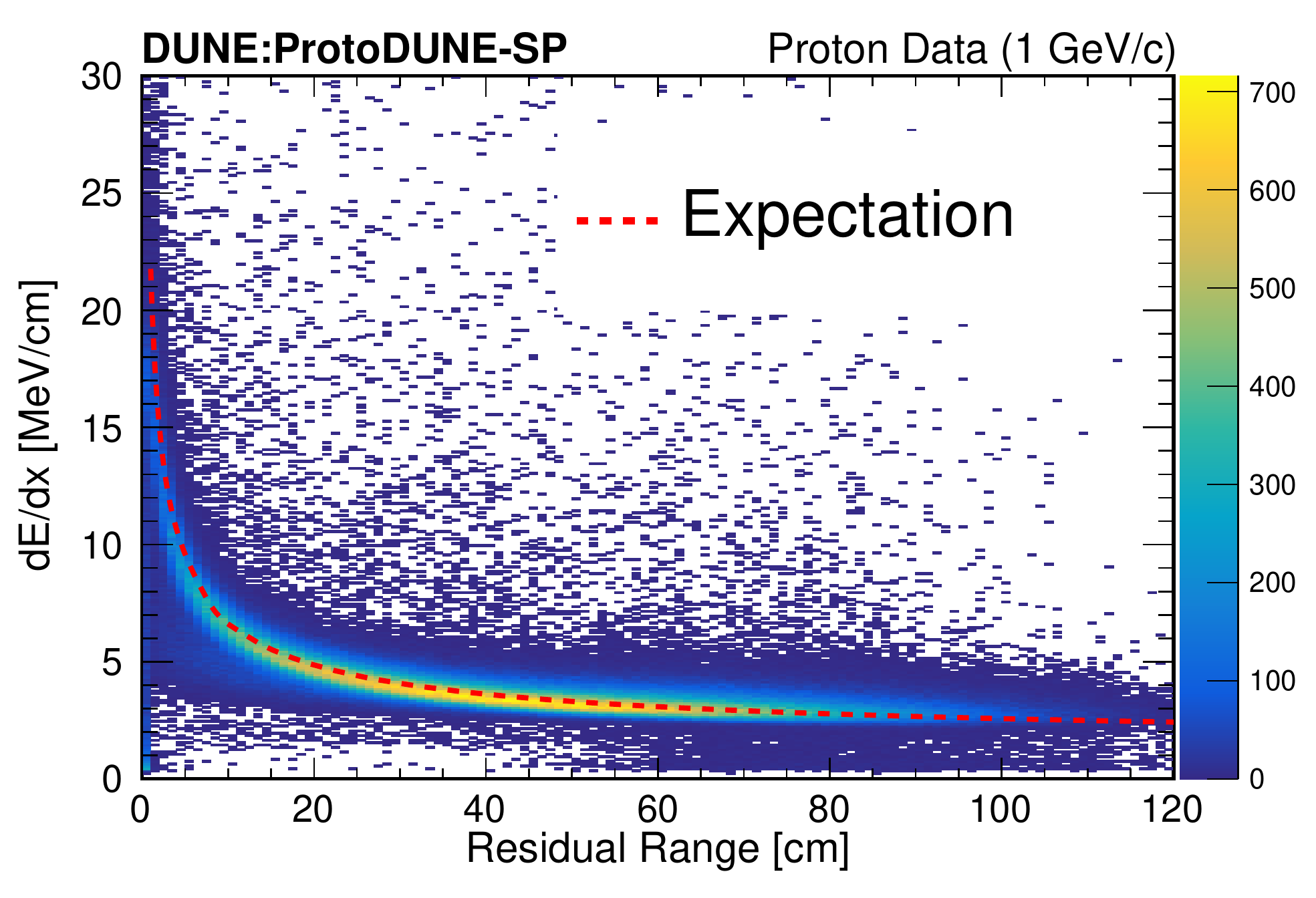}
    \caption{}
    \label{fig:proton_dedx_rr_data}
  \end{subfigure}
  \hfill
  \begin{subfigure}[b]{0.5\textwidth}
    \centering
    \includegraphics[width=\textwidth]{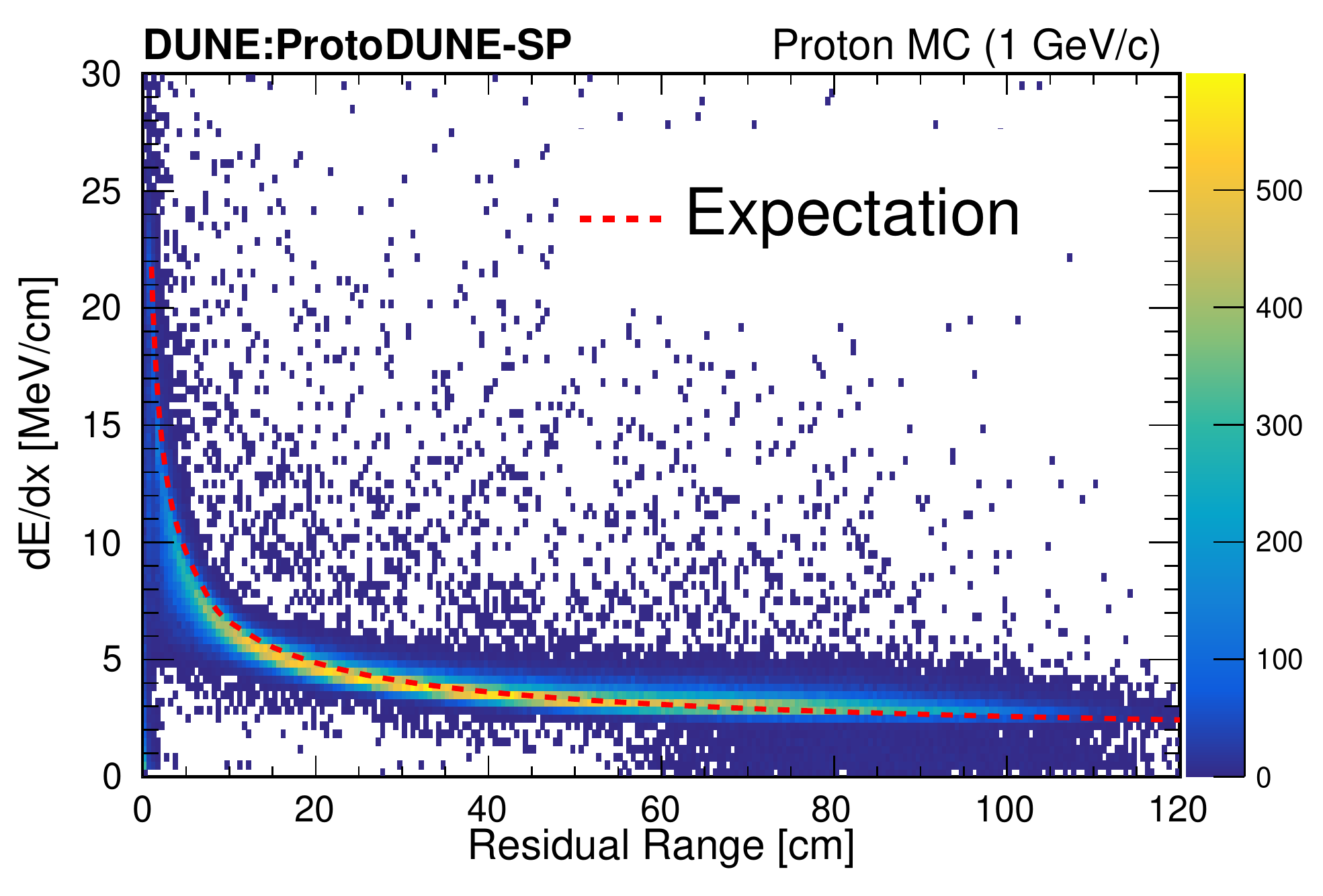}
    \caption{}
      \label{fig:proton_dedx_rr_mc}
  \end{subfigure}
  \hfill
  \begin{subfigure}[b]{\textwidth}
    \centering
    \includegraphics[width=0.5\textwidth]{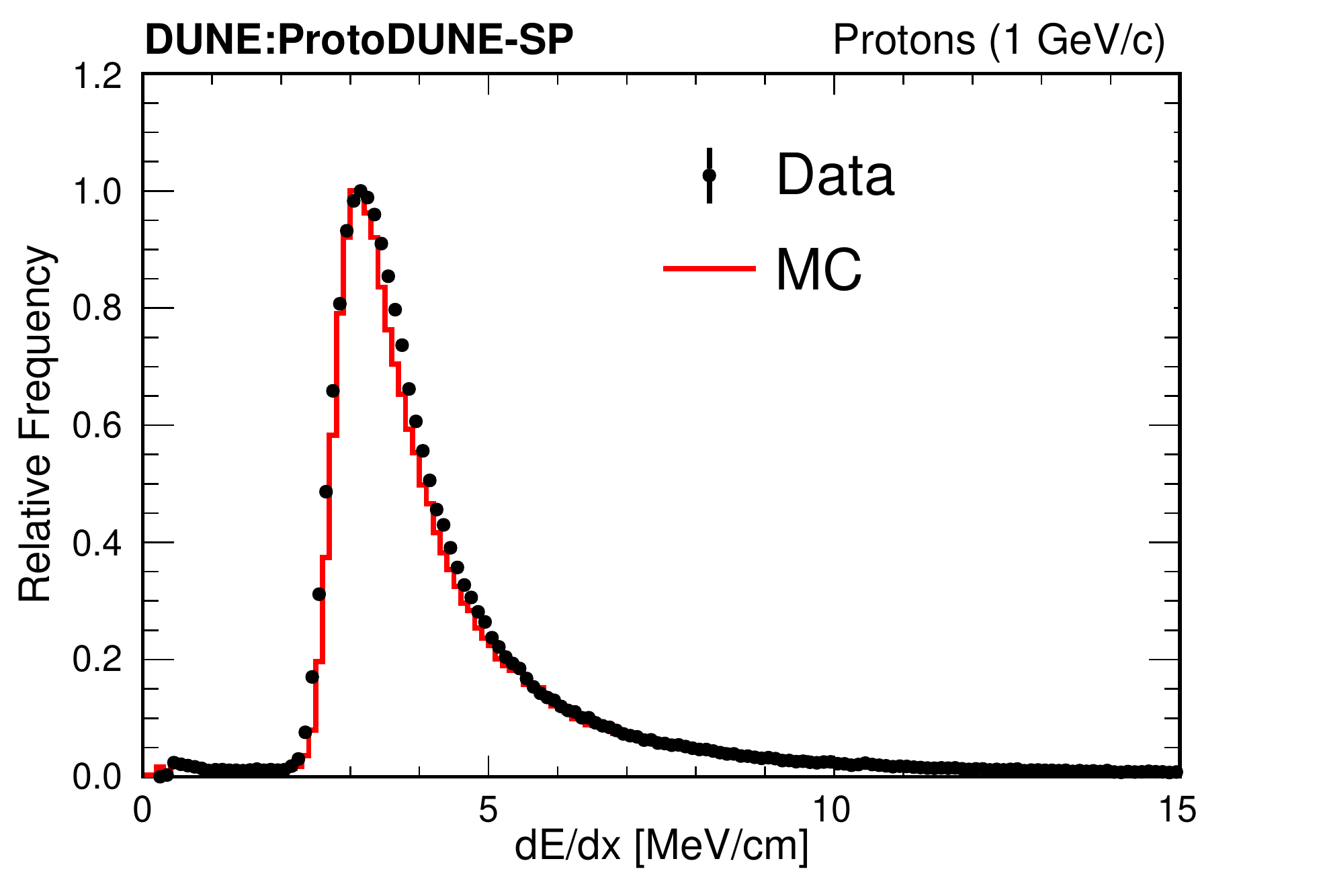}
    \caption{}
    \label{fig:proton_com_dedx}
  \end{subfigure}
  \caption{Stopping proton $dE/dx$ versus residual range distributions for the ProtoDUNE-SP beam data~\protect\subref{fig:proton_dedx_rr_data} and MC~\protect\subref{fig:proton_dedx_rr_mc}, the expected most probable values are shown in red. The $dE/dx$ distributions after the SCE corrections of data and MC are shown in~\protect\subref{fig:proton_com_dedx}. The histograms in ~\protect\subref{fig:proton_com_dedx} are normalized such that the maximum frequency is one.
  }
  \label{fig:proton_dedx_rr}
\end{figure}

Figures~\ref{fig:proton_dedx_rr_data} and~\ref{fig:proton_dedx_rr_mc} show the energy loss versus the residual range of the stopping proton candidates for data and MC, respectively. The data and MC after the calibration procedure show good agreement with the expected MPVs. The distributions of $dE/dx$ in the data and the MC are shown in figure~\ref{fig:proton_com_dedx}. The MPVs of the $dE/dx$ distributions between data and MC agree to better than 1\%. 

\subsubsection{\texorpdfstring{$dE/dx$}{dE/dx} for 1 \texorpdfstring{GeV/$c$}{GeV/c} electrons}
\label{sec:electron}
It is important to understand the LArTPC response to electromagnetic showers since DUNE will measure electrons coming from oscillated neutrinos, produced via charged current interactions. Accurate measurement of the calorimetric response of electrons in the ProtoDUNE-SP TPC allows a more precise understanding of $e/\gamma$ separation and estimation of electron neutrino energy. 
The $dE/dx$ for a photon-induced shower is expected to be twice the $dE/dx$ of a single electron at the beginning of the shower, due to the photon conversion into an electron-positron pair.  This has been verified in a LArTPC by the ArgoNeuT collaboration~\cite{PhysRevD.95.072005}. To successfully select $\nu_{e}$ charged-current interactions in DUNE, a $dE/dx$ metric can be used to remove electromagnetic-like background from interactions such as neutral-current $\pi^{0}$ production where the photons from $\pi^0$ decay can mimic an electron shower. 

Electrons are selected in Run 5809 taken on Nov. 8, 2018 using the same beam-TPC matching criteria described in section~\ref{sec:PionMuon}. For the 1 GeV/$c$ beam momentum runs considered here, the beam line PID conditions for electrons can be found in table~\ref{tab:PID}.


The reconstruction of electron-induced showers in the detector follows the same procedure as in track-like events.  Signal processing (including deconvolution and noise removal) is followed by hit finding and 2D cluster formation. The reconstruction framework Pandora~\cite{Marshall:2015rfa} is used to reconstruct 3D~showers. 
The position and the direction of the shower are used to define the beginning of the shower, which is before the electromagnetic cascade develops. To ensure that the electron candidate has not developed a cascade shower before entering the active TPC volume, a completeness cut was required.  The completeness is defined as the reconstructed shower energy divided by the incoming electron's momentum.  Based on MC studies we expect to have losses due to energy loss upstream and due to signal processing thresholds. If the completeness is at least 80\%, the event is selected. 

To measure $dE/dx$, first, the charge deposition per unit length $dQ/dx$ is measured on a single wire at the collection plane. To calculate the effective pitch $dx$ between hits, the direction of the shower is used to measure the actual distance that the electron traverses in the TPC between adjacent wires. Then, following the discussion in section~\ref{sec:SCE} the SCE corrections are applied. The conversion from $dQ/dx$ to $dE/dx$ uses the calibration constants described in section~\ref{sec:muonCal}. To measure $dE/dx$ at the beginning of the shower, only hits within 4~cm along the direction of the shower and 1~cm perpendicular to the shower are considered and the median $dE/dx$ is computed. The same analysis procedure is applied to the Monte Carlo sample.

\begin{figure}[!htp]
\begin{center}
\includegraphics[width=9cm,height=6.cm]{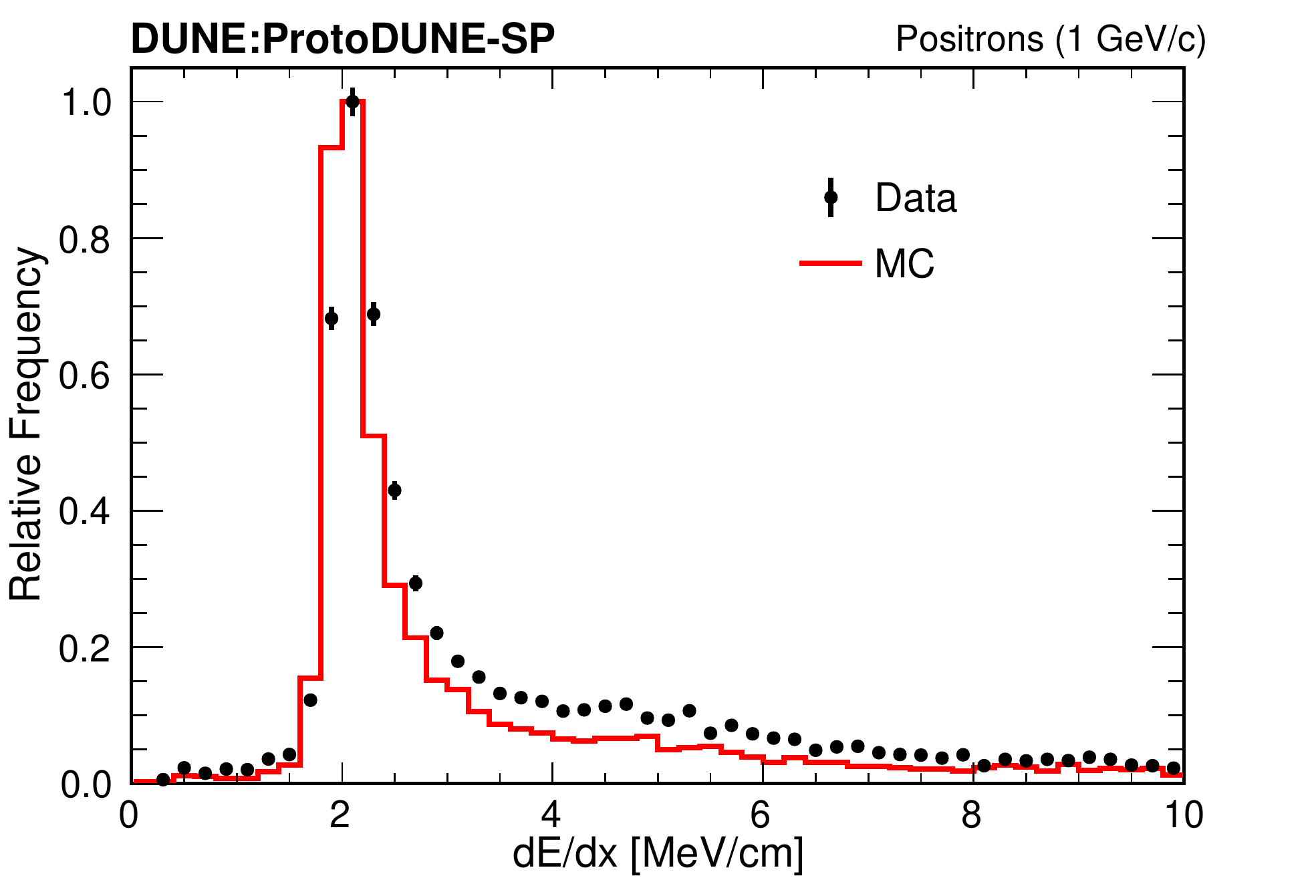}
\end{center}
\caption{$dE/dx$ at the beginning of the shower. The histograms are normalized such that the maximum frequency is one.}
\label{fig:dEdx}
\end{figure}

The $dE/dx$ distributions for 1 GeV/$c$ electron candidates are shown in figure~\ref{fig:dEdx}.
The $dE/dx$ distributions in figure~\ref{fig:dEdx} follow the expected Gaussian-convolved Landau distribution with the $dE/dx$ peak value corresponding to one single ionizing particle. This results demonstrates an understanding of the $dE/dx$ metric for electrons that would be a valuable input for future analyses.
Electrons are selected using the same beam-TPC matching criteria described in section~\ref{sec:PionMuon}. For the 1 GeV/$c$ beam momentum runs considered here, the beam line PID conditions for electrons can be found in table~\ref{tab:PID}.

The $dE/dx$ distributions for various particle all show a good agreement between data and MC in the peak. However, the resolution of $dE/dx$ is slightly overestimated in the MC. This could be due to the imperfect modeling of physics processes and/or detector effects. 

\subsubsection{Particle Identification: Protons and Muons}\label{sec:P_Mu_PID}

\begin{figure}[!ht]
  \begin{subfigure}[b]{0.5\textwidth}
    \centering
    \includegraphics[width=\textwidth]{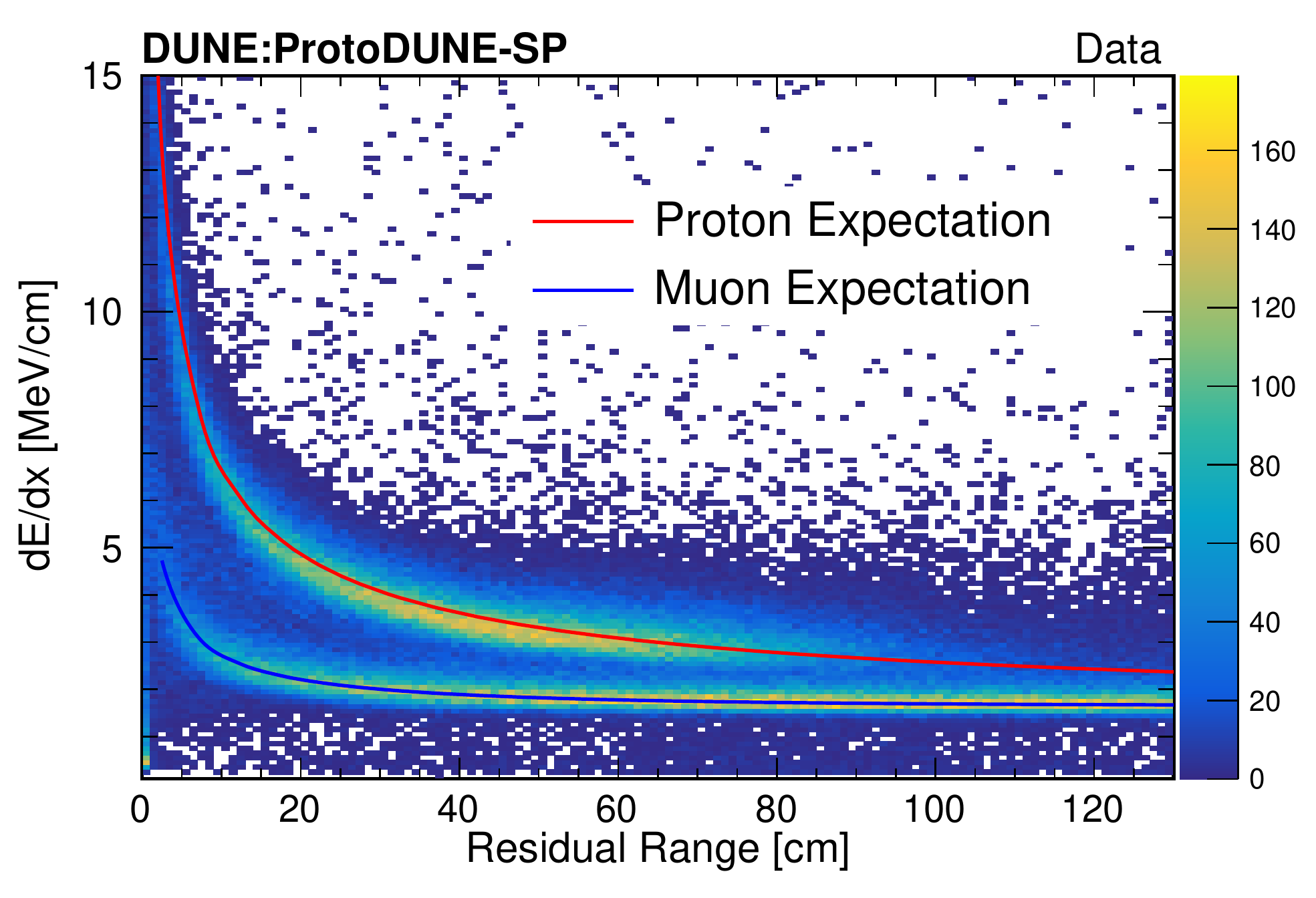}
    \caption{}
    \label{fig:pid_p_mu}
  \end{subfigure}
  \hfill
  \begin{subfigure}[b]{0.5\textwidth}
    \centering
    \includegraphics[width=\textwidth]{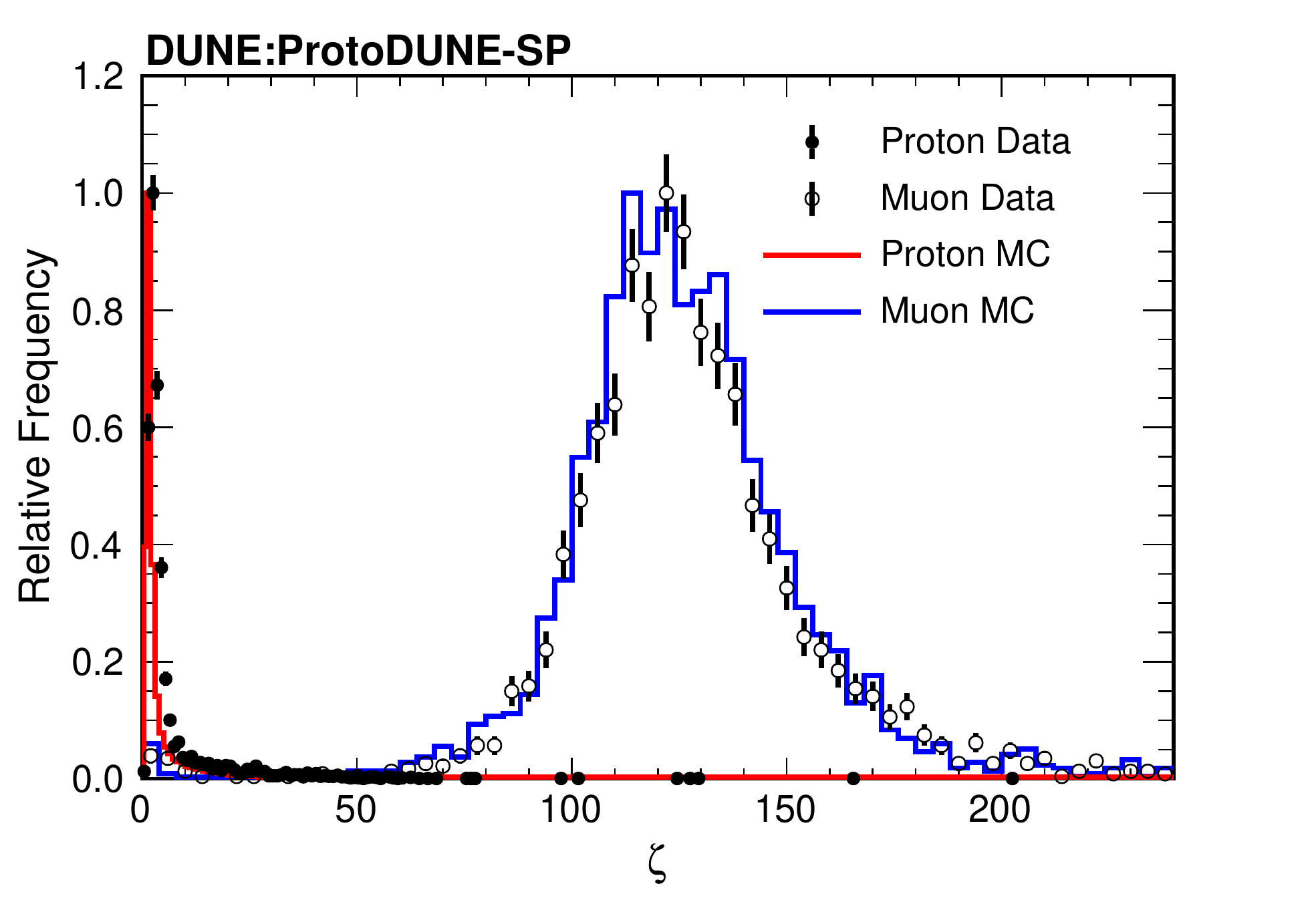}
    \caption{}
    \label{fig:pid_p_chi2}
  \end{subfigure}
  \caption{\protect\subref{fig:pid_p_mu} $dE/dx$ versus residual range after the SCE corrections for the stopping protons (upper band) and the muons (lower band). The solid lines  represent the expected most probable values for the protons (red) and the muons (blue). \protect\subref{fig:pid_p_chi2} The $\zeta$ distributions of the stopping protons and muons. The histograms are normalized such that the maximum frequency is one.
  }
  \label{fig:proton_dedx_rr_pid}
\end{figure}

Robust particle identification (PID) is of fundamental importance for the physics goals of ProtoDUNE-SP and the future DUNE experiment. The calorimetric-based PID method in a LArTPC uses the reconstructed energy deposits as a function of residual range for the stopping particles. Event selections of the stopping muons and the stopping protons are described in  section~\ref{sec:Ecali} and section~\ref{sec:Proton}, respectively.
The stopping protons and muons are shown in figure~\ref{fig:pid_p_mu}. The highly ionizing protons are clearly separated from the muons over the entire range from their stopping points. 

Based on the obtained calorimetry information, a likelihood-based parameter, $\zeta$, is adopted to quantify the PID performance of the ProtoDUNE-SP detector. The method is to compare the particle species with respect to the stopping proton hypothesis. The parameter $\zeta$ is defined to be: 

\begin{equation}
\label{eqn:proton_pid}
\zeta=
\frac{1}{n_{\rm{T}}}\sum_{j}\frac{\left[(\frac{dE}{dx})_{j}({\rm Data})-{(\frac{dE}{dx})_{j}({\rm MC ~Proton})}\right]^{2}}{\sqrt{\left[\sigma (\frac{dE}{dx})_{j}({\rm Data})\right]^{2}+{\left[ \sigma(\frac{dE}{dx})_{j}({\rm MC~Proton})\right]^{2}}}}, 
\end{equation}
where $j$ is the $j$-th measurement before the end of the track, $\sigma(\frac{dE}{dx})_{j}$ is the associated $\frac{dE}{dx}$ error of the $j$-th hit, $j$ covers the measurements over the last 26 cm of the track, and $n_{\rm{T}}$ is the total number of hits. Only the collection plane information is used.

Figure~\ref{fig:pid_p_chi2} shows the $\zeta$ distributions of the stopping protons and the stopping muons. The protons and the muons are well-separated. The PID performance for pions is expected to be similar to that of muons.

\section{Photon detector response}
\label{sec:pdresponse}

Homogeneous calorimeters are instrumented targets where the kinetic energy ($E$) of incident particles is absorbed and transformed into a detectable signal. The deposited energy is typically detected in the form of charge or light. When the  energy is large enough, a shower of secondary particles is produced (through electromagnetic or strong processes) with progressively reduced energy. If the shower is fully contained and the output signal is efficiently collected, the calorimetric energy resolution is expected to be good, improving with energy as $1/\sqrt E$.
Calorimeters can also provide information on shower position, direction and size as well as arrival time $t_0$ of the particle.

A LArTPC is a sophisticated version of a homogeneous calorimeter, with additional imaging and particle identification capabilities. Energy deposition in the liquid argon target yields free charge from ionization and it also yields fast scintillation light. The best energy resolution would be obtained by collecting both the charge and the light signals, which are anticorrelated by the randomness of the recombination processes. With detectors based on LArTPC technology, calorimetry typically relies only on the charge signal collection, while the use of the light signal is limited to $t_0$ determination and triggering purposes. A first attempt to extend the use of scintillation light for calorimetry was recently performed in a low energy range with a small sized LArTPC~\cite{Foreman:2019dzm}.
Operating protoDUNE-SP on the H4-VLE charged particle test beam offers the opportunity to directly probe with light the calorimetric response of liquid argon to fully contained EM and hadronic showers in the sub- to few-GeV energy range. 

\subsection{Calorimetric energy reconstruction from scintillation light and energy resolution}
\label{sec:calorimetry-light}

As described in section \ref{sec:PDSDescription}, in ProtoDUNE-SP the photon detection system comprises a series of optical modules positioned inside the APA frames, behind the TPC wire planes and the grounded mesh. The active liquid argon volume is only on one side of the APA's, on the side facing the central cathode.  The total photo-sensitive area is $\sim 1.5$\% ~ of the boundary surface of the LAr volume. The relatively modest optical coverage and the one-sided geometry of the PD system, compared for example to the 4$\pi$ coverage of scintillation or Cherenkov detectors, are expected to limit the light yield and the uniformity of the calorimetric response along the drift direction. In this section, beam electrons and data from the ARAPUCA module in the beam side of the PDS are utilized to investigate the light yield and resolution of the ProtoDUNE-SP PD system.  

\begin{figure}[!htbp]
\centering
\includegraphics[height=8cm]{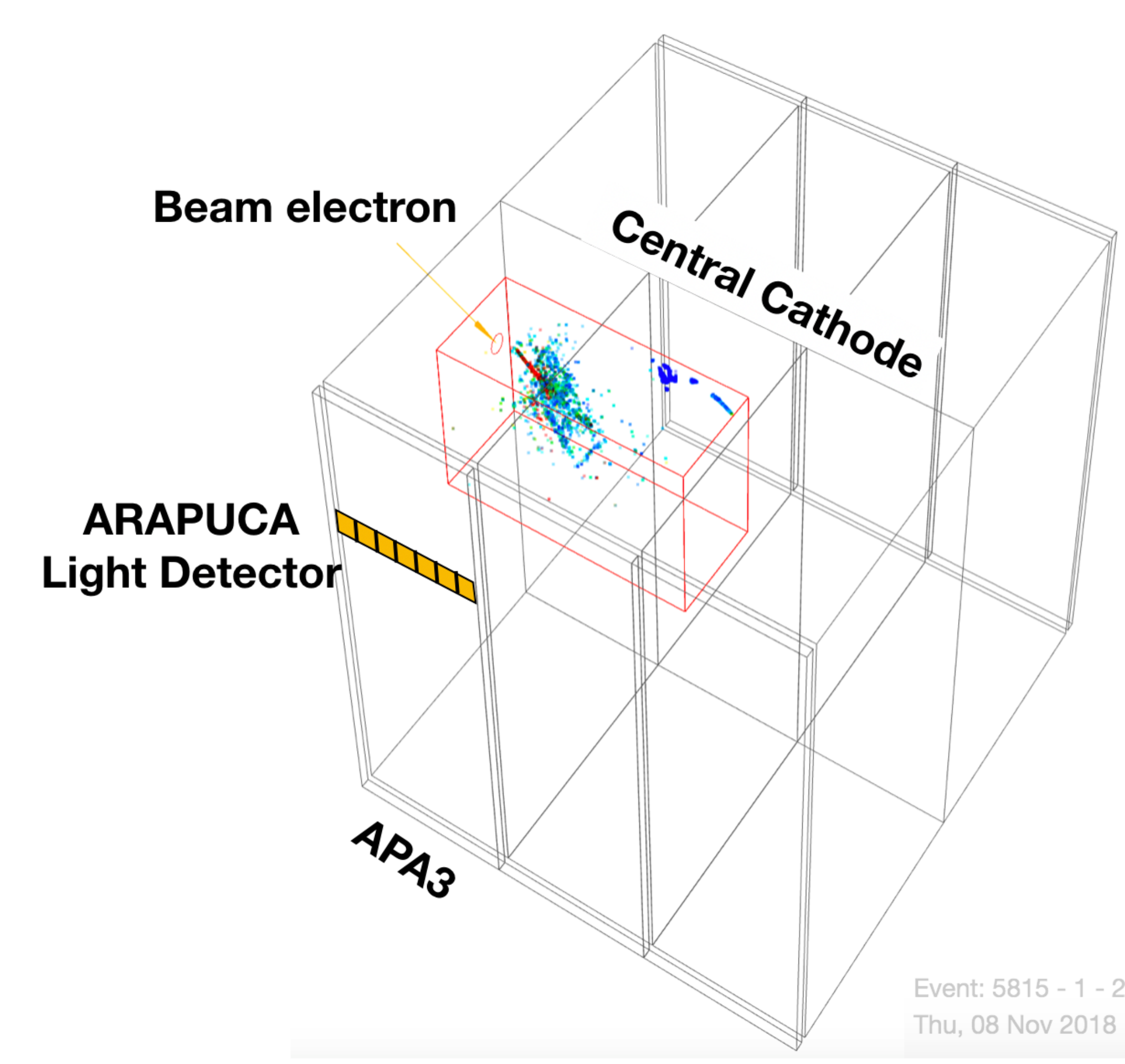}
\caption{3D event display of a real 7~GeV beam electron in the TPC volume [only tracks inside a predefined sub-volume (red box) are shown]. The beam electron enters near the cathode and an EM shower develops in LAr along the beam direction, about 10$^\circ$ downward and 11$^\circ$ toward the anode plane, where the PDS beam side modules are located inside the APA frames. Scintillation photons from energy deposits along the shower are detected by the ARAPUCA module.}
\label{fig:3D-EM-showewr}
\end{figure}
\subsubsection{Beam electrons and EM showers }
\begin{figure}[tbh]
\begin{minipage}[t]{0.52\textwidth}
\includegraphics[width=1.0\textwidth]{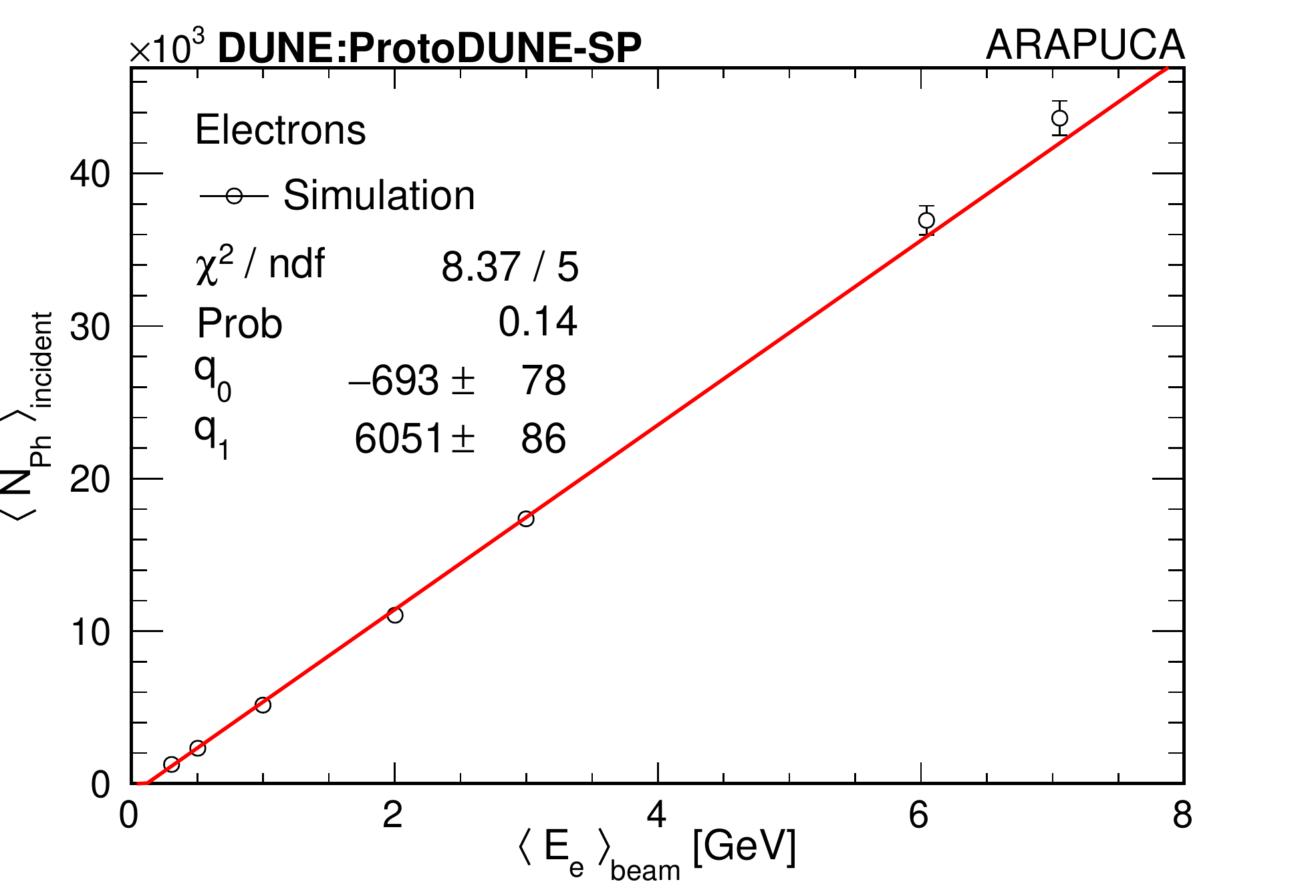}
\end{minipage}
\begin{minipage}[t]{0.52\textwidth}
\includegraphics[width=1.0\textwidth]{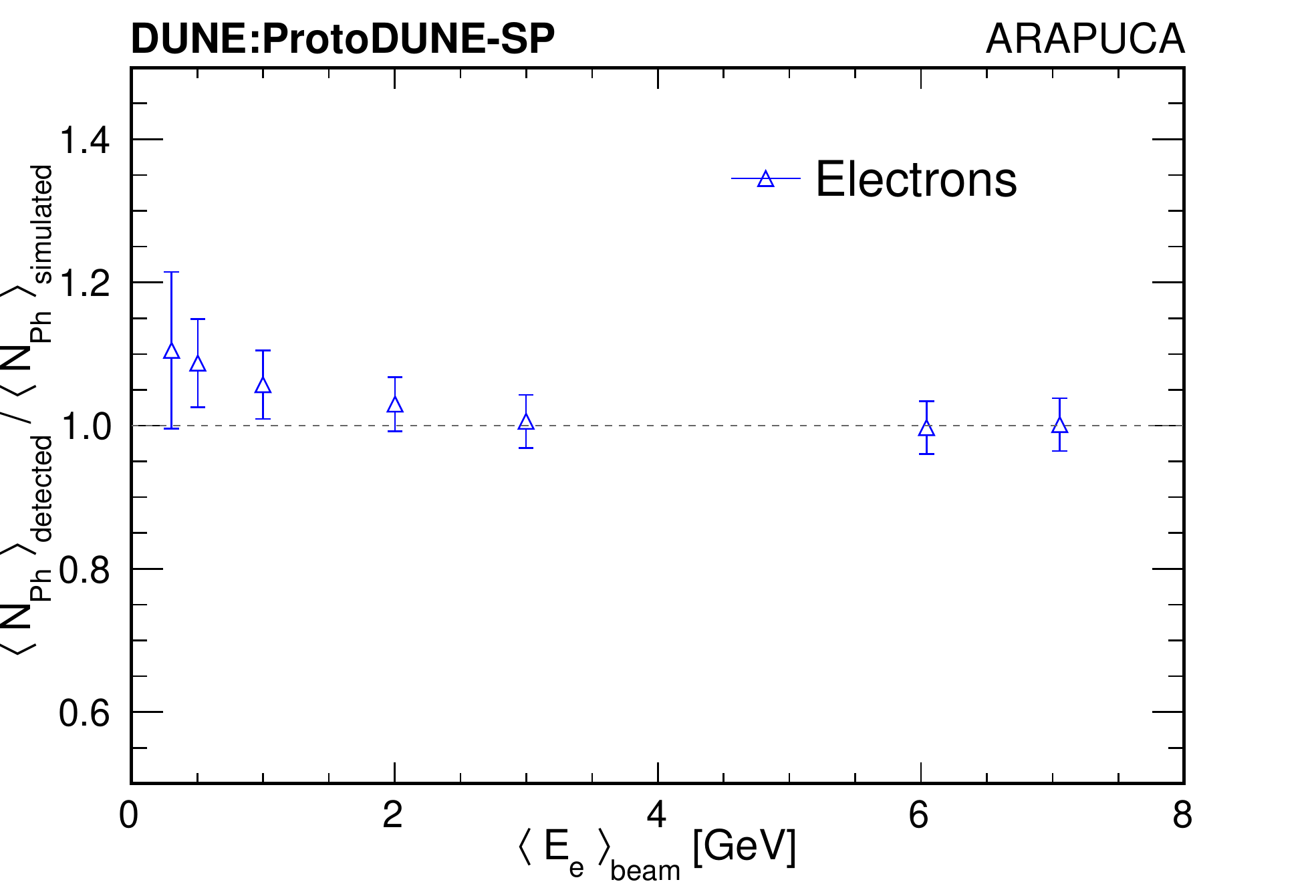}
\end{minipage}
\caption{Monte Carlo simulation of light response to EM showers:  average number of incident photons at the ARAPUCA bar surface from simulation vs average electron beam energy (left) and Data/MC comparison (right) by the ratio of average number of photons detected to the corresponding number of photons from MC simulation, at the different beam energies.}
\label{fig:MC-cfr}
\end{figure}

In a sequence of beam runs, data were collected with incident positive electrons ($e^+$) with energies of 0.3, 0.5, 1, 2, 3, 6 and 7~GeV, providing a large sample of EM showers developing in the LAr volume. The beam is delivered with a typical momentum spread of $\pm 5\%$ around the nominal setting. The beam line instrumentation (see section 
\ref{subsec:momSpec}) provides an event-by-event particle identification and momentum  measurement with a precision of $\Delta p/p \simeq 2.5\%$~\cite{PhysRevAccelBeams.22.061003}. 
Light signal from the single beam side ARAPUCA module is used for this calorimetric response study. The ARAPUCA module is positioned inside the upstream APA3 frame, nearly at the same height $y$ of the entering beam and oriented along the $z$ axis. 
In figure \ref{fig:3D-EM-showewr} a 3D display of a 7~GeV electron event from the ProtoDUNE-SP data sample is shown as reconstructed by the TPC. The EM shower develops immediately downstream of the beam entry point in the LArTPC volume in front of the ARAPUCA module, at $\sim$3~m distance in the $x$ (drift) direction, and propagates longitudinally along the beam direction, about 10$^\circ$ downward and 11$^\circ$ toward the anode plane. Scintillation light is emitted isotropically from every location at which ionization occurs along the shower. The total photo-sensitive area of the ARAPUCA module is $\sim 0.5\times 10^{-3}$ 
of the surface surrounding the LArTPC active volume (beam side). 
\begin{figure}[tbh]
\includegraphics[width=1.0\textwidth]{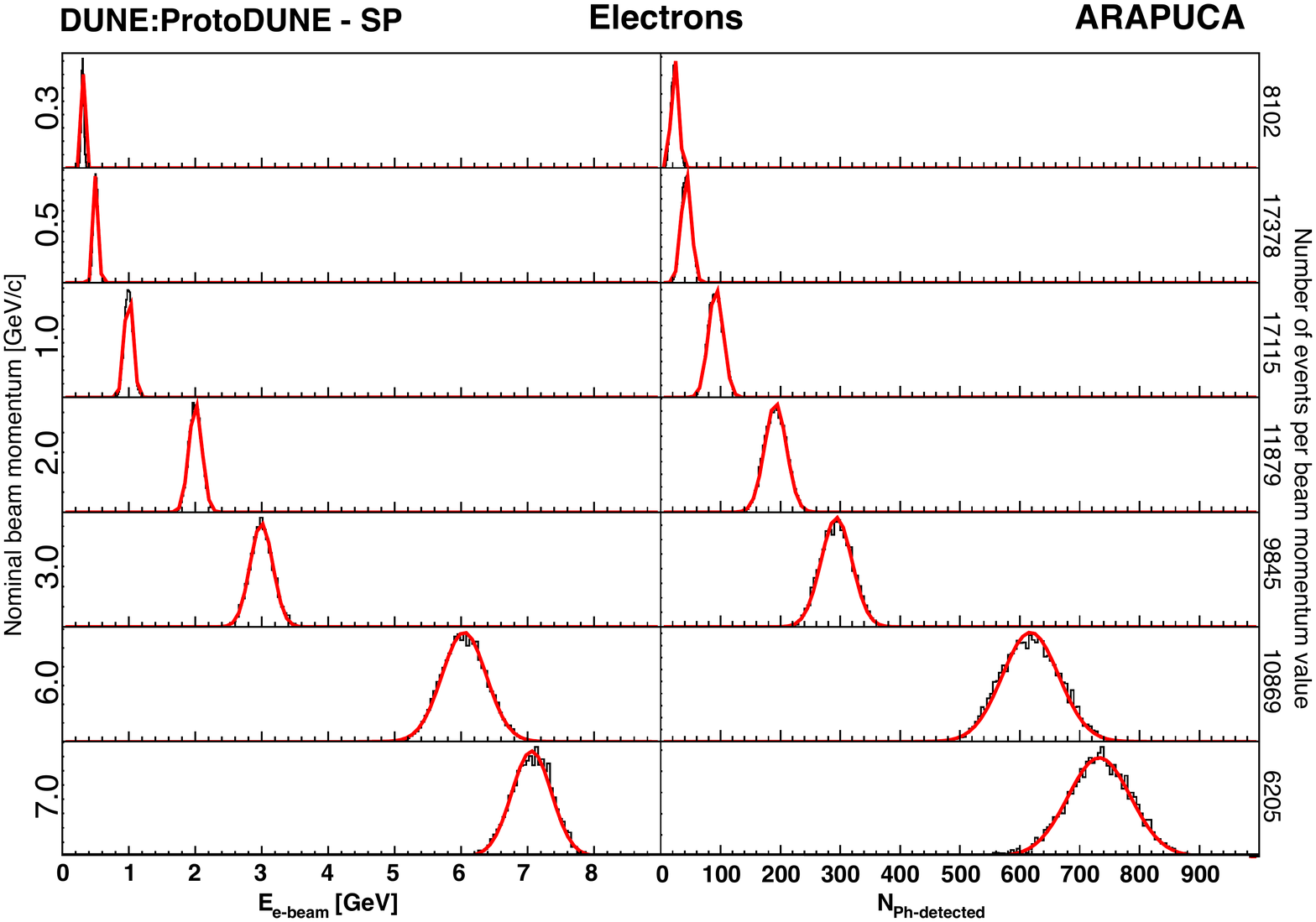}
\caption{Distributions of the incident beam electron energies (left) and corresponding detected photon spectra (right) for the collected seven nominal beam energies (Gaussian fit superimposed - red line). Fit parameters are used in figure \ref{fig:ElectronsPHvsKE}. The photon spectra are relative to the sum of photons detected by the 12 cells of the ARAPUCA PD module.}
\label{fig:ebeam-phdet}
\end{figure}
Summing over the twelve cells in the ARAPUCA module, the total number of photons detected $N_{\rm Ph}=\sum N^{\rm Det}_j$ is evaluated event by event, for each beam run. 
Monte Carlo events, already used for the efficiency study, are also available,  generated as described in section \ref{sec:PDPerf} with incident electrons as in the beam runs (same energy distributions and direction). Scintillation light from the EM shower development in LAr is propagated to the photo-detector(s) and the total number of photons incident  on the surface of the ARAPUCA module is evaluated for each event of the MC simulated momentum run.
In figure \ref{fig:ebeam-phdet}, the energy distribution (left) of the incident electrons from the beam spectrometer of the CERN H4-VLE beam line for each beam run and the corresponding calorimetric response from the ARAPUCA detector (right) expressed by the number of detected photons are shown. 
A Gaussian fit of both distributions (red lines in figure \ref{fig:ebeam-phdet}) gives the average electron energy $\langle E_{\rm e}\rangle$ and the corresponding average photon counting $\langle N_{\rm Ph}\rangle$, and their spreads ($\sigma_E,~\sigma_{N}$), for each run.
The average number of detected photons as a function of the beam energy is shown in figure~\ref{fig:ElectronsPHvsKE}. This relationship gives the calorimetric energy response from light data. Correspondingly, the response as obtained from the MC simulation, expressed by the average number of detectable photons (incident at the detector surface) from EM showers at given electron beam energy, is presented in figure \ref{fig:MC-cfr} (left).

To a first approximation, the average light response is a linear function of the energy over the entire range of tested beam energies as shown in figure \ref{fig:ElectronsPHvsKE} (left) with the result of the linear fit. The slope of the fit $p_1$ gives the light yield $Y_{\rm{light}}=102.1$~photons/GeV. 
The quoted $Y_{\rm{light}}$ is relative to a diffuse light source (EM shower) at a distance of about 3~m (see figure \ref{fig:3D-EM-showewr}). The non-zero (negative) intercept ($p_0=-8.4$ photons from the fit) corresponds to an incident energy offset of $-82\pm 14$ MeV from the nominal value for all beam energies. From the CERN H4-VLE beam line MC simulations (section 3.2 and 3.3) beam electrons are expected to release  10-20~MeV in the material in the portion of the beam line downstream the spectrometer and additional $\sim$ 20-30 MeV while crossing the materials inside the cryostat from the end of the beam pipe and the active volume of the TPC (cryostat insulation and membrane, beam tube and a thin LAr layer in between). The observed energy offset from the linear fit of the light response provides direct evidence, though in slight excess, of the expected energy loss of beam electrons before entering the LArTPC.   
A slight deviation from linearity is observed in the light response at higher incident energies, both in the data (figure \ref{fig:ElectronsPHvsKE} - left) and in the Monte Carlo (figure \ref{fig:MC-cfr} - left). This is due to the light response dependence on source-to-detector distance. At higher energies, the longitudinal shower profile extends deeper in the LAr volume along the beam direction and closer to the ARAPUCA module, with some increase of visibility. Based on the reconstruction of the shower profile at the different incident electron energies (see figure \ref{fig:PhL-PhD-eff} - top-right), a geometry correction to the cells' acceptance has been calculated and a normalization factor applied to the data at different energies. Most of the nonlinearity was then removed. 
The intercept of the linear fit after correction indicates an energy offset of -56$~\pm~$14 MeV, in better agreement with the expected beam electron energy loss in the materials before entering the TPC - details of this study can be found in \cite{ref:Dante-FLC-LightAna}. 

\begin{figure}[tbh]
\begin{minipage}[t]{0.52\textwidth}
\includegraphics[width=1.0\textwidth]{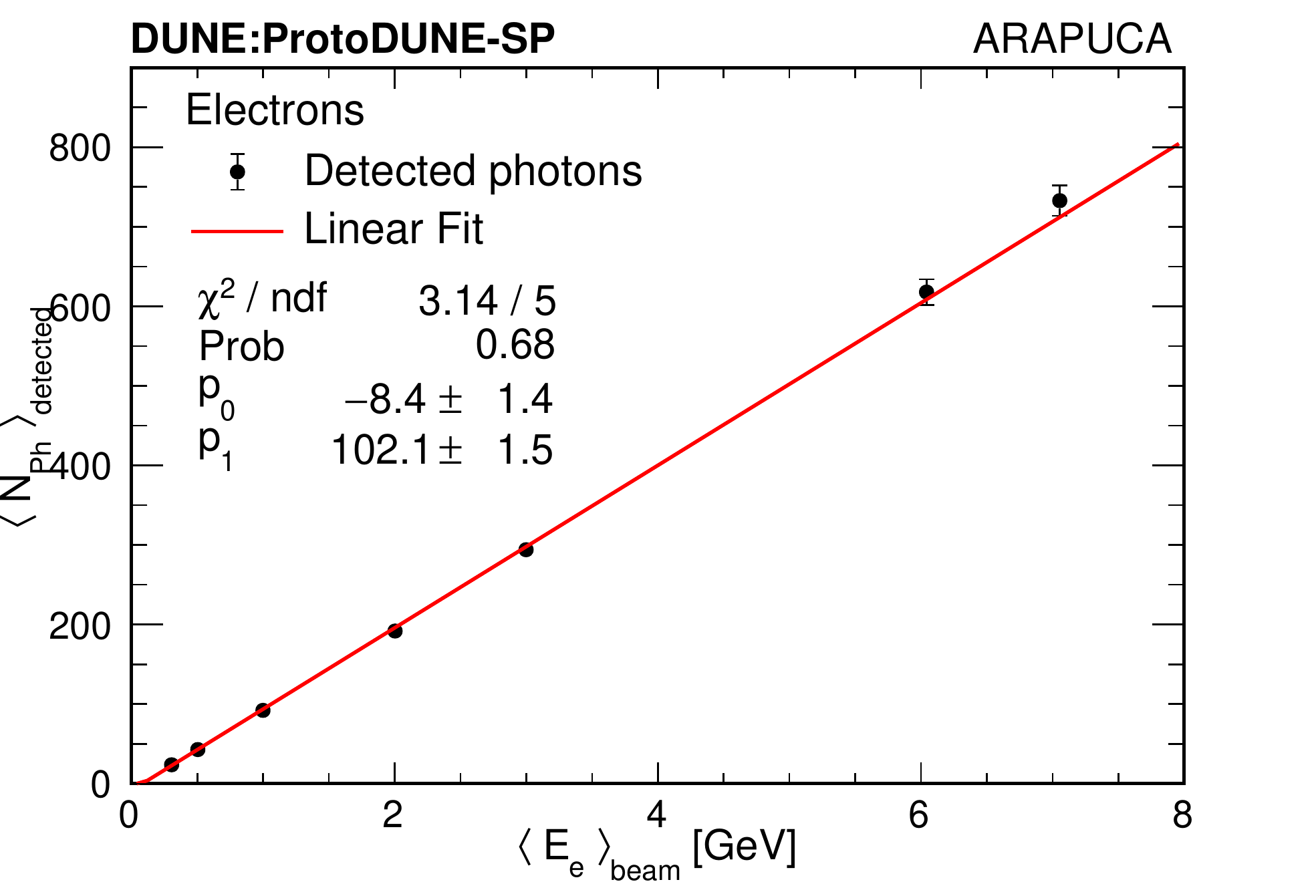}
\end{minipage}
\begin{minipage}[t]{0.52\textwidth}
\includegraphics[width=1.0\textwidth]{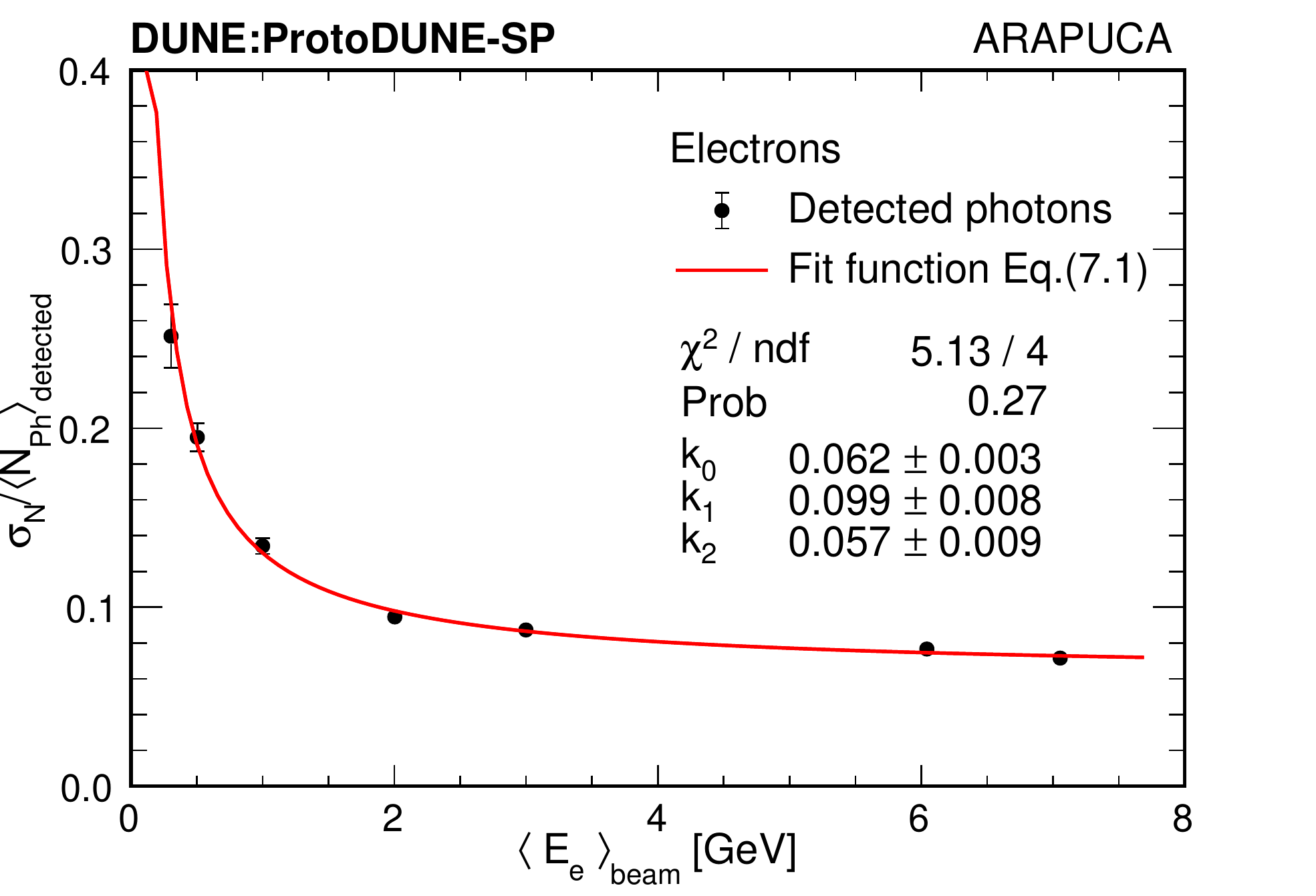}
\end{minipage}
\caption{Number of detected photons (Gaussian fit mean value, from  fig.\ref{fig:ebeam-phdet} right) as a function of incident electron energy (Gaussian fit mean value, from fig.\ref{fig:ebeam-phdet} - left) (left panel). Reconstructed energy resolution from the detected photon distributions (Gaussian fit standard deviation to the mean ratio) (right panel).  A line of slope $p_1$ and intercept $p_0$ is fit to the data in the left-hand plot, and the function in equation~\eqref{eq:resolution} is fit to the data in the right-hand plot. 
}
\label{fig:ElectronsPHvsKE}
\end{figure}
Based on the linearity of the light response, the relative calorimetric energy resolution $\sigma_E/E_{\rm e}$ is obtained from the $\sigma_N/N_{\rm Ph}$ ratio of the light response (figure \ref{fig:ElectronsPHvsKE} - right). The energy resolution, as for a homogeneous calorimeter, can be expressed in a general form \cite{Fabjan:2003aq} depending on three different contributions: 
\begin{equation}
\frac{\sigma_E}{E}~=~k_0\oplus \frac{k_1}{\sqrt E}\oplus\frac{k_2}{E}
\label{eq:resolution}
\end{equation}
where the symbol $\oplus$ indicates a quadratic sum and $E$ is in GeV. The terms on the right-hand side are referred to below as the ``constant term'', the ``stochastic term'' and the  ``noise term.'' The relative weight of the three terms depends on the energy of the incident particle.
The stochastic term contribution to the energy resolution comes from the statistical fluctuations in the number of photons detected. The relatively large value $({k_1}=9.9\%)$ - when compared to typical homogeneous calorimeters - is ascribed to the limited photo-sensitive coverage of the ARAPUCA module. The noise term comes from the electronic noise of the readout chain. Its value (${k_2}=0.057$~GeV) is exactly as expected from the measured signal-to-noise-ratio
 of the ARAPUCA readout (section \ref{sec:StoN}). The constant term is large (${k_0}=6.2\%$) and due to different contributions. The main one comes from the incident beam electron energy spread (figure \ref{fig:ebeam-phdet} - left), depending on the actual beam line configuration (collimators aperture at different momentum setting). The mean value of the relative energy spread  $\sigma_E/E_e=(5.8\pm 0.4)\%$. An additional contribution comes from fluctuations in the energy loss in the materials before electrons enter the TPC. Since the energy loss occurs downstream of the momentum spectrometer, this energy degradation and its fluctuation do not appear in the incident beam energy spectra and it is evaluated by simulations ($\sim 2\%$). Other contributions to the resolution, such as non-uniformity on the detector illumination, channel to channel response variation and possible shower leakage across the cathode, have been investigated and shown to be negligible \cite{ref:Dante-FLC-LightAna}. 
 The light yield and resolution response obtained from a single ARAPUCA module is adequate. An extrapolation of the performance to a PDS system consisting entirely of ARAPUCA modules, obtained by scaling the detected photon response of the light guide bars with the ratio of the ARAPUCA-to-light guide efficiencies (presented in section~\ref{Sec:Efficiency}), indicates that it can perform calorimetric energy reconstruction with an expected light yield of 1.9~photons/MeV. This performance exceeds the specifications of the DUNE Far Detector~\cite{Abi:2020loh} by almost a factor of four.

A comparison of electron data and the corresponding MC simulation can be used to validate the simulation of the light propagation and collection. The average number of photons incident on the surface of the ARAPUCA module from the MC (shown in figure \ref{fig:MC-cfr} - left) is scaled by a normalization factor $\eta$, an average value over the ARAPUCA cells' efficiency shown in section \ref{Sec:Efficiency}, to give the simulated detected photons $\langle N_{\rm Ph}\rangle_{\rm simulated}=\eta~ \langle N_{\rm Ph}\rangle_{\rm incident}$.
The ratio $\langle N_{\rm Ph}\rangle_{\rm detected}/\langle N_{\rm Ph}\rangle_{\rm simulated}$ is shown as a function of incident beam energy in figure \ref{fig:MC-cfr} (right).
A systematic deviation, between 5 and $\le10$\%, is found for electrons with energies between 0.3~GeV and 1~GeV. This deviation is attributed to the statistical limitations on the visibility values of the photon library used in the Monte Carlo for converting the energy deposited along the EM shower into the number of photons impinging upon the ARAPUCA module.  

\section{Conclusions}\label{conclusions}

This paper summarizes the first results on the performance of the ProtoDUNE-SP LArTPC using large samples of data from a test-beam run at the CERN Neutrino Platform. The dedicated H4-VLE beam line delivers electrons, pions, protons and kaons in the 0.3 -- 7 GeV/$c$ momentum range, which are crucial to the study of detector performance and the measurement of particle-argon cross sections.   In table \ref{tab:pDUNE-Perf} the detector's high-level performance parameters from studies and findings presented in this report are shown and they are compared with the corresponding DUNE SP Far Detector design specifications.  For each of the categories shown, the ProtoDUNE-SP performance meets or exceeds the DUNE specification, in several cases by a large margin.  This successful performance demonstrates the effectiveness of the single-phase detector design and the execution of the fabrication, assembly, installation, commissioning, and operations phases~\cite{protodunetechnicalpaper}.
\begin{table}[ht]
\centering
\caption{ProtoDUNE-SP performance for main parameters and corresponding DUNE specifications.} 
\begin{tabular}[t]{lccc} 
\hline
\hline
&{\sf Detector parameter}&{\sf ProtoDUNE-SP performance}&{\sf DUNE specification} \\
\hline
\hline
&Average drift electric field & 500 V/cm                      & $\>$ 250 V/cm  (min) \\
& & & ~~~~~~~500 V/cm (nominal) \\
\hline
& LAr e-lifetime                 & $>$ 20 ms                   & $>$ 3 ms          \\
\hline
\hline
& TPC+CE & &  \\
& Noise     &  (C) 550 e,  (I) 650 e ENC (raw)                   & $<$ 1000 e ENC          \\
& Signal-to-noise $\langle$SNR$\rangle$    &   (C) 48.7, (I) 21.2  (w/CNR)                 &                \\
\hline
&CE dead channels                  &0.2\%                   & $<$ 1\%         \\
\hline
&PDS light yield                      & 1.9 photons/MeV                   & $>$ 0.5   photons/MeV     \\
& &($@$ 3.3 m distance) & ($@$ cathode distance - 3.6 m) \\
\hline
&PDS time resolution                      &  14 ns                   & $<$  100 ns  \\
\hline
\hline
\end{tabular}
\label{tab:pDUNE-Perf}
\end{table}

The electric field in the TPC drift volume was stable at the nominal level of 500 V/cm  with $>$99.5\% of uptime during the data taking periods with beam and cosmics. 

A drift electron lifetime in LAr in excess of 20~ms has been achieved and it was sustained for an extended period of data-taking.  It reached approximately $(89\pm 22)$~ms for the last day of beam data-taking. This corresponds to a concentration of impurity in the liquid argon of 3.4$\pm$0.7~ppt oxygen equivalent. The DUNE Far Detector specification is for the impurity concentration to be less than 100~ppt O$_2$ equivalent.

The TPC and cold electronics show excellent signal-to-noise performance.
  The signal-to-noise ratios corresponding to the most-probable-value ionization of a minimum ionizing particle are 40.3, 15.1 and 18.6, for collection, U and V wires, respectively, after noise filtering and signal processing.  

The number of solidly unresponsive TPC channels was initially 29 out of 15360 and rose to 36 over the course of a year and five months of operations.  Approximately 105 additional channels are noisy or have other issues with the electronics so that they were not included in the analyses presented in this report.

Three different photon detection technologies were implemented in the PDS and characterized with muon and electron beam data. The ARAPUCA technology showed 2\% efficiency, the highest among the three, with a light response to EM energy deposit linear over the entire range of beam energies. A PDS system consisting entirely of ARAPUCA modules, with an expected light yield of 1.9~photons/MeV, will exceeds the specifications of the DUNE Far Detector.

Space-charge effects were predicted to be prominent in ProtoDUNE-SP.  Spatial distortions in the apparent positions of tracks of up to 40~cm are observed in the data, based on the points of entry and exit into the TPC.  Changes in the magnitude of the electric field by up to 25\% are inferred from the spatial distortion measurements.   Data-based, three-dimensional maps of spatial offsets and electric field strengths are made and are used to correct the observed data positions and ionization strengths for use in precision analyses.

The measured resolutions of the calibrated TPC and photon detector responses to protons, muons, electrons, and charged pions are similar to those in the simulations that are used to predict the performance of the first DUNE far detector module. 

The data collected by ProtoDUNE-SP during beam runs and cosmic-ray runs will allow detailed studies of detector characteristics such as fluid flow, and they will also allow the measurement of argon-hadron cross sections.  The results of these studies and measurements will be reported in future publications.


\acknowledgments

%
%
The ProtoDUNE-SP detector was constructed and operated on the CERN Neutrino Platform.
We thank the CERN management for providing the infrastructure for this experiment and gratefully acknowledge the support of the CERN EP, BE, TE, EN, IT and IPT Departments for NP04/Proto\-DUNE-SP.
%
%
This document was prepared by the DUNE collaboration using the
resources of the Fermi National Accelerator Laboratory 
(Fermilab), a U.S. Department of Energy, Office of Science, 
HEP User Facility. Fermilab is managed by Fermi Research Alliance, 
LLC (FRA), acting under Contract No. DE-AC02-07CH11359.
%
%
This work was supported by
CNPq, FAPERJ, FAPEG and FAPESP,              Brazil;
CFI, IPP and NSERC,                          Canada;
CERN;
M\v{S}MT,	                                 Czech Republic;
ERDF, H2020-EU and MSCA,                     European Union;
CNRS/IN2P3 and CEA,                          France;
INFN,                                        Italy;
FCT,                                         Portugal;
NRF,                                         South Korea;
CAM, Fundaci\'{o}n ``La Caixa'' and MICINN,  Spain;
SERI and SNSF,                               Switzerland;
T\"UB\.ITAK,                                 Turkey;
The Royal Society and UKRI/STFC,             United Kingdom;
DOE and NSF,                                 United States of America.
%
%
This research used resources of the 
National Energy Research Scientific Computing Center (NERSC), 
a U.S. Department of Energy Office of Science User Facility 
operated under Contract No. DE-AC02-05CH11231.



\bibliographystyle{JHEP}
\bibliography{citedb}

\end{document}